\RequirePackage{fix-cm}
\documentclass{iopjournal}
\usepackage{booktabs}
\usepackage{graphicx}
\usepackage{amsmath,amssymb}
\usepackage{xcolor}
\usepackage{hyperref}
\usepackage{url}
\usepackage{longtable}
\usepackage{booktabs}
\usepackage{pdflscape}
\usepackage{xltabular}
\usepackage{array}
\keepXColumns

\begin{document}

\articletype{Paper}

\title{From Circuits to Hardware: Benchmarking Standard and Qubit-Efficient
Quantum Optimization on Real Hardware}

\author{Monit Sharma$^1$\orcid{0009-0000-7191-5548} and Hoong Chuin Lau$^{1,2*}$\orcid{0000-0002-5326-411X} }

\affil{$^1$School of Computing and Information Systems,  Singapore Management University, Singapore}

\affil{$^2$Institute of High Performance Computing, A*STAR, Singapore}

\affil{$^*$Author to whom any correspondence should be addressed.}

\email{{monitsharma,hclau}@smu.edu.sg}


\begin{abstract}
  Despite rapid progress in quantum optimization, the field lacks broad
  real-hardware benchmarks comparing multiple algorithmic families across
  diverse classically hard combinatorial problems under one protocol. We present
  a hardware-aware benchmark of gate-based quantum optimization across four
  NP-hard problems: the multi-dimensional knapsack problem (MDKP), maximum
  independent set (MIS), quadratic assignment problem (QAP), and market-share
  problem (MSP), spanning variational methods (VQE, CVaR-VQE), standard,
  multi-angle, and warm-start QAOA, and qubit-efficient encodings (PCE, QRAO),
  executed on IBM Heron r1/r2 processors under resilience-level-2 mitigation. To
  our knowledge this includes the first real-hardware QRAO results and the first
  multi-problem PCE hardware benchmark.

  Across 247 method--instance combinations we report transpiled circuit size, an
  independent-error gate-count fidelity proxy $F_{\mathrm{est}}$, and hardware
  outcomes. An empirical operating point near $F_{\mathrm{est}}\approx0.1$
  ($\sim$770 two-qubit gates at the median Heron-r2 CZ error rate) marks the
  transition to noise-dominated execution in the MDKP and MIS regimes.

  Two limitations emerge. QAP couples dense one-hot encodings with an
  exponentially sparse feasible manifold (feasible fraction $10!/2^{100}$ at
  $n{=}10$); no tested hardware method returns a feasible assignment. The tested
  QAOA-family circuits become noise-dominated after compilation, and a matched
  uniform-random control shows most feasible low-fidelity outcomes lie within the
  random range, with one MIS warm-start result reported as a finite-sample
  exception. A compilation counterfactual (SWAP-aware, fractional-gate,
  Nighthawk-topology) reduces two-qubit counts but moves no circuit above
  $F_{\mathrm{est}}=10^{-3}$; conclusions therefore apply to the tested
  implementations, not QAOA in general. Qubit-efficient methods extend runnable
  instance sizes but gain only within the empirical fidelity budget.\footnote{Code
    available at: \href{https://github.com/SMU-Quantum/quantum-optimization-benchmarks}
  {https://github.com/SMU-Quantum/quantum-optimization-benchmarks}}
\end{abstract}

\section{Introduction}

Benchmarking quantum optimization methods under realistic execution conditions remains an
open challenge~\cite{abbas2023quantum}. Although gate-based quantum algorithms such as the Variational Quantum
Eigensolver (VQE)~\cite{Peruzzo_2014}, Conditional Value-at-Risk VQE (CVaR-VQE)~\cite{Barkoutsos_2020}, and the Quantum Approximate
Optimization Algorithm (QAOA)~\cite{farhi2014qaoa} have been widely studied for combinatorial optimization,
existing evaluations are often limited to a single problem class, a single algorithm
family, or simulator-based settings that do not fully reflect the practical constraints
of present-day quantum hardware. As a result, it remains difficult to assess how different
quantum optimization strategies compare under a unified protocol when compilation overhead,
qubit connectivity, gate noise, and execution variability are taken into account.

This limitation is particularly important because the empirical performance of quantum
optimization methods depends on more than the nominal algorithm alone. In practice,
performance is shaped by a layered interaction between problem formulation, encoding
strategy, circuit structure, transpilation overhead, and backend-specific noise processes.
Methods that appear competitive before transpilation may degrade substantially
after routing and compilation, while approaches that reduce algorithm-level
circuit width may introduce different costs in circuit depth, classical
post-processing overhead, or hardware robustness.
A meaningful comparison therefore requires a benchmark that goes beyond idealized resource
estimates and instead evaluates methods under realistic execution conditions.

These issues are especially relevant for comparisons between standard variational methods
and qubit-efficient approaches. Variational energy-minimization methods such as VQE and
CVaR-VQE operate directly on problem Hamiltonians, while QAOA-style methods exploit
structured alternating operators. In contrast, encoding or reduction techniques such as
Pauli Correlation Encoding (PCE)~\cite{Sciorilli_2025} and Quantum Random Access Optimization (QRAO)~\cite{fuller2021approximatesolutionscombinatorialproblems} aim to
compress the problem representation or transform it into a more hardware-manageable form.
Such methods can expand the range of instances that are runnable on current
devices by reducing algorithm-level circuit width, but the practical value of
this compression depends on whether the resulting circuits remain executable
after transpilation and whether the recovered solutions remain robust under
hardware noise.Consequently, qubit count alone
is not a sufficient indicator of practical performance.

Despite growing interest in quantum optimization benchmarks~\cite{abbas2023quantum,
  koch2025quantum,Lubinski_2023,Dupont_2025,Brandhofer_2022,
Li_2023,Shaydulin_2024,mcdowall2026crossplatformbenchmarkingneartermquantum}, broad real-hardware
comparisons across multiple hard combinatorial problems~(such as~\cite{chekuri2004multidimensional,lawler1980generating, sahni1976qap}) and multiple algorithmic families
remain limited. Much of the existing literature focuses either on a single optimization
method, a narrow benchmark family, or simulator-based assessments that do not expose the
full gap between algorithmic design and hardware-realized behavior. This makes it difficult to
separate algorithmic effects from encoding effects, and to determine when qubit-efficient
formulations offer a genuine practical advantage rather than only a nominal reduction in
width. It also limits the ability of the community to compare methods using consistent
metrics and reproducible evaluation criteria.

In this work, we present a hardware-aware benchmarking study of gate-based quantum
optimization methods across four classically hard combinatorial optimization problem classes: the
Multi-Dimensional Knapsack Problem (MDKP)~\cite{mdkp}, Maximum Independent Set (MIS)~\cite{mis},
Quadratic Assignment Problem (QAP)~\cite{qap}, and Market Share Problem (MSP)~\cite{market_share}. These problems were chosen
to span structurally diverse binary optimization regimes, including packing-type
constraints, graph-based independence structure, strongly coupled quadratic assignment
structure, and application-driven balancing constraints. Our benchmark includes
variational energy-minimization methods (VQE and CVaR-VQE), QAOA-style approaches,
and qubit-efficient encoding or reduction methods, including PCE and QRAO\@. Rather
than focusing only on final objective values, we evaluate a broader set of metrics
that affect performance on current hardware, including qubit requirements, transpiled
circuit depth, total and two-qubit gate counts, compilation overhead, feasibility of
execution, and execution robustness across real quantum processors.

The goal of this study is not to claim a practical quantum advantage, but to establish a
systematic empirical baseline for assessing quantum optimization methods under realistic
constraints. In particular, we seek to understand three questions: how different
algorithmic families behave across structurally different hard problem classes, how
algorithmic resource savings translate into hardware-realized behavior after transpilation,
and how strongly simulator-based expectations align with real-device outcomes. This
perspective is important for near-term benchmarking because conclusions drawn from
idealized simulations may not persist once routing overhead and backend noise are taken
into account.

Our empirical results show that practical performance is strongly shaped by the
interaction between encoding choice and hardware constraints. Across the tested
problems and backends, lower qubit count does not consistently imply better
final performance on hardware. In several cases, qubit-efficient methods extend the
range of runnable instances, but the resulting gains are partly offset by transpilation
overhead, increased effective circuit depth, or degraded robustness during execution.
Conversely, methods that are competitive in simulation may become substantially less
effective after mapping to hardware-native gate sets and coupling graphs. These findings
reinforce the need for benchmark protocols that report not only solution quality but also
the compilation and execution characteristics that determine whether a method is
practically deployable on current devices.

The main contributions of this paper are as follows:
\begin{itemize}
  \item We present the first unified benchmark of multiple gate-based quantum optimization families (variational, QAOA-style, and qubit-efficient) evaluated across four structurally distinct NP-hard combinatorial optimization problems under a common consistent experimental protocol. This breadth enables direct cross-family and cross-problem comparison that has not previously been reported in the literature.
  \item We evaluate these methods under a hardware-aware protocol that goes beyond solution quality to include qubit requirements, transpiled circuit depth, gate counts, compilation overhead, and execution robustness on real quantum hardware (IBM Heron processors). By treating these compilation and execution characteristics as primary benchmark dimensions rather than engineering footnotes, we provide a more complete picture of practical deployability on current devices.
  \item We provide a quantitative noise model derived from backend calibration data and use it to establish an empirically grounded fidelity threshold that distinguishes signal-preserving from noise-dominated execution. This threshold clarifies when qubit-efficient encodings provide genuine practical benefit and when compilation overhead erodes their advantage, yielding actionable guidance for method selection on near-term hardware.
  \item We establish a fully reproducible empirical baseline, namely software stack, backend metadata, transpilation policy, and per-instance execution diagnostics, for future benchmarking studies of quantum optimization on realistic devices. All code and benchmark instances are publicly available, enabling direct comparison with methods developed as hardware continues to improve.
\end{itemize}

The remainder of the paper is organized as follows.
Section~\ref{sec:background} reviews the relevant quantum
optimization methods and benchmarking context.
Section~\ref{sec:benchmark_design} describes the benchmark problems,
evaluation metrics, and experimental protocol.
Section~\ref{sec:methods} presents the optimization methods,
encodings, and hardware execution workflow.
Section~\ref{sec:results} reports the benchmark results on
simulators and real quantum hardware.
Section~\ref{sec:discussion} discusses the main empirical insights
and their implications for near-term quantum optimization.
Section~\ref{sec:conclusion} concludes the paper.

\section{Background and Related Work}
\label{sec:background}

\subsection{Gate-based quantum optimization methods}

Gate-based quantum optimization for combinatorial problems is commonly studied through
variational hybrid algorithms, where a parameterized quantum circuit prepares candidate
states and a classical optimizer updates the circuit parameters based on measurement
outcomes. Among the most widely used approaches are the Variational Quantum Eigensolver
(VQE) and the Quantum Approximate Optimization Algorithm (QAOA), together with several
variants designed to improve robustness, trainability, or sampling quality~\cite{Peruzzo_2014,farhi2014qaoa,Cerezo_2021,Blekos_2024}.

In the optimization setting, the goal is to prepare low-energy states of a
problem Hamiltonian whose ground-state structure encodes high-quality classical
solutions.
VQE typically uses a hardware-efficient or structured ans\"atze and minimizes the expectation value of the Hamiltonian through iterative classical optimization. This formulation is flexible and can be applied across a wide range of problems, but its empirical behavior depends strongly on the chosen ans\"atze, optimizer, measurement budget, and hardware noise~\cite{Tilly_2022,Cerezo_2021_barren_plateaus}.

QAOA adopts a more structured strategy based on alternating applications of a problem
Hamiltonian and a mixing Hamiltonian~\cite{farhi2014qaoa,Blekos_2024}. Its
problem-informed circuit structure has made it one of the central methods in
quantum optimization, particularly for graph and constraint-based binary problems.
Several variants have been proposed to improve practical performance, including
warm-start approaches, multi-angle parameterizations, and risk-sensitive objectives
such as Conditional Value-at-Risk (CVaR) optimization~\cite{Barkoutsos_2020,Egger_2021,herrman2022multi}. In practice, however, both VQE- and QAOA-style methods face similar near-term limitations: circuit depth grows under transpilation, two-qubit gates dominate hardware error, and performance can degrade substantially when mapped to sparse-connectivity devices.

A key point for benchmarking is that these methods should not be compared only at the
level of idealized pre-transpiled circuits. Their practical behavior depends on how the
encoded Hamiltonian interacts with ans\"atze structure, routing overhead, hardware-native
gate sets, and sampling noise. This motivates the hardware-aware perspective adopted in
this work.

\subsection{Qubit-efficient encodings and reduction strategies}

Alongside standard variational approaches, a growing line of work studies qubit-efficient
encodings and reduction methods that attempt to make larger optimization problems
accessible on current hardware by reducing qubit requirements~\cite{Sciorilli_2025,fuller2021approximatesolutionscombinatorialproblems}. Two representative examples considered in this paper are Pauli Correlation Encoding (PCE) and Quantum Random Access Optimization (QRAO).

PCE reformulates the original binary optimization problem into a reduced quantum
representation by embedding binary decision structure into correlations among Pauli
observables, typically after an intermediate reduction to a related quadratic form
such as weighted Max-Cut~\cite{Sciorilli_2025}. The central motivation is to trade algorithm-level circuit width against more
involved encoding, correlator reconstruction, rounding, and classical
post-processing. QRAO
follows a related compression philosophy, using quantum random access codes to encode
multiple classical bits into fewer qubits and then recover approximate classical solutions
through rounding procedures~\cite{fuller2021approximatesolutionscombinatorialproblems}. These approaches are attractive for near-term hardware because qubit availability
remains a dominant bottleneck. Their practical value, however, must be assessed
through the full execution stack, including encoding, compilation, measurement,
correlator reconstruction or QRAC rounding, feasibility checks, and any
method-specific local search.

For this reason, qubit-efficient methods are evaluated using the same practical
criteria as standard variational methods: final objective quality, backend
eligibility, transpiled circuit size, compilation overhead, and execution
robustness. A central purpose of the present benchmark is to assess these
trade-offs empirically across multiple problem classes.

\subsection{Benchmarking quantum optimization}
\label{sec:rw_benchmarks}

Contemporary efforts to establish broader benchmarks for quantum optimization
fall into three complementary categories. First, instance-library initiatives
such as the Quantum Optimization Benchmarking Library (QOBLIB)~\cite{koch2025quantum}
have assembled curated collections of classically hard combinatorial problems
to support reproducible cross-study comparison. Our benchmark draws on problem
families consistent with QOBLIB but extends the evaluation to multiple method
families executed on real quantum processors. Second, cross-platform hardware
studies such as McDowall~\textit{et~al.}~\cite{mcdowall2025practical} compare
near-term quantum optimization algorithms across multiple processor
architectures, focusing on how algorithmic performance depends on backend
choice. Our study is complementary in scope: we focus on a single processor
family (IBM Heron r1/r2) but span a wider range of method families and problem
classes, and we additionally provide a quantitative noise model linking
benchmark outcomes to backend calibration data.
Method-specific benchmarking studies further show that reported VQA
performance can depend materially on ansatz construction, hyperparameter
selection, optimizer cost, and the choice of fixed-resource classical
comparators. In particular, Turati \textit{et al.}~\cite{Turati_2023}
benchmark adaptive VQAs and QAOA on QUBO instances, while
Schwägerl \textit{et al.}~\cite{Schwaegerl_2024} study VQA performance for
combinatorial optimization under fixed practical resource budgets relative to
sampling and greedy baselines.

Third, protocol-level
methodology work such as Abbas~\textit{et~al.}~\cite{abbas2023quantum} and
Lubinski~\textit{et~al.}~\cite{Lubinski_2023} proposes evaluation standards
for quantum optimization. The present study implements several of the practices
these works advocate; backend metadata logging, transpiled resource
accounting, shot-budget reporting, and feasibility tracking, and treats them
as first-class benchmark dimensions rather than engineering details.

Despite these efforts, broad real-hardware comparisons across multiple hard
combinatorial problems and multiple algorithmic families remain limited.
Much of the existing literature focuses either on a single optimization
method, a narrow benchmark family, or simulator-based assessments that do
not expose the full gap between algorithmic design and hardware-realized
behaviour. This makes it difficult to separate algorithmic effects from
encoding effects, and to determine when qubit-efficient formulations offer
a genuine practical advantage rather than only a nominal reduction in width.
Relative to these efforts, the present paper contributes the deepest
hardware-facing diagnostic analysis to date across multiple algorithmic
families on structurally distinct NP-hard binary problems, tied to an
explicit and reproducible process-fidelity model.

\subsection{Position of the present work}

The present paper is designed to address this gap through a unified, hardware-aware
benchmark spanning multiple algorithmic families and multiple structurally distinct
NP-hard problems. Rather than isolating one method or one problem class, we compare
standard variational energy-minimization methods, QAOA-style methods, and qubit-efficient
encodings within a common experimental protocol. Our emphasis is not only on solution
quality, but also on the practical execution characteristics that determine whether a
method is usable on current devices, including qubit requirements, transpiled depth,
gate counts, compilation overhead, and execution robustness.

This perspective places the paper closer to a benchmarking and methodology contribution
than to a single-algorithm performance study. In particular, it complements
simulator-based comparative studies by quantifying the gap between algorithmic-level design and
hardware-realized behavior, and it extends narrower hardware studies by broadening both
the algorithmic and problem-level scope. The resulting benchmark is intended to serve as
a reproducible baseline for future work on quantum optimization under realistic execution
constraints.

\section{Benchmark Design}
\label{sec:benchmark_design}

\subsection{Design objectives}

The benchmark is designed to evaluate quantum optimization methods under a unified and
hardware-aware protocol across multiple classes of hard binary optimization problems. Our
objective is not merely to compare final objective values, but to assess how algorithmic
families behave when differences in problem structure, encoding strategy, and hardware
execution are taken into account. To this end, the benchmark was constructed around four
design principles.

First, the selected problems should be classically hard and structurally diverse. A
benchmark restricted to a single family may reflect problem-specific biases rather than
general algorithmic behavior. We therefore include problem classes with substantially
different constraint and interaction structure, ranging from packing-type constraints to
graph-based exclusivity, strongly coupled quadratic assignment structure, and
application-driven balancing constraints.

Second, the benchmark should include method families with distinct resource trade-offs.
In particular, we compare standard variational methods, QAOA-style approaches, and
qubit-efficient encoding or reduction strategies. This makes it possible to assess not
only which methods perform well on a given problem, but also how algorithm-level width
reduction interacts with compilation and noise.

Third, the evaluation should reflect realistic execution conditions. Metrics such as
algorithmic-level circuit width and noiseless simulator objective value are informative, but
insufficient for understanding performance on current devices. We therefore include
hardware-facing diagnostics such as transpiled circuit depth, total and two-qubit gate
counts, compilation overhead, backend eligibility, and execution robustness.

Fourth, the benchmark should support fair comparison across methods and problem
instances. All methods are evaluated on common benchmark families with consistent
reporting of solution quality, feasibility, and execution characteristics. This allows
algorithmic and encoding effects to be interpreted within a shared experimental framework
rather than through isolated case studies.

\subsection{Terminology and reporting conventions}
\label{sec:terminology}

We use \emph{algorithm-level circuit width}, to mean the number of circuit wires in the encoded
circuit before hardware-aware transpilation. For direct QUBO formulations,
number of qubits normally equals the number of encoded binary variables.
For PCE and QRAO, it is the number of circuit wires after the respective
compression mapping.

This quantity is not a fault-tolerant logical-qubit count and should not be
interpreted as such. When a physical-qubit count is
reported, it refers to the device qubits occupied by the transpiled layout and
may differ from pre-transpiled qubits, because layout, routing, and ancillary wires
can alter the hardware realization.

We use \emph{classical post-processing} as the umbrella term for the
method-specific operations that turn measurement outcomes into reported
candidates. For direct VQE, CVaR-VQE, and QAOA-style methods, this consists of
bitstring interpretation, original-problem scoring, and best-candidate
selection. For PCE, the corresponding operation is correlator reconstruction;
for QRAO, it is QRAC rounding. We use \emph{feasibility repair} only when a
procedure explicitly modifies a candidate to restore constraint satisfaction,
and \emph{local search} only for an explicit objective-improving neighborhood
search. These operations are reported separately rather than being collectively
called ``decoding.''

\subsection{Benchmark problem classes and instance scope}
\label{sec:benchmark_problems}

We consider four NP-hard 0--1 combinatorial optimization problem classes:
the Multi-Dimensional Knapsack Problem (MDKP)~\cite{mdkp,mdkp_reference},
Maximum Independent Set (MIS)~\cite{mis,mis_reference},
Quadratic Assignment Problem (QAP)~\cite{qap,qap_reference,qaplib},
and Market Share Problem (MSP)~\cite{market_share,market_split_reference}.
These problems were selected to cover complementary structural regimes:
packing constraints, graph-induced exclusivity, dense assignment interactions,
and target-allocation balancing. This diversity is important because near-term
quantum optimization performance depends not only on nominal problem size, but
also on constraint density, interaction structure, encoding overhead, and
compiled circuit complexity.

\paragraph{Multi-Dimensional Knapsack Problem (MDKP).}
MDKP generalizes the classical knapsack problem by imposing multiple resource
constraints~\cite{mdkp,mdkp_reference,puchinger2010multidimensional}. Given item
profits and several capacity dimensions, the goal is to select a subset of items
that maximizes total value while satisfying all resource limits. In this
benchmark, MDKP represents a packing problem in which feasibility and constraint
handling are central to performance evaluation.

\paragraph{Maximum Independent Set (MIS).}
MIS asks for the largest subset of mutually nonadjacent vertices in a
graph~\cite{mis,mis_reference}. Its graph-induced exclusivity structure maps
naturally to quadratic binary objectives and provides a contrast to
capacity-constrained packing formulations such as MDKP\@.

\paragraph{Quadratic Assignment Problem (QAP).}
QAP assigns facilities to locations so as to minimize a quadratic flow-distance
cost~\cite{qap,qap_reference,qaplib}. Its dense pairwise couplings and strict
assignment structure make it the most demanding family in the benchmark from the
standpoint of encoded Hamiltonian density and compiled circuit growth.

\paragraph{Market Share Problem (MSP).}
MSP is an application-driven binary optimization problem in which selected
decisions must match prescribed market-share or allocation
targets~\cite{market_share,market_split_reference}. It differs from the other
three families because solution quality is naturally measured by target
deviation rather than by a conventional profit or cost gap alone.

For each problem class, we select instances that are nontrivial relative to
current quantum optimization methods while remaining within the executable
regime of at least some compared approaches after encoding and compilation.
The reference ``Optimal'' or ``BKS'' values reported in the result tables are
taken from the corresponding benchmark sources~\cite{dataset,mis_datasets,qaplib,market_split_reference}
and are used only as evaluation targets for gap or deviation reporting. This
intermediate benchmark regime reveals how practical viability changes with
encoded width, coupling density, transpilation burden, and hardware noise.

\subsection{Compared method families}
\label{sec:compared_families}

The benchmark includes three broad categories of gate-based quantum optimization methods.

\paragraph{Variational energy-minimization methods.}
These include VQE and CVaR-VQE, which optimize parameterized circuits with respect to the expectation value or a risk-sensitive tail objective derived from the measured energy distribution. These methods provide a flexible baseline and are representative of a large portion of near-term quantum optimization literature.

\paragraph{QAOA-style methods.}
This category includes standard QAOA, warm-start QAOA, and multi-angle QAOA.
The benchmark evaluates these three concrete implementations under a fixed
depth-selection, parameter-optimization, compilation, and hardware-execution
protocol. The results should therefore not be interpreted as a universal
statement about all QAOA variants. In particular, routing-specialized QAOA
compilation, SWAP-network constructions, fixed-schedule linear-ramp QAOA, and
spectral-gap-informed parameter schedules are distinct protocol families with
different parameter-selection and compilation assumptions
\cite{montanez2025ptc_qaoa,montanez2024lrqaoa,mcdowall2026sgirqaoa}.

\paragraph{Qubit-efficient encodings and reductions.}
This category includes PCE and QRAO, which aim to reduce qubit requirements by compressing or reformulating the optimization problem before quantum execution. These methods are especially relevant in the present benchmarking regime because qubit availability is often the first practical bottleneck on current hardware.

The purpose of this classification is not to impose strict boundaries between all algorithmic variants, but to compare representative optimization strategies with meaningfully different width-depth-decoding trade-offs.

\subsection{Evaluation metrics}

To support a hardware-aware comparison, we report metrics from four complementary categories.

\paragraph{Solution quality.}
We evaluate the best solution obtained by each method relative to a classical
reference, using feasibility status and a problem-appropriate quality metric.
For MDKP, MIS, and QAP, solution quality is reported through the optimality gap
to the best-known or classically certified reference value:
\begin{equation}
  \label{eq:optimality_gap}
  \mathrm{Gap}(\%) =
  100 \times
  \frac{|z_{\mathrm{ref}} - z_{\mathrm{alg}}|}{|z_{\mathrm{ref}}|},
\end{equation}
where $z_{\mathrm{ref}}$ is the classical reference value and
$z_{\mathrm{alg}}$ is the objective value recovered by the quantum method.

For the Market Share Problem (MSP), however, the classical optimal objectives
are small integer deviation values. In this setting, percentage optimality gaps
can be misleading or unstable: a change of only a few deviation units may
produce a large percentage gap, even though the operational interpretation is
better captured directly by the deviation from the target allocation. We
therefore report MSP quality using the total absolute deviation
\begin{equation}
  \label{eq:tdev}
  \mathrm{TDev}
  =
  \sum_{i} \left| A_i(\mathbf{x}) - T_i \right|,
\end{equation}
where $A_i(\mathbf{x})$ denotes the achieved allocation for target category
$i$ under solution $\mathbf{x}$, and $T_i$ is the corresponding prescribed
target. Lower TDev indicates a better market-share match, with
$\mathrm{TDev}=0$ corresponding to an exact target match when such a solution
is attainable. For MSP tables, we also report the maximum single-category
deviation, denoted MDev, to indicate whether the total deviation is spread
across many categories or concentrated in one large imbalance.

To permit comparison across MSP instances with different target scales, we
additionally report the range-normalized TDev approximation ratio
\begin{equation}
  \label{eq:msp_ar_tdev}
  \mathrm{AR}_{\mathrm{TDev}}(\mathbf{x})
  =
  \frac{
    \mathrm{TDev}^{\max}_{I}
    -
    \mathrm{TDev}(\mathbf{x})
  }{
    \mathrm{TDev}^{\max}_{I}
    -
    \mathrm{TDev}^{\star}_{I}
  }.
\end{equation}
where $\mathrm{TDev}^{\star}_{I}$ and
$\mathrm{TDev}^{\max}_{I}$ are respectively the classically certified minimum
and maximum TDev values over the original feasible decision domain for MSP
instance $I$. Thus, $\mathrm{AR}_{\mathrm{TDev}}=1$ denotes the certified
minimum-TDev solution, whereas $\mathrm{AR}_{\mathrm{TDev}}=0$ denotes the
worst feasible target mismatch. TDev remains the primary application-level
metric because the normalized ratio alone does not convey the absolute
magnitude of the market-share deviation.

For constrained problems, feasibility is reported explicitly because
near-optimal but infeasible solutions are not operationally comparable to
feasible ones. Where relevant, additional measures of violation magnitude are
also reported.

\paragraph{Algorithm-level circuit width.}
We record the algorithm-level circuit width, number of qubits, required by
each encoding or method. This is the pre-transpilation width defined in
Sec.~\ref{sec:terminology}, not a fault-tolerant logical-qubit count. It is
particularly important when comparing standard formulations against
qubit-efficient strategies, since reduced algorithm-level width is one of the
main claimed advantages of encoding-based methods.

\paragraph{Compiled circuit characteristics.}
Algorithm-level resource counts alone do not capture hardware difficulty. We
therefore also report post-transpilation metrics, including circuit depth, total
gate count, and two-qubit gate count after mapping to the target backend. These
quantities provide a more realistic picture of execution cost and noise
exposure.

\paragraph{Compiled circuit characteristics.}
Algorithm-level resource counts alone do not capture hardware difficulty. We
therefore record both pre-transpilation and post-transpilation circuit metrics:
algorithm-level depth and two-qubit-gate count before backend mapping, together
with transpiled depth, total gate count, and two-qubit-gate count after routing,
native-basis decomposition, and scheduling on the target processor. The
difference between these two descriptions quantifies compilation expansion
under hardware connectivity constraints. For the representative QAOA-family
circuits shown in Figure~\ref{fig:qaoa_prepost_2q}, historical transpilation
increases two-qubit-gate counts by a factor of $4.47$–$6.24$ across all four
problem families.

\paragraph{Execution robustness.}
Finally, we assess practical hardware behavior through metrics such as the number of successful jobs, evaluation counts, shot budgets, latency summaries, and overall execution stability across repeated runs. These diagnostics are important because current-device benchmarking is affected not only by nominal circuit size, but also by queueing, job failures, and backend-specific variability.

Taken together, these metrics allow us to distinguish between methods that look
attractive at the algorithm-level design stage and methods that remain
competitive after compilation and hardware execution.

\subsection{Fairness and comparison protocol}
\label{sec:fairness_protocol}

A central challenge in benchmarking quantum optimization methods is ensuring that
comparisons remain meaningful across methods with different operating
principles. In this work, fairness is enforced at the level of benchmark
instances, evaluation criteria, quantum-evaluation budget, and final
candidate-processing protocol rather than by forcing every method into an
artificially identical parameterization.

All compared methods are evaluated on the same benchmark families and judged
using the same core outcome measures: solution quality, feasibility,
algorithm-level circuit width, compiled circuit characteristics, and execution
robustness. Classical reference solutions are used to anchor performance
interpretation. Where a method produces probabilistic outputs, the reported
solution is obtained through the method-specific classical post-processing
procedure defined in Sec.~\ref{sec:terminology}. This makes reconstruction,
rounding, repair, and local-search stages explicit rather than conflating them
under the single term ``decoding.''

At the same time, we avoid conflating fundamentally different notions of
resource use. In particular, algorithm-level circuit width, transpiled gate
counts, and execution success statistics are reported separately rather than
merged into a single composite score. This makes it possible to identify the
source of a method's strengths or limitations more precisely.

This distinction is important for qubit-efficient methods. PCE’s
correlator-sign reconstruction and QRAO’s rounding rule are native decoding
steps required to map their compressed quantum representations back to binary
decision variables. They are distinct from the subsequent shared local-improvement
routine. We therefore do not attribute the final performance of PCE or QRAO
solely to their quantum circuits; the reported values reflect their complete
documented pipeline, as do the results for all other method families.

At the same time, we avoid conflating fundamentally different notions of
resource use. In particular, algorithm-level circuit width, transpiled gate
counts, and execution success statistics are reported separately rather than
merged into a single composite score.

\subsection{Simulator and hardware scope}

The benchmark includes both simulator-based and real-hardware evaluations,
since the two provide complementary information. Simulators offer a controlled
setting for comparing algorithmic behavior without device-specific noise,
queueing, or availability effects, whereas hardware runs reveal how those
expectations change after compilation, routing, and execution on present-day
processors. This distinction is particularly important for width-reduction
strategies, whose apparent advantages at the encoding level may weaken or
change after transpilation.

The benchmark is therefore empirical in scope. It is not intended to establish
asymptotic superiority among quantum optimization families or to claim
practical quantum advantage over classical methods. Rather, it provides a
reproducible baseline for characterizing the current operating regime of
representative gate-based quantum optimization methods on classically hard
benchmark problems, including executability, solution quality, and the
simulator--hardware gap under realistic resource constraints.

\section{Methods}
\label{sec:methods}

\subsection{Problem-to-QUBO workflow}

All benchmark instances are first expressed as binary optimization models and then converted into Quadratic Unconstrained Binary Optimization (QUBO) form. This QUBO--Ising workflow follows the standard equivalence between binary quadratic optimization and Ising Hamiltonian formulations, which has become a common interface between combinatorial optimization and quantum optimization methods~\cite{lucas2014ising,kochenberger2014qubo}. This common representation enables the application of multiple gate-based quantum optimization methods within a unified pipeline. For constrained problems, the QUBO is obtained by incorporating constraint violations into the objective through quadratic penalty terms, yielding an unconstrained binary model whose low-energy states correspond to high-quality solutions of the original problem.

The use of a common QUBO representation serves two purposes. First, it standardizes the input format across structurally different benchmark problems, allowing multiple algorithmic families to be compared within the same evaluation framework. Second, it enables a consistent transition from classical binary models to quantum Hamiltonians. Each QUBO is mapped to an Ising-type cost Hamiltonian, which is then used either directly in variational energy-minimization methods or indirectly through encoding and reduction techniques such as PCE and QRAO\@.

\subsection{Structural descriptors of the benchmark QUBOs}
\label{sec:qubo_structure}

The common QUBO--Ising workflow produces structurally different optimization
landscapes across the four benchmark families. Table~\ref{tab:structural_summary}
therefore reports the QUBO width, coupling density, coefficient dynamic range,
and penalty-to-objective scale ratio for the actual encoded instances. Here,
$\rho_Q$ denotes the fraction of nonzero off-diagonal QUBO couplings,
$R_Q$ is the ratio of the largest to smallest absolute nonzero QUBO
coefficient, and $R_\lambda$ is the ratio between the largest penalty scale
and the largest objective-term magnitude. Exact definitions and full
instance-level structural records are provided in
Appendix~\ref{app:structural_metrics}.

The families occupy distinct structural regimes. MIS remains comparatively
sparse because its couplings are inherited from the input graph. In contrast,
MDKP and MSP acquire moderate-to-high density through capacity, slack, and
target-balancing penalties. QAP is the most restrictive case: its direct
one-hot assignment encoding combines dense interactions, the broadest
coefficient ranges, and an extremely small feasible permutation manifold.
For the selected $n=10$ QAP instances, only
$10!/2^{100}\approx10^{-23.54}$ of binary strings encode valid assignments;
for the $n=12$ instances, this fraction falls to
$12!/2^{144}\approx10^{-34.67}$. These descriptors are not universal
hardness scores, but they identify formulation-level features that influence
variational conditioning, compilation overhead, and feasibility recovery.

\begin{table*}[t]
  \centering
  \caption{Compact structural profile of the benchmark QUBOs. $N_Q$ is QUBO
    width, $\rho_Q$ is coupling density, $R_Q$ is the coefficient dynamic range,
    and $R_\lambda$ is the penalty-to-objective scale ratio. $R_Q$ is reported as
    an across-instance range; $\rho_Q$ and $R_\lambda$ are reported as family
    medians. Full per-instance values are provided in
  Appendix~\ref{app:structural_metrics}.}
  \label{tab:structural_summary}
  \footnotesize
  \setlength{\tabcolsep}{2.5pt}
  \renewcommand{\arraystretch}{1.08}
  \begin{tabularx}{\textwidth}{@{}lrrrrr>{\raggedright\arraybackslash}X@{}}
    \toprule
    Problem & Inst. & $N_Q$ range & Median $\rho_Q$ & $R_Q$ range
    & Median $R_\lambda$ & Dominant encoded structure \\
    \midrule
    MDKP & 12 & 45--122 & 0.678 &
    $8.53\times10^{3}$--$6.75\times10^{5}$ &
    4.78 &
    Capacity and slack-induced couplings \\

    MIS & 7 & 8--128 & 0.181 &
    $9$--$129$ &
    65.0 &
    Sparse graph-induced edge couplings \\

    QAP & 11 & 100--144 & 0.818 &
    $1.07\times10^{5}$--$2.21\times10^{10}$ &
    389.3 &
    Dense row--column one-hot assignment encoding \\

    MSP & 8 & 48--156 & 0.589 &
    $9.86\times10^{4}$--$1.70\times10^{6}$ &
    14.7 &
    Target-allocation and balancing penalties \\
    \bottomrule
  \end{tabularx}
\end{table*}

\subsection{Constraint penalties and coefficient scaling}
\label{sec:penalty_selection}

All constrained benchmark instances are converted to QUBO form using quadratic
penalties generated by the Qiskit Optimization converter pipeline. Let the
pre-penalty binary objective be written as
\begin{equation}
  f(x)=f_{\mathrm{lin}}(x)+f_{\mathrm{quad}}(x),
\end{equation}
and let the converted equality or slack-expanded inequality constraints be
represented by integer-valued residuals $g_r(x,s)$. The resulting penalized
QUBO has the generic form
\begin{equation}
  E_{\lambda}(x,s)
  = f(x)
  + \lambda\sum_r g_r(x,s)^2.
  \label{eq:generic_penalty_qubo}
\end{equation}
For every benchmark instance, the penalty coefficient was selected before any
quantum execution by the Qiskit automatic-bound rule
\begin{equation}
  \lambda
  = 1+
  \left(
    U_{\mathrm{lin}}-L_{\mathrm{lin}}
  \right)
  +
  \left(
    U_{\mathrm{quad}}-L_{\mathrm{quad}}
  \right),
  \label{eq:qiskit_auto_penalty}
\end{equation}
where $L_{\mathrm{lin}}$, $U_{\mathrm{lin}}$, $L_{\mathrm{quad}}$, and
$U_{\mathrm{quad}}$ are lower and upper bounds on the linear and quadratic
pre-penalty objective contributions at the converter stage. This construction
is analytical rather than hardware-tuned or grid-tuned. Because the binary and
slack-expanded constraint residuals are integer valued, any violation has
$g_r(x,s)^2\geq1$. The selected penalty therefore exceeds the pre-penalty
objective bound range used by the converter and makes a unit constraint
violation energetically more expensive than the full modeled objective range.

The resulting penalty construction differs by problem family only through the
constraint residual. For MDKP, binary-expanded integer slack variables enforce
capacity inequalities,
\begin{equation}
  E_{\mathrm{MDKP}}(x,s)
  = -\sum_i p_i x_i
  + \lambda_{\mathrm{K}}
  \sum_k
  \left(
    \sum_i w_{ki}x_i+s_k-C_k
  \right)^2.
  \label{eq:mdkp_penalty}
\end{equation}
For MIS, edge-conflict inequalities are penalized as
\begin{equation}
  E_{\mathrm{MIS}}(x)
  = -\sum_{i\in V}x_i
  + \lambda_{\mathrm{MIS}}
  \sum_{(i,j)\in E}x_ix_j.
  \label{eq:mis_penalty}
\end{equation}
For the unit-weight MIS instances, the automatic bound gives
$\lambda_{\mathrm{MIS}}=|V|+1$, which is conservative: any
$\lambda_{\mathrm{MIS}}>1$ is sufficient to prefer removing an endpoint of a
violating edge over retaining the conflict penalty. For QAP, the direct
assignment encoding uses row and column one-hot penalties,
\begin{equation}
  E_{\mathrm{QAP}}(x)
  = \sum_{i,k}\sum_{j,l}
  f_{ik}d_{jl}x_{ij}x_{kl}
  + \lambda_{\mathrm{Q}}
  \left[
    \sum_i
    \left(
      1-\sum_jx_{ij}
    \right)^2
    +
    \sum_j
    \left(
      1-\sum_ix_{ij}
    \right)^2
  \right].
  \label{eq:qap_penalty}
\end{equation}
For MSP, target-allocation equalities are enforced after binary expansion of
the deviation variables,
\begin{equation}
  E_{\mathrm{MSP}}(x,s)
  = f_{\mathrm{TDev}}(x,s)
  + \lambda_{\mathrm{M}}
  \sum_g
  \left(
    A_g(x)+s_g-T_g
  \right)^2,
  \label{eq:msp_penalty}
\end{equation}
where $A_g(x)$ is the allocation expression, $s_g$ is the corresponding
binary-expanded deviation variable, and $T_g$ is the target.

No coefficient normalization is applied after QUBO conversion. The resulting
penalized QUBO is passed directly to \texttt{to\_ising()} for VQE- and
QAOA-family methods or to the subsequent PCE and QRAO encodings. Thus, all
methods begin from the same instance-specific penalized QUBO, preserving the
relative objective-to-penalty scale through the common construction stage;
method-specific differences arise only during encoding, variational
optimization, sampling, and post-processing. The resulting family-level
structural and coefficient scales are summarized in
Table~\ref{tab:structural_summary}. Appendix~\ref{app:structural_metrics}
defines the structural descriptors and reports the detailed QAP instance-level
analysis, while Appendix~\ref{app:penalty_audit} documents the
instance-specific penalty values, automatic selection rule, and representative
penalty-sensitivity audit.

\begin{table*}[t]
  \centering
  \caption{Penalty construction summary for the benchmark QUBOs. All penalties
    are selected by the automatic pre-penalty objective-bound rule in
    Eq.\eqref{eq:qiskit_auto_penalty}; they are fixed before quantum execution,
    shared across all converted constraints of an instance, and not normalized
    before Hamiltonian construction. Full instance-level structural metrics and
  penalty scales are reported in Appendix~\ref{app:penalty_audit}.}
  \label{tab:penalty_summary}
  \footnotesize
  \setlength{\tabcolsep}{3.0pt}
  \renewcommand{\arraystretch}{1.08}
  \begin{tabularx}{\textwidth}{@{}l>{\raggedright\arraybackslash}Xc>{\raggedright\arraybackslash}p{2.7cm}>{\raggedright\arraybackslash}X@{}}
    \toprule
    Problem & Constraint representation & Penalty symbol &
    Tested penalty range & Selection basis \\
    \midrule
    MDKP &
    Capacity inequalities with binary slack &
    $\lambda_{\mathrm{K}}$ &
    $4{,}022$–$182{,}685$ &
    Automatic objective-bound rule \\
    MIS &
    Edge-conflict inequalities &
    $\lambda_{\mathrm{MIS}}$ &
    $9$–$129$ &
    Automatic objective-bound rule;
    $\lambda_{\mathrm{MIS}}=|V|+1$ \\
    QAP &
    Row and column one-hot equalities &
    $\lambda_{\mathrm{Q}}$ &
    $1.07185\times10^{5}$–$1.102967\times10^{10}$ &
    Automatic objective-bound rule \\
    MSP &
    Target-allocation equalities with binary deviation variables &
    $\lambda_{\mathrm{M}}$ &
    $1{,}963$–$20{,}613$ &
    Automatic objective-bound rule \\
    \bottomrule
  \end{tabularx}
\end{table*}

\subsection{Method instantiation details}
\label{sec:method_specific_encoding}
The compared method families are defined in Sec.~\ref{sec:compared_families}. Here we specify only
their experimental instantiations. The variational family consists of VQE and
CVaR-VQE, with CVaR evaluated using the fixed confidence level $\alpha=0.25$.
The QAOA-style family consists of standard QAOA together with warm-start QAOA
and multi-angle QAOA\@. The qubit-efficient family consists of PCE and QRAO,
which differ from the direct formulations through problem compression and
method-specific classical recovery structure rather than through ans\"atze choice
alone. Additional
ans\"atze, optimizer, and encoding settings are reported in the following
subsections.

\subsection{Variational optimization and candidate-selection protocol}
\label{sec:optimization_protocol}

All method--instance configurations were compared under a common nominal
shot-based variational-optimization protocol. The direct variational methods
use their native parameterized circuits and objectives: VQE and CVaR-VQE use
hardware-efficient ans"atze, while QAOA, MA-QAOA, and WS-QAOA use layered
cost--mixer constructions at method- and instance-dependent algorithmic depth
$p$. QRAO likewise uses a parameterized hardware-efficient circuit after its
compressed encoding, whereas PCE uses a correlation-encoding circuit comprising
single-qubit rotations and entangling $R_{XX}$ layers. These method-specific
representations determine the circuit and parameterization, but all methods are
evaluated within the same nominal optimizer budget, shot allocation, and
candidate-selection framework. The additional encoding and native recovery
steps for PCE and QRAO are specified in
Section~\ref{sec:method_specific_encoding}.

The benchmark protocol uses COBYLA through
\texttt{scipy.optimize.minimize} in SciPy~1.15.3, with a maximum of
200 optimizer iterations and 200 objective evaluations. Each objective
evaluation uses 1{,}000 shots. Final candidate sampling also uses 1{,}000
shots where a distinct final-sampling record is retained. For runs with
recorded initialization metadata, variational parameters are initialized
independently and uniformly over $[0,2\pi]$. The runner retains the parameter
vector associated with the best observed objective value and then applies the
method-specific reconstruction or rounding step, followed by the common
feasibility-preserving local-improvement procedure described in
Section~\ref{sec:fairness_protocol}.

The 200-evaluation specification is an upper bound, rather than a requirement
that every execution consume exactly 200 objective evaluations. To document
realized execution behaviour, we audited 262 retained benchmark records.
Among these, 149 reached the planned evaluation cap and 113 returned an
\texttt{optimizer\_success} status before reaching the cap. The retained
records do not contain explicit function, parameter, or gradient tolerance
values. We therefore use \texttt{optimizer\_success} only as the status
returned by the optimization framework; it is not interpreted as evidence of
tolerance-based convergence, local optimality, or global optimality.

The benchmark reports one primary protocol execution for each method--instance
configuration. It is consequently a fixed-protocol comparison rather than an
exhaustive hyperparameter sweep or an estimate averaged over all possible
random initializations. A reduced simulator study nevertheless shows that
decoded quality can be materially initialization-sensitive for selected
nontrivial VQE and warm-start-QAOA cases. No additional method-specific tuning
or re-optimization was introduced after inspection of individual headline
outcomes. The reported values should therefore be interpreted as outcomes of
the stated common-budget protocol, not as optimizer-independent best-case
performance for each algorithmic family.

Table~\ref{tab:optimizer_protocol_audit} in
Appendix~\ref{app:optimization_protocol} summarizes the recovered optimizer,
shot-budget, stopping-status, and realized-evaluation information, while
Table~\ref{tab:optimizer_seed_sensitivity} reports the reduced
initialization-sensitivity analysis. Problem-wise compiled circuit depths,
gate counts, and trainable-parameter counts are reported in
Appendices~\ref{app:hw_mdkp}--\ref{app:hw_msp}.

\subsection{Method-specific encoding settings}

Qubit-efficient methods require additional compression choices beyond the ans\"atze selection. For QRAO, we use compression parameter $k=3$, corresponding to the highest
compression level in the QRAC-based formulation used here~\cite{fuller2021approximatesolutionscombinatorialproblems},
and the highest level that remained practically usable under our circuit-depth constraints.. This configuration yields substantial qubit savings while avoiding the more severe depth growth associated with stronger compression. For PCE, we use two-body Pauli correlations, corresponding to quadratic compression~\cite{Sciorilli_2025}.
Higher-order compression is possible in principle, but it requires deeper circuits and was not practical within the present execution regime.

For PCE, we employ a QUBO-based cost formulation together with controlled
perturbation and correlator-sign reconstruction from the compressed
Pauli-correlation representation. The recovered benchmark artifacts indicate
that the reported PCE runs use a single-pass COBYLA protocol with a planned
upper bound of 200 objective evaluations and no recursive multi-round
re-optimization.

After correlator-sign reconstruction, the PCE candidate receives the same
single one-pass feasible local-improvement routine applied to every quantum
method after its native decoding or rounding step. Thus, the reported PCE
values represent the prescribed end-to-end hybrid quantum--classical pipeline
under the common fixed-budget and common post-processing protocol, rather than
the strongest attainable performance of a recursively tuned PCE procedure or
of a PCE-exclusive local-search procedure. PCE's method-specific contribution
is its compressed encoding and native correlator reconstruction; the subsequent
local refinement is shared across VQE, CVaR-VQE, QAOA, MA-QAOA, WS-QAOA, PCE,
and QRAO, as described in Section~\ref{sec:fairness_protocol}.

Realized evaluation counts, termination statuses, and replayability metadata
are reported in Appendix~\ref{app:optimization_protocol}. Full mathematical
details of the PCE formulation, correlator reconstruction, and shared
post-processing protocol are reported in Appendix~\ref{app:pce}.

\subsection{Simulator environment and MPS validation}
\label{sec:simulator_environment}

Simulator-based experiments were conducted with Qiskit Aer 0.17.2 using the
\texttt{matrix\_product\_state} method in double precision on CPU. Production
runs used adaptive uncapped bond dimension
(\texttt{matrix\_product\_state\_max\_bond\_dimension=None}) with truncation
enabled at threshold $10^{-16}$. The simulations were executed on an Apple M3
Pro system with 12 CPU cores and 36\,GB physical memory; no GPU acceleration
was used. Shot-based execution was retained in simulation to preserve the
sampling, candidate-selection, and post-processing workflow used for hardware
experiments.

To test whether MPS truncation affects the MDKP simulator reference, we
performed a targeted fixed-final-parameter stability audit on two large
VQE-family circuits that contribute to the negative MDKP simulator--hardware
gap discussed in Section~\ref{sec:noise_analysis}: a 99-qubit CVaR-VQE circuit
for \texttt{pet2} and a 60-qubit VQE circuit for \texttt{hp1}. For each
circuit, the final parameter vector, QUBO formulation, circuit
construction, final-shot budget, and decoding procedure were held fixed. The
only changed variables were the MPS bond-dimension cap and truncation threshold.
Each setting used the same 20 independent sampling seeds, with 1,000 shots per
seed.

For the 99-qubit CVaR-VQE circuit, the production uncapped run reached maximum
observed bond dimension $\chi_{\mathrm{obs}}=64$. Explicit caps of $64$ and
$128$ reproduced the production empirical output distribution and decoded
solution quality exactly. For the 60-qubit VQE circuit, the production run
reached $\chi_{\mathrm{obs}}=66$; explicit caps of $128$ and $256$ likewise
reproduced the production result. Relaxing the truncation threshold from
$10^{-16}$ to $10^{-12}$ produced no change in either case. At threshold
$10^{-8}$, the empirical output-distribution change remained small and the
decoded solution quality was unchanged. In contrast, deliberately restrictive
bond caps produced measurable distribution deviations, confirming that the
audit is sensitive to harmful bond restriction. Full settings and results are
reported in Appendix~\ref{app:mps_validation}.

\subsection{Quantum hardware execution}

To complement simulation-based results, we re-ran the benchmarking pipeline on real
quantum processors accessed through provider-supported runtime interfaces. Hardware
experiments are treated as execution-realistic evaluations rather than as noisy
replications of simulator runs. Accordingly, backend identity, execution window,
and job-level status information were recorded as part of the experimental context.

A fully controlled campaign in which every method–instance combination is
executed on one fixed processor was not practical within the available hardware
access window. Queue times and backend availability varied substantially during
the campaign, so runs were submitted to compatible IBM backends when they were
available rather than being delayed indefinitely for a single device. This
choice was made to complete the planned benchmark under realistic access
constraints; it was not intended to treat the available processors as
interchangeable noise sources.

Backend heterogeneity can contribute to observed differences because processor
topology, calibration state, native gate set, routing overhead, and execution
window affect the realized circuit. We therefore do not interpret
cross-backend hardware differences as controlled estimates of an isolated
algorithmic effect. Instead, they are reported as end-to-end
hardware-realized outcomes under the stated practical access protocol. Each
result is tied to its backend identity, execution window, transpiled resource
metrics, and available calibration metadata in the appendix, so that backend
context remains visible rather than being absorbed into an algorithm-only
comparison.

\subsection{Hardware configuration and software stack}
\label{sec:hardware_config}

The hardware experiments reported in this study were executed on superconducting
quantum processors accessed through provider-supported runtime interfaces. Across
the reported runs, the primary IBM backends were \texttt{ibm\_fez},
\texttt{ibm\_torino}, and \texttt{ibm\_marrakesh}. Backend identity,
transpiled circuit resources, and execution-level diagnostics for each reported
method–instance configuration are provided in the problem-wise
hardware-diagnostics appendices,
Appendices~\ref{app:hw_mdkp}–\ref{app:hw_msp}. These processors were treated as
distinct execution environments rather than interchangeable noise sources, since
differences in processor family, qubit count, topology, calibration state, and
routing overhead can materially affect realized circuit performance. Multiple
backends were used because queue times and device availability prevented a
practical single-backend campaign across the full benchmark; consequently, the
hardware results are not presented as a same-device controlled ranking of
algorithm families.

For each backend used in the benchmark, we recorded the processor identity, total
available qubit count, connectivity structure, and native gate basis exposed by
the backend configuration at execution time. Compilation targeted the backend’s
native one- and two-qubit gate set and coupling map supplied through the provider
interface. When available, we additionally logged representative backend property
summaries during the experimental window, including readout error and two-qubit
gate error information, so that observed degradation can be interpreted in the
context of hardware state rather than attributed solely to the optimization method.

Because calibration values drift over time, hardware performance is not treated as
a timeless property of a named backend. Instead, each hardware result is tied to
an execution window and backend-specific metadata. We therefore report backend
identity in the main benchmark tables and provide backend-facing circuit and
execution summaries in the appendix.

All reported experiments were implemented in Qiskit~\cite{qiskit2024}, with the exact software
versions fixed throughout the benchmarking campaign. Unless otherwise stated,
hardware circuits were transpiled using a fixed compilation policy with
transpiler optimization level~3. Compilation maps each pre-transpilation circuit to the backend's native gate
set and coupling graph, thereby determining the realized depth, two-qubit gate
count, physical-qubit layout, and routing overhead. Runtime resilience and error
mitigation are applied after this mapping stage and are reported separately.

\subsection{Execution and mitigation pipeline}
\label{sec:error_mitigation}

The hardware workflow separates compilation, expectation-value estimation,
final bitstring sampling, and classical solution recovery. Transpiler
\texttt{optimization\_level=3} is used for backend-aware mapping, gate
decomposition, routing, and circuit optimization before execution. This is a
compilation setting, not an error-mitigation setting.

Variational objective evaluations use IBM Runtime EstimatorV2 with
\texttt{resilience\_level=2}. No custom TREX, ZNE noise-factor, extrapolator,
twirling, or dynamical-decoupling options were manually configured or retained
as part of the benchmark protocol. The managed Runtime preset is
therefore treated as a provider-controlled mitigation setting for
expectation-value estimates, rather than as a fully specified custom
mitigation protocol.

Final candidate bitstrings are obtained through separate SamplerV2 jobs with
raw/default options and no custom sampler noise-management settings. These
SamplerV2 outputs are then passed to the method-specific decoding procedure,
original-problem feasibility checks, and the shared one-round local refinement.
Consequently, runtime-managed mitigation affects the Estimator expectation
values used by the COBYLA objective evaluations; it does not directly mitigate
the final bitstrings used to decode the reported solutions. Table~\ref{tab:mitigation_pipeline} summarizes the four-stage execution path
and distinguishes the backend-aware compilation step, Runtime-managed
mitigation of Estimator objective evaluations, raw/default final sampling, and
the subsequent classical recovery procedure.

\begin{table}[t]
  \centering
  \caption{Execution and mitigation pipeline. Compilation, Estimator mitigation,
    raw final sampling, and classical recovery are distinct stages of the reported
  benchmark protocol.}
  \label{tab:mitigation_pipeline}
  \footnotesize
  \setlength{\tabcolsep}{3.0pt}
  \renewcommand{\arraystretch}{1.08}
  \begin{tabularx}{\textwidth}{@{}>{\raggedright\arraybackslash}p{0.23\textwidth}>{\raggedright\arraybackslash}p{0.29\textwidth}>{\raggedright\arraybackslash}X@{}}
    \toprule
    Stage & Primitive / configuration & Role in reported result \\
    \midrule
    Circuit compilation &
    Transpiler \texttt{optimization\_level=3} &
    Backend-aware mapping, decomposition, routing, and circuit optimization;
    not error mitigation \\
    \addlinespace
    Variational objective evaluation &
    EstimatorV2,
    \texttt{resilience\_level=2} &
    Runtime-managed mitigation applied to expectation-value estimates used by
    COBYLA \\
    \addlinespace
    Final candidate sampling &
    Separate SamplerV2, raw/default options &
    Unmitigated/default bitstring distribution used for decoding and candidate
    selection \\
    \addlinespace
    Decoding and local refinement &
    Classical post-processing &
    Method-specific reconstruction, feasibility checking, and shared one-round
    local improvement \\
    \bottomrule
  \end{tabularx}
\end{table}

Runtime-managed mitigation can increase Estimator execution cost through
additional internally managed circuit ensembles, but the exact overhead was
not isolated from backend queue time, calibration variation, and ordinary
runtime variability. The complete execution
provenance is reported in Appendix~\ref{app:mitigation_provenance}.

\subsection{Compilation and transpilation protocol}

Compilation is treated as a first-class component of the benchmarking stack.
For each circuit–backend pair, we transpile the parameterized circuit using the
backend-native gate basis and coupling map under the fixed compilation policy
described in Section~\ref{sec:hardware_config}. This mapping stage determines
the realized circuit after layout selection, routing, native-basis
decomposition, and transpiler optimization. Pre-transpilation circuit
descriptions can therefore differ substantially from the backend-native circuits
ultimately executed on the processor.

For every evaluated circuit–backend configuration, we record both
algorithm-level and post-transpilation resource characteristics. The
algorithm-level description captures the circuit implied by the selected problem
encoding and ansatz before backend mapping. The transpiled description records
the circuit depth, total gate count, two-qubit gate count, and backend identity
after routing and native-gate decomposition. This distinction is necessary
because connectivity constraints can induce substantial routing overhead,
including SWAP insertion, and can amplify depth and two-qubit-gate burden even
when the original encoded circuit is relatively compact.

The complete backend-specific transpilation records are reported in the
problem-wise hardware-diagnostics appendices,
Appendices~\ref{app:hw_mdkp}–\ref{app:hw_msp}. These records provide
problem- and method-level circuit-resource information together with the
associated execution context. Representative empirical evidence of the
resulting pre-/post-transpilation expansion for bound QAOA-family circuits is
presented later in Figure~\ref{fig:qaoa_prepost_2q} and discussed in
Section~\ref{sec:qaoa_prepost_connectivity}.

We therefore treat transpiled circuit statistics as primary evaluation metadata
rather than as secondary implementation details. They quantify the
backend-native execution burden relevant to noise exposure and hardware
deployability, whereas algorithm-level resource counts describe only the
abstract problem encoding and ansatz construction.

Runtime resilience and error-mitigation settings are not folded into these
transpilation statistics. Compilation, Runtime-managed Estimator mitigation,
raw/default final sampling, and classical candidate recovery are reported as
separate stages of the hardware-execution protocol.

\subsection{Sampling and objective estimation}

All quantum evaluations use shot-based primitives. Simulator and hardware
protocols use the shot budgets recorded in the execution artifact.
For hardware runs, EstimatorV2 evaluations are used to obtain expectation-value
objectives during variational optimization, while final candidate bitstrings
are obtained separately from raw/default SamplerV2 jobs.

For expectation-based methods, the Estimator output defines the objective
supplied to the optimizer. For CVaR-based methods, the objective is computed
from the lower tail of the measured energy distribution using the fixed
confidence level $\alpha=0.25$.

The final reported solution is not decoded from the mitigated Estimator
expectation value. Instead, it is reconstructed from the separate SamplerV2
bitstring distribution and then evaluated in the original combinatorial
problem domain. This distinction is important: runtime-managed Estimator
mitigation may alter the optimization landscape seen by COBYLA, but it does
not directly correct the bitstrings used for final decoding, feasibility
assessment, or local refinement.

\subsection{Hardware noise model and process fidelity estimation}
\label{sec:noise_model}

The quantum processors used in this study are IBM Heron-family superconducting
transmon devices: \texttt{ibm\_fez} and \texttt{ibm\_marrakesh} (Heron~r2,
156~qubits) and \texttt{ibm\_torino} (Heron~r1, 133~qubits). All three use a
heavy-hexagonal lattice topology with fixed-frequency transmon qubits and
tunable couplers, operating near 15~mK in dilution refrigerators. The native
two-qubit gate is CZ, with basis gate set $\{\mathrm{CZ}, \mathrm{ID},
\mathrm{RZ}, \mathrm{SX}, \mathrm{X}\}$ on Heron~r2 and
$\{\mathrm{CZ}, \mathrm{RZZ}, \mathrm{ID}, \mathrm{RZ}, \mathrm{SX},
\mathrm{X}\}$ on Heron~r1. Representative backend characteristics during the
experimental window are summarized in Table~\ref{tab:backend_calibration}.

\begin{table}[ht]
  \centering
  \caption{Representative calibration characteristics for the backends used in
    this study, consistent with published Heron-family
    specifications~\cite{IBMHeronR2} and backend calibration snapshots during the
    experimental window. Exact per-job calibration values are logged in the
  reproducibility appendix.}
  \label{tab:backend_calibration}
  \small
  \begin{tabular}{lccccccc}
    \toprule
    Backend & Processor & Qubits & Topology & $\tilde{T_{1}}$ ($\mu$s) & $\tilde{T_{2}}$ ($\mu$s) & $\tilde{\varepsilon}_{2Q}$ & $\tilde{\varepsilon}_{\mathrm{ro}}$ \\
    \midrule
    \texttt{ibm\_fez}       & Heron~r2 & 156 & heavy-hex & 100--300 & 80--200 & $3\times10^{-3}$ & $1.45\times10^{-2}$ \\
    \texttt{ibm\_marrakesh} & Heron~r2 & 156 & heavy-hex & 100--300 & 80--200 & $3\times10^{-3}$ & $1.25\times10^{-2}$ \\
    \texttt{ibm\_torino}    & Heron~r1 & 133 & heavy-hex & 100--300 & 80--200 & $4\times10^{-3}$ & $2.41\times10^{-2}$ \\
    \bottomrule
  \end{tabular}
\end{table}

The Heron~r2 processors benefit from two-level system (TLS) mitigation
circuitry that actively manages material defects contributing to
decoherence, resulting in improved coherence stability relative to earlier
generations~\cite{IBMHeronR2}. Published Heron-family specifications report
that updated calibration procedures introduced in 2024 reduced the median CZ
error from approximately $5\times10^{-3}$ to $3\times10^{-3}$, and that
circuits with up to approximately 5\,000 two-qubit gate operations remain
within the signal-preserving regime on this
architecture~\cite{IBMHeronR2,IBMRoadmap2025}. These are the two quantities
we use to build our noise model.

\subsubsection{Process fidelity estimate.}
\label{subsec:fidel-est}

For a transpiled circuit with $N_{2Q}$ two-qubit gates executed on a backend
with median two-qubit gate error $\varepsilon_{2Q}$, we estimate the process
fidelity as
\begin{equation}
  \label{eq:fidelity}
  F_{\mathrm{est}} \;\approx\; {\bigl(1 - \varepsilon_{2Q}\bigr)}^{N_{2Q}}.
\end{equation}

This expression assumes independent, identically distributed depolarizing
errors on each two-qubit gate and neglects coherent error accumulation,
crosstalk, single-qubit gate errors, readout effects, and calibration drift.
Accordingly, $F_{\mathrm{est}}$ is an independent-error gate-count proxy
rather than a direct measurement of a circuit's process fidelity or final
output distribution.

The proxy tracks the dominant two-qubit-noise contribution under the
representative calibration conditions used in this study and provides a useful
heuristic for distinguishing signal-preserving from strongly noise-dominated
execution regimes. However, an extremely small value of $F_{\mathrm{est}}$
does not by itself prove that measured output samples are uniformly random:
structured residual correlations, device-specific noise effects, and
problem-dependent post-processing can remain relevant. The thresholds derived
from $F_{\mathrm{est}}$ should therefore be interpreted as benchmark-specific
diagnostic reference points rather than universal hardware thresholds.

Under this model, two reference values are informative for interpreting the
Heron-family hardware outcomes:
\begin{itemize}
  \item $F_{\mathrm{est}} \approx 0.1$ (equivalently
    $N_{2Q} \approx 770$ at $\varepsilon_{2Q}=3\times10^{-3}$): a
    benchmark-specific operating point at which the present data begin to
    show a transition toward strongly noise-dominated execution; see
    Figures~\ref{fig:complexity_vs_quality} and
    \ref{fig:fidelity_vs_quality}. This value is not a universal ZNE
    threshold. ZNE estimates noiseless expectation values by evaluating
    deliberately amplified noise levels and extrapolating to zero noise
    \cite{temme2017error_mitigation,Endo_2018}; whether it is useful for a given circuit
    cannot be inferred from $F_{\mathrm{est}}$ alone.
  \item $F_{\mathrm{est}} \approx 0.01$ (equivalently
    $N_{2Q} \approx 1540$): a more conservative reference value at which the
    independent-error proxy attenuates a unit-scale ideal signal to roughly
    one percent of its noiseless magnitude.
\end{itemize}
These numerical values are benchmark-specific diagnostic reference points,
derived from the representative Heron-family calibration rate and the
shot-budget regime used in this study; they would shift with different
hardware, circuits, calibration windows, or shot budgets.

\subsubsection{Representative fidelity estimates.}

Table~\ref{tab:representative_fidelity} reports estimated process fidelity for
representative method--instance--backend combinations from each benchmark
family, chosen to span the signal-preserving and noise-dominated regimes.
Full per-instance fidelity proxies, hardware outcomes, and simulator outcomes
for all evaluated method–instance–backend combinations are reported in the
master fidelity table in Appendix~\ref{app:reproducibility},
Table~\ref{tab:master_fidelity}.

\begin{table}[ht]
  \centering
  \caption{Representative gate-count fidelity proxies by problem family.
    $N_{2Q}$ is the transpiled two-qubit gate count from the hardware-diagnostics
    appendices; $F_{\mathrm{est}}$ is computed from
    Eq.~(\ref{eq:fidelity}) under the simplified independent-error model.
    It is an execution diagnostic rather than a direct measurement of the final
    output distribution. Sim gap and HW gap are the respective optimality gaps to
    BKS for MDKP, MIS, and QAP. For MSP, we report TDev instead of percentage
    optimality gap, as defined in Eq.~\eqref{eq:tdev}, because MSP objectives are
    small integer target-deviation values for which percentage gaps are less
    informative. Feas: Y if the hardware run returned a feasible solution, N
  otherwise.}
  \label{tab:representative_fidelity}
  \small
  \begin{tabular}{lllrrrrrc}
    \toprule
    Problem & Method & Instance & $N_{2Q}$ & $F_{\mathrm{est}}$ & Sim gap \% & HW gap \% & Feas \\
    \midrule
    MDKP & VQE        & hp1    & 180    & $5.8\times10^{-1}$ & 39.8 & 27.0 & Y \\
    MDKP & CVaR-VQE   & pet2   & 347    & $2.5\times10^{-1}$ & 30.4 & 1.5  & Y \\
    MDKP & QAOA       & hp1    & 22\,501 & $4.4\times10^{-30}$ & -- & 36.2 & Y \\
    MDKP & PCE        & hp1    & 168    & $5.1\times10^{-1}$ & 20.9 & 13.3 & Y \\
    MDKP & QRAO       & hp1    & 111    & $7.2\times10^{-1}$ & --   & 5.9  & Y \\
    MIS  & VQE        & 1tc.32 & 96     & $7.5\times10^{-1}$ & 8.3  & --   & N \\
    MIS  & QAOA       & 1dc.128 & 48\,702 & $2.8\times10^{-64}$ & -- & -- & N \\
    MIS  & PCE        & 1tc.16 & 6      & $9.8\times10^{-1}$ & 50.0 & 62.5 & Y \\
    QAP  & VQE        & chr12a & 1\,323 & $1.9\times10^{-2}$ & --  & --   & N \\
    QAP  & QAOA       & chr12a & 84\,111 & $1.8\times10^{-110}$ & -- & -- & N \\
    QAP  & PCE        & nug12  & 20     & $9.4\times10^{-1}$ & 36.0 & --   & N \\
    MSP  & VQE        & ms20   & 144    & $6.5\times10^{-1}$ & --   & TDev=4 & Y \\
    MSP  & QAOA       & ms40   & 76\,599 & $1.1\times10^{-100}$ & -- & TDev=42 & Y \\
    MSP  & PCE        & ms20   & 168    & $5.1\times10^{-1}$ & --   & TDev=45 & Y \\
    \bottomrule
  \end{tabular}
\end{table}

Three patterns stand out in Table~\ref{tab:representative_fidelity}. First,
VQE and CVaR-VQE circuits remain in the proxy range $0.25$--$0.6$ at the
tested scales, consistent with their ability to return useful solutions on
MDKP and MSP. Second, QAOA-style circuits have transpiled two-qubit gate
counts exceeding 20\,000 for all representative instances and therefore
extremely small gate-count fidelity proxies. This places them well inside the
strongly noise-dominated regime, but does not by itself establish that their
measured output distributions are uniform.

To test the practical consequence of this regime, we performed a matched
uniform-random best-shot control over all 106 low-fidelity QAOA-family
hardware runs. For each run, the control matches the decision-variable
dimension, optimizer-trajectory candidate budget, batch-selection rule,
within-batch tie-breaking rule, and shared one-round local refinement used by
the hardware pipeline. Of the 106 runs, 43 hardware outputs are infeasible
and are treated as feasibility-only comparisons. Among the remaining feasible
runs with a finite matched random-quality distribution, 51 lie within the
empirical random range and 11 are worse than the matched random baseline. One
warm-start-QAOA MIS run returned a feasible candidate while none of the 300
matched random replicates did; this is reported as a finite-sample feasibility
exception rather than as a direct optimality-gap comparison. Thus, the
dominant low-fidelity QAOA-family pattern is compatible with best-of-budget
random candidate selection under the benchmark protocol, while the feasibility
exception shows that low $F_{\mathrm{est}}$ alone must not be treated as proof
of fully uniform-random output. Full results are reported in
Appendix~\ref{app:random_baseline}.

Third, for QAP, even the shallowest VQE ans\"atze at 144 qubits produces
transpiled circuits exceeding 1\,200 two-qubit gates, with estimated proxy
below $0.02$, consistent with the complete hardware infeasibility reported in
Section~\ref{sec:results_qap}. These patterns support the use of the
benchmark-specific fidelity thresholds derived in Sec.~\ref{sec:noise_model}
as diagnostic reference points rather than universal predictive laws.

\subsubsection{Compilation counterfactuals for QAOA-family circuits.}
\label{sec:qaoa_counterfactuals}

\begin{table}[htbp]
  \centering
  \caption{Compilation-only counterfactual summary for 12 representative bound
    QAOA-family circuits. Reductions are measured relative to reconstructed
    historical transpilation. The Nighthawk result uses topology-only compilation;
    the fractional-gate result is a gate-resource counterfactual and was not
    executed under the original resilience configuration. No audited circuit
    crosses $F_{\mathrm{est}}\geq10^{-3}$ or $F_{\mathrm{est}}\geq10^{-2}$ under
  any condition.}
  \label{tab:qaoa_counterfactual_summary}
  \footnotesize
  \setlength{\tabcolsep}{2.6pt}
  \renewcommand{\arraystretch}{1.08}
  \begin{tabular}{lrrr}
    \toprule
    Compilation condition &
    Median transpiled $N_{2Q}$ &
    Median change &
    Circuits with $F_{\mathrm{est}}\geq10^{-3}$ \\
    \midrule
    Historical compilation & 31,025 & -- & 0 / 12 \\
    SWAP-aware Heron surrogate & 34,443 & $-12.0\%$ & 0 / 12 \\
    Fractional-gate Heron surrogate & 29,411 & $+4.1\%$ & 0 / 12 \\
    SWAP-aware Nighthawk surrogate & 24,866 & $+19.8\%$ & 0 / 12 \\
    \bottomrule
  \end{tabular}
\end{table}

The negative QAOA-family hardware results should be interpreted at the level of
the tested implementations rather than as a statement about QAOA in general.
To assess whether plausible compilation improvements alone could move the
reported circuits back into the signal-preserving regime, we performed a
compilation-only counterfactual audit on 12 representative bound circuits:
QAOA, MA-QAOA, and WS-QAOA for MDKP \texttt{hp1}, MIS \texttt{1tc.32}, QAP
\texttt{tai10a}, and MSP \texttt{ms20}. The pre-transpilation circuits and final bound
parameter vectors were fixed; no parameters were retrained and no additional
hardware executions were performed.

Each circuit was compared under four conditions: (i) the reconstructed
historical compilation recorded in the saved hardware artifacts; (ii) a
connectivity-aware SABRE mapping and routing configuration on a Heron-like
heavy-hex surrogate; (iii) a bound-circuit fractional-gate Heron
counterfactual that preserves $R_{ZZ}$ structure where possible; and
(iv) the same connectivity-aware routing approach on a topology-only
square-lattice Nighthawk surrogate. The latter is a topology study only and
does not use Nighthawk calibration data or hardware execution.

Table~\ref{tab:qaoa_counterfactual_summary} summarizes the results. The
square-lattice Nighthawk surrogate produces the largest median reduction in
transpiled two-qubit gates, $19.8\%$, followed by fractional-gate Heron
compilation at $4.1\%$. The particular SWAP-aware Heron surrogate used here
increases median two-qubit-gate count by $12.0\%$ relative to the saved
historical routes. This should not be interpreted as evidence that all
routing-aware methods fail; rather, it shows that the historical routing was
already competitive with this specific SABRE-based surrogate configuration.

Most importantly, none of the 12 representative bound circuits crosses either
$F_{\mathrm{est}}=10^{-3}$ or $F_{\mathrm{est}}=10^{-2}$ under any tested
counterfactual. The strongest topology-only reduction therefore remains
insufficient to restore these circuits to the benchmark’s
signal-preserving regime. Fractional-gate results are reported only as
gate-resource counterfactuals: they were not executed under the original
runtime resilience configuration because that workflow is not directly
compatible with the original ZNE/probabilistic-error-amplification setting.

The resulting conclusion is deliberately narrow. The historical standard
QAOA, MA-QAOA, and WS-QAOA implementations remain strongly noise-dominated at
the reported scales, even after the examined compilation counterfactuals.
This result does not exclude improved performance from other QAOA
formulations, deeper architecture-specific compilation, alternative
parameter schedules, different mixers, or future hardware generations.

\subsection{Classical post-processing and solution recovery}
\label{sec:postprocessing}

The output of each quantum method is ultimately interpreted at the level of the
original combinatorial optimization problem. For the direct QUBO-based methods
(VQE, CVaR-VQE, QAOA, MA-QAOA, and WS-QAOA), each measured bitstring is treated
directly as a candidate binary decision vector and scored in the original
problem domain.

For PCE, measured Pauli correlators are converted into a binary candidate by
correlator reconstruction, as defined in Appendix~\ref{app:pce}. For QRAO,
measurement outcomes are converted into a binary candidate through the QRAC
rounding rule. These operations are distinct and are collectively referred to
as method-specific classical post-processing.

After a candidate is obtained, its feasibility is checked against the original
constrained problem rather than only against the unconstrained QUBO objective.
When an explicit procedure modifies a candidate to restore constraint
satisfaction, we call that procedure feasibility repair. The PCE pipeline also
uses the separately reported, feasibility-preserving multi-bit local search
described in Appendix~\ref{app:pce}; this is classical improvement rather than
measurement interpretation.

For each instance and method, the reported solution is the best candidate
returned by the prescribed sampling and classical post-processing pipeline.
This distinction is especially important for constrained benchmarks such as
MDKP, QAP, and MSP, where a low penalized objective does not necessarily imply
an operationally valid solution.

\section{Results}
\label{sec:results}

\subsection{Overview of the empirical comparison}

We now report the benchmark results under the protocol described in
Sections~\ref{sec:benchmark_design} and~\ref{sec:methods}. Our goal is not only
to compare final objective quality across method families, but also to assess how
those comparisons change once realistic execution constraints are taken into
account. Accordingly, we evaluate each method family both through
problem-level solution metrics and through hardware-facing diagnostics such as
algorithmic level width, transpiled circuit growth, and execution robustness.

To keep the discussion interpretable, we organize the main results around the four
benchmark problems and compare three broad method families in each case:
(i) VQE-style energy-minimization methods (VQE and CVaR-VQE),
(ii) QAOA-style methods (QAOA, MA-QAOA, and WS-QAOA), and
(iii) qubit-efficient encoding or reduction methods (PCE and QRAO).
The main text focuses on benchmark-level patterns, while detailed
circuit-compilation and execution-robustness summaries are reported in the appendix.

Across the full benchmark, performance is heterogeneous across problem classes
and method families. The detailed cross-problem synthesis is given in
Sec.~\ref{sec:results_cross_problem}; here we first examine each problem family
separately.

\subsection{MDKP: balanced comparison across all method families}
\label{sec:results_mdkp}

MDKP is the most balanced family in the benchmark because all compared methods
return feasible solutions on hardware, allowing a genuine comparison of solution
quality under realistic execution constraints.
The VQE-style results in Table~\ref{tab:mdkp_vqe_family_results} show that both
VQE and CVaR-VQE remain feasible on all instances, with CVaR-VQE achieving the
lower gap on most cases, including especially strong gains on \texttt{pet2},
\texttt{pb2}, \texttt{pb5}, \texttt{pet5}, and \texttt{pet6}.
This indicates that the tail-focused objective can improve recovery of better
samples for constrained packing instances, although the advantage is not uniform.

\begin{table}[ht]
  \caption{\label{tab:mdkp_vqe_family_results}
    Combined MDKP VQE and CVaR-VQE results summary across instances. Entries report the primary
    performance metrics (Optimal/BKS, Qubits, Feasibility, and Gap). The Qubits column is reported as
    VQE / CVaR-VQE when the two encodings require different algorithmic-level widths. The lower gap per instance
  is boldfaced.}
  \centering
  \begin{tabular*}{\columnwidth}{@{\extracolsep{\fill}}lcccccc}
    \toprule
    \textbf{Instance} & \textbf{Optimal} & \textbf{Qubits} &
    \multicolumn{2}{c}{\textbf{VQE}} & \multicolumn{2}{c}{\textbf{CVaR-VQE}} \\
    \cmidrule(lr){4-5} \cmidrule(lr){6-7}
    & & & \textbf{Feasible} & \textbf{Gap (\%)} & \textbf{Feasible} & \textbf{Gap (\%)} \\
    \midrule
    hp1  & 3418  & 60  & Yes & \textbf{26.95} & Yes & 33.85 \\
    hp2  & 3186  & 67  & Yes & 17.92          & Yes & \textbf{15.88} \\
    pb1  & 3090  & 59  & Yes & 22.62          & Yes & \textbf{14.89} \\
    pb2  & 3186  & 66  & Yes & 26.77          & Yes & \textbf{11.93} \\
    pb4  & 95168 & 45  & Yes & \textbf{33.63} & Yes & 34.97 \\
    pb5  & 2139  & 116 & Yes & 29.73          & Yes & \textbf{18.14} \\
    pet2 & 87061 & 99  & Yes & 16.98          & Yes & \textbf{1.50} \\
    pet3 & 4015  & 102 & Yes & \textbf{32.38} & Yes & 36.61 \\
    pet4 & 6120  & 107 & Yes & 44.69          & Yes & \textbf{41.91} \\
    pet5 & 12400 & 122 & Yes & 25.65          & Yes & \textbf{13.83} \\
    pet6 & 10618 & 86  & Yes & 20.24          & Yes & \textbf{13.88} \\
    pet7 & 16537 & 100 & Yes & \textbf{23.28} & Yes & 24.52 \\
    \bottomrule
  \end{tabular*}
\end{table}

The QAOA-family results in Table~\ref{tab:mdkp_qaoa_family_results} show that all
three variants are also feasible on every tested instance, but the best variant is
instance-dependent.
MA-QAOA attains the lowest gap on several instances, including \texttt{hp1},
\texttt{hp2}, \texttt{pb1}, \texttt{pb4}, and \texttt{pet4}, whereas WS-QAOA is
best on \texttt{pb2}, \texttt{pb5}, \texttt{pet2}, \texttt{pet5}, and
\texttt{pet7}. Standard QAOA is occasionally best, for example on
\texttt{pet3} and \texttt{pet6}, but is less consistently competitive than the
structured variants. Thus, for MDKP, modifying the operator schedule or warm-start
structure often improves performance relative to vanilla QAOA\@.

\begin{table*}[t]
  \caption{\label{tab:mdkp_qaoa_family_results}
    Combined MDKP results for QAOA, MA-QAOA, and WS-QAOA across instances. Entries report
    the primary performance metrics (Optimal/BKS, Qubits, Feasibility, and Gap). The lower
  gap per instance among the three variants is boldfaced.}
  \centering
  \small
  \setlength{\tabcolsep}{4pt}
  \begin{tabular*}{\textwidth}{@{\extracolsep{\fill}}lcccccccc}
    \toprule
    \textbf{Instance} & \textbf{Optimal} & \textbf{Qubits} &
    \multicolumn{2}{c}{\textbf{QAOA}} &
    \multicolumn{2}{c}{\textbf{MA-QAOA}} &
    \multicolumn{2}{c}{\textbf{WS-QAOA}} \\
    \cmidrule(lr){4-5} \cmidrule(lr){6-7} \cmidrule(lr){8-9}
    & & & \textbf{Feas.} & \textbf{Gap} & \textbf{Feas.} & \textbf{Gap} & \textbf{Feas.} & \textbf{Gap} \\
    \midrule
    hp1  & 3418  & 60  & Yes & 36.22 & Yes & \textbf{21.06} & Yes & 23.61 \\
    hp2  & 3186  & 67  & Yes & 19.68 & Yes & \textbf{15.66} & Yes & 29.19 \\
    pb1  & 3090  & 59  & Yes & 36.67 & Yes & \textbf{27.86} & Yes & 28.45 \\
    pb2  & 3186  & 66  & Yes & 42.18 & Yes & 44.51 & Yes & \textbf{33.93} \\
    pb4  & 95168 & 45  & Yes & 37.53 & Yes & \textbf{27.63} & Yes & 43.13 \\
    pb5  & 2139  & 116 & Yes & 23.47 & Yes & 28.00 & Yes & \textbf{20.80} \\
    pet2 & 87061 & 99  & Yes & 14.70 & Yes & 30.97 & Yes & \textbf{1.28} \\
    pet3 & 4015  & 102 & Yes & \textbf{20.05} & Yes & 27.52 & Yes & 36.74 \\
    pet4 & 6120  & 107 & Yes & 33.33 & Yes & \textbf{14.38} & Yes & 25.65 \\
    pet5 & 12400 & 122 & Yes & 37.30 & Yes & 26.21 & Yes & \textbf{13.79} \\
    pet6 & 10618 & 86  & Yes & \textbf{23.09} & Yes & 39.87 & Yes & 30.45 \\
    pet7 & 16537 & 100 & Yes & 49.76 & Yes & 28.91 & Yes & \textbf{27.62} \\
    \bottomrule
  \end{tabular*}
\end{table*}

The encoding-based comparison in
Table~\ref{tab:mdkp_encoding_family_results} reveals the clearest width--quality
trade-off in the paper.
PCE uses only 6--10 qubits across the tested instances, whereas QRAO uses
32--62 qubits and the direct formulations require substantially larger widths.
This compression is substantial and materially expands the set of backends on which
the problem can be executed.
However, PCE is not uniformly best in objective quality: QRAO attains the lower
gap on 8 of the 12 MDKP instances, while PCE is best on the remaining 4.
MDKP therefore provides a clear
positive example of the benchmark thesis: qubit-efficient encodings improve executability, but the
strongest compression does not by itself determine the best final hardware outcome. In
particular, the benchmarked quadratic PCE configuration is competitive on several
instances under the same shared one-pass local-improvement rule applied to all
quantum methods. Its results should not be interpreted as an upper bound on what
a more aggressive multi-pass or higher-order PCE variant might achieve.

\begin{table*}[t]
  \caption{\label{tab:mdkp_encoding_family_results}
    Combined MDKP results for PCE and QRAO across instances. Entries report
    Optimal/BKS, Qubits, shown as PCE/QRAO when the encodings differ,
  feasibility, and gap to BKS\@. The lower gap per instance is boldfaced.}
  \centering
  \small
  \setlength{\tabcolsep}{4pt}
  \begin{tabular*}{\textwidth}{@{\extracolsep{\fill}}lcccccc}
    \toprule
    \textbf{Instance} & \textbf{Optimal} & \textbf{Qubits} &
    \multicolumn{2}{c}{\textbf{PCE}} &
    \multicolumn{2}{c}{\textbf{QRAO}} \\
    \cmidrule(lr){4-5} \cmidrule(lr){6-7}
    & & & \textbf{Feas.} & \textbf{Gap} & \textbf{Feas.} & \textbf{Gap} \\
    \midrule
    hp1  & 3418  & 7 / 38  & Yes & 13.25          & Yes & \textbf{5.85} \\
    hp2  & 3186  & 8 / 45  & Yes & \textbf{14.56} & Yes & 16.67 \\
    pb1  & 3090  & 7 / 37  & Yes & 22.78          & Yes & \textbf{15.40} \\
    pb2  & 3186  & 8 / 44  & Yes & 21.37          & Yes & \textbf{13.65} \\
    pb4  & 95168 & 6 / 32  & Yes & \textbf{21.30} & Yes & 21.71 \\
    pb5  & 2139  & 10 / 57 & Yes & \textbf{17.72} & Yes & 18.00 \\
    pet2 & 87061 & 9 / 44  & Yes & 1.69           & Yes & \textbf{0.64} \\
    pet3 & 4015  & 9 / 44  & Yes & \textbf{7.97}  & Yes & 34.12 \\
    pet4 & 6120  & 9 / 47  & Yes & 20.42          & Yes & \textbf{18.63} \\
    pet5 & 12400 & 10 / 58 & Yes & 37.50          & Yes & \textbf{25.52} \\
    pet6 & 10618 & 9 / 52  & Yes & 19.92          & Yes & \textbf{15.23} \\
    pet7 & 16537 & 9 / 62  & Yes & 25.27          & Yes & \textbf{16.79} \\
    \bottomrule
  \end{tabular*}
\end{table*}

\begin{figure}
  \centering
  \includegraphics[width=\textwidth]{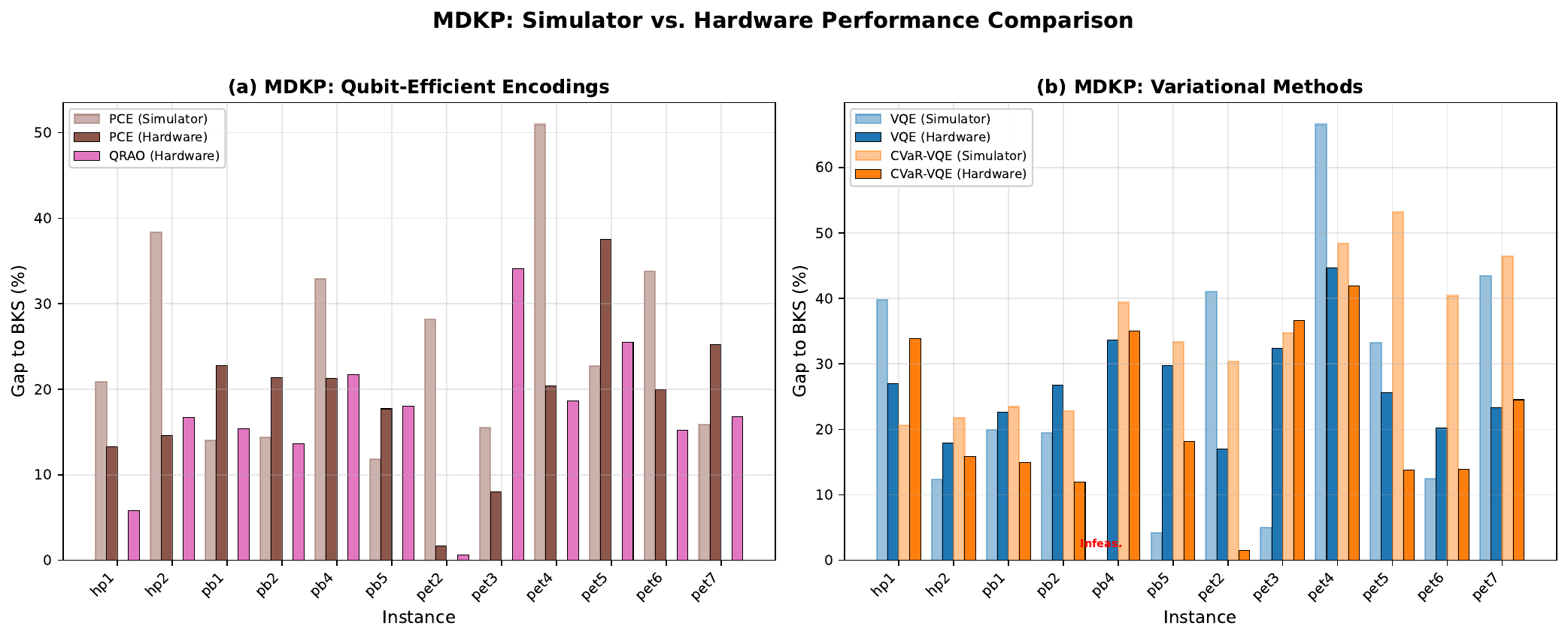}
  \caption{MDKP simulator versus hardware performance across qubit-efficient and
    variational methods. Panel (a) compares PCE and QRAO on MDKP instances
    using gap to the optimal solution, while panel (b) reports
    the corresponding comparison for VQE and CVaR-VQE\@. The figure shows
    that simulator-side rankings do not transfer uniformly to hardware:
    some methods retain moderate performance degradation, whereas others
    exhibit larger hardware-induced gap inflation or infeasibility on
    selected instances. Overall, the plot highlights a nontrivial
    simulator–hardware gap even at these small benchmark sizes and shows
  that hardware execution can materially change relative method quality.}
  \label{fig1}
\end{figure}

The simulator results reported in the appendix provide a controlled reference
for this comparison, while the hardware results show how the final ranking
changes after compilation and execution.
Taken together, MDKP is the clearest example in the benchmark where all method
families remain viable and meaningful trade-offs between quality, width, and
hardware overhead can be observed directly. Figure~\ref{fig1} contrasts the simulator and hardware gaps for these method
families and shows that simulator-side rankings do not transfer uniformly to
hardware.

\subsection{MIS: a sharp feasibility threshold under hardware execution}
\label{sec:results_mis}

The MIS results reveal a different behavior from MDKP\@.
Rather than a balanced quality comparison across all methods, the dominant pattern
is a sharp feasibility breakdown as instance size increases.
The VQE-style results in Table~\ref{tab:mis_vqe_family_results} show that both VQE
and CVaR-VQE solve the smallest instances well, with zero gap on \texttt{1tc.8}
and near-optimal performance on \texttt{1tc.16}.
However, this behavior does not persist: from \texttt{1tc.32} onward, feasibility
begins to collapse, and by the 64- and 128-variable instances both methods become
infeasible.

\begin{table*}[t]
  \caption{\label{tab:mis_vqe_family_results}
    Combined MIS VQE and CVaR-VQE results across instances. Entries report the known
    optimum/BKS, Qubits, feasibility, and gap to BKS\@. The lower gap
    per instance among feasible methods is boldfaced; infeasible runs are reported
  with gap $\infty$.}
  \centering
  \small
  \setlength{\tabcolsep}{4pt}
  \begin{tabular*}{\textwidth}{@{\extracolsep{\fill}}lccccccc}
    \toprule
    \textbf{Instance} & \textbf{Optimal} & \textbf{Qubits} &
    \multicolumn{2}{c}{\textbf{VQE}} &
    \multicolumn{2}{c}{\textbf{CVaR-VQE}} \\
    \cmidrule(lr){4-5} \cmidrule(lr){6-7}
    & & & \textbf{Feas.} & \textbf{Gap} & \textbf{Feas.} & \textbf{Gap} \\
    \midrule
    1tc.8   & 4  & 8   & Yes & \textbf{0.00}   & Yes & \textbf{0.00} \\
    1tc.16  & 8  & 16  & Yes & \textbf{0.00}   & Yes & 12.50 \\
    1tc.32  & 12 & 32  & No  & $\infty$        & Yes & \textbf{16.67} \\
    1tc.64  & 20 & 64  & No  & $\infty$        & No  & $\infty$ \\
    1et.64  & 18 & 64  & No  & $\infty$        & No  & $\infty$ \\
    1dc.64  & 10 & 64  & No  & $\infty$        & No  & $\infty$ \\
    1dc.128 & 16 & 128 & No  & $\infty$        & No  & $\infty$ \\
    \bottomrule
  \end{tabular*}
\end{table*}

A similar pattern appears for the QAOA family in
Table~\ref{tab:mis_qaoa_family_results}.
The structured variants remain strong on the smallest graphs, and WS-QAOA is the
best-performing QAOA-style method on several of the larger feasible cases.
Most notably, WS-QAOA remains feasible on \texttt{1tc.64}, where the other
QAOA-style methods fail.
Even so, the larger MIS instances still expose a clear feasibility cliff:
performance deteriorates rapidly once the width and compiled circuit burden exceed
what current hardware can support reliably.

\begin{table*}[t]
  \caption{\label{tab:mis_qaoa_family_results}
    Combined MIS results for QAOA, MA-QAOA, and WS-QAOA across instances. Entries
    report the known optimum/BKS, Qubits, feasibility, and gap to
    BKS\@. The lower gap per instance among feasible variants is boldfaced; infeasible
  runs are reported with gap $\infty$.}
  \centering
  \small
  \setlength{\tabcolsep}{4pt}
  \begin{tabular*}{\textwidth}{@{\extracolsep{\fill}}lccccccccc}
    \toprule
    \textbf{Instance} & \textbf{Optimal} & \textbf{Qubits} &
    \multicolumn{2}{c}{\textbf{QAOA}} &
    \multicolumn{2}{c}{\textbf{MA-QAOA}} &
    \multicolumn{2}{c}{\textbf{WS-QAOA}} \\
    \cmidrule(lr){4-5} \cmidrule(lr){6-7} \cmidrule(lr){8-9}
    & & & \textbf{Feas.} & \textbf{Gap} & \textbf{Feas.} & \textbf{Gap} & \textbf{Feas.} & \textbf{Gap} \\
    \midrule
    1tc.8   & 4  & 8   & Yes & \textbf{0.00} & Yes & \textbf{0.00} & Yes & \textbf{0.00} \\
    1tc.16  & 8  & 16  & Yes & 25.00         & Yes & 12.50         & Yes & \textbf{0.00} \\
    1tc.32  & 12 & 32  & Yes & 25.00         & No  & $\infty$      & Yes & \textbf{16.67} \\
    1tc.64  & 20 & 64  & No  & $\infty$      & No  & $\infty$      & Yes & \textbf{35.00} \\
    1et.64  & 18 & 64  & No  & $\infty$      & No  & $\infty$      & No  & $\infty$ \\
    1dc.64  & 10 & 64  & No  & $\infty$      & No  & $\infty$      & No  & $\infty$ \\
    1dc.128 & 16 & 128 & No  & $\infty$      & No  & $\infty$      & No  & $\infty$ \\
    \bottomrule
  \end{tabular*}
\end{table*}

\begin{figure}
  \centering
  \includegraphics[width=\textwidth]{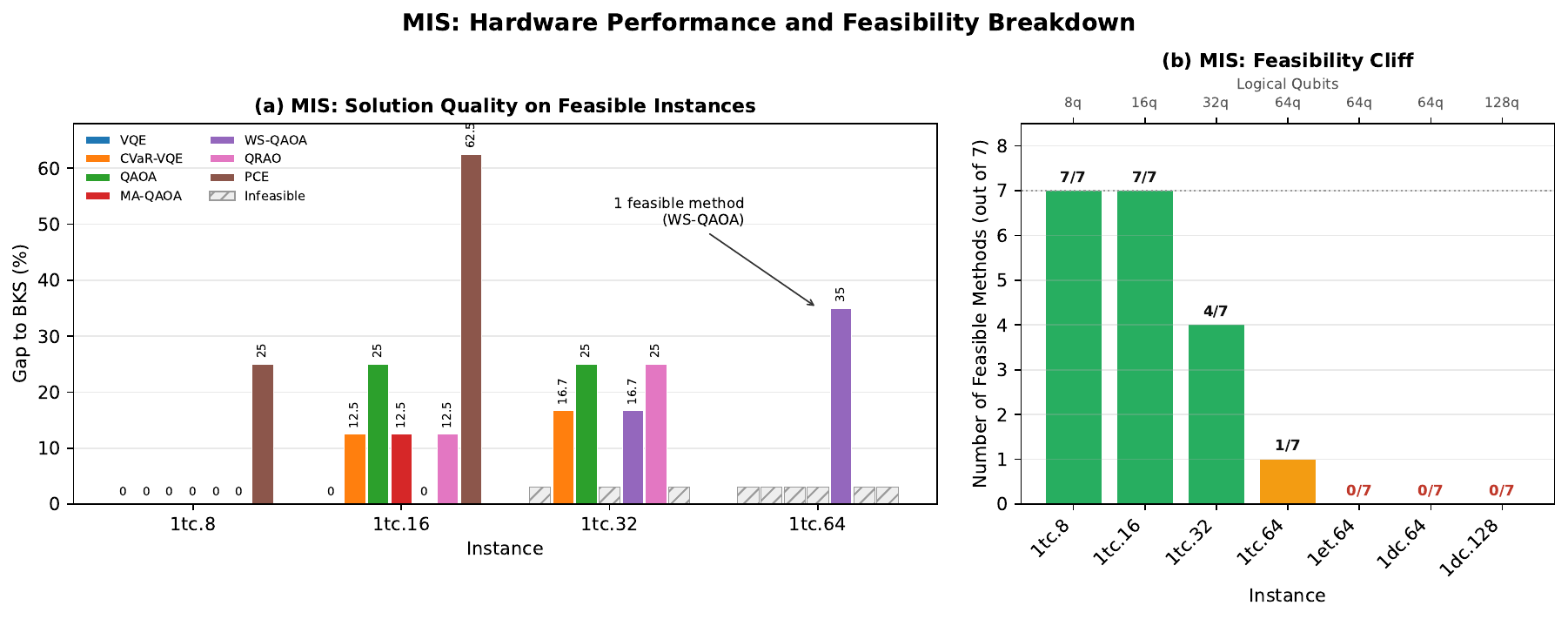}
  \caption{Hardware performance and feasibility breakdown for MIS.
    Panel (a) reports finite optimality gaps for feasible
    method--instance outcomes; hatched bars denote infeasible outcomes.
    WS-QAOA is the only feasible method on \texttt{1tc.64}. Panel (b)
    reports the number of feasible methods out of seven for each
    instance: $7/7$, $7/7$, $4/7$, $1/7$, $0/7$, $0/7$, and $0/7$ from
    \texttt{1tc.8} through \texttt{1dc.128}. Feasibility therefore
    collapses sharply with problem size, even though an isolated feasible
  outcome remains at 64 algorithm-level qubits.}

  \label{fig2}
\end{figure}

The encoding-based MIS comparison in
Table~\ref{tab:mis_pce_qrao_results} confirms that compression helps, but only to a
point.
QRAO outperforms PCE on the smaller and intermediate feasible instances, achieving
zero gap on \texttt{1tc.8} and better recovery than PCE on \texttt{1tc.16} and
\texttt{1tc.32}.
However, neither encoding family can maintain feasibility indefinitely as the
instances grow.
Thus, MIS does not simply reward lower width; rather, it exposes a practical
boundary beyond which current hardware noise and compilation overhead dominate. The feasibility breakdown is summarized in Figure~\ref{fig2}.

\begin{table*}[t]
  \caption{\label{tab:mis_pce_qrao_results}
    Combined MIS results for PCE and QRAO across instances. Entries report the known
    optimum/BKS, qubits used, feasibility, and gap to BKS\@. The lower gap per instance among feasible
  methods is boldfaced; infeasible runs are reported with gap $\infty$.}
  \centering
  \small
  \setlength{\tabcolsep}{4pt}
  \begin{tabular*}{\textwidth}{@{\extracolsep{\fill}}lccccccc}
    \toprule
    \textbf{Instance} & \textbf{Optimal} &
    \multicolumn{3}{c}{\textbf{PCE}} &
    \multicolumn{3}{c}{\textbf{QRAO}} \\
    \cmidrule(lr){3-5} \cmidrule(lr){6-8}
    & & \textbf{Qubits} & \textbf{Feas.} & \textbf{Gap} & \textbf{Qubits} & \textbf{Feas.} & \textbf{Gap} \\
    \midrule
    1tc.8   & 4  & 3 & Yes & 25.00     & 4 & Yes & \textbf{0.00} \\
    1tc.16  & 8  & 4 & Yes & 62.50     & 6 & Yes & \textbf{12.50} \\
    1tc.32  & 12 & 6 & No  & $\infty$  & 13 & Yes & \textbf{25.00} \\
    1tc.64  & 20 & 8 & No  & $\infty$  & 23 & No  & $\infty$ \\
    1et.64  & 18 & 8 & No  & $\infty$  & 24 & No  & $\infty$ \\
    1dc.64  & 10 & 8 & No  & $\infty$  & 18 & No  & $\infty$ \\
    1dc.128 & 16 & 10 & No  & $\infty$  & 46 & No  & $\infty$ \\
    \bottomrule
  \end{tabular*}
\end{table*}

This makes MIS one of the most informative families in the benchmark from a
hardware perspective.
At small scale, several methods perform well and even reach zero gap.
At moderate scale, the benchmark becomes sensitive to architecture and encoding
choice.
At larger scale, feasibility itself becomes the central issue.
The main lesson from MIS is therefore not just that gaps worsen with size, but that
backend-realized feasibility can collapse sharply even when the problem size
still appears moderate.

\subsection{QAP: the clearest negative result in the benchmark}
\label{sec:results_qap}

QAP is the most structurally restrictive family in the study and provides the
clearest negative result under real-hardware execution. The difficulty is not
explained by circuit width alone. The benchmark uses standard QAPLIB instances
in the $n=10$–$12$ range; the smallest selected instances,
\texttt{tai10a} and \texttt{tai10b}, already require 100 binary variables
under the direct one-hot formulation. A valid assignment corresponds to a
permutation, so only $10!$ of the $2^{100}$ binary strings are feasible:
\begin{equation}
  \frac{10!}{2^{100}}
  \approx 10^{-23.54}.
\end{equation}
For the $n=12$ instances, the direct formulation contains 144 binary variables
and the feasible fraction contracts to
\begin{equation}
  \frac{12!}{2^{144}}
  \approx 10^{-34.67}.
\end{equation}

This sparse permutation-feasible manifold is combined with dense
flow-distance interactions and row-and-column one-hot penalties. Across the
tested QAP instances, the median QUBO density is $0.818$, the median
interaction degree is $95.4$, and coefficient dynamic range reaches as high as
$2.21\times10^{10}$. Thus, even the smallest standard QAPLIB instances
selected for this benchmark are already structurally very different from a
100-variable sparse graph or packing QUBO.

The VQE-style results in Table~\ref{tab:qap_vqe_family_results} show that both
VQE and CVaR-VQE are infeasible on all tested QAP instances. The same conclusion
holds for the QAOA family in Table~\ref{tab:qap_qaoa_family_results}, where
QAOA, MA-QAOA, and WS-QAOA all fail to recover feasible solutions on the
hardware-tested instances. Even the encoding-based methods in
Table~\ref{tab:qap_pce_qrao_results} do not resolve this difficulty: PCE and
QRAO remain infeasible on the tested QPU runs.

\begin{table*}[t]
  \caption{\label{tab:qap_vqe_family_results}
    Combined QAP VQE and CVaR-VQE results across instances. Entries report the known
    optimum/BKS, circuit width (Qubits), feasibility, and gap to BKS\@. The lower gap
    per instance among feasible methods is boldfaced; infeasible runs are reported
  with gap $\infty$.}
  \centering
  \small
  \setlength{\tabcolsep}{4pt}
  \begin{tabular*}{\textwidth}{@{\extracolsep{\fill}}lccccccc}
    \toprule
    \textbf{Instance} & \textbf{Optimal} & \textbf{Qubits} &
    \multicolumn{2}{c}{\textbf{VQE}} &
    \multicolumn{2}{c}{\textbf{CVaR-VQE}} \\
    \cmidrule(lr){4-5} \cmidrule(lr){6-7}
    & & & \textbf{Feas.} & \textbf{Gap} & \textbf{Feas.} & \textbf{Gap} \\
    \midrule
    chr12a & 9552     & 144 & No & $\infty$ & No & $\infty$ \\
    chr12b & 9742     & 144 & No & $\infty$ & No & $\infty$ \\
    chr12c & 11156    & 144 & No & $\infty$ & No & $\infty$ \\
    nug12  & 578      & 144 & No & $\infty$ & No & $\infty$ \\
    had12  & 1652     & 144 & No & $\infty$ & No & $\infty$ \\
    rou12  & 235528   & 144 & No & $\infty$ & No & $\infty$ \\
    scr12  & 31410    & 144 & No & $\infty$ & No & $\infty$ \\
    tai10a & 135028   & 100 & No & $\infty$ & No & $\infty$ \\
    tai10b & 1183760  & 100 & No & $\infty$ & No & $\infty$ \\
    tai12a & 224416   & 144 & No & $\infty$ & No & $\infty$ \\
    tai12b & 39464925 & 144 & No & $\infty$ & No & $\infty$ \\
    \bottomrule
  \end{tabular*}
\end{table*}

\begin{table*}[t]
  \caption{\label{tab:qap_qaoa_family_results}
    Combined QAP results for QAOA, MA-QAOA, and WS-QAOA across instances. Entries report
    the known optimum/BKS, circuit width (Qubits), feasibility, and gap to BKS\@. The
    lower gap per instance among feasible variants is boldfaced; infeasible runs are
    reported with gap $\infty$. A dash indicates that the method was not run or not
  reported for that instance.}
  \centering
  \small
  \setlength{\tabcolsep}{4pt}
  \begin{tabular*}{\textwidth}{@{\extracolsep{\fill}}lccccccccc}
    \toprule
    \textbf{Instance} & \textbf{Optimal} & \textbf{Qubits} &
    \multicolumn{2}{c}{\textbf{QAOA}} &
    \multicolumn{2}{c}{\textbf{MA-QAOA}} &
    \multicolumn{2}{c}{\textbf{WS-QAOA}} \\
    \cmidrule(lr){4-5} \cmidrule(lr){6-7} \cmidrule(lr){8-9}
    & & & \textbf{Feas.} & \textbf{Gap} & \textbf{Feas.} & \textbf{Gap} & \textbf{Feas.} & \textbf{Gap} \\
    \midrule
    chr12a & 9552     & 144 & No & $\infty$ & No & $\infty$ & No & $\infty$ \\
    chr12b & 9742     & 144 & No & $\infty$ & No & $\infty$ & No & $\infty$ \\
    chr12c & 11156    & 144 & No & $\infty$ & No & $\infty$ & No & $\infty$ \\
    nug12  & 578      & 144 & No & $\infty$ & No & $\infty$ & No & $\infty$ \\
    had12  & 1652     & 144 & No & $\infty$ & No & $\infty$ & No & $\infty$ \\
    rou12  & 235528   & 144 & No & $\infty$ & No & $\infty$ & No & $\infty$ \\
    scr12  & 31410    & 144 & No & $\infty$ & No & $\infty$ & No & $\infty$ \\
    tai10a & 135028   & 100 & No & $\infty$ & No & $\infty$ & No & $\infty$ \\
    tai10b & 1183760  & 100 & No & $\infty$ & No & $\infty$ & No & $\infty$ \\
    tai12a & 224416   & 144 & No & $\infty$ & No & $\infty$ & No & $\infty$ \\
    tai12b & 39464925 & 144 & No & $\infty$ & No & $\infty$ & No & $\infty$ \\
    \bottomrule
  \end{tabular*}
\end{table*}

\begin{table*}[t]
  \caption{\label{tab:qap_pce_qrao_results}
    Combined QAP results for PCE and QRAO across instances. Entries report the known
    optimum/BKS, the number of qubits required by each encoding, feasibility, and gap
    to BKS\@. The lower gap per instance among feasible methods is boldfaced; infeasible
    runs are reported with gap $\infty$. A dash indicates that the method was not run
  or not reported for that instance.}
  \centering
  \small
  \setlength{\tabcolsep}{4pt}
  \begin{tabular*}{\textwidth}{@{\extracolsep{\fill}}lccccccc}
    \toprule
    \textbf{Instance} & \textbf{Optimal} &
    \multicolumn{3}{c}{\textbf{PCE}} &
    \multicolumn{3}{c}{\textbf{QRAO}} \\
    \cmidrule(lr){3-5} \cmidrule(lr){6-8}
    & & \textbf{Qubits} & \textbf{Feas.} & \textbf{Gap} &
    \textbf{Qubits} & \textbf{Feas.} & \textbf{Gap} \\
    \midrule
    chr12a & 9552     & 11 & No & $\infty$ & 56  & No & $\infty$ \\
    chr12b & 9742     & 11 & No & $\infty$ & 62  & No & $\infty$ \\
    chr12c & 11156    & 11 & No & $\infty$ & 54  & No & $\infty$ \\
    nug12  & 578      & 11 & No & $\infty$ & 63  & No & $\infty$ \\
    had12  & 1652     & 11 & No & $\infty$ & 144 & No & $\infty$ \\
    rou12  & 235528   & 11 & No & $\infty$ & 132 & No & $\infty$ \\
    scr12  & 31410    & 11 & No & $\infty$       & 60  & No & $\infty$ \\
    \bottomrule
  \end{tabular*}
\end{table*}

This result is scientifically important and should not be treated as merely a
failure case. QAP combines three adverse properties in the direct benchmark
encoding: dense flow-distance interactions, strict row-and-column one-hot
assignment constraints, and an exponentially sparse feasible subset of the
binary search space. The resulting QUBOs have widths of 100–144 variables,
median density $0.818$, and coefficient dynamic ranges extending to
$2.21\times10^{10}$ across the tested instances.

The simulator results in Appendix~\ref{app:sim_qap} provide an instructive
contrast. VQE and CVaR-VQE remain infeasible on the tested QAP instances even
under simulator execution, whereas PCE returns feasible candidates on several
instances, albeit with large optimality gaps of approximately $15\%$–$235\%$.
This indicates that compressed PCE reconstruction can retain partial access to
the permutation-feasible manifold under noise-free execution, but does not
remove the underlying structural difficulty of the dense QAP formulation.

\begin{figure}
  \centering
  \includegraphics[width=\textwidth]{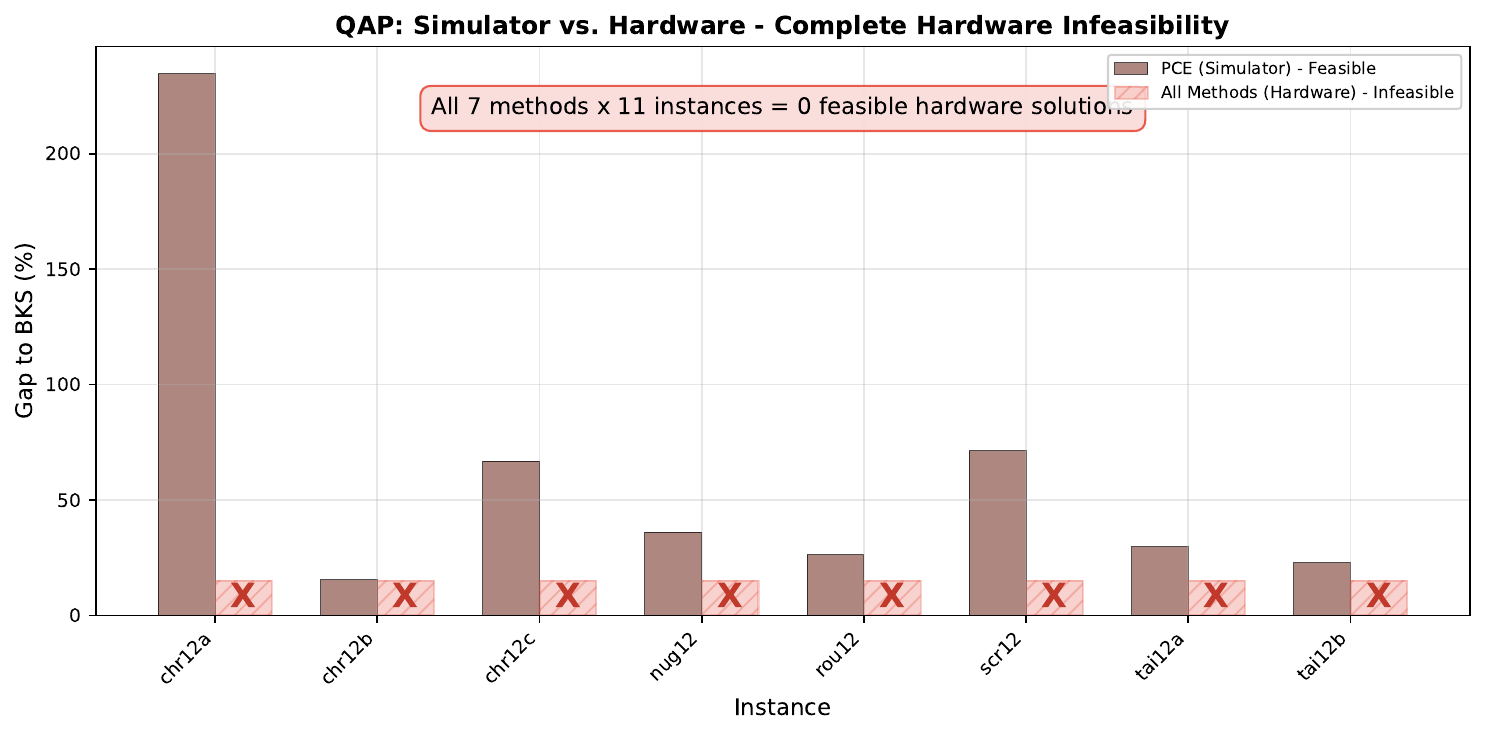}
  \caption{QAP simulator versus hardware comparison. The direct QAP encoding
    combines dense flow-distance couplings with row-and-column one-hot assignment
    constraints; for the tested $n=10$ and $n=12$ instances, valid assignments
    occupy approximately $10^{-23.54}$ to $10^{-34.67}$ of the full binary
    hypercube. PCE returns feasible simulator candidates on several instances,
    although with large optimality gaps, whereas none of the seven
    hardware-executed methods returns a feasible QAP assignment. The comparison
    therefore illustrates the interaction of restrictive QAP feasibility geometry
  with compilation overhead, execution noise, and recovery sensitivity.}
  \label{fig3}
\end{figure}

The complete hardware infeasibility should therefore not be attributed to
qubit count or compilation cost in isolation. The observed barrier reflects the
combined effect of dense interactions, broad coefficient scales, one-hot
penalty structure, a combinatorially sparse feasible assignment manifold,
hardware compilation overhead, execution noise, and recovery sensitivity. QAP
therefore serves as a stress test that separates partial representability in
simulation from reliable feasible-solution recovery on current hardware. Figure~\ref{fig3} contrasts the feasible PCE simulator candidates with the
uniform hardware infeasibility across all seven methods.

\subsection{MSP: strong compression but weak recovered hardware quality}
\label{sec:results_msp}

MSP exhibits a markedly different pattern from both MDKP and MIS\@.
Here, the most important observation is that qubit-efficient encodings achieve the
strongest compression, but not the best recovered hardware solution quality.
The VQE-style results in Table~\ref{tab:market_share_vqe_family_results} show that
the relative performance of VQE and CVaR-VQE is instance-dependent.
VQE is better on some instances, including \texttt{ms20}, \texttt{ms30},
\texttt{ms31}, \texttt{ms51}, while CVaR-VQE improves on others such as
\texttt{ms40}, \texttt{ms50}, \texttt{ms21}, and \texttt{ms41}.
Thus, unlike MDKP, there is no consistent advantage of the CVaR objective for MSP\@.

\begin{table*}[t]
  \caption{\label{tab:market_share_vqe_family_results}
    Market Share results for VQE and CVaR-VQE\@. Because the classically
    certified MSP optima are small integers (0--3), percentage gaps are
    unstable and uninformative. We report total absolute deviation
    (\textbf{TDev}), maximum single-product deviation (\textbf{MDev}), and
    the range-normalized TDev approximation ratio
    ($\mathbf{AR}_{\mathrm{TDev}}$). TDev remains the primary
    application-level quality measure. No reported hardware run achieved an
  exact target match.}
  \centering
  \footnotesize
  \setlength{\tabcolsep}{4pt}
  \begin{tabular*}{\textwidth}{@{\extracolsep{\fill}}lcccccccc}
    \toprule
    & & & \multicolumn{3}{c}{\textbf{VQE}} & \multicolumn{3}{c}{\textbf{CVaR-VQE}} \\
    \cmidrule(lr){4-6} \cmidrule(lr){7-9}
    \textbf{Inst.} & \textbf{Opt.} & \textbf{Q} &
    \textbf{TDev} & \textbf{MDev} & $\mathbf{AR}_{\mathrm{TDev}}$ &
    \textbf{TDev} & \textbf{MDev} & $\mathbf{AR}_{\mathrm{TDev}}$ \\
    \midrule
    ms20 & 3 & 48  & \textbf{4}   & \textbf{2}   & \textbf{0.998} & 27  & 21  & 0.951 \\
    ms30 & 2 & 84  & \textbf{127} & \textbf{89}  & \textbf{0.919} & 347 & 214 & 0.777 \\
    ms40 & 1 & 118 & 188          & 85           & 0.936 & \textbf{150} & \textbf{91}  & \textbf{0.949} \\
    ms50 & 0 & 150 & 644          & 214          & 0.865 & \textbf{476} & \textbf{165} & \textbf{0.900} \\
    ms21 & 3 & 50  & 215          & 121          & 0.586 & \textbf{9}   & \textbf{7}   & \textbf{0.988} \\
    ms31 & 3 & 84  & \textbf{12}  & \textbf{7}   & \textbf{0.994} & 51  & 21  & 0.968 \\
    ms41 & 1 & 118 & 273          & 135          & 0.916 & \textbf{48}  & \textbf{16}  & \textbf{0.985} \\
    ms51 & 1 & 156 & \textbf{439} & \textbf{152} & \textbf{0.915} & 565 & 177 & 0.891 \\
    \bottomrule
  \end{tabular*}
\end{table*}

The QAOA-style results in Table~\ref{tab:market_share_qaoa_family_results} are
again highly instance-dependent.
QAOA is strongest on some cases, WS-QAOA is best on others, and MA-QAOA is
competitive only on a subset.
This reinforces that MSP has a different optimization structure from the graph-based
or packing-based benchmarks and cannot be interpreted through the same intuition as
MIS or MDKP\@.

\begin{table*}[t]
  \caption{\label{tab:market_share_qaoa_family_results}
    Market Share results for QAOA, MA-QAOA, and WS-QAOA\@.
    \textbf{TDev}: total absolute deviation from target split.
    \textbf{MDev}: maximum deviation on any single product.
    $\mathbf{AR}_{\mathrm{TDev}}$: range-normalized TDev approximation
  ratio. No reported hardware run achieved an exact target match.}
  \centering
  \footnotesize
  \setlength{\tabcolsep}{3pt}
  \begin{tabular*}{\textwidth}{@{\extracolsep{\fill}}lccccccccccc}
    \toprule
    & & & \multicolumn{3}{c}{\textbf{QAOA}} &
    \multicolumn{3}{c}{\textbf{MA-QAOA}} &
    \multicolumn{3}{c}{\textbf{WS-QAOA}} \\
    \cmidrule(lr){4-6} \cmidrule(lr){7-9} \cmidrule(lr){10-12}
    \textbf{Inst.} & \textbf{Opt.} & \textbf{Q} &
    \textbf{TDev} & \textbf{MDev} & $\mathbf{AR}_{\mathrm{TDev}}$ &
    \textbf{TDev} & \textbf{MDev} & $\mathbf{AR}_{\mathrm{TDev}}$ &
    \textbf{TDev} & \textbf{MDev} & $\mathbf{AR}_{\mathrm{TDev}}$ \\
    \midrule
    ms20 & 3 & 48  & 21   & 15  & 0.963 & \textbf{10}   & \textbf{8}   & \textbf{0.986} & 41   & 34  & 0.922 \\
    ms30 & 2 & 84  & \textbf{59}   & \textbf{35}  & \textbf{0.963} & 203  & 101 & 0.870 & 296  & 121 & 0.810 \\
    ms40 & 1 & 118 & \textbf{42}   & \textbf{20}  & \textbf{0.986} & 619  & 213 & 0.790 & 901  & 285 & 0.694 \\
    ms50 & 0 & 150 & 846  & 287 & 0.822 & 2440 & 576 & 0.488 & \textbf{318}  & \textbf{92}  & \textbf{0.933} \\
    ms21 & 3 & 50  & 15   & 14  & 0.977 & 19   & 13  & 0.969 & \textbf{9}    & \textbf{7}   & \textbf{0.988} \\
    ms31 & 3 & 84  & 170  & 65  & 0.889 & 54   & 31  & 0.966 & \textbf{36}   & \textbf{26}  & \textbf{0.978} \\
    ms41 & 1 & 118 & 594  & 279 & 0.817 & \textbf{85}   & \textbf{60}  & \textbf{0.974} & 348  & 183 & 0.893 \\
    ms51 & 1 & 156 & 1242 & 317 & 0.759 & \textbf{495}  & \textbf{218} & \textbf{0.904} & 1024 & 270 & 0.801 \\
    \bottomrule
  \end{tabular*}
\end{table*}

The encoding-based results in
Table~\ref{tab:market_share_encoding_family_results} provide the most important
benchmark message for MSP\@.
PCE and QRAO compress the instances to only 7--11 qubits, which is a
dramatic reduction relative to the 48--156 qubits required by the direct
formulations.
However, this compression comes at a substantial cost in recovered solution
quality.
Across all tested MSP instances, the gaps obtained by PCE and QRAO are much larger
than those of the best full-width methods, often by one or more orders of
magnitude.
In addition, feasibility remains limited despite the compression, showing that low
width alone does not guarantee robust recovery for balancing-type problems.


\begin{table*}[t]
  \caption{\label{tab:market_share_encoding_family_results}
    Market Share results for PCE and QRAO\@.
    \textbf{TDev}: total absolute deviation from target split.
    \textbf{MDev}: maximum deviation on any single product.
    $\mathbf{AR}_{\mathrm{TDev}}$: range-normalized TDev approximation
    ratio. Despite substantial algorithm-level width reduction
    (7--11 qubits versus 48--156 for full-width methods), recovered
    deviations remain large. No reported hardware run achieved an exact
  target match.}
  \centering
  \footnotesize
  \setlength{\tabcolsep}{3pt}
  \begin{tabular*}{\textwidth}{@{\extracolsep{\fill}}lccccccccc}
    \toprule
    & & \multicolumn{4}{c}{\textbf{PCE}} &
    \multicolumn{4}{c}{\textbf{QRAO}} \\
    \cmidrule(lr){3-6} \cmidrule(lr){7-10}
    \textbf{Inst.} & \textbf{Opt.} &
    \textbf{Q} & \textbf{TDev} & \textbf{MDev} & $\mathbf{AR}_{\mathrm{TDev}}$ &
    \textbf{Q} & \textbf{TDev} & \textbf{MDev} & $\mathbf{AR}_{\mathrm{TDev}}$ \\
    \midrule
    ms20 & 3 & 7  & \textbf{45}  & \textbf{31}  & \textbf{0.914} & 11 & 81   & 62  & 0.840 \\
    ms30 & 2 & 8  & \textbf{167} & \textbf{82}  & \textbf{0.893} & 11 & 291  & 133 & 0.813 \\
    ms40 & 1 & 10 & \textbf{296} & \textbf{95}  & \textbf{0.900} & 11 & 485  & 156 & 0.836 \\
    ms50 & 0 & 11 & \textbf{822} & \textbf{207} & \textbf{0.827} & 11 & 1120 & 301 & 0.765 \\
    ms21 & 3 & 7  & \textbf{52}  & \textbf{34}  & \textbf{0.904} & 11 & 99   & 71  & 0.812 \\
    ms31 & 3 & 8  & \textbf{185} & \textbf{90}  & \textbf{0.879} & 11 & 310  & 145 & 0.797 \\
    ms41 & 1 & 10 & \textbf{317} & \textbf{102} & \textbf{0.902} & 11 & 509  & 164 & 0.843 \\
    ms51 & 1 & 11 & \textbf{845} & \textbf{215} & \textbf{0.836} & 11 & 1150 & 310 & 0.777 \\
    \bottomrule
  \end{tabular*}
\end{table*}

MSP is therefore the strongest counterexample in the benchmark to any simple
``fewer qubits is better'' narrative.
The compressed methods make the instances far easier to fit on current devices, but
the best hardware outcomes are generally obtained by full-width VQE/CVaR-VQE or
QAOA-style methods, depending on the instance.
This problem family most clearly demonstrates that qubit count should be treated as
an important but incomplete metric of practical performance.

\subsection{Cross-problem benchmark observations}
\label{sec:results_cross_problem}

When viewed together, the four benchmark families yield a problem-dependent
picture of near-term quantum optimization. MDKP permits a balanced comparison
across all method families, MIS exposes a sharp feasibility boundary, QAP is the
most demanding dense-assignment case, and MSP shows that strong compression does
not automatically imply strong recovered solutions. This variation justifies the
use of multiple structurally distinct benchmark families rather than relying on a
single canonical optimization problem.

The problem-wise hardware diagnostics in
Appendices~\ref{app:hw_mdkp}–\ref{app:hw_msp} show that these differences are
closely tied to backend-native circuit growth, two-qubit-gate burden, and
execution robustness. The dedicated QAOA-family pre-/post-transpilation audit
in Appendix~\ref{app:qaoa_prepost_expansion} further shows how routing,
native-basis decomposition, and connectivity constraints can substantially
increase the realized two-qubit-gate count relative to the algorithm-level
circuit. Complete method–instance–backend fidelity and outcome records are
provided in Appendix~\ref{app:reproducibility},
Table~\ref{tab:master_fidelity}.

Methods that appear only modestly different at the algorithmic level can therefore
diverge substantially after transpilation, particularly when dense encoded
interactions or nonlocal entangling patterns increase effective depth and
two-qubit-gate burden. The benchmark should consequently be interpreted as a
full-stack evaluation of encoding, compilation, execution, and solution
recovery, rather than as a comparison of abstract encodings alone.

\subsection{Connectivity-induced compilation expansion}
\label{sec:qaoa_prepost_connectivity}

\begin{figure}[htbp]
  \centering
  \includegraphics[width=0.98\linewidth]{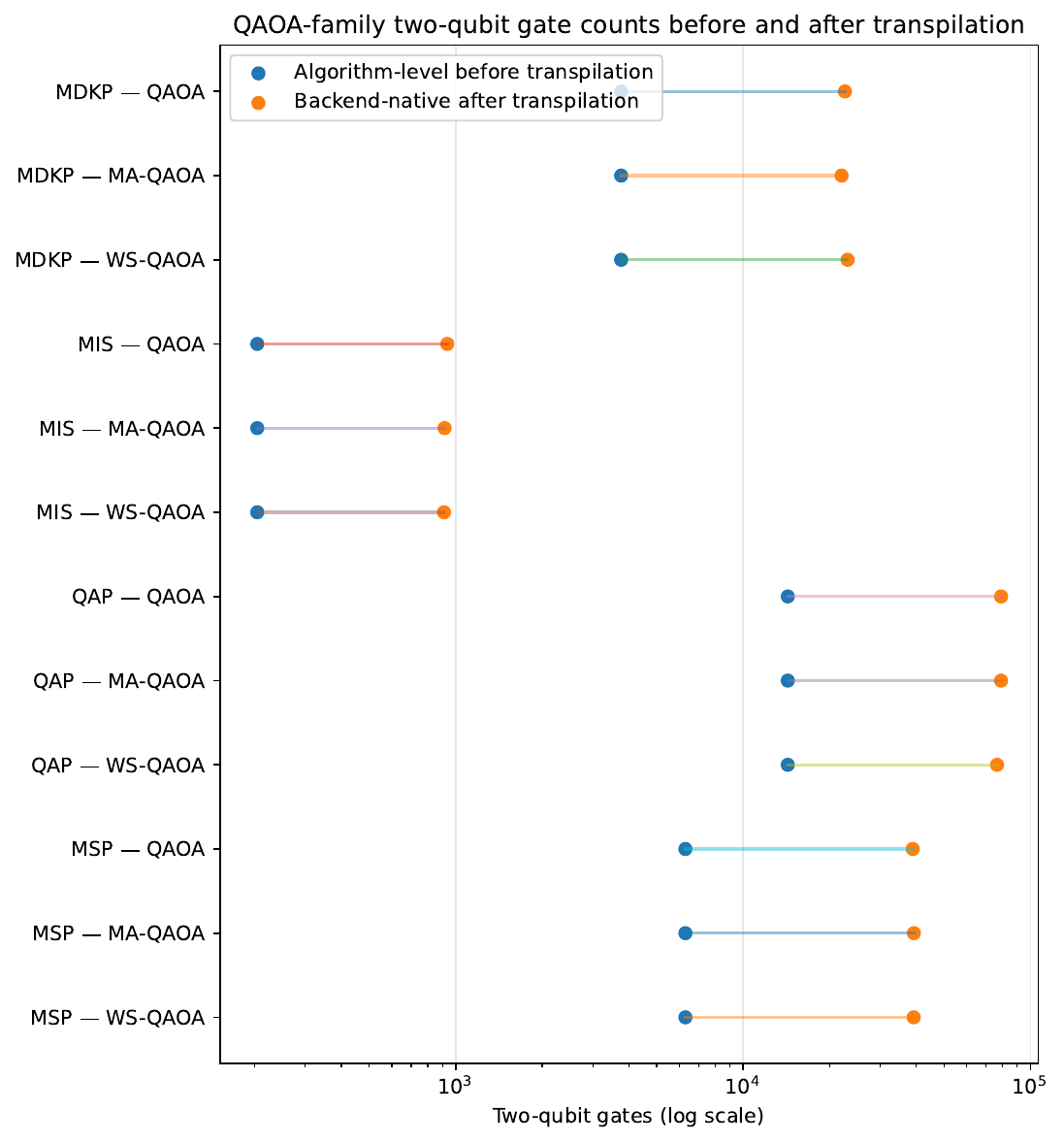}
  \caption{Historical QAOA-family two-qubit-gate counts before and after
    transpilation. Each row compares the same representative bound depth-$3$
    circuit before backend mapping with the corresponding historically executed
    backend-native circuit. The four problem families are represented by MDKP
    \texttt{hp1}, MIS \texttt{1tc.32}, QAP \texttt{tai10a}, and MSP \texttt{ms20}.
    The horizontal axis is logarithmic. Expansion reflects the combined effects of
    routing, SWAP insertion, native-basis decomposition, scheduling, and
    post-routing circuit simplification; it should not be attributed to SWAP gates
  alone.}
  \label{fig:qaoa_prepost_2q}
\end{figure}

The effect of hardware connectivity is most visible when comparing
algorithm-level QAOA-family circuits with the backend-native circuits that are
actually executed. Figure~\ref{fig:qaoa_prepost_2q} reports this comparison for
12 representative bound depth-$3$ circuits: QAOA, MA-QAOA, and WS-QAOA on MDKP
\texttt{hp1}, MIS \texttt{1tc.32}, QAP \texttt{tai10a}, and MSP \texttt{ms20}.
The lower point in each row is the algorithm-level two-qubit-gate count before
transpilation; the upper point is the two-qubit-gate count recorded after
historical backend-native transpilation.

All 12 representative circuits expand substantially after transpilation, with
post-transpilation to algorithm-level two-qubit-gate ratios ranging from
$4.47$ to $6.24$. The family-level median expansion is $6.01\times$ for MDKP,
$4.49\times$ for MIS, $5.52\times$ for QAP, and $6.23\times$ for MSP.
The largest absolute burden occurs for QAP: the median representative
QAOA-family circuit grows from 14,310 algorithm-level two-qubit gates to
79,027 transpiled two-qubit gates. The largest relative median expansion
occurs for MSP, where the median count grows from 6,300 to 39,227.

The historical compilation artifacts also retain routing metadata. Median
recorded SWAP counts are 5,030 for MDKP, 169 for MIS, 16,802 for QAP, and
8,876 for MSP. These values reinforce that sparse or structured processor
connectivity materially affects the realized circuit burden. They are not used
as an additive accounting identity because native-gate decomposition and later
transpiler simplifications can change the relation between recorded SWAP
operations and final two-qubit-gate count.

This evidence supports the interpretation that circuit feasibility and fidelity
cannot be inferred from algorithm-level QAOA gate counts alone. The relevant
hardware quantity is the post-transpilation circuit produced for the selected
backend and execution protocol.

\subsection{Noise-induced performance degradation: a quantitative analysis}
\label{sec:noise_analysis}

\begin{figure}[t]
  \centering
  \includegraphics[width=0.98\linewidth]{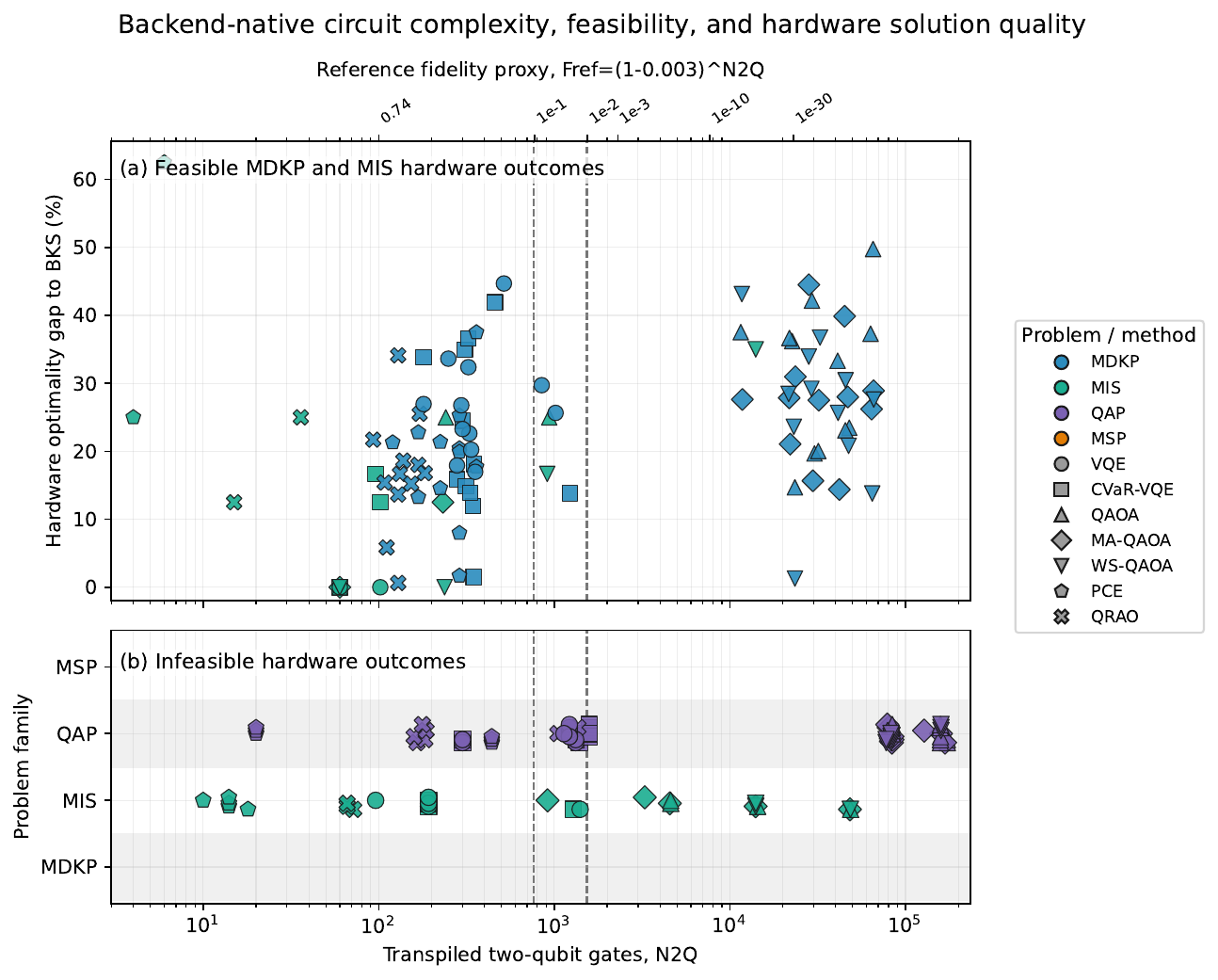}
  \caption{Backend-native circuit complexity, feasibility, and hardware solution
    quality. \textbf{(a)} Feasible MDKP and MIS hardware outcomes, plotted as
    optimality gap to BKS against the transpiled two-qubit-gate count
    $N_{2Q}$. \textbf{(b)} Infeasible hardware outcomes, grouped into shaded
    problem-family lanes; marker shape denotes method and color denotes problem
    family in both panels. MSP feasible outcomes are omitted from panel (a)
    because they use TDev rather than percentage optimality gap, but MSP
    infeasible outcomes are retained in panel (b). QAP appears only in panel (b)
    because all reported hardware QAP outcomes are infeasible. The lower x-axis
    shows backend-native $N_{2Q}$; the upper x-axis gives the common reference
    mapping $F_{\mathrm{ref}}=(1-0.003)^{N_{2Q}}$. This upper axis is a visual
    reference under the median Heron-r2 CZ error and is not a substitute for the
    backend-specific $F_{\mathrm{est}}$ values reported in the appendix. Vertical
    dashed lines mark the reference thresholds
    $F_{\mathrm{ref}}=0.1$ and $F_{\mathrm{ref}}=0.01$. The matched
    uniform-random control for low-fidelity QAOA-family runs is reported
  separately in Appendix~\ref{app:random_baseline}.}
  \label{fig:complexity_vs_quality}
\end{figure}

The problem-wise hardware diagnostics in
Appendices~\ref{app:hw_mdkp}–\ref{app:hw_msp}, combined with the fidelity
model of Section~\ref{sec:noise_model}, enable a quantitative comparison
between backend-native circuit burden, feasibility, and recovered hardware
quality. Figure~\ref{fig:complexity_vs_quality} separates this comparison into
two views. Panel~(a) plots feasible MDKP and MIS hardware outcomes as
optimality gap to BKS against the transpiled two-qubit-gate count. Panel~(b)
shows all infeasible hardware outcomes in problem-family lanes, including QAP
runs, for which no reported hardware execution returns a feasible assignment.
Feasible MSP outcomes are not included in panel~(a) because they are evaluated
using total absolute deviation rather than percentage optimality gap.

Several observations emerge from Figure~\ref{fig:complexity_vs_quality}.
First, feasible low-gate-count outcomes span a substantial range of optimality
gaps, from near-zero values to roughly $45\%$. This spread shows that remaining
solution-quality variation cannot be attributed to two-qubit-gate burden alone:
ansatz choice, encoding, optimization trajectory, sampling, and classical
recovery remain relevant within the lower-(N\_{2Q}) regime. Second, the
feasible QAOA-family outcomes at $N_{2Q}>10^{4}$ lie in a markedly different
backend-native circuit regime, with gate-count fidelity proxies below
$10^{-30}$. The matched uniform-random control shows that the dominant pattern
among these feasible low-fidelity QAOA-family outcomes is compatibility with,
or underperformance relative to, random candidate pools processed by the same
trajectory-level selection and one-round local-refinement rule. This supports a
best-of-budget sampling interpretation for most of the tested QAOA-family runs,
while not establishing that every measured output distribution is fully
uniform.

Third, panel~(b) shows that infeasibility is not confined to a single
two-qubit-gate threshold. MIS failures emerge across an expanding range of
compiled circuit sizes, whereas QAP is infeasible throughout the reported
hardware set despite substantial variation in its compiled resource burden.
The QAP pattern is therefore consistent with a combined structural and
execution barrier: dense assignment interactions, one-hot feasibility
constraints, sparse feasible support, compilation overhead, and hardware noise
all contribute to the observed failure of feasible-solution recovery.

\begin{figure}[t]
  \centering
  \includegraphics[width=0.98\linewidth]{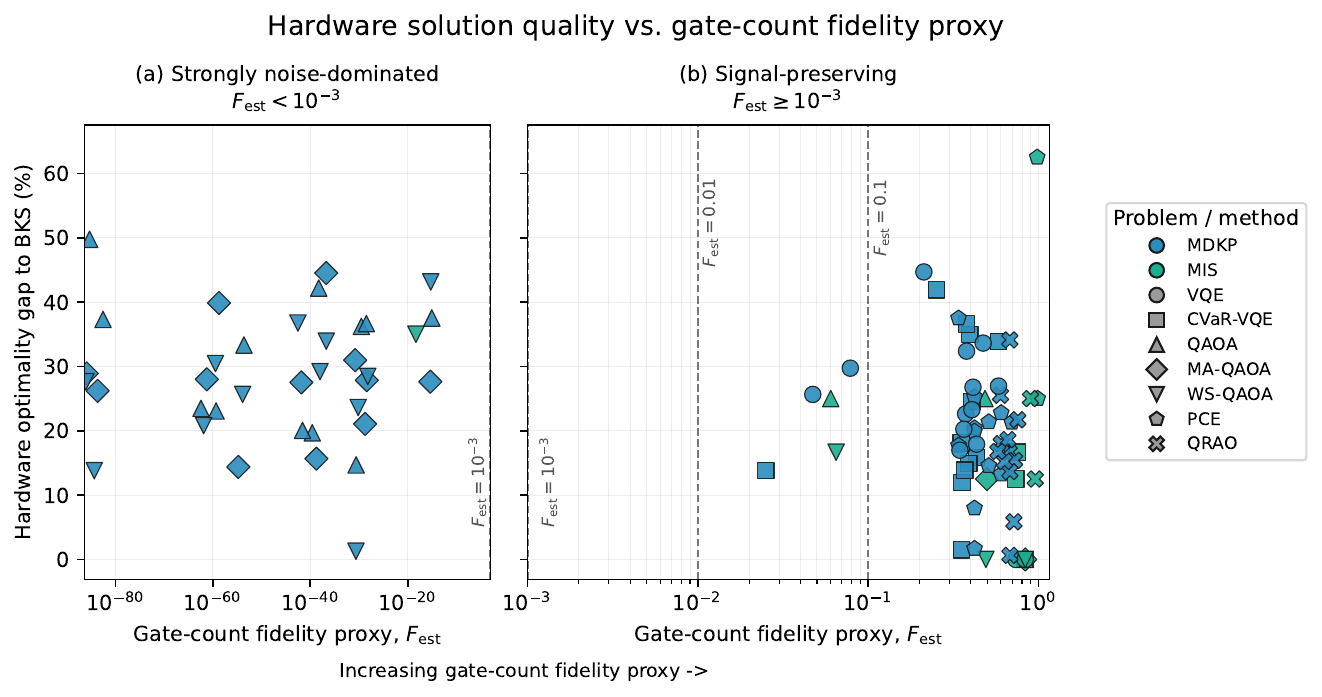}
  \caption{Feasible MDKP and MIS hardware outcomes versus the independent-error
    gate-count fidelity proxy $F_{\mathrm{est}}$. \textbf{(a)} Strongly
    noise-dominated regime, $F_{\mathrm{est}}<10^{-3}$. \textbf{(b)}
    Signal-preserving regime, $F_{\mathrm{est}}\geq10^{-3}$. The panels are
    ordered from lower to higher fidelity to make the transition across the
    $F_{\mathrm{est}}=10^{-3}$ boundary visually continuous. The dashed lines in
    panel (b) mark the benchmark-specific heuristic reference values
    $F_{\mathrm{est}}=0.01$ and $F_{\mathrm{est}}=0.1$. MSP outcomes are omitted
    because they are evaluated using TDev rather than percentage optimality gap;
    QAP outcomes are omitted because no reported hardware QAP run is feasible.
    The matched uniform-random control for the low-fidelity QAOA-family subset is
  reported separately in Appendix~\ref{app:random_baseline}.}
  \label{fig:fidelity_vs_quality}
\end{figure}

Figure~\ref{fig:fidelity_vs_quality} presents the feasible MDKP and MIS
hardware outcomes as a function of the independent-error gate-count fidelity
proxy $F_{\mathrm{est}}$. The figure separates the strongly noise-dominated
regime, $F_{\mathrm{est}}<10^{-3}$, in panel~(a) from the
signal-preserving regime, $F_{\mathrm{est}}\geq10^{-3}$, in panel~(b).
This separation distinguishes the deeply compiled QAOA-family executions from
the lower-depth variational and qubit-efficient circuits that remain within the
higher-fidelity regime. The dashed reference lines in panel~(b) mark the
benchmark-specific values $F_{\mathrm{est}}=0.01$ and
$F_{\mathrm{est}}=0.1$, which are used as diagnostic reference points rather
than universal performance thresholds. MSP outcomes are excluded because they
use total absolute deviation rather than percentage optimality gap, while QAP
outcomes are excluded because no reported hardware QAP execution returns a
feasible solution.

Table~\ref{tab:noise_penalty} reports the mean noise penalty, defined as
the per-run difference between hardware gap and simulator gap, for each
method family and problem where both gaps are available.

\begin{table}[t]
  \centering
  \caption{Mean noise-induced gap inflation per method family and problem.
    The noise penalty is defined as (HW gap $-$ Sim gap) for each instance where
    both are available; positive values indicate that hardware execution worsens
    the gap, negative values indicate that hardware execution improves on the
  simulator-measured gap. Reported errors are standard errors across instances.}
  \label{tab:noise_penalty}
  \small
  \begin{tabular}{llrrr}
    \toprule
    Problem & Method family & $N$ & Median $N_{2Q}$ & Mean noise penalty (\%) \\
    \midrule
    MDKP & VQE-family     & 22 & 310 & $-7.1 \pm 3.5$ \\
    MDKP & Qubit-efficient & 12 & 260 & $-6.3 \pm 4.6$ \\
    MIS  & VQE-family     & 4  & 120 & $+4.2 \pm 2.6$ \\
    MIS  & Qubit-efficient & 4  & 13  & $+6.2 \pm 6.3$ \\
    \bottomrule
  \end{tabular}
\end{table}

Two features of Table~\ref{tab:noise_penalty} warrant explicit discussion.
First, the MIS noise penalty is small and positive, as expected: hardware
execution on small MIS instances degrades solution quality relative to
simulation. Second, the MDKP noise penalty is \textit{negative}: under the
fixed-shot and fixed-evaluation protocol, hardware runs obtain lower
optimality gaps than the corresponding shot-based simulator runs on average.

This observation should not be interpreted as evidence that hardware noise is
intrinsically beneficial or that it generically improves variational
optimization. It may reflect a combination of finite-shot stochasticity,
best-candidate selection, optimizer-trajectory variability, and
backend-specific execution effects. To test whether MPS bond truncation could
explain this pattern, we performed targeted stability checks on the 99-qubit
CVaR-VQE \texttt{pet2} circuit and the 60-qubit VQE \texttt{hp1} circuit,
both of which contribute to the negative MDKP gap. As reported in
Appendix~\ref{app:mps_validation}, the production uncapped MPS results were
reproduced by explicit converged and conservative bond caps, while relaxed
truncation thresholds left decoded solution quality unchanged. Thus, for these
representative VQE-family MDKP cases, we find no evidence that MPS
bond-dimension truncation explains the negative simulator--hardware gap.

The result remains an empirical property of the present protocol rather than a
causal demonstration of noise-assisted escape from local minima. The stability
audit is deliberately scoped to the selected VQE-family cases. It does not
separately validate a historical MDKP PCE result because the archived
182-parameter PCE circuit cannot be reconstructed exactly from the currently
available 42-parameter implementation and no serialized legacy circuit artifact
is available.

\begin{figure}[t]
  \centering
  \includegraphics[width=0.9\linewidth]{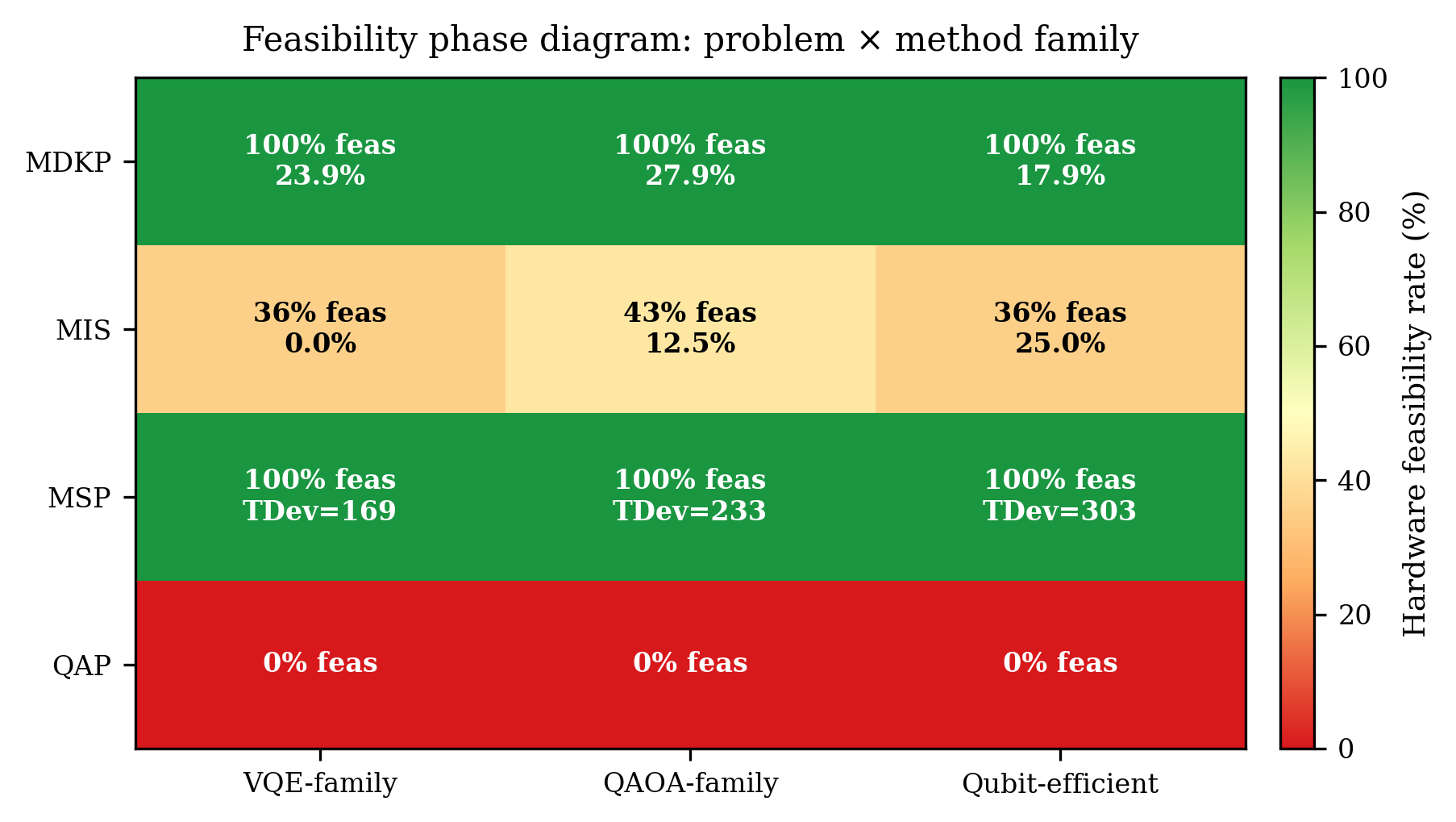}
  \caption{Feasibility phase diagram. Cell colour indicates the fraction of
    hardware runs that returned feasible solutions for each (problem,
    method-family) pair. Cell annotation reports the median hardware gap to BKS
    among feasible runs (TDev for MSP, for which a target-deviation metric is
    appropriate). MDKP and MSP remain universally feasible because their QUBO
    formulations are unconstrained and every bitstring is classically feasible;
    MIS exhibits the sharp feasibility cliff documented in
    Section~\ref{sec:results_mis}; QAP is uniformly infeasible across all seven
  methods and all eleven instances.}
  \label{fig:feasibility_phase}
\end{figure}

Figure~\ref{fig:feasibility_phase} summarizes the feasibility landscape across
the whole benchmark.

\subsubsection{Implications.}

Taken together, the fidelity analysis, Figures~\ref{fig:complexity_vs_quality}--\ref{fig:feasibility_phase},
and Table~\ref{tab:noise_penalty} support four quantitative conclusions.
\begin{enumerate}
  \item \textit{The fidelity thresholds derived in Sec.~\ref{sec:noise_model}
    serve as useful empirical diagnostics within this benchmark.} They should be
    read as execution-aware reference points under the observed Heron calibration
    regime, not as first-principles predictive laws.

  \item \textit{QAP provides the clearest example of a structural feasibility
    barrier under current hardware constraints.} The direct QAP encodings combine
    100–144 binary variables, dense quadratic interactions, row-and-column
    one-hot assignment penalties, broad coefficient ranges, and feasible subsets
    occupying only approximately $10^{-23.54}$ to $10^{-34.67}$ of the
    corresponding binary hypercubes. The QAP failure is therefore not attributable
    to gate count alone: restrictive feasibility geometry, dense compilation
    overhead, execution noise, and recovery sensitivity jointly place the tested
    end-to-end pipelines outside the reliable hardware operating regime.

  \item \textit{Deeply compiled QAOA-style methods are predominantly
    random-baseline-limited under the tested protocol.} Their compiled circuits
    fall far below the benchmark-specific gate-count fidelity regime identified
    above. Across the 106 low-fidelity QAOA-family hardware runs, matched
    uniform-random controls show that most feasible outcomes lie within the
    empirical random range or are worse than random best-shot selection under the
    same trajectory-level selection and one-round local-refinement procedure. This
    does not prove that every low-fidelity output distribution is uniform; one MIS
    warm-start-QAOA case is reported separately as a finite-sample feasibility
    exception.

  \item \textit{Qubit efficiency is beneficial only when compression survives
    compilation and decoding.} PCE gives short compiled circuits for MIS and MDKP,
    but the MSP results show that circuit width alone does not guarantee strong
    recovered solutions.
\end{enumerate}

\section{Discussion}
\label{sec:discussion}

\subsection{Problem-dependent method performance}

The problem-dependent variation summarized in Sec.~\ref{sec:results_cross_problem}
suggests that near-term quantum optimization benchmarks should be interpreted as
problem--method--hardware triples rather than as method rankings alone. The
qualitative behavior changes across MDKP, MIS, QAP, and MSP: different problem
structures expose different bottlenecks in the quantum workflow, including
constraint handling, graph structure, interaction density, and decoding
sensitivity. As a result, a method that appears competitive on one benchmark
family may fail to retain that position once the problem structure changes.

This has an important consequence for benchmark design. Claims about practical
method quality should be based on portfolios of structurally distinct problems
rather than on isolated demonstrations. In particular, the present results
support difficulty-at-small-sizes benchmarking, where instances remain
classically meaningful while still exposing different failure modes of
near-term quantum methods. From this perspective, the value of a benchmark
suite such as QOBLIB is not only that it contains hard instances, but that
it prevents overgeneralization from a narrow class of favorable formulations.

A second implication is methodological. Since method rankings are conditional
rather than absolute, comparisons should be framed less as searches for a u
niversally best NISQ optimizer and more as attempts to identify which
algorithmic features remain effective under which structural regimes.
This shifts the benchmarking goal from winner selection to regime
characterization, which is more informative for both hardware-facing
evaluation and future method development.

\subsection{Qubit compression versus practical hardware performance}

\begin{figure}
  \centering
  \includegraphics[width=\textwidth]{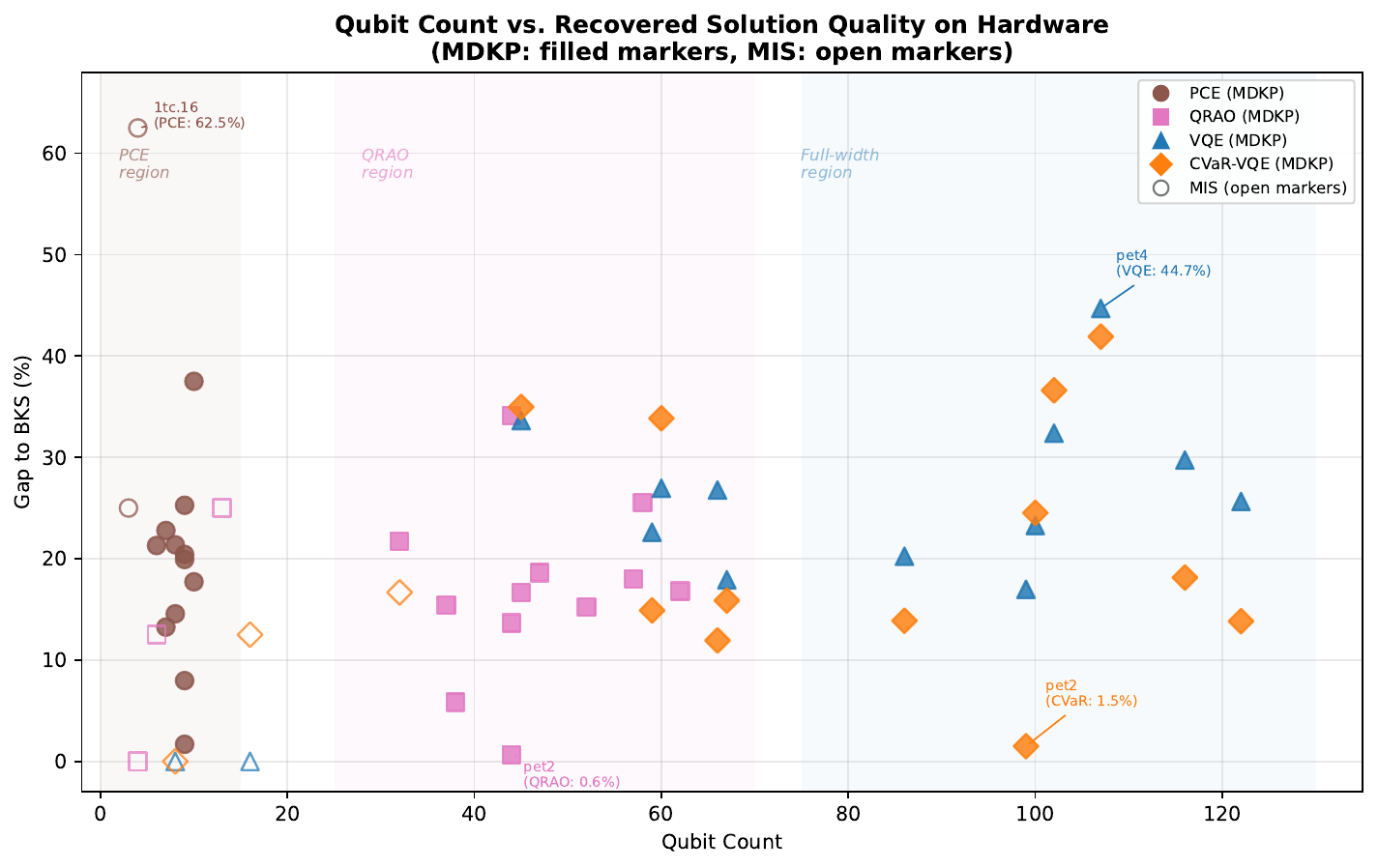}
  \caption{Qubit count versus recovered hardware solution quality for
    MDKP and MIS\@. Filled markers denote MDKP results and open markers denote
    MIS results, with regions indicating PCE, QRAO, and full-width formulations.
    The scatter shows that lower qubit count does not by itself guarantee better
    recovered hardware performance: qubit-efficient methods often enable access
    to smaller-width executions, but their achieved gaps vary substantially,
    while some larger-width formulations remain competitive or superior on
    selected instances. The plot thus illustrates that qubit compression
    is an access-enabling resource reduction, not a standalone predictor of
  final hardware solution quality.}
  \label{fig4}
\end{figure}

The cross-problem results show that compression is useful but not sufficient.
PCE and QRAO reduce circuit width substantially for several problem families,
but practical performance depends on whether the compressed representation also
leads to favorable compiled circuits and reliable recovery. Figure~\ref{fig4} plots circuit width against recovered hardware quality and
shows that lower qubit count does not by itself predict better outcomes.

The main implication is that width reduction must be evaluated jointly
with the distortions it may introduce elsewhere. Compressed encodings
can alter the effective optimization landscape, increase decoding
sensitivity, or shift difficulty from qubit count to recovery quality
after measurement. They may also remain vulnerable to compilation-induced
overhead once the reduced pre-transpiled circuit is mapped to a specific backend.
As a result, the relevant practical question is not whether a method uses
fewer qubits, but whether the width reduction survives the full path from
encoding to executed circuit to recovered solution.

This point matters because qubit count is often the most visible headline
metric in quantum optimization studies. The present benchmark suggests that
such reporting can be misleading when presented in isolation. A lower-width
method may be more runnable yet less competitive in recovered hardware
quality, while a larger-width method may still be preferable if its encoding,
circuit structure, and decoding pipeline are better aligned with the backend.
For this reason, claims of practical advantage from qubit-efficient
formulations should be supported by hardware-realized solution quality,
compilation diagnostics, and recovery robustness rather than by width
reduction alone.

\subsection{The simulator--hardware gap}

The simulator results provide controlled algorithmic references, but the hardware
runs show how those references change after compilation, routing, calibration
variability, and finite-shot execution are introduced.

This has an important methodological consequence. Simulator results should
not be interpreted as direct evidence of near-term deployability, but as
one stage in a layered evaluation workflow. In that workflow, simulation
can help rule out clearly weak formulations, estimate baseline sensitivity,
and guide candidate selection, but claims about practical robustness should
be reserved for methods that remain competitive after execution on real
hardware. The relevant object of evaluation is therefore not the algorithmic
circuit alone, but the executed circuit together with its backend-specific
realization.

More broadly, the simulator--hardware gap changes how progress should be
interpreted in quantum optimization. Improvement in idealized simulation
is scientifically useful, but it is not sufficient evidence that a method
has advanced the practical frontier. A formulation that performs well before
transpilation yet fails systematically after mapping to hardware should be
understood as algorithmically interesting but operationally immature. For
this reason, simulator studies remain necessary, but they should be paired
with hardware-facing diagnostics whenever the claim concerns practical
near-term capability.

\subsection{Compilation and execution as first-class benchmark dimensions}

A central implication of the benchmark is that compilation and execution
should be treated as transformation stages that can materially change the
identity of a quantum optimization method in practice. Two approaches that
appear similar at the algorithmic level may cease to be comparable after
transpilation if one incurs substantially greater routing overhead, deeper
compiled circuits, or higher exposure to backend-specific error channels.
From this perspective, practical benchmarking should evaluate not only the
nominal formulation, but also how strongly that formulation degrades when
mapped to hardware.

This shifts the interpretation of algorithmic performance. Algorithmic-level
resource counts remain useful because they describe the intended structure
of a method, but they are not sufficient to characterize deployability. The
more informative question is how much of the algorithmic advantage survives
compilation and execution. In near-term settings, this survival rate may be
more consequential than the original circuit specification itself, especially
when differences in compiled depth, two-qubit gate counts, or job robustness
dominate the final quality of recovered solutions.

The broader methodological consequence is that benchmark studies should
distinguish clearly between algorithmic design efficiency and hardware-realized
efficiency. Conflating the two can lead to overstated claims, particularly
for methods whose nominal compactness weakens after routing or whose execution
behavior is unstable across backends. Treating compilation and execution as
first-class benchmark dimensions therefore does more than add engineering
detail: it changes what counts as convincing evidence in near-term quantum
optimization.

\subsection{Implications for future quantum optimization benchmarks}

Beyond the specific methods and problems studied here, the benchmark has broader
methodological implications.
First, future benchmark studies should move beyond reporting only the best
objective value obtained by a method.
For constrained combinatorial problems, feasibility must be reported explicitly.
For hardware studies, circuit width (qubits) alone is insufficient; transpiled circuit size,
two-qubit gate burden, and execution robustness are also necessary for meaningful
interpretation.
Without these additional dimensions, it is difficult to distinguish between nominal
algorithmic promise and practical hardware viability.

Second, future benchmarks should continue to include structurally diverse problem
families.
The present results show that conclusions derived from one class do not transfer
reliably to others.
A method that appears strong on graph-based problems may behave very differently on
dense assignment problems or on balancing-type formulations.
This diversity is not incidental; it is essential for assessing whether a
quantum optimization strategy is genuinely general-purpose or only well matched to
a narrow structural regime.

Third, the results suggest that the most informative near-term benchmarking
question is not which method is globally best, but which trade-offs are currently
dominant.
In the present study, these trade-offs include width versus recovered quality,
algorithmic design versus compiled execution cost, and simulator promise versus
hardware robustness.
Framing benchmarks in terms of these trade-offs may be more scientifically useful
than searching for a single winning algorithm family in a regime where hardware
constraints still dominate.

\subsection{Practitioner guidelines for near-term deployment}
\label{sec:guidelines}

The empirical patterns documented in Sections~\ref{sec:results_mdkp}--\ref{sec:noise_analysis}
support a compact decision table that practitioners can use to select quantum
optimization methods under current hardware constraints.
Table~\ref{tab:guidelines} is not intended as a definitive recommendation, but
as a summary of the trade-offs we observed; it should be updated as hardware
improves.

\begin{table}[t]
  \centering
  \caption{Summary decision table for near-term gate-based quantum optimization,
  based on the empirical regimes documented in this benchmark.}
  \label{tab:guidelines}
  \small
  \begin{tabularx}{\textwidth}{>{\raggedright\arraybackslash}X>{\raggedright\arraybackslash}X>{\raggedright\arraybackslash}X}
    \toprule
    If your problem has\ldots & Consider\ldots & Avoid\ldots \\
    \midrule
    $<60$ qubits with sparse quadratic couplings (small packing, MIS) &
    VQE or CVaR-VQE with EfficientSU2 of moderate repetition count; Runtime
    Estimator objective evaluation at \texttt{resilience\_level=2}, with final
    candidate recovery evaluated separately from raw/default sampling&
    The tested standard, MA-QAOA, and WS-QAOA implementations when compiled
    two-qubit-gate counts exceed the empirical fidelity budget
    \\
    \midrule
    $60$--$120$ qubits &
    PCE for width reduction; full-width VQE when depth permits &
    QAOA if exact assignment constraints are needed ($p\geq 2$ pushes $N_{2Q}$
    above 20\,000) \\
    \midrule
    Large permutation or assignment structure (QAP, QAP-like) &
    Classical solver (Gurobi, CPLEX, Concorde) &
    All tested quantum methods at current scale; hardware infeasibility is
    structural \\
    \midrule
    Balancing or target-matching constraints (MSP-style) &
    Full-width VQE or CVaR-VQE; retain the richer ans\"atze &
    Aggressive qubit compression via PCE or QRAO at large instance sizes
    (recovery quality degrades) \\
    \midrule
    A need to demonstrate hardware execution &
    Always report transpiled $N_{2Q}$, median backend $\varepsilon_{2Q}$,
    estimated $F$, Estimator mitigation settings, final-sampling configuration,
    and classical recovery protocol alongside the optimization result &
    Simulator-only claims about practical advantage; hardware execution can
    reshape the ranking \\
    \bottomrule
  \end{tabularx}
\end{table}

Two meta-level observations follow from Table~\ref{tab:guidelines}. First,
the most useful decision is rarely between two quantum methods but between a
quantum method and a strong classical baseline on the specific instance size
of interest. At all sizes tested in this benchmark, classical solvers such
as CPLEX reach provable optimality for all MDKP, MIS, and MSP instances
within seconds to minutes, and near-optimal feasible QAP solutions in
minutes. The value of the quantum benchmark is not in producing a better
objective value but in characterizing the execution pipeline. Second, once
the fidelity budget is fixed by backend calibration, circuit width (qubits) is only
one component of deployability: the product $N_{2Q}\cdot\varepsilon_{2Q}$, the cumulative noise exposure, is the more informative figure of merit,
and practitioners should optimize for it rather than for qubit count alone.

\subsection{Hardware roadmap and forward-looking implications}
\label{sec:roadmap}

The fidelity thresholds derived in Sec.~\ref{sec:noise_model} are anchored to
the current generation of IBM Heron processors and should move as gate
fidelities, connectivity, and supported circuit depths improve. IBM's published
roadmap projects deeper single-processor circuits through continued
gate-fidelity improvements and richer connectivity through long-range
couplers~\cite{IBMHeronR2,IBMRoadmap2025}. Such progress could expand the
region in which QAOA-family circuits can be tested. The counterfactual audit in
Sec.~\ref{sec:qaoa_counterfactuals} shows, however, that topology-level
two-qubit-gate reductions of approximately $20\%$ are insufficient for the
representative historical circuits studied here. Progress for these instances
will require hardware improvement together with more substantial reductions in
algorithmic cost-layer complexity, routing overhead, or parameterized-circuit
depth.

However, dense assignment problems such as QAP are unlikely to become accessible
through hardware scaling alone. Their compiled two-qubit-gate counts remain far
above the present execution regime, so progress on this class will require both
hardware improvements and encoding or compilation strategies that reduce
compiled circuit size by more than a constant factor.

These observations underscore that the numerical thresholds reported here are
not universal. The broader evaluation protocol---circuit width, transpiled
depth, transpiled $N_{2Q}$, estimated fidelity, error-mitigation settings, and
hardware-realized gap---is intended to remain useful as the underlying hardware
numbers move.

\subsection{Scope and limitations}

This study is intentionally empirical and should be interpreted within that scope.
The benchmark does not establish asymptotic superiority of one quantum optimization
family over another, nor does it claim a practical quantum advantage over strong
classical optimization methods. The tested instances remain limited by the
capabilities of current hardware, and the reported results are shaped by
present-day device noise, limited circuit depth, and finite execution budgets.

The MPS simulations provide an execution-consistent shot-based reference under
the stated Aer configuration; they are not intended as an exhaustive comparison
against all scalable classical simulation approaches. Alternative methods,
including other tensor-network strategies, Pauli-propagation techniques,
Gibbs-state approximations, and symmetry-aware simulators, may be advantageous
in different circuit, entanglement, and noise regimes. The targeted validation
in Appendix~\ref{app:mps_validation} establishes bond-cap and
truncation-threshold stability for selected large MDKP VQE-family circuits, but
does not constitute a universal convergence guarantee across every benchmark
family, ans\"atze, or optimizer trajectory.

In addition, the benchmark compares representative fixed-budget implementations
of each method family rather than exhaustively optimized versions of every
possible variant. The recovered artifacts use a common upper cap of 200
objective evaluations, but realized stopping behavior differs across methods:
many VQE-family, PCE, and QRAO runs reach the cap, whereas some QAOA-style runs
terminate earlier with an optimizer-success status. Because the archived
artifacts do not retain function, parameter, or gradient tolerance values, this
study does not claim that every reported run reaches a common numerical
convergence tolerance.

The real-hardware campaign also involves backend heterogeneity. Because queue
times and backend availability made a full single-device campaign impractical,
some method–instance combinations were executed on different compatible IBM
processors. Backend identity, transpiled resource metrics, and execution
metadata are reported for each run, but residual differences in calibration,
topology, routing overhead, and execution window may contribute to
cross-backend variation. The hardware results should therefore be read as
execution-realistic end-to-end outcomes rather than as a controlled
same-device estimate of the isolated algorithmic effect.

A reduced multi-seed simulator check further shows that initialization can
materially affect decoded quality for representative VQE and warm-start QAOA
instances. The headline tables should therefore be interpreted as outcomes of
the documented fixed-budget protocol rather than as optimizer-independent
best-case performance for each method family. Different ans\"atze, optimizers, or decoding strategies could improve
performance in specific cases. Constraint penalties were not tuned against
hardware outcomes in this study: they were fixed analytically at QUBO
construction time using Eq.~\eqref{eq:qiskit_auto_penalty}. Alternative
penalty constructions may alter conditioning and practical variational
optimization behavior, even when they preserve the same constrained optimum. The protocol audit and
seed-sensitivity evidence are reported in
Appendix~\ref{app:optimization_protocol}.

With these caveats in mind, the main conclusion of the discussion is clear:
practical performance in gate-based quantum optimization is governed by a coupled
interaction between algorithm family, encoding strategy, compilation overhead, and
hardware noise.
Benchmarking studies that ignore any one of these layers risk overstating the
practical value of algorithmic-level improvements.

\section{Conclusion}
\label{sec:conclusion}

We presented a comparative benchmark of gate-based quantum optimization methods
across four classically hard 0--1 combinatorial problem families: MDKP, MIS,
QAP, and MSP\@. By evaluating variational, QAOA-style, and qubit-efficient
approaches under a shared protocol spanning both simulation and real hardware,
the study provides a unified view of current near-term performance across
structurally distinct problem classes.

The main empirical conclusion is that practical performance is determined by
the interaction between problem structure, encoding strategy, compilation
overhead, backend noise, and decoding. Algorithmic qubit count and simulator
performance are therefore incomplete indicators of hardware deployability. A
method that is attractive at the algorithmic level may become ineffective after
routing and basis decomposition, while a compressed encoding is useful only when
the resulting compiled circuit and recovery procedure remain robust.

Beyond the specific benchmark outcomes, the study argues that compilation and
execution diagnostics should be treated as first-class evaluation dimensions in
quantum optimization. Reporting circuit width (qubits), transpiled depth, two-qubit gate
count, estimated fidelity, feasibility, and hardware-realized solution quality
makes it possible to distinguish algorithmic promise from operational
deployability.

The fidelity analysis developed here is hardware-window dependent, but the
evaluation framework is intended to remain useful as devices improve. As gate
fidelities, connectivity, and supported circuit depths increase, the numerical
boundary between executable and noise-dominated regimes will move. The same
protocol can therefore be used to track whether future algorithmic and hardware
advances translate into measurable improvements on structurally diverse
optimization problems.

Overall, the benchmark provides a reproducible empirical baseline and a
diagnostic framework for interpreting near-term quantum optimization results
under realistic execution constraints.

\appendix
\textbf{Appendix}

The appendices provide the detailed formulation, protocol, simulator,
compilation, hardware-diagnostic, classical-baseline, and reproducibility
evidence supporting the benchmark. They are organized in the same progression
as the main paper: formulation and penalty construction; optimization and
method-specific recovery; simulator validation and results; compilation and
hardware diagnostics; matched controls; and reproducibility and classical
reference data.

Appendix~\ref{app:structural_metrics} reports instance-level QUBO structural
descriptors, including encoded width, coupling density, mean interaction degree,
coefficient dynamic range, source metadata, and the direct-QAP feasible-space
geometry. Appendix~\ref{app:penalty_audit} documents the automatic
instance-specific penalty construction, the selected penalty values, and a
representative classical penalty-sensitivity audit.

Appendix~\ref{app:optimization_protocol} verifies the nominal fixed-budget
variational-optimization protocol against retained benchmark records and reports
the reduced initialization-sensitivity and current-PCE optimizer-resource
diagnostics. Appendix~\ref{app:pce} gives the mathematical formulation of
Pauli Correlation Encoding (PCE), including its correlator-based
reconstruction rule and the shared feasibility-preserving local-refinement
procedure used after native method-specific recovery. The appendix describes
available PCE enhancement mechanisms for methodological completeness, but the
reported benchmark PCE runs use the single-pass protocol documented in
Appendix~\ref{app:optimization_protocol}.

Appendix~\ref{app:mps_validation} documents the Aer MPS simulator
configuration, execution environment, and targeted bond-cap and
truncation-threshold stability audit for representative large MDKP VQE-family
circuits. Appendix~\ref{app:qaoa_counterfactuals} reports the
compilation-only counterfactual audit for representative bound QAOA-family
circuits under historical, SWAP-aware, fractional-gate, and topology-only
conditions. Appendix~\ref{app:qaoa_prepost_expansion} provides the numerical
pre-/post-transpilation source data underlying
Figure~\ref{fig:qaoa_prepost_2q}. Appendix~\ref{app:acronyms} provides a
compact reference for the acronyms and table notation used throughout the
manuscript.

Appendix~\ref{app:simulator_results} reports the full shot-based simulator
results for MDKP, MIS, QAP, and MSP. Appendices~\ref{app:hw_mdkp}–
\ref{app:hw_msp} provide the corresponding problem-wise hardware diagnostics,
including backend identity, algorithmic-level and transpiled circuit resources, and
execution-robustness records for the VQE-style, QAOA-style, and
encoding-based method families. Appendix~\ref{app:random_baseline} reports the
matched uniform-random best-shot control for low-fidelity QAOA-family hardware
runs, preserving the recorded trajectory-level candidate-selection and shared
one-round local-refinement procedure.

Appendix~\ref{app:reproducibility} records the software stack, Runtime
mitigation provenance, final-sampling configuration, and the master
method–instance–backend fidelity and outcome table. Appendix~\ref{app:classical_baselines}
reports the classical Gurobi and CPLEX baseline results used to interpret the
quantum solution-quality metrics. Finally,
Appendix~\ref{app:resource_usage} reports simulator resource usage by problem
and method, including algorithm-level qubit count, circuit depth, total gate
count, two-qubit-gate count, parameter count, and runtime.

\section{Instance-Level Structural Metrics}
\label{app:structural_metrics}

This appendix reports formulation-level descriptors for the QUBOs used in the
benchmark. The purpose is to make explicit that nominal problem size alone does
not capture the structural differences among MDKP, MIS, QAP, and MSP. For each instance, we characterize QUBO width, coupling density, interaction
degree, coefficient dynamic range, penalty scale, source metadata, and, where
analytically available, feasible-space geometry. The family-level summaries
below and the instance-level tables reported in this appendix provide the
structural descriptors used in the analysis.

For a QUBO with $N_Q$ binary variables and $m_Q$ unique nonzero off-diagonal
couplings, coupling density is defined as
\begin{equation}
  \rho_Q =
  \frac{2m_Q}{N_Q(N_Q-1)}.
  \label{eq:qubo_density}
\end{equation}
The corresponding mean interaction degree is
\begin{equation}
  \bar d_Q =
  \frac{2m_Q}{N_Q},
  \label{eq:qubo_mean_degree}
\end{equation}
and coefficient dynamic range is defined over the absolute values of nonzero
QUBO coefficients as
\begin{equation}
  R_Q =
  \frac{\max_{i\leq j:Q_{ij}\neq0}|Q_{ij}|}
  {\min_{i\leq j:Q_{ij}\neq0}|Q_{ij}|}.
  \label{eq:qubo_dynamic_range}
\end{equation}

\subsection{Problem-family summary}

The benchmark contains qualitatively different encoded structures. MIS is
sparse because its quadratic interactions are induced by the input graph.
MDKP and MSP contain moderate-to-high density introduced by capacity, slack,
and target-balancing penalties. QAP is structurally distinct because its
direct assignment encoding combines dense flow-distance interactions with
row-and-column one-hot constraints.

\begin{table*}[ht]
  \centering
  \caption{Family-level structural summary of the QUBOs used in the benchmark.
  Dynamic range is computed from absolute nonzero QUBO coefficients.}
  \label{tab:structural_summary_appendix}
  \footnotesize
  \setlength{\tabcolsep}{3.5pt}
  \renewcommand{\arraystretch}{1.08}
  \begin{tabular}{lrrrrrr}
    \toprule
    Problem & Instances & Width range & Density range & Median density &
    Median degree & Dynamic-range range \\
    \midrule
    MDKP & 12 & 45–122 & 0.260–0.756 & 0.678 & 42.5 &
    $8.53\times10^{3}$–$6.75\times10^{5}$ \\
    MIS & 7 & 8–128 & 0.095–0.269 & 0.181 & 6.0 &
    $9$–$129$ \\
    QAP & 11 & 100–144 & 0.293–1.000 & 0.818 & 95.4 &
    $1.07\times10^{5}$–$2.21\times10^{10}$ \\
    MSP & 8 & 48–156 & 0.549–0.681 & 0.589 & 58.7 &
    $9.86\times10^{4}$–$1.70\times10^{6}$ \\
    \bottomrule
  \end{tabular}
\end{table*}

We do not report a single cross-family landscape-hardness scalar based on
solution degeneracy or local-minimum counts. Exact enumeration depends on the
problem-specific objective and neighborhood definition, and it is not directly
comparable across packing, graph, permutation, and target-allocation
formulations. Instead, this appendix reports directly computable formulation
descriptors for all instances and the exact feasible-space fraction for QAP,
where the latter is directly relevant to the observed feasibility barrier.

\subsection{QAP feasibility geometry}

The direct QAP formulation uses binary assignment variables
$x_{ij}\in{0,1}$, where $x_{ij}=1$ indicates assignment of facility $i$ to
location $j$. An $n$-facility QAP therefore has $n^2$ binary variables, while
exactly $n!$ bitstrings correspond to valid permutations. The feasible fraction
of the binary hypercube is
\begin{equation}
  \frac{n!}{2^{n^2}}.
  \label{eq:qap_feasible_fraction}
\end{equation}

For the smallest QAPLIB instances selected for this benchmark,
\texttt{tai10a} and \texttt{tai10b}, this fraction is
$10!/2^{100}\approx10^{-23.54}$. For the tested $n=12$ instances, it is
$12!/2^{144}\approx10^{-34.67}$. Thus, even the smallest selected direct-QAP
instances have an extremely sparse feasible assignment manifold.

\begin{table*}[t]
  \centering
  \caption{QAP direct-encoding feasibility geometry and structural descriptors.
    All listed QAP hardware runs are infeasible. Simulator outcomes differ:
    PCE returns feasible candidates on several instances, whereas VQE and
  CVaR-VQE remain infeasible; see Appendix~\ref{app:sim_qap}.}
  \label{tab:qap_feasibility_geometry}
  \footnotesize
  \setlength{\tabcolsep}{3.2pt}
  \renewcommand{\arraystretch}{1.08}
  \begin{tabular}{lrrrrrr}
    \toprule
    Instance & $n$ & $N_Q$ &
    $\log_{10}(|\mathcal F|/2^{N_Q})$ &
    Density & Mean degree & Dynamic range \\
    \midrule
    chr12a & 12 & 144 & $-34.668$ & 0.293 & 41.9 & $1.19\times10^{6}$ \\
    chr12b & 12 & 144 & $-34.668$ & 0.293 & 41.9 & $1.19\times10^{6}$ \\
    chr12c & 12 & 144 & $-34.668$ & 0.293 & 41.9 & $1.19\times10^{6}$ \\
    had12  & 12 & 144 & $-34.668$ & 1.000 & 143.0 & $2.49\times10^{5}$ \\
    nug12  & 12 & 144 & $-34.668$ & 0.731 & 104.5 & $1.07\times10^{5}$ \\
    rou12  & 12 & 144 & $-34.668$ & 0.987 & 141.2 & $5.09\times10^{6}$ \\
    scr12  & 12 & 144 & $-34.668$ & 0.513 & 73.3 & $2.18\times10^{5}$ \\
    tai10a & 10 & 100 & $-23.543$ & 0.964 & 95.4 & $1.68\times10^{7}$ \\
    tai10b & 10 & 100 & $-23.543$ & 0.818 & 81.0 & $1.37\times10^{8}$ \\
    tai12a & 12 & 144 & $-34.668$ & 0.974 & 139.3 & $4.13\times10^{7}$ \\
    tai12b & 12 & 144 & $-34.668$ & 0.838 & 119.8 & $2.21\times10^{10}$ \\
    \bottomrule
  \end{tabular}
\end{table*}

These structural values do not imply that QAP is impossible in simulation.
The simulator results show a meaningful distinction among method families:
PCE returns feasible candidates on several QAP instances, although with large
optimality gaps, whereas VQE and CVaR-VQE remain infeasible. This suggests that
compressed PCE reconstruction can retain partial access to the assignment
manifold under noise-free execution, but it does not remove the underlying
difficulty caused by dense couplings, broad coefficient scales, and sparse
permutation feasibility.

\section{QUBO Penalty Provenance and Sensitivity Audit}
\label{app:penalty_audit}

This appendix documents the constraint-penalty construction used for every
benchmark instance. The purpose is to distinguish analytical constraint
encoding from hardware-level parameter tuning. Penalties were selected once at
QUBO-construction time, before quantum execution, and were not modified by
method family, backend, optimizer outcome, or hardware performance.

\subsection{Automatic penalty selection}

All constrained models are converted through the Qiskit Optimization
\texttt{QuadraticProgramToQubo} pipeline, using
\texttt{LinearEqualityToPenalty} or \texttt{LinearInequalityToPenalty} as
appropriate. The converter selects a single instance-specific penalty
\begin{equation}
  \lambda
  = 1+
  \left(
    U_{\mathrm{lin}}-L_{\mathrm{lin}}
  \right)
  +
  \left(
    U_{\mathrm{quad}}-L_{\mathrm{quad}}
  \right),
\end{equation}
where the bounds are evaluated on the objective before the corresponding
constraint-elimination stage. The added unit margin ensures that the selected
penalty exceeds the modeled pre-penalty objective range. For the integer-valued
constraint residuals used in the present binary and slack-expanded
formulations, any nonzero violation has squared penalty at least one.

No QUBO normalization is applied after conversion. The resulting penalized
QUBO is passed directly to \texttt{to\_ising()} for direct methods or to the
method-specific PCE and QRAO encoders. Consequently, all methods receive the
same original penalized QUBO, and the relative scale between objective and
constraint terms is preserved.

\begin{table*}[t]
  \centering
  \caption{Instance-level penalty values selected by the automatic converter
    rule. Every listed value equals one plus the corresponding pre-penalty
    objective bound range. No post-conversion coefficient normalization is
  applied.}
  \label{tab:penalty_values}
  \footnotesize
  \setlength{\tabcolsep}{2.5pt}
  \renewcommand{\arraystretch}{1.04}
  \begin{tabular}{llr@{\qquad}llr}
    \toprule
    MDKP instance & $\lambda_{\mathrm{K}}$ &
    Value &
    MIS instance & $\lambda_{\mathrm{MIS}}$ &
    Value \\
    \midrule
    hp1  & $\lambda_{\mathrm{K}}$ & 5,124 &
    1tc.8   & $\lambda_{\mathrm{MIS}}$ & 9 \\
    hp2  & $\lambda_{\mathrm{K}}$ & 6,451 &
    1tc.16  & $\lambda_{\mathrm{MIS}}$ & 17 \\
    pb1  & $\lambda_{\mathrm{K}}$ & 4,796 &
    1tc.32  & $\lambda_{\mathrm{MIS}}$ & 33 \\
    pb2  & $\lambda_{\mathrm{K}}$ & 5,326 &
    1tc.64  & $\lambda_{\mathrm{MIS}}$ & 65 \\
    pb4  & $\lambda_{\mathrm{K}}$ & 182,685 &
    1dc.64  & $\lambda_{\mathrm{MIS}}$ & 65 \\
    pb5  & $\lambda_{\mathrm{K}}$ & 4,022 &
    1et.64  & $\lambda_{\mathrm{MIS}}$ & 65 \\
    pet2 & $\lambda_{\mathrm{K}}$ & 125,895 &
    1dc.128 & $\lambda_{\mathrm{MIS}}$ & 129 \\
    pet3 & $\lambda_{\mathrm{K}}$ & 5,166 & & & \\
    pet4 & $\lambda_{\mathrm{K}}$ & 8,656 & & & \\
    pet5 & $\lambda_{\mathrm{K}}$ & 15,496 & & & \\
    pet6 & $\lambda_{\mathrm{K}}$ & 14,724 & & & \\
    pet7 & $\lambda_{\mathrm{K}}$ & 22,498 & & & \\
    \bottomrule
  \end{tabular}
\end{table*}

\begin{table*}[t]
  \centering
  \caption{Instance-level QAP and MSP penalty values selected by the automatic
  converter rule.}
  \label{tab:penalty_values_qap_msp}
  \footnotesize
  \setlength{\tabcolsep}{2.5pt}
  \renewcommand{\arraystretch}{1.04}
  \begin{tabular}{llr@{\qquad}llr}
    \toprule
    QAP instance & $\lambda_{\mathrm{Q}}$ &
    Value &
    MSP instance & $\lambda_{\mathrm{M}}$ &
    Value \\
    \midrule
    chr12a & $\lambda_{\mathrm{Q}}$ & $5.955985\times10^{6}$ &
    ms\_seed0\_prod2 & $\lambda_{\mathrm{M}}$ & 1,963 \\
    chr12b & $\lambda_{\mathrm{Q}}$ & $5.955985\times10^{6}$ &
    ms\_seed0\_prod3 & $\lambda_{\mathrm{M}}$ & 6,193 \\
    chr12c & $\lambda_{\mathrm{Q}}$ & $5.955985\times10^{6}$ &
    ms\_seed0\_prod4 & $\lambda_{\mathrm{M}}$ & 11,775 \\
    had12  & $\lambda_{\mathrm{Q}}$ & $2.492410\times10^{5}$ &
    ms\_seed0\_prod5 & $\lambda_{\mathrm{M}}$ & 19,045 \\
    nug12  & $\lambda_{\mathrm{Q}}$ & $1.071850\times10^{5}$ &
    ms\_seed1\_prod2 & $\lambda_{\mathrm{M}}$ & 2,061 \\
    rou12  & $\lambda_{\mathrm{Q}}$ & $4.073476\times10^{7}$ &
    ms\_seed1\_prod3 & $\lambda_{\mathrm{M}}$ & 6,053 \\
    scr12  & $\lambda_{\mathrm{Q}}$ & $7.845993\times10^{6}$ &
    ms\_seed1\_prod4 & $\lambda_{\mathrm{M}}$ & 12,947 \\
    tai10a & $\lambda_{\mathrm{Q}}$ & $1.682441\times10^{7}$ &
    ms\_seed1\_prod5 & $\lambda_{\mathrm{M}}$ & 20,613 \\
    tai10b & $\lambda_{\mathrm{Q}}$ & $2.054462\times10^{8}$ &
    & & \\
    tai12a & $\lambda_{\mathrm{Q}}$ & $4.125248\times10^{7}$ &
    & & \\
    tai12b & $\lambda_{\mathrm{Q}}$ & $1.102967\times10^{10}$ &
    & & \\
    \bottomrule
  \end{tabular}
\end{table*}

\subsection{Classical sensitivity audit}

A formulation-level sensitivity audit was performed on representative
instances MDKP \texttt{hp1}, MIS \texttt{1tc.32}, QAP \texttt{tai10a}, and
MSP \texttt{ms\_seed0\_prod3}. The reported penalty was multiplied by
\[
  \kappa
  \in
  \{0.25,\ 0.5,\ 0.75,\ 1.0,\ 1.25,\ 2.0,\ 4.0\},
\]
while holding the objective, binary encoding, variable ordering, and
normalization policy fixed. Each scaled QUBO was solved classically using
Gurobi with a 10-second time limit, and the decoded candidate was checked
against the original constraints independently of QUBO energy.

For MIS \texttt{1tc.32}, every tested multiplier, including
$\kappa=0.25$, yields a certified feasible QUBO optimum matching the classical
reference independent-set size of 12. This is consistent with the direct
unit-weight MIS sufficiency condition $\lambda_{\mathrm{MIS}}>1$ and confirms
that the automatic $\lambda_{\mathrm{MIS}}=33$ choice is conservative.

For MDKP \texttt{hp1}, QAP \texttt{tai10a}, and MSP
\texttt{ms\_seed0\_prod3}, the 10-second classical solves return feasible
incumbents at the reported penalty but do not certify QUBO optimality.
Accordingly, these rows are not used as evidence that the reported penalties
are minimal, tight, or uniquely optimal. They serve only as a transparent
diagnostic showing that the sensitivity sweep did not reveal an immediate
feasibility failure at the tested incumbent level. The principal justification
for the reported penalties remains the converter’s analytical pre-penalty
objective-bound construction.

\section{Optimization Protocol Verification and Sensitivity Analysis}
\label{app:optimization_protocol}

This appendix documents the fixed-budget variational-optimization protocol used
throughout the benchmark, verifies its realized execution against retained
benchmark records, and reports reduced diagnostics for initialization
sensitivity and optimizer-resource interpretation. The primary benchmark is a
single-execution, fixed-protocol comparison for each method--instance
configuration; it is not an exhaustive hyperparameter search or an estimate
averaged over all possible initializations. Table~\ref{tab:optimizer_protocol_audit}
summarizes the protocol and realized execution records,
Table~\ref{tab:optimizer_seed_sensitivity} reports the reduced
initialization-sensitivity study, and
Table~\ref{tab:pce_slsqp_diagnostic} separates optimizer-iteration counts from
realized objective-evaluation cost for a current-PCE diagnostic.

\subsection{Nominal protocol and execution-record verification}

All benchmark method--instance configurations were evaluated under the nominal
shot-based variational-optimization protocol described in
Section~\ref{sec:optimization_protocol}. To provide an auditable account of
its realized use, we reviewed 262 retained benchmark result records. For each
recoverable record, the audit captures the problem family, instance, execution
mode, method, ansatz metadata, optimizer, initialization metadata,
objective-shot count, final-sampling-shot count, planned budget, realized
objective evaluations, termination status, final-parameter selection rule,
post-processing rule, and replayability status.

The retained records consistently identify COBYLA, invoked through
\texttt{scipy.optimize.minimize} in SciPy~1.15.3, as the benchmark optimizer.
The nominal protocol specifies a maximum of 200 optimizer iterations and 200
objective evaluations. Objective-evaluation metadata record 1{,}000 shots per
evaluation, and separately retained final-sampling metadata record 1{,}000
shots where available. Recorded initialization metadata specify uniform
parameter initialization over $[0,2\pi]$. The benchmark runner retains the
parameter vector associated with the best observed objective value before
applying method-specific reconstruction or rounding, feasibility handling, and
post-processing.

Table~\ref{tab:optimizer_protocol_audit} summarizes the aggregate audit.
The planned 200-evaluation budget is an upper bound rather than a requirement
that every execution consume exactly 200 objective evaluations.

\begin{table}[t]
  \centering
  \caption{Aggregate verification of the nominal optimization protocol using
    262 retained benchmark records. The planned 200-evaluation limit is an
  upper bound, not a fixed realized cost for every execution.}
  \label{tab:optimizer_protocol_audit}
  \footnotesize
  \setlength{\tabcolsep}{4pt}
  \renewcommand{\arraystretch}{1.10}
  \begin{tabular}{lr}
    \toprule
    Audit quantity & Value \\
    \midrule
    Saved benchmark records audited & 262 \\
    Records with \texttt{budget\_reached} & 149 \\
    Records with \texttt{optimizer\_success} & 113 \\
    Recorded optimizer & COBYLA \\
    Optimizer implementation & \texttt{scipy.optimize.minimize} \\
    SciPy version & 1.15.3 \\
    Planned optimizer-iteration cap & 200 \\
    Planned objective-evaluation cap & 200 \\
    Objective shots per evaluation & 1{,}000 \\
    Final-sampling shots, where retained & 1{,}000 \\
    Function/parameter/gradient tolerances & not retained \\
    \bottomrule
  \end{tabular}
\end{table}

Of the 262 audited records, 149 reached the planned evaluation cap and 113
returned an \texttt{optimizer\_success} status before reaching that cap.
Because the retained records do not preserve explicit function, parameter, or
gradient tolerances, \texttt{optimizer\_success} is used only as the status
returned by the optimization framework. It is not interpreted as evidence of
tolerance-based convergence, local optimality, or global optimality.

The audit includes 33 retained PCE hardware records. Their stored
optimizer-budget fields specify the same nominal cap of 200, while realized
objective/circuit-evaluation counts are 124, 184, or 200 according to whether
the run returned before the cap or exhausted the available budget. The retained
PCE benchmark source path uses COBYLA.

\subsection{Reduced initialization-sensitivity study}

The main benchmark reports one primary protocol execution per method--instance
configuration. To assess the practical sensitivity of this choice to parameter
initialization, we conducted a reduced simulator study using three independent
parameter initializations for three representative cases:

\begin{itemize}
  \item MDKP \texttt{hp1} with VQE;
  \item MIS \texttt{1tc.32} with warm-start QAOA;
  \item MIS \texttt{1tc.16} with the current reproducible PCE implementation.
\end{itemize}

The study uses initialization seeds 1103, 4409, and 7703. Each run uses the
same Aer MPS configuration, nominal 200-evaluation cap, 1{,}000-shot objective
budget, reconstruction or rounding procedure, and post-processing pipeline as
the corresponding benchmark protocol. Each seed produces one final
production-style sampling result; the study therefore measures
initialization-induced variation under the stated protocol, rather than
separately estimating variance from repeated final-sampling draws.

Table~\ref{tab:optimizer_seed_sensitivity} reports the resulting best, median,
and worst decoded gaps across the three tested initializations. For MDKP and
MIS, lower decoded optimality gap is better.

\begin{table*}[t]
  \centering
  \caption{Reduced initialization-sensitivity study. Each row summarizes three
    independent parameter initializations under the nominal 200-evaluation
    production protocol. For MDKP and MIS, lower decoded optimality gap is
  better.}
  \label{tab:optimizer_seed_sensitivity}
  \scriptsize
  \setlength{\tabcolsep}{2pt}
  \renewcommand{\arraystretch}{1.10}
  \begin{tabular*}{\textwidth}{@{\extracolsep{\fill}}llrllrrr@{}}
    \toprule
    Problem & Inst./method & Seeds & Evals. & Status
    & Best gap (\%) & Median gap (\%) & Worst gap (\%) \\
    \midrule
    MDKP & \texttt{hp1} / VQE & 3 & 200, 200, 200 &
    budget reached & 20.07 & 24.40 & 24.63 \\
    MIS & \texttt{1tc.32} / WS-QAOA & 3 & 63, 70, 75 &
    optimizer success & 8.33 & 16.67 & 50.00 \\
    MIS & \texttt{1tc.16} / current PCE & 3 & 200, 200, 200 &
    budget reached & 62.50 & 62.50 & 62.50 \\
    \bottomrule
  \end{tabular*}
\end{table*}

The VQE and warm-start-QAOA cases show that decoded solution quality can be
materially initialization-sensitive. In particular, the three warm-start-QAOA
runs returned \texttt{optimizer\_success} after 63, 70, and 75 objective
evaluations, but their decoded gaps ranged from $8.33\%$ to $50.00\%$. The
selected current-PCE case produced identical decoded gaps across the three
tested initializations, although all three runs exhausted the nominal
200-evaluation budget. These results support interpreting the primary benchmark
values as representative outcomes of a stated fixed-budget protocol rather than
as optimizer-independent best-case values.

\subsection{PCE optimizer-resource diagnostic}

Optimizer iteration counts are not directly comparable across method families,
because different optimizers can require different numbers of objective
evaluations per iteration. To make this distinction concrete, we performed a
reduced SLSQP diagnostic using the current reproducible PCE implementation on
MIS \texttt{1tc.8}, MIS \texttt{1tc.16}, and QAP \texttt{nug12}. This
diagnostic is not an exact replay of the historical Brickwork-PCE benchmark
runs and is not used to validate their reported performance. Its sole purpose
is to illustrate why realized objective evaluations, rather than nominal
optimizer iterations, are the appropriate cross-method cost metric.

Table~\ref{tab:pce_slsqp_diagnostic} reports the median number of objective
evaluations, optimizer-success frequency, cap-hit frequency, decoded quality,
and feasibility rate for SLSQP limits of 100 and 200 iterations.

\begin{table*}[t]
  \centering
  \caption{Reduced current-PCE SLSQP diagnostic. Realized objective evaluations,
    rather than nominal optimizer iterations, are reported as the relevant
    cross-method cost metric. This diagnostic uses the current reproducible PCE
    implementation and is not an exact replay of the historical Brickwork-PCE
  hardware artifacts.}
  \label{tab:pce_slsqp_diagnostic}
  \scriptsize
  \setlength{\tabcolsep}{1.8pt}
  \renewcommand{\arraystretch}{1.08}
  \begin{tabular*}{\textwidth}{@{\extracolsep{\fill}}llrrrrrrrr@{}}
    \toprule
    Problem & Instance & Max.\ iter. & Median $n_{\mathrm{fev}}$
    & Succ. & Cap hit & Best gap & Median gap & Worst gap & Feas. \\
    \midrule
    MIS & \texttt{1tc.8.txt} & 100 & 616 & 1.00 & 0.00
    & 25.0 & 25.0 & 25.0 & 1.00 \\
    MIS & \texttt{1tc.8.txt} & 200 & 616 & 1.00 & 0.00
    & 25.0 & 25.0 & 25.0 & 1.00 \\
    MIS & \texttt{1tc.16.txt} & 100 & 2,590 & 0.00 & 1.00
    & 62.5 & 62.5 & 62.5 & 1.00 \\
    MIS & \texttt{1tc.16.txt} & 200 & 5,105 & 0.00 & 1.00
    & 62.5 & 62.5 & 62.5 & 1.00 \\
    QAP & \texttt{nug12.dat} & 100 & 7,690 & 0.00 & 1.00
    & -- & -- & -- & 0.00 \\
    QAP & \texttt{nug12.dat} & 200 & 15,317 & 0.00 & 1.00
    & -- & -- & -- & 0.00 \\
    \bottomrule
  \end{tabular*}
\end{table*}

For MIS \texttt{1tc.8}, SLSQP terminates successfully at both iteration limits,
with median 616 objective evaluations and an unchanged median decoded gap of
$25.0\%$. For MIS \texttt{1tc.16}, increasing the SLSQP cap from 100 to 200
iterations increases the median objective-evaluation count from 2,590 to 5,105
without changing the median decoded gap of $62.5\%$. For QAP \texttt{nug12},
the corresponding median objective-evaluation counts increase from 7,690 to
15,317, while no feasible decoded solution is recovered at either cap. These
results reinforce the use of realized objective evaluations, rather than
optimizer iterations alone, as the relevant cross-method budget measure.

\subsection{Scope and interpretation}

The analyses in Tables~\ref{tab:optimizer_protocol_audit},
\ref{tab:optimizer_seed_sensitivity}, and
\ref{tab:pce_slsqp_diagnostic} do not constitute a complete multi-seed or
budget-doubling rebenchmark over all method--instance combinations. They
instead provide targeted evidence on three issues raised by the benchmark
protocol: realized budget consumption, sensitivity to initialization, and the
non-comparability of optimizer iterations across optimization methods.

Accordingly, the manuscript reports realized objective evaluations and
termination statuses where available, avoids tolerance-based convergence claims
when the corresponding metadata were not retained, and interprets headline
results as outcomes of the documented fixed-budget protocol. The results should
not be read as optimizer-independent best-case performance estimates for each
algorithmic family.

\section{Pauli Correlation Encoding: Formulation and Enhancements}
\label{app:pce}

This appendix provides the mathematical details of the Pauli Correlation
Encoding (PCE) method used in the benchmark, including the QUBO-based loss
formulation, the parameterized quantum ans\"atze, and the dynamic perturbation
and multi-reoptimization strategy that we introduce as enhancements to the
original PCE framework~\cite{Sciorilli_2025}.

\subsection{Encoding principle}

PCE maps a combinatorial optimization problem with
$m = \mathcal{O}(n^k)$ binary variables onto only $n$ qubits, where $k$
is a user-selected integer controlling the compression order.  The
encoding leverages $k$-body Pauli correlations rather than assigning one
qubit per variable as in standard QAOA or VQE encodings.

Concretely, the $m$ binary decision variables
$x = {\{x_i\}}_{i \in [m]}$ are associated with a subset of Pauli strings
$\Pi_i$ (excluding the $n$-fold identity) via
\begin{equation}
  x_i = \operatorname{sgn}\!\bigl(\langle \Pi_i \rangle\bigr)
  \quad \text{for all } i \in [m],
  \label{eq:sgn_app}
\end{equation}
where $\langle \Pi_i \rangle = \langle \Psi | \Pi_i | \Psi \rangle$ is
the expectation value of $\Pi_i$ with respect to the parameterized quantum
state $|\Psi(\theta)\rangle$.

The parameters $\theta$ are optimized variationally to minimize a
non-linear loss function whose first term relaxes the binary sign function
into a smooth hyperbolic tangent:
\begin{equation}
  \label{eq:pce_loss_app}
  \mathcal{L} = \sum_{(i,j)\in E}
  W_{ij}\,\tanh\!\bigl(\alpha \langle \Pi_i\rangle\bigr)\,
  \tanh\!\bigl(\alpha \langle \Pi_j\rangle\bigr)
  + \mathcal{L}^{(\mathrm{reg})}.
\end{equation}
The regularization term
\begin{equation}
  \mathcal{L}^{(\mathrm{reg})} = \beta\,\nu
  {\left[\frac{1}{m} \sum_{i=1}^{m}
      {\tanh\!\bigl(\alpha \langle \Pi_i \rangle\bigr)}^{2}
  \right]}^{\!2}
\end{equation}
penalizes large correlator magnitudes, keeping the optimizer within the
correlator domain where all bitstring solutions are representable.  The
parameter $\nu$ is set using the Polj\'{a}k--Turz\'{\i}k
bound~\cite{POLJAK198699} for Max-Cut problems, while $\beta$ is
typically fixed at $\tfrac{1}{2}$.

\subsection{QUBO-based loss formulation}

For general QUBO problems (beyond Max-Cut), we replace the graph-based loss
with a formulation native to the QUBO objective
$\min_{\mathbf{x}\in{\{0,1\}}^n} \mathbf{x}^\top Q\,\mathbf{x}
+ \mathbf{c}^\top \mathbf{x} + \mathrm{offset}$,
where $Q\in\mathbb{R}^{n\times n}$ is a symmetric cost matrix and
$\mathbf{c}\in\mathbb{R}^n$ contains linear coefficients.
The updated loss function is
\begin{align}
  \mathcal{L}
  &= \sum_{(i,j)\in E}
  Q_{ij}\,\tanh\!\bigl(\alpha\langle\pi_i\rangle\bigr)\,
  \tanh\!\bigl(\alpha\langle\pi_j\rangle\bigr) \notag\\
  &\quad
  + \sum_{i=1}^{m} c_i\,
  {\tanh\!\bigl(\alpha\langle\pi_i\rangle\bigr)}^{2}
  + \mathcal{L}^{(\mathrm{reg})},
  \label{eq:qubo_loss_app}
\end{align}
with the regularization parameter $\nu$ set via the Frobenius norm of $Q$:
\begin{equation}
  \nu = c \cdot \sqrt{\textstyle\sum_{i,j} Q_{ij}^2}\,.
\end{equation}
This formulation avoids the intermediate conversion to weighted Max-Cut
that is otherwise required to apply the original PCE loss, thereby
broadening the method's applicability to arbitrary QUBO instances.

\subsection{Quantum ans\"atze}

The parameterized circuit $|\Psi(\boldsymbol{\theta})\rangle$ uses a
\emph{Brickwork} architecture consisting of alternating layers of
single-qubit rotation gates ($R_X$, $R_Y$, $R_Z$) and two-qubit
$R_{XX}$ entangling gates arranged in a staggered brick-like pattern.
This structure ensures full qubit connectivity within each layer pair
while keeping the circuit depth linear in the number of layers.

\subsection{Dynamic perturbation and multi-reoptimization}
\label{app:pce_reopt}

To reduce the risk of premature convergence to local minima, we introduce
a multi-phase re-optimization strategy that iteratively perturbs the
trained circuit parameters.

At each re-optimization round $r$, a candidate parameter vector is
generated as
\begin{equation}
  \tilde{\theta} = \theta + \Delta_r + \Delta_d\,,
\end{equation}
where the random component
$\Delta_r \sim \mathcal{N}\bigl(0,\,{[P'(1+f/5)]}^{2}\,I\bigr)$ is scaled
by both a perturbation factor $P'$ and a failure counter $f$ that tracks
consecutive rounds without improvement, and the directional component
$\Delta_d = P'\cdot\operatorname{sgn}(T)$ biases the search along the
historical trend $T$.

The perturbation factor is amplified every third round
($P' = E\cdot P$, with exploration factor $E$) and decayed by a factor
$\delta < 1$ upon improvement.  If the failure counter exceeds a restart
threshold $f_{\mathrm{restart}}$, the parameters are reinitialized
uniformly in $[-\pi,\pi]$.  For intermediate stagnation levels, the
algorithm applies either weighted blending toward the best-known solution
or a stronger adaptive perturbation.

\subsection{Native PCE reconstruction and shared local refinement}

PCE converts its compressed quantum representation into a binary decision
vector through the correlator-sign reconstruction rule in
Eq.~\eqref{eq:sgn_app}. This reconstruction is specific to PCE and is the
method’s native decoding operation; it is not a separate classical
local-search advantage.

After native decoding, the benchmark applies the same single one-pass feasible
local-improvement routine to the decoded candidate produced by every quantum
method. Thus, PCE, VQE, CVaR-VQE, QAOA, MA-QAOA, WS-QAOA, and QRAO all receive
the same final refinement stage after their respective native decoding or
rounding procedures. The routine evaluates the prescribed local-swap
neighborhood for the underlying problem formulation and accepts only moves that
improve the original objective while preserving feasibility.

For PCE, the shared local-improvement stage starts from the binary candidate
obtained through Eq.~\eqref{eq:sgn_app}. It does not introduce recursive PCE
re-optimization, additional variational training rounds, or a PCE-exclusive
search budget. The benchmark therefore reports PCE as part of the same
end-to-end hybrid quantum–classical evaluation protocol used for all compared
quantum methods. The distinction between PCE-specific reconstruction and the
shared local-improvement stage is important: the former is required to decode
the compressed correlator representation, whereas the latter is a common
post-processing rule applied uniformly across method families.

\section{MPS Validation and Simulator Reproducibility}
\label{app:mps_validation}

This appendix documents the matrix-product-state (MPS) simulator configuration
and a targeted stability audit for the large MDKP VQE-family circuits most
relevant to the negative simulator--hardware gap discussed in the main text.
The purpose of the audit is to test whether the production MPS reference is
sensitive to bond-dimension restriction or truncation-threshold choices.

\subsection{Software and execution environment}

All validation runs used Python 3.10.16, Qiskit 2.3.0, and Qiskit Aer 0.17.2.
The calculations were performed on an Apple M3 Pro CPU system with 12 CPU
cores and 36\,GB physical memory; no GPU acceleration was used. Aer was run
with the \texttt{matrix\_product\_state} method on CPU in double precision.
The production MPS configuration used adaptive uncapped bond dimension,
\begin{equation}
  \texttt{matrix\_product\_state\_max\_bond\_dimension=None},
\end{equation}
with truncation enabled at threshold
\begin{equation}
  \epsilon_{\mathrm{trunc}}=10^{-16}.
\end{equation}

\subsection{Validation protocol}

We selected two large MDKP circuits that contribute directly to the negative
simulator--hardware gap: the 99-qubit CVaR-VQE circuit for \texttt{pet2} and
the 60-qubit VQE circuit for \texttt{hp1}. The validation used the archived
final parameter vectors and the legacy $R_Y$ plus circular-entanglement circuit
specification reflected by the stored run artifacts. For each selected case,
the variational parameters, QUBO instance, circuit construction, final
sampling budget, candidate-selection procedure, and decoder were fixed to the
original protocol. Only the MPS bond-dimension cap and truncation threshold
were changed.

Each setting used 20 final-sampling repetitions with 1,000 shots per
repetition, giving 20,000 shots per setting. The same set of 20 sampling seeds
was used across all settings within a case. For each setting, we recorded the
empirical sampled total-variation distance (TVD) relative to the production
uncapped MPS setting,
\begin{equation}
  \operatorname{TVD}(\hat p,\hat q)
  =
  \frac{1}{2}\sum_z
  \left|
  \hat p(z)-\hat q(z)
  \right|,
  \label{eq:mps_sampled_tvd}
\end{equation}
where $\hat p$ and $\hat q$ are the empirical distributions formed by
aggregating the 20 sampled repetitions. We also recorded raw feasible-sample
mass before recovery, decoded optimality-gap statistics, best recovered gap,
and the fraction of final-sampling repetitions that produced a feasible decoded
solution.

For \texttt{pet2}, the production uncapped MPS run reached
$\chi_{\mathrm{obs}}=64$. We compared the production setting against a
deliberately restrictive cap of $32$, a converged cap of $64$, a conservative
cap of $128$, and uncapped threshold relaxations to $10^{-12}$ and $10^{-8}$.
For \texttt{hp1}, the production run reached $\chi_{\mathrm{obs}}=66$. We
compared against a deliberately restrictive cap of $33$, a converged cap of
$128$, a conservative cap of $256$, and the same two threshold relaxations.

\subsection{MDKP MPS stability results}

Table~\ref{tab:mdkp_mps_stability} reports the complete validation
results. The explicitly converged and conservative caps reproduce the
production sampled distributions exactly for both cases. The $10^{-12}$
threshold relaxation also reproduces the production result exactly. The more
aggressive $10^{-8}$ threshold causes only small empirical TVD values,
$0.00135$ for CVaR-VQE/\texttt{pet2} and $0.00035$ for VQE/\texttt{hp1},
without changing decoded solution quality. All 20 decoded repetitions were
feasible for every listed setting.

The restrictive-cap controls demonstrate that the audit can detect harmful
bond restriction. For CVaR-VQE/\texttt{pet2}, cap $32$ produces sampled TVD
$0.04635$ relative to the production setting. For VQE/\texttt{hp1}, cap $33$
produces sampled TVD $0.00585$. These deviations disappear once the cap is
increased to the converged or conservative setting. Thus, for the two
representative large MDKP VQE-family circuits examined here, the production
adaptive MPS configuration is stable to the tested bond-cap and truncation
settings.

\begin{table*}[t]
  \centering
  \caption{Targeted MPS stability audit for representative large MDKP VQE-family
    circuits. Each setting uses 20 independent final-sampling repetitions with
    1,000 shots per repetition. TVD is the empirical total-variation distance
    relative to the production uncapped MPS setting, computed from the aggregated
    20,000-shot empirical distributions. $\chi_{\max}$ denotes the requested MPS
    bond-dimension cap; ``uncapped'' denotes
    \texttt{matrix\_product\_state\_max\_bond\_dimension=None}. Raw feasible mass
  is measured before classical recovery. All decoded repetitions were feasible.}
  \label{tab:mdkp_mps_stability}
  \scriptsize
  \setlength{\tabcolsep}{2pt}
  \renewcommand{\arraystretch}{1.08}
  \begin{tabular*}{\textwidth}{@{\extracolsep{\fill}}llrrrrrr@{}}
    \toprule
    Case & Setting & $\chi_{\max}$ & $\epsilon_{\mathrm{trunc}}$
    & TVD & Raw feas. mass
    & Mean gap (\%) & Best gap (\%) \\
    \midrule
    CVaR-VQE/\texttt{pet2}
    & production & uncapped & $10^{-16}$ & $0$
    & $0.6419$ & $11.27 \pm 6.15$ & $0.21$ \\
    & restrictive cap & $32$ & $10^{-16}$ & $0.04635$
    & $0.6418$ & $11.27 \pm 6.15$ & $0.21$ \\
    & converged cap & $64$ & $10^{-16}$ & $0$
    & $0.6419$ & $11.27 \pm 6.15$ & $0.21$ \\
    & conservative cap & $128$ & $10^{-16}$ & $0$
    & $0.6419$ & $11.27 \pm 6.15$ & $0.21$ \\
    & uncapped threshold test & uncapped & $10^{-12}$ & $0$
    & $0.6419$ & $11.27 \pm 6.15$ & $0.21$ \\
    & uncapped threshold test & uncapped & $10^{-8}$ & $0.00135$
    & $0.6421$ & $11.27 \pm 6.15$ & $0.21$ \\
    \midrule
    VQE/\texttt{hp1}
    & production & uncapped & $10^{-16}$ & $0$
    & $0.6036$ & $18.69 \pm 4.93$ & $9.98$ \\
    & restrictive cap & $33$ & $10^{-16}$ & $0.00585$
    & $0.6034$ & $18.69 \pm 4.93$ & $9.98$ \\
    & converged cap & $128$ & $10^{-16}$ & $0$
    & $0.6036$ & $18.69 \pm 4.93$ & $9.98$ \\
    & conservative cap & $256$ & $10^{-16}$ & $0$
    & $0.6036$ & $18.69 \pm 4.93$ & $9.98$ \\
    & uncapped threshold test & uncapped & $10^{-12}$ & $0$
    & $0.6036$ & $18.69 \pm 4.93$ & $9.98$ \\
    & uncapped threshold test & uncapped & $10^{-8}$ & $0.00035$
    & $0.6036$ & $18.69 \pm 4.93$ & $9.98$ \\
    \bottomrule
  \end{tabular*}
\end{table*}

\subsection{Scope of the validation}

The establishes stability of the MPS reference for the selected large
MDKP VQE-family circuits. It does not constitute an exact-statevector
comparison, which is not practical at 60 and 99 qubits, nor does it establish
equivalent convergence for every problem family, ans\"atze, or optimizer
trajectory in the benchmark. It tests the more specific question relevant to
the MDKP noise-penalty discussion: whether the final simulator references for
representative high-width VQE-family cases are sensitive to plausible
bond-dimension or truncation choices. For these cases, the answer is negative.

\section{QAOA-Family Compilation Counterfactual Audit}
\label{app:qaoa_counterfactuals}

This appendix evaluates whether compilation changes alone could move
representative QAOA-family circuits from the strongly noise-dominated regime
into the empirical signal-preserving regime identified in the main text. The
analysis is compilation-only: algorithmic problem encodings, QAOA depth, and final
bound parameter vectors are fixed, no parameters are retrained, and no
additional quantum-hardware experiments are performed.

\subsection{Representative circuits and compilation conditions}

The audit includes 12 representative bound circuits: standard QAOA, MA-QAOA,
and WS-QAOA for MDKP \texttt{hp1}, MIS \texttt{1tc.32}, QAP \texttt{tai10a},
and MSP \texttt{ms20}. The MSP experiment is mapped to the generated
\texttt{ms\_seed0\_prod3} instance retained in the saved benchmark artifacts.

Each circuit is compiled under four conditions:
\begin{enumerate}
  \item \textbf{Historical baseline:} reconstructed from saved hardware-artifact
    metadata using the historical heavy-hex compilation workflow.
  \item \textbf{SWAP-aware Heron surrogate:} connectivity-aware SABRE initial
    layout and routing on a Heron-like heavy-hex surrogate.
  \item \textbf{Fractional-gate Heron surrogate:} bound-circuit compilation that
    preserves $R_{ZZ}$ structure where possible on the same Heron-like topology.
  \item \textbf{SWAP-aware Nighthawk surrogate:} the same connectivity-aware
    routing procedure on a topology-only square-lattice Nighthawk surrogate.
\end{enumerate}

The SWAP-aware surrogate uses a heuristic connectivity-aware SABRE layout and
routing configuration. It is not an optimal-routing or SAT-proved lower bound.
The Nighthawk compilation is topology-only: no Nighthawk calibration data,
backend noise model, or hardware execution is used. Fractional-gate rows are
also resource counterfactuals only. They are not directly comparable with the
original mitigated hardware campaign because they were not executed under the
original ZNE/probabilistic-error-amplification workflow.

\subsection{Metrics}

For each compilation condition, we record algorithmic-level two-qubit-gate count,
transpiled two-qubit-gate count, SWAP count, two-qubit depth, total depth, and
the same conservative gate-count fidelity proxy used elsewhere in the paper.
For fractional-gate rows, this quantity is denoted
$\widetilde F_{\mathrm{gate}}$ because it is a resource proxy and not a
calibrated fractional-gate process-fidelity estimate.

\subsection{Results}

Table~\ref{tab:qaoa_counterfactual_detail} reports the full representative
audit. Across all 12 circuits, square-lattice Nighthawk topology gives the
largest median two-qubit-gate reduction, $19.8\%$, relative to reconstructed
historical compilation. Fractional-gate Heron compilation yields a median
reduction of $4.1\%$. The tested SWAP-aware Heron surrogate instead increases
median two-qubit-gate count by $12.0\%$, illustrating that a generic
connectivity-aware routing heuristic does not necessarily improve on the
historical compiled routes.

\begin{figure*}[t]
  \centering
  \includegraphics[width=\textwidth]{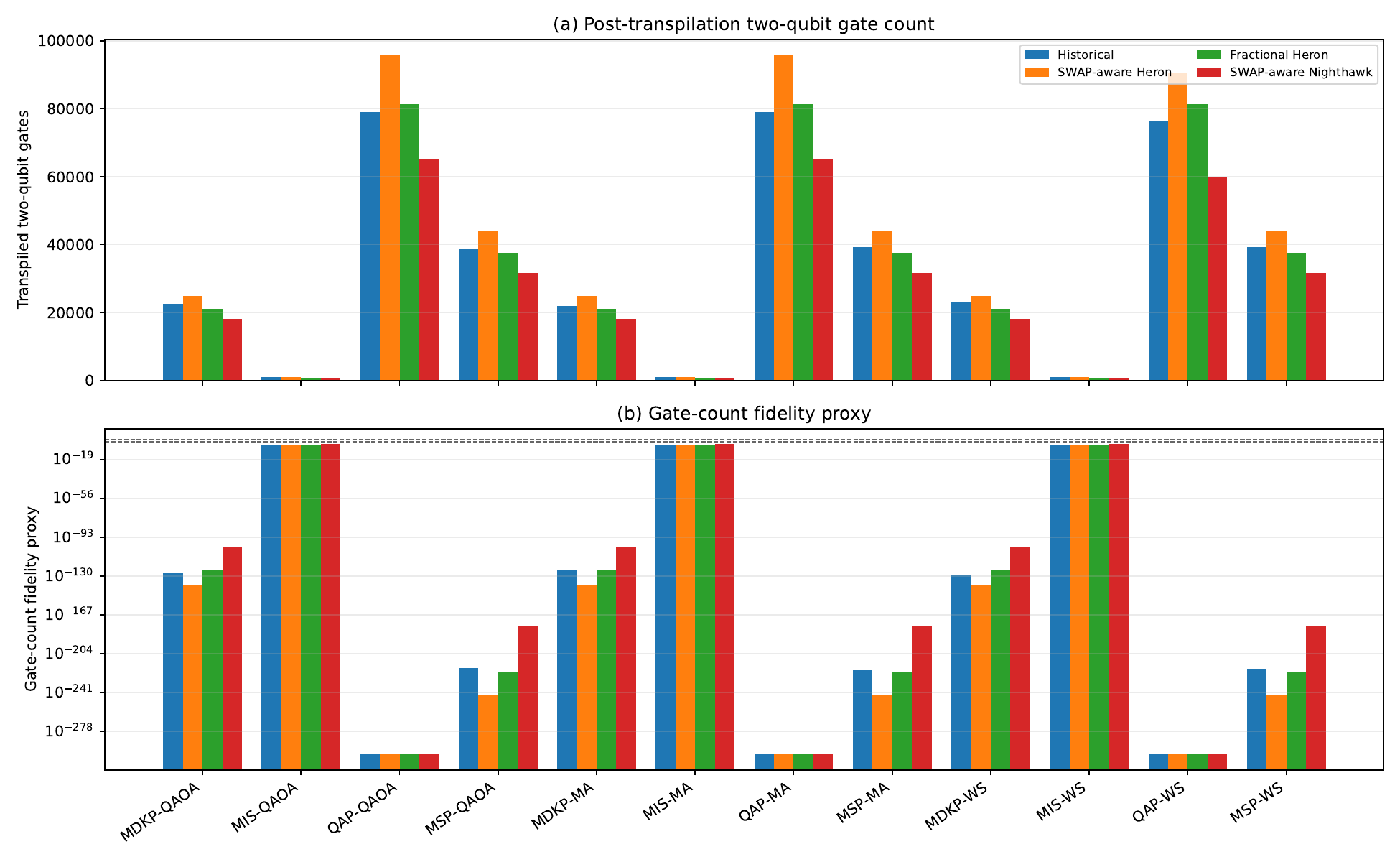}
  \caption{Compilation-only counterfactual audit for 12 representative bound
    QAOA-family circuits. \textbf{Top:} post-transpilation two-qubit-gate counts
    under reconstructed historical compilation, a SWAP-aware Heron surrogate,
    fractional-gate Heron compilation, and a topology-only SWAP-aware Nighthawk
    surrogate. \textbf{Bottom:} corresponding conservative gate-count fidelity
    proxies. The dashed reference lines indicate the benchmark diagnostic
    thresholds $F_{\mathrm{est}}=10^{-2}$ and $F_{\mathrm{est}}=10^{-3}$.
    No circuit crosses either threshold under any tested counterfactual.
    Fractional-gate values are resource proxies only and were not executed under
  the original ZNE/probabilistic-error-amplification workflow.}
  \label{fig:qaoa_counterfactuals}
\end{figure*}

Most importantly, no circuit crosses either
$F_{\mathrm{est}}\geq10^{-3}$ or $F_{\mathrm{est}}\geq10^{-2}$ under any
counterfactual. The observed reductions are therefore insufficient to return
the selected historical QAOA-family circuits to the empirical
signal-preserving regime. This supports a conclusion about the tested standard
QAOA, MA-QAOA, and WS-QAOA implementations and their reported circuit scales;
it does not rule out other mixers, fixed-schedule QAOA variants, specialized
routing methods, or future hardware. Figure~\ref{fig:qaoa_counterfactuals}
visualizes the corresponding post-transpilation two-qubit-gate counts and
gate-count fidelity proxies for all four compilation conditions.

\begin{table*}[t]

  \centering
  \caption{Representative bound QAOA-family compilation counterfactuals.
    $N_{2Q}$ denotes transpiled two-qubit-gate count. H is reconstructed historical
    heavy-hex compilation; S is the SWAP-aware Heron surrogate; F is fractional-gate
    Heron compilation; N is the topology-only square-lattice Nighthawk surrogate.
    No listed circuit reaches $F_{\mathrm{est}}\geq10^{-3}$ or
  $F_{\mathrm{est}}\geq10^{-2}$ under any condition.}
  \label{tab:qaoa_counterfactual_detail}
  \footnotesize
  \setlength{\tabcolsep}{2.4pt}
  \renewcommand{\arraystretch}{1.05}
  \begin{tabular}{lllrrrrrr}
    \toprule
    Problem & Method & Instance & $N_{2Q}^{\mathrm{H}}$ &
    $N_{2Q}^{\mathrm{S}}$ & $N_{2Q}^{\mathrm{F}}$ &
    $N_{2Q}^{\mathrm{N}}$ & Best reduction & Best proxy \\
    \midrule
    MDKP & QAOA    & hp1    & 22,621 & 24,918 & 21,153 & 18,075 & 20.1\% & $1.77\times10^{-102}$ \\
    MDKP & MA-QAOA & hp1    & 22,019 & 24,918 & 21,153 & 18,075 & 17.9\% & $1.77\times10^{-102}$ \\
    MDKP & WS-QAOA & hp1    & 23,114 & 24,918 & 21,153 & 18,075 & 21.8\% & $1.71\times10^{-102}$ \\
    MIS  & QAOA    & 1tc.32 & 935 & 969 & 765 & 720 & 23.0\% & $2.91\times10^{-5}$ \\
    MIS  & MA-QAOA & 1tc.32 & 915 & 969 & 765 & 720 & 21.3\% & $2.91\times10^{-5}$ \\
    MIS  & WS-QAOA & 1tc.32 & 911 & 969 & 765 & 720 & 21.0\% & $2.91\times10^{-5}$ \\
    QAP  & QAOA    & tai10a & 79,041 & 95,754 & 81,444 & 65,244 & 17.5\% & $0$ \\
    QAP  & MA-QAOA & tai10a & 79,027 & 95,754 & 81,444 & 65,244 & 17.4\% & $0$ \\
    QAP  & WS-QAOA & tai10a & 76,532 & 90,610 & 81,444 & 60,100 & 21.5\% & $0$ \\
    MSP  & QAOA    & ms20 & 38,936 & 43,968 & 37,668 & 31,656 & 18.7\% & $2.21\times10^{-178}$ \\
    MSP  & MA-QAOA & ms20 & 39,302 & 43,968 & 37,668 & 31,656 & 19.5\% & $2.21\times10^{-178}$ \\
    MSP  & WS-QAOA & ms20 & 39,227 & 43,968 & 37,668 & 31,656 & 19.3\% & $2.21\times10^{-178}$ \\
    \bottomrule
  \end{tabular}
\end{table*}

\section{QAOA-Family Transpilation Expansion}
\label{app:qaoa_prepost_expansion}

\begin{figure}[t]
  \centering
  \includegraphics[width=0.98\linewidth]{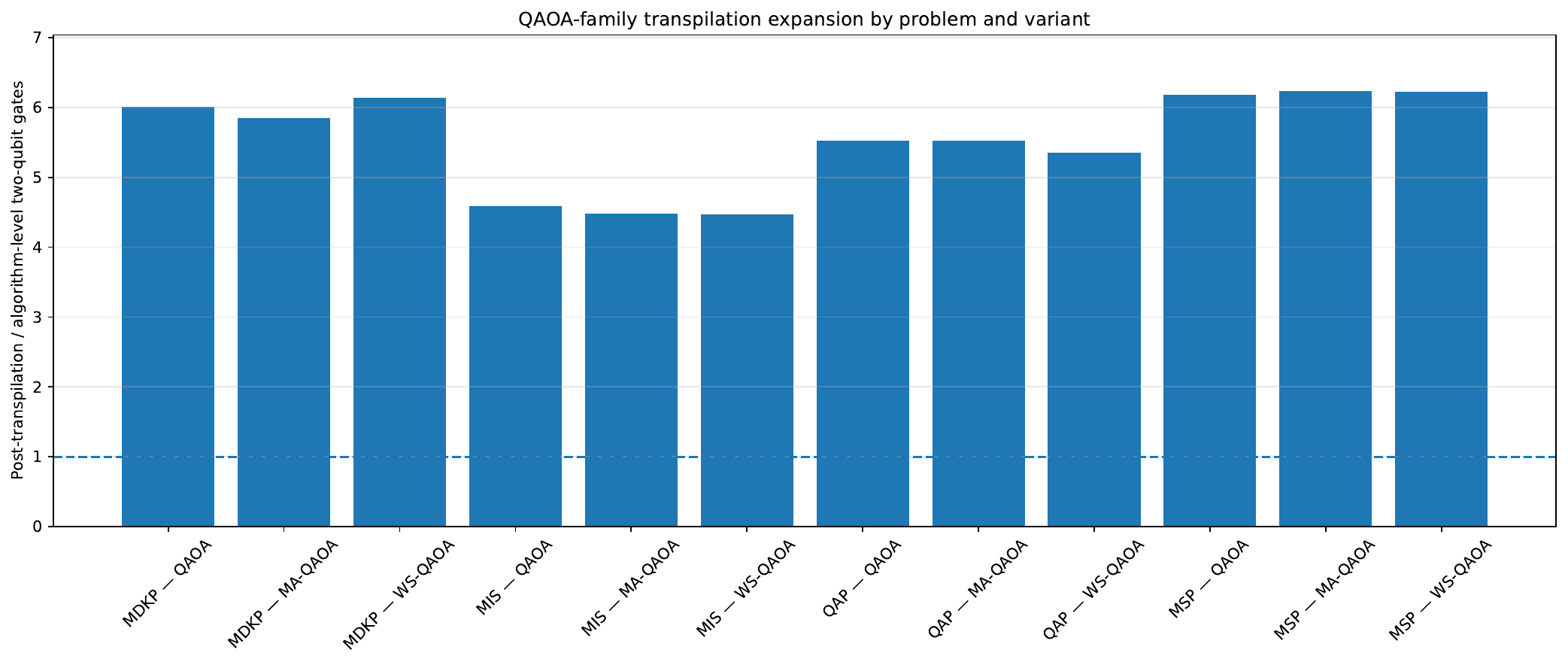}
  \caption{QAOA-family transpilation expansion factors for the 12
    representative circuits in Figure~\ref{fig:qaoa_prepost_2q}. The dashed line at
    one indicates no expansion. Every circuit shows a post-transpilation
  two-qubit-gate count between $4.47$ and $6.24$ times its algorithm-level count.}
  \label{fig:qaoa_transpilation_expansion}
\end{figure}

This appendix provides the numerical source data behind
Figure~\ref{fig:qaoa_prepost_2q}. It separates algorithm-level two-qubit-gate
counts from the backend-native transpiled counts for representative
bound depth-$3$ QAOA-family circuits. The analysis uses the same historical
compilation rows as the QAOA counterfactual audit, but considers only the
original reported hardware-transpilation pathway. Figure~\ref{fig:qaoa_transpilation_expansion}
visualizes the circuit-level expansion factors across all 12 representative
QAOA-family circuits.

\begin{table}[t]

  \centering
  \caption{Pre- and post-transpilation two-qubit-gate counts.
    Values are medians across QAOA, MA-QAOA, and WS-QAOA for the stated
    representative instance. The expansion factor is
    $N_{2Q}^{\mathrm{transpiled}}/N_{2Q}^{\mathrm{algorithm}}$. Recorded SWAP
    counts are routing metadata and should not be interpreted as an additive
  decomposition of final two-qubit-gate count.}
  \label{tab:qaoa_prepost_summary}
  \footnotesize
  \setlength{\tabcolsep}{2.8pt}
  \renewcommand{\arraystretch}{1.08}
  \begin{tabular}{llrrrr}
    \toprule
    Problem & Instance &
    $N_{2Q}^{\mathrm{alg}}$ &
    $N_{2Q}^{\mathrm{trans}}$ &
    Expansion &
    Median recorded SWAPs \\
    \midrule
    MDKP & hp1 & 3,765 & 22,621 & $6.01\times$ & 5,030 \\
    MIS  & 1tc.32 & 204 & 915 & $4.49\times$ & 169 \\
    QAP  & tai10a & 14,310 & 79,027 & $5.52\times$ & 16,802 \\
    MSP  & ms20 & 6,300 & 39,227 & $6.23\times$ & 8,876 \\
    \bottomrule
  \end{tabular}
\end{table}

The complete circuit-level data, including circuit width, QAOA depth,
algorithm-level two-qubit-gate count, transpiled two-qubit-gate count, recorded
SWAP count, backend, optimization level, and historical-compilation metadata,
are included in the reproducibility artifact.

\section{Acronym and Table-Notation Reference}
\label{app:acronyms}

Table~\ref{tab:acronym_reference} provides a quick reference for the
abbreviations and compact table notation used throughout the manuscript and
appendices. Acronyms are grouped by their role in the benchmark rather than
listed alphabetically, so that problem classes, algorithm families, execution
settings, and result-table quantities can be identified quickly.

\begin{table*}[t]
  \centering
  \caption{\label{tab:acronym_reference}
    Acronym and table-notation reference. Full terms are given at first use in the
    main text; this table is intended as a compact reader aid for the benchmark
  results and supplementary diagnostics.}
  \footnotesize
  \setlength{\tabcolsep}{3pt}
  \renewcommand{\arraystretch}{1.04}
  \begin{tabularx}{\textwidth}{@{}>{\raggedright\arraybackslash}p{0.23\textwidth}>{\raggedright\arraybackslash}p{0.16\textwidth}>{\raggedright\arraybackslash}X@{}}
    \toprule
    \textbf{Category} & \textbf{Abbreviation} & \textbf{Meaning and use in this study} \\
    \midrule

    Problem class
    & MDKP
    & Multi-Dimensional Knapsack Problem: a packing problem with multiple capacity constraints. \\

    Problem class
    & MIS
    & Maximum Independent Set: selection of the largest set of mutually nonadjacent graph vertices. \\

    Problem class
    & QAP
    & Quadratic Assignment Problem: assignment of facilities to locations under quadratic flow--distance costs. \\

    Problem class
    & MSP
    & Market Share Problem: allocation problem evaluated against prescribed market-share targets. \\

    Problem formulation
    & QUBO
    & Quadratic Unconstrained Binary Optimization: the common binary quadratic representation used before conversion to an Ising-type cost Hamiltonian. \\

    Reference metric
    & BKS
    & Best-known solution or value used as the classical reference target in the result tables. \\

    MSP metric
    & TDev
    & Total absolute deviation from the prescribed market-share targets; lower values indicate better target matching. \\

    MSP metric
    & MDev
    & Maximum single-category deviation from the prescribed market-share targets. \\

    Variational method
    & VQE
    & Variational Quantum Eigensolver. \\

    Risk-sensitive method
    & CVaR
    & Conditional Value-at-Risk: lower-tail objective used here with confidence level $\alpha=0.25$. \\

    Risk-sensitive method
    & CVaR-VQE
    & Conditional Value-at-Risk Variational Quantum Eigensolver. \\

    Alternating-operator method
    & QAOA
    & Quantum Approximate Optimization Algorithm. \\

    QAOA variant
    & MA-QAOA
    & Multi-Angle Quantum Approximate Optimization Algorithm. \\

    QAOA variant
    & WS-QAOA
    & Warm-Start Quantum Approximate Optimization Algorithm. \\

    Qubit-efficient method
    & PCE
    & Pauli Correlation Encoding: a compressed representation that reconstructs binary variables from measured Pauli correlators. \\

    Qubit-efficient method
    & QRAO
    & Quantum Random Access Optimization: a quantum random-access-code-based optimization method with classical rounding and recovery. \\

    Encoding primitive
    & QRAC
    & Quantum Random Access Code: the compressed coding primitive underlying the QRAO implementation. \\

    Simulation method
    & MPS
    & Matrix Product State: the tensor-network simulation method used for the shot-based AerSimulator reference runs. \\

    Hardware term
    & QPU
    & Quantum Processing Unit. \\

    Compilation metric
    & 2Q / $N_{2Q}$
    & Two-qubit gates / transpiled two-qubit-gate count. $N_{2Q}$ is used in the process-fidelity diagnostic. \\

    Fidelity diagnostic
    & $F_{\mathrm{est}}$
    & Estimated process-fidelity proxy computed from transpiled two-qubit-gate count and representative backend error data. \\

    Error mitigation
    & ZNE
    & Zero-Noise Extrapolation: estimator-side gate-noise mitigation used through IBM Runtime resilience settings. \\

    Error mitigation
    & TREX
    & Twirled Readout Error eXtinction: readout-error mitigation used through IBM Runtime resilience settings. \\

    Classical optimizer
    & SLSQP
    & Sequential Least Squares Programming: the classical optimizer used for the PCE variational objective. \\

    Benchmark resource
    & QOBLIB
    & Quantum Optimization Benchmarking Library. \\

    Table notation
    & Q / D / G / P
    & Respectively, circuit qubit count, circuit depth, total gate count, and number of variational parameters in the simulator-resource tables. \\

    Solver status
    & OPT
    & Solver status indicating that a proven optimal solution was obtained. \\

    Near-term regime
    & NISQ
    & Noisy Intermediate-Scale Quantum: the current regime of non-fault-tolerant quantum processors considered in this study. \\

    \bottomrule
  \end{tabularx}
\end{table*}

\clearpage

\section{Simulator-Based Results}
\label{app:simulator_results}

This appendix collects the full simulator-based performance tables for each
benchmark problem family.  These results serve as controlled reference points
against which hardware-induced degradation can be measured: because the
Qiskit AerSimulator (matrix product state backend, shot-based evaluation)
removes queueing, calibration drift, and device-specific noise, performance
differences between simulator and hardware runs isolate the cost of
compilation and execution on real devices.

\subsection{MDKP simulator results}
\label{app:sim_mdkp}

Table~\ref{tab:results_mdkp} reports simulator-based performance for PCE,
VQE, and CVaR-VQE on the MDKP benchmark instances.  QAOA-based methods were
excluded from the simulator comparison because their evaluation cost was
prohibitively high even at minimal circuit depths on the dense MDKP QUBO
instances.  Under controlled simulation, all three methods return feasible
solutions on every instance except VQE on \texttt{pb4}.  Comparing these
gaps with the hardware results in Section~\ref{sec:results_mdkp} reveals
how transpilation and backend noise reshape the method ranking.

\begin{table*}[htbp]
\caption{\label{tab:results_mdkp}
MDKP simulator results for PCE, VQE, and CVaR-VQE\@.
QAOA-based methods were too computationally expensive to simulate
even at minimal depths. \textbf{Vars.}:\ number of binary variables
in the QUBO formulation. For \texttt{pb4}, VQE does not return a
feasible solution (gap reported as $\infty$).}
\centering
\footnotesize
\setlength{\tabcolsep}{3pt}
\begin{tabular*}{\textwidth}{@{\extracolsep{\fill}}lccccccccc}
\toprule
& & & \multicolumn{3}{c}{\textbf{PCE}} &
\multicolumn{2}{c}{\textbf{VQE}} &
\multicolumn{2}{c}{\textbf{CVaR-VQE}} \\
\cmidrule(lr){4-6} \cmidrule(lr){7-8} \cmidrule(lr){9-10}
\textbf{Inst.} & \textbf{Opt.} & \textbf{Vars.} &
\textbf{Q} & \textbf{Feas.} & \textbf{Gap} &
\textbf{Feas.} & \textbf{Gap} &
\textbf{Feas.} & \textbf{Gap} \\
\midrule
hp1  & 3418  & 60  & 7  & Yes & 20.88 & Yes & 39.76    & Yes & 20.59 \\
hp2  & 3186  & 67  & 8  & Yes & 38.38 & Yes & 12.34    & Yes & 21.78 \\
pb1  & 3090  & 59  & 7  & Yes & 14.04 & Yes & 19.94    & Yes & 23.43 \\
pb2  & 3186  & 66  & 8  & Yes & 14.37 & Yes & 19.49    & Yes & 22.78 \\
pb4  & 95168 & 45  & 6  & Yes & 32.91 & No  & $\infty$ & Yes & 39.38 \\
pb5  & 2139  & 116 & 10 & Yes & 11.87 & Yes & 4.25     & Yes & 33.34 \\
pet2 & 87061 & 99  & 9  & Yes & 28.19 & Yes & 41.07    & Yes & 30.37 \\
pet3 & 4015  & 102 & 9  & Yes & 15.56 & Yes & 4.98     & Yes & 34.74 \\
pet4 & 6120  & 107 & 9  & Yes & 50.98 & Yes & 66.58    & Yes & 48.44 \\
pet5 & 12400 & 122 & 10 & Yes & 22.74 & Yes & 33.23    & Yes & 53.15 \\
pet6 & 10618 & 86  & 9  & Yes & 33.84 & Yes & 12.50    & Yes & 40.41 \\
pet7 & 16537 & 100 & 9  & Yes & 15.87 & Yes & 43.46    & Yes & 46.47 \\
\bottomrule
\end{tabular*}
\end{table*}

\subsection{MIS simulator results}
\label{app:sim_mis}

Table~\ref{tab:results_mis} reports simulator-based performance on the MIS
instances.  In contrast to the hardware results
(Section~\ref{sec:results_mis}), most methods achieve zero or near-zero gap
on the smallest instances and retain feasibility across a wider range.  This
indicates that the sharp feasibility cliff observed on hardware is driven
primarily by compilation overhead and backend noise rather than by
algorithmic limitations at these scales.

\begin{table*}[htbp]
\caption{\label{tab:results_mis}
MIS simulator results for PCE, VQE, CVaR-VQE, QAOA, and MA-QAOA\@.
In contrast to the hardware results (Section~\ref{sec:results_mis}),
most methods perform well on small instances in simulation, indicating
that the hardware feasibility breakdown is driven by compilation and
noise rather than algorithmic limitations at these scales.}
\centering
\footnotesize
\setlength{\tabcolsep}{3pt}
\begin{tabular*}{\textwidth}{@{\extracolsep{\fill}}lcccccccccccc}
\toprule
& & \multicolumn{3}{c}{\textbf{PCE}} &
\multicolumn{2}{c}{\textbf{VQE}} &
\multicolumn{2}{c}{\textbf{CVaR-VQE}} &
\multicolumn{2}{c}{\textbf{QAOA}} \\
\cmidrule(lr){3-5} \cmidrule(lr){6-7} \cmidrule(lr){8-9} \cmidrule(lr){10-11}
\textbf{Inst.} & \textbf{Opt.} &
\textbf{Q} & \textbf{Feas.} & \textbf{Gap} &
\textbf{Feas.} & \textbf{Gap} &
\textbf{Feas.} & \textbf{Gap} &
\textbf{Feas.} & \textbf{Gap} \\
\midrule
1tc.8   & 4  & 3  & Yes & 25.00 & Yes & 0.00  & Yes & 0.00  & Yes & 0.00 \\
1tc.16  & 8  & 4  & Yes & 50.00 & Yes & 0.00  & Yes & 0.00  & Yes & 0.00 \\
1tc.32  & 12 & 6  & Yes & 41.67 & Yes & 8.33  & Yes & 8.33  & Yes & 16.67 \\
1tc.64  & 20 & 8  & No  & $\infty$ & No & $\infty$ & No & $\infty$ & No & $\infty$ \\
1et.64  & 18 & 8  & No  & $\infty$ & No & $\infty$ & No & $\infty$ & No & $\infty$ \\
1dc.64  & 10 & 8  & No  & $\infty$ & No & $\infty$ & No & $\infty$ & No & $\infty$ \\
1dc.128 & 16 & 10 & No  & $\infty$ & No & $\infty$ & No & $\infty$ & No & $\infty$ \\
\bottomrule
\end{tabular*}
\end{table*}

\subsection{QAP simulator results}
\label{app:sim_qap}

Table~\ref{tab:results_qap} reports simulator-based performance on the QAP
instances.  The results provide an instructive contrast to the complete
hardware infeasibility reported in Section~\ref{sec:results_qap}: PCE is
the only method that returns feasible solutions, albeit with large
optimality gaps (15--235\%), while VQE and CVaR-VQE remain infeasible even
under noise-free conditions.  This confirms that QAP's dense permutation
structure is inherently difficult for standard variational methods
regardless of hardware effects.

\begin{table*}[htbp]
\caption{\label{tab:results_qap}
QAP simulator results for VQE, PCE, and CVaR-VQE\@.
VQE and CVaR-VQE are infeasible on all instances (gap reported as --).
PCE returns feasible solutions on all tested instances, albeit with
large optimality gaps.}
\centering
\footnotesize
\setlength{\tabcolsep}{3pt}
\begin{tabular*}{\textwidth}{@{\extracolsep{\fill}}lccccccccc}
\toprule
& & & \multicolumn{2}{c}{\textbf{VQE}} &
\multicolumn{3}{c}{\textbf{PCE}} &
\multicolumn{2}{c}{\textbf{CVaR-VQE}} \\
\cmidrule(lr){4-5} \cmidrule(lr){6-8} \cmidrule(lr){9-10}
\textbf{Inst.} & \textbf{Opt.} & \textbf{Vars.} &
\textbf{Feas.} & \textbf{Gap} &
\textbf{Q} & \textbf{Feas.} & \textbf{Gap} &
\textbf{Feas.} & \textbf{Gap} \\
\midrule
chr12a & 9552     & 144 & No & $\infty$ & 11 & Yes & 234.75 & No & $\infty$ \\
chr12b & 9742     & 144 & No & $\infty$ & 11 & Yes & 15.47  & No & $\infty$ \\
chr12c & 11156    & 144 & No & $\infty$ & 11 & Yes & 66.82  & No & $\infty$ \\
nug12  & 578      & 144 & No & $\infty$ & 11 & Yes & 35.98  & No & $\infty$ \\
rou12  & 235528   & 144 & No & $\infty$ & 11 & Yes & 26.30  & No & $\infty$ \\
scr12  & 31410    & 144 & No & $\infty$ & 11 & Yes & 71.30  & No & $\infty$ \\
tai12a & 224416   & 144 & No & $\infty$ & 11 & Yes & 29.78  & No & $\infty$ \\
tai12b & 39464925 & 144 & No & $\infty$ & 11 & Yes & 22.90  & No & $\infty$ \\
\bottomrule
\end{tabular*}
\end{table*}

\subsection{MSP simulator results}
\label{app:sim_msp}

Table~\ref{tab:results_msp} reports simulator-based performance on the
Market Share Problem instances.  Even in simulation, residual deviations
from the target allocation remain non-trivial, reflecting the inherent
difficulty of the balancing constraints rather than hardware noise alone.

\begin{table*}[htbp]
\caption{\label{tab:results_msp}
MSP simulator results for PCE, VQE, and CVaR-VQE\@.
\textbf{TDev}: total absolute deviation from target split.
\textbf{MDev}: maximum single-product deviation.
Even in simulation, large residual deviations are observed, partly
because the optimal objectives are small integers.}
\centering
\footnotesize
\setlength{\tabcolsep}{3pt}
\begin{tabular*}{\textwidth}{@{\extracolsep{\fill}}lccccccccccc}
\toprule
& & & \multicolumn{3}{c}{\textbf{PCE}} &
\multicolumn{2}{c}{\textbf{VQE}} &
\multicolumn{2}{c}{\textbf{CVaR-VQE}} \\
\cmidrule(lr){4-6} \cmidrule(lr){7-8} \cmidrule(lr){9-10}
\textbf{Inst.} & \textbf{Opt.} & \textbf{Q} &
\textbf{TDev} & \textbf{MDev} & \textbf{Match} &
\textbf{TDev} & \textbf{MDev} &
\textbf{TDev} & \textbf{MDev} \\
\midrule
ms20 & 3 & 7  & 3   & 3   & No & 3   & 2   & 5   & 3 \\
ms30 & 2 & 8  & 18  & 12  & No & 12  & 8   & 15  & 10 \\
ms40 & 1 & 10 & 35  & 15  & No & 25  & 12  & 20  & 9 \\
ms50 & 0 & 11 & 72  & 28  & No & 55  & 22  & 48  & 18 \\
ms21 & 3 & 7  & 5   & 4   & No & 4   & 3   & 3   & 2 \\
ms31 & 3 & 8  & 20  & 14  & No & 10  & 6   & 12  & 8 \\
ms41 & 1 & 10 & 38  & 16  & No & 28  & 14  & 22  & 10 \\
ms51 & 1 & 11 & 78  & 30  & No & 60  & 25  & 50  & 20 \\
\bottomrule
\end{tabular*}
\end{table*}

\section{Hardware Diagnostics: MDKP}
\label{app:hw_mdkp}

This appendix reports circuit compilation statistics and execution
robustness diagnostics for all seven methods evaluated on the MDKP
benchmark.  For each method we provide two tables:
a \emph{circuit compilation} table reporting algorithmic-level and post-transpilation
metrics (qubits, depth, total gates, two-qubit gates, and parameters) per
backend, and an \emph{execution robustness} table reporting shot budget,
evaluation count, job success rates, latency statistics, and wall-clock
time.  These diagnostics contextualize the solution quality results in
Section~\ref{sec:results_mdkp} by quantifying the hardware-induced
overheads that shape practical performance.

\subsection{MDKP: VQE-style methods}
\label{sec:appendix_mdkp_vqe_family}

Tables~\ref{tab:mdkp_vqe_circuit_compilation}
and~\ref{tab:mdkp_vqe_execution_robustness} report compilation and
execution diagnostics for VQE\@.
Tables~\ref{tab:mdkp_cvar_vqe_circuit_compilation}
and~\ref{tab:mdkp_cvar_vqe_execution_robustness} report the corresponding
diagnostics for CVaR-VQE\@.

\begin{table*}[htbp]
\caption{\label{tab:mdkp_vqe_circuit_compilation}
MDKP (VQE): circuit and compilation statistics. 
Logical metrics (qubits, depth, gates, two-qubit gates, parameters) and post-transpilation metrics per backend.}
\centering
\footnotesize
\setlength{\tabcolsep}{3pt}
\begin{tabular*}{\textwidth}{@{\extracolsep{\fill}}llccccccccc}
\toprule
 & & \multicolumn{5}{c}{\textbf{Logical}} & \multicolumn{3}{c}{\textbf{Transpiled}} \\
\cmidrule(lr){3-7} \cmidrule(lr){8-10}
\textbf{Inst.} & \textbf{Backend} & \textbf{Q} & \textbf{D} & \textbf{G} & \textbf{2Q} & \textbf{P} & \textbf{D} & \textbf{G} & \textbf{2Q} \\
\midrule
hp1 & ibm\_fez & 60 & 185 & 480 & 180 & 240 & 731 & 1740 & 180 \\
hp1 & ibm\_torino & 60 & 185 & 480 & 180 & 240 & 731 & 1740 & 180 \\
hp2 & ibm\_fez & 67 & 206 & 536 & 201 & 268 & 1068 & 2356 & 338 \\
hp2 & ibm\_torino & 67 & 206 & 536 & 201 & 268 & 934 & 2161 & 279 \\
pb1 & ibm\_fez & 59 & 182 & 472 & 177 & 236 & 1015 & 2247 & 363 \\
pb1 & ibm\_torino & 59 & 182 & 472 & 177 & 236 & 985 & 2160 & 328 \\
pb2 & ibm\_fez & 66 & 203 & 528 & 198 & 264 & 1074 & 2356 & 344 \\
pb2 & ibm\_torino & 66 & 203 & 528 & 198 & 264 & 964 & 2196 & 295 \\
pb4 & ibm\_fez & 45 & 140 & 360 & 135 & 180 & 818 & 1707 & 270 \\
pb4 & ibm\_torino & 45 & 140 & 360 & 135 & 180 & 735 & 1636 & 249 \\
pb5 & ibm\_fez & 116 & 353 & 928 & 348 & 464 & 1403 & 3364 & 348 \\
pb5 & ibm\_torino & 116 & 353 & 928 & 348 & 464 & 2083 & 4766 & 847 \\
pet2 & ibm\_fez & 99 & 302 & 792 & 297 & 396 & 1236 & 2962 & 326 \\
pet2 & ibm\_torino & 99 & 302 & 792 & 297 & 396 & 1269 & 3032 & 354 \\
pet3 & ibm\_fez & 102 & 311 & 816 & 306 & 408 & 1284 & 3078 & 347 \\
pet3 & ibm\_torino & 102 & 311 & 816 & 306 & 408 & 1268 & 3009 & 324 \\
pet4 & ibm\_fez & 107 & 326 & 856 & 321 & 428 & 1405 & 3349 & 410 \\
pet4 & ibm\_torino & 107 & 326 & 856 & 321 & 428 & 1519 & 3640 & 516 \\
pet5 & ibm\_fez & 122 & 371 & 976 & 366 & 488 & 1508 & 3589 & 384 \\
pet5 & ibm\_torino & 122 & 371 & 976 & 366 & 488 & 2380 & 5355 & 1016 \\
pet6 & ibm\_fez & 86 & 267 & 1032 & 258 & 688 & 1035 & 3391 & 390 \\
pet6 & ibm\_torino & 86 & 267 & 1032 & 258 & 688 & 935 & 3237 & 336 \\
pet7 & ibm\_fez & 100 & 309 & 1200 & 300 & 800 & 919 & 3497 & 300 \\
pet7 & ibm\_torino & 100 & 309 & 1200 & 300 & 800 & 919 & 3497 & 300 \\
\bottomrule
\end{tabular*}
\end{table*}
\begin{table*}[htbp]
\caption{\label{tab:mdkp_vqe_execution_robustness}
MDKP (VQE): execution robustness summary. 
Shots per evaluation, number of evaluations, job success counts, latency statistics, and wall-clock time.}
\centering
\footnotesize
\setlength{\tabcolsep}{4pt}
\begin{tabular*}{\textwidth}{@{\extracolsep{\fill}}lcccccc}
\toprule
\textbf{Inst.} & \textbf{Shots} & \textbf{Evals} & \textbf{Jobs (ok/fail)} & \textbf{Med.\ Lat.\ (s)} & \textbf{P95 Lat.\ (s)} & \textbf{Time (min)} \\
\midrule
hp1 & 1000 & 200 & 200/0 & 8.534 & 13.958 & 43.128 \\
hp2 & 1000 & 200 & 200/0 & 8.500 & 14.251 & 42.111 \\
pb1 & 1000 & 200 & 200/0 & 8.492 & 13.914 & 36.385 \\
pb2 & 1000 & 200 & 200/0 & 8.505 & 14.894 & 41.330 \\
pb4 & 1000 & 200 & 200/0 & 8.546 & 14.318 & 37.562 \\
pb5 & 1000 & 200 & 200/0 & 8.949 & 19.649 & 44.451 \\
pet2 & 1000 & 200 & 200/0 & 8.548 & 18.904 & 42.780 \\
pet3 & 1000 & 200 & 200/0 & 8.670 & 14.169 & 38.936 \\
pet4 & 1000 & 200 & 200/0 & 9.000 & 52.542 & 87.149 \\
pet5 & 1000 & 200 & 200/0 & 9.106 & 104.800 & 104.926 \\
pet6 & 1000 & 200 & 200/0 & 8.700 & 14.409 & 41.173 \\
pet7 & 1000 & 200 & 200/0 & 8.712 & 14.797 & 44.246 \\
\bottomrule
\end{tabular*}
\end{table*}
\begin{table*}[htbp]
\caption{\label{tab:mdkp_cvar_vqe_circuit_compilation}
MDKP (CVaR-VQE): circuit and compilation statistics. 
Logical metrics (qubits, depth, gates, two-qubit gates, parameters) and post-transpilation metrics per backend.}
\centering
\footnotesize
\setlength{\tabcolsep}{3pt}
\begin{tabular*}{\textwidth}{@{\extracolsep{\fill}}llccccccccc}
\toprule
 & & \multicolumn{5}{c}{\textbf{Logical}} & \multicolumn{3}{c}{\textbf{Transpiled}} \\
\cmidrule(lr){3-7} \cmidrule(lr){8-10}
\textbf{Inst.} & \textbf{Backend} & \textbf{Q} & \textbf{D} & \textbf{G} & \textbf{2Q} & \textbf{P} & \textbf{D} & \textbf{G} & \textbf{2Q} \\
\midrule
hp1 & ibm\_torino & 60 & 185 & 480 & 180 & 240 & 731 & 1740 & 180 \\
hp2 & ibm\_marrakesh & 67 & 206 & 536 & 201 & 268 & 1068 & 2356 & 338 \\
hp2 & ibm\_torino & 67 & 206 & 536 & 201 & 268 & 929 & 2162 & 279 \\
pb1 & ibm\_marrakesh & 59 & 182 & 472 & 177 & 236 & 972 & 2153 & 327 \\
pb1 & ibm\_torino & 59 & 182 & 472 & 177 & 236 & 972 & 2124 & 314 \\
pb2 & ibm\_marrakesh & 66 & 203 & 528 & 198 & 264 & 1080 & 2354 & 344 \\
pb2 & ibm\_torino & 66 & 203 & 528 & 198 & 264 & 964 & 2196 & 295 \\
pb4 & ibm\_marrakesh & 45 & 140 & 360 & 135 & 180 & 970 & 1993 & 369 \\
pb4 & ibm\_torino & 45 & 140 & 360 & 135 & 180 & 812 & 1803 & 309 \\
pb5 & ibm\_marrakesh & 116 & 353 & 928 & 348 & 464 & 1403 & 3364 & 348 \\
pb5 & ibm\_torino & 116 & 353 & 928 & 348 & 464 & 2104 & 4838 & 876 \\
pet2 & ibm\_torino & 99 & 302 & 792 & 297 & 396 & 1259 & 3016 & 347 \\
pet3 & ibm\_torino & 102 & 311 & 816 & 306 & 408 & 1268 & 3009 & 324 \\
pet4 & ibm\_torino & 107 & 326 & 856 & 321 & 428 & 1452 & 3485 & 459 \\
pet5 & ibm\_torino & 122 & 371 & 976 & 366 & 488 & 2619 & 5963 & 1227 \\
pet6 & ibm\_torino & 86 & 263 & 688 & 258 & 344 & 1170 & 2720 & 332 \\
pet7 & ibm\_torino & 100 & 305 & 800 & 300 & 400 & 1211 & 2900 & 300 \\
\bottomrule
\end{tabular*}
\end{table*}
\begin{table*}[htbp]
\caption{\label{tab:mdkp_cvar_vqe_execution_robustness}
MDKP (CVaR-VQE): execution robustness summary. 
Shots per evaluation, number of evaluations, job success counts, latency statistics, and wall-clock time.}
\centering
\footnotesize
\setlength{\tabcolsep}{4pt}
\begin{tabular*}{\textwidth}{@{\extracolsep{\fill}}lcccccc}
\toprule
\textbf{Inst.} & \textbf{Shots} & \textbf{Evals} & \textbf{Jobs (ok/fail)} & \textbf{Med.\ Lat.\ (s)} & \textbf{P95 Lat.\ (s)} & \textbf{Time (min)} \\
\midrule
hp1 & 1000 & 200 & 200/0 & 19.270 & 21.626 & 57.458 \\
hp2 & 1000 & 200 & 200/0 & 19.692 & 26.366 & 66.246 \\
pb1 & 1000 & 200 & 200/0 & 19.580 & 21.897 & 59.023 \\
pb2 & 1000 & 200 & 200/0 & 19.790 & 45.706 & 75.479 \\
pb4 & 1000 & 200 & 200/0 & 8.630 & 24.464 & 42.742 \\
pb5 & 1000 & 200 & 200/0 & 8.974 & 25.054 & 52.700 \\
pet2 & 1000 & 200 & 200/0 & 8.763 & 41.798 & 66.312 \\
pet3 & 1000 & 200 & 200/0 & 8.470 & 13.788 & 35.619 \\
pet4 & 1000 & 200 & 200/0 & 8.828 & 14.489 & 43.222 \\
pet5 & 1000 & 200 & 200/0 & 8.894 & 14.097 & 41.606 \\
pet6 & 1000 & 200 & 200/0 & 8.678 & 14.505 & 39.838 \\
pet7 & 1000 & 200 & 200/0 & 8.504 & 9.147 & 37.156 \\
\bottomrule
\end{tabular*}
\end{table*}

\newpage

\subsection{MDKP: QAOA-style methods}
\label{sec:appendix_mdkp_qaoa_family}

Tables~\ref{tab:mdkp_qaoa_circuit_compilation}--\ref{tab:mdkp_qaoa_execution_robustness}
report diagnostics for QAOA,
Tables~\ref{tab:mdkp_maqaoa_circuit_compilation}--\ref{tab:mdkp_maqaoa_execution_robustness}
for MA-QAOA, and
Tables~\ref{tab:mdkp_wsqaoa_circuit_compilation}--\ref{tab:mdkp_wsqaoa_execution_robustness}
for WS-QAOA\@.
The most notable compilation effect in this family is the rapid growth of
transpiled two-qubit gate counts for WS-QAOA on larger instances, where
the warm-start operator introduces substantial additional entangling
structure.

\begin{table*}[htbp]
\caption{\label{tab:mdkp_qaoa_circuit_compilation}
MDKP (QAOA): circuit and compilation statistics. 
Logical metrics (qubits, depth, gates, two-qubit gates, parameters) and post-transpilation metrics per backend.}
\centering
\footnotesize
\setlength{\tabcolsep}{3pt}
\begin{tabular*}{\textwidth}{@{\extracolsep{\fill}}llccccccccc}
\toprule
 & & \multicolumn{5}{c}{\textbf{Logical}} & \multicolumn{3}{c}{\textbf{Transpiled}} \\
\cmidrule(lr){3-7} \cmidrule(lr){8-10}
\textbf{Inst.} & \textbf{Backend} & \textbf{Q} & \textbf{D} & \textbf{G} & \textbf{2Q} & \textbf{P} & \textbf{D} & \textbf{G} & \textbf{2Q} \\
\midrule
hp1 & ibm\_fez & 60 & 220 & 4245 & 3765 & 6 & 11330 & 86546 & 22501 \\
hp1 & ibm\_torino & 60 & 220 & 4245 & 3765 & 6 & 12280 & 86578 & 22621 \\
hp2 & ibm\_fez & 67 & 248 & 5552 & 5016 & 6 & 13857 & 111513 & 28967 \\
hp2 & ibm\_torino & 67 & 248 & 5552 & 5016 & 6 & 14514 & 115588 & 30329 \\
pb1 & ibm\_fez & 59 & 216 & 4060 & 3588 & 6 & 12461 & 83382 & 21821 \\
pb1 & ibm\_torino & 59 & 216 & 4060 & 3588 & 6 & 10922 & 80985 & 20909 \\
pb2 & ibm\_fez & 66 & 244 & 5346 & 4818 & 6 & 16222 & 111749 & 29320 \\
pb2 & ibm\_torino & 66 & 244 & 5346 & 4818 & 6 & 13674 & 108542 & 28059 \\
pb4 & ibm\_fez & 45 & 177 & 2316 & 1956 & 6 & 7176 & 44412 & 11522 \\
pb4 & ibm\_torino & 45 & 177 & 2316 & 1956 & 6 & 6179 & 43837 & 11119 \\
pb5 & ibm\_fez & 116 & 383 & 8443 & 7515 & 6 & 22923 & 178516 & 46275 \\
pb5 & ibm\_torino & 116 & 383 & 8443 & 7515 & 6 & 28035 & 181487 & 47838 \\
pet2 & ibm\_fez & 99 & 321 & 4581 & 3789 & 6 & 15014 & 92729 & 24261 \\
pet2 & ibm\_torino & 99 & 321 & 4581 & 3789 & 6 & 14955 & 90280 & 23440 \\
pet3 & ibm\_fez & 102 & 335 & 5958 & 5142 & 6 & 17066 & 124046 & 32232 \\
pet3 & ibm\_torino & 102 & 335 & 5958 & 5142 & 6 & 18158 & 122519 & 31868 \\
pet4 & ibm\_fez & 107 & 355 & 7405 & 6549 & 6 & 18534 & 158092 & 41325 \\
pet4 & ibm\_torino & 107 & 355 & 7405 & 6549 & 6 & 20703 & 157900 & 41078 \\
pet5 & ibm\_fez & 122 & 408 & 10777 & 9801 & 6 & 30564 & 239760 & 63249 \\
pet5 & ibm\_torino & 122 & 408 & 10777 & 9801 & 6 & 28907 & 237033 & 61641 \\
pet6 & ibm\_fez & 86 & 311 & 8329 & 7641 & 6 & 23359 & 178654 & 47056 \\
pet6 & ibm\_torino & 86 & 311 & 8329 & 7641 & 6 & 21612 & 175309 & 45464 \\
pet7 & ibm\_fez & 100 & 364 & 11774 & 10974 & 6 & 25821 & 254155 & 65365 \\
pet7 & ibm\_torino & 100 & 364 & 11774 & 10974 & 6 & 30778 & 251117 & 64817 \\
\bottomrule
\end{tabular*}
\end{table*}
\begin{table*}[htbp]
\caption{\label{tab:mdkp_qaoa_execution_robustness}
MDKP (QAOA): execution robustness summary. 
Shots per evaluation, number of evaluations, job success counts, latency statistics, and wall-clock time.}
\centering
\footnotesize
\setlength{\tabcolsep}{4pt}
\begin{tabular*}{\textwidth}{@{\extracolsep{\fill}}lcccccc}
\toprule
\textbf{Inst.} & \textbf{Shots} & \textbf{Evals} & \textbf{Jobs (ok/fail)} & \textbf{Med.\ Lat.\ (s)} & \textbf{P95 Lat.\ (s)} & \textbf{Time (min)} \\
\midrule
hp1 & 1000 & 56 & 56/0 & 16.565 & 22.136 & 19.152 \\
hp2 & 1000 & 60 & 60/0 & 16.688 & 51.128 & 28.525 \\
pb1 & 1000 & 49 & 49/0 & 16.204 & 126.202 & 25.526 \\
pb2 & 1000 & 49 & 49/0 & 16.470 & 94.612 & 25.196 \\
pb4 & 1000 & 57 & 57/0 & 15.213 & 112.305 & 29.595 \\
pb5 & 1000 & 53 & 53/0 & 17.840 & 33.536 & 19.755 \\
pet2 & 1000 & 63 & 63/0 & 15.675 & 17.464 & 18.786 \\
pet3 & 1000 & 53 & 53/0 & 16.759 & 36.663 & 23.444 \\
pet4 & 1000 & 53 & 53/0 & 17.232 & 65.362 & 24.957 \\
pet5 & 1000 & 56 & 56/0 & 24.119 & 43.775 & 66.174 \\
pet6 & 1000 & 51 & 51/0 & 17.597 & 27.909 & 23.111 \\
pet7 & 1000 & 58 & 58/0 & 23.737 & 57.358 & 31.400 \\
\bottomrule
\end{tabular*}
\end{table*}
\begin{table*}[htbp]
\caption{\label{tab:mdkp_maqaoa_circuit_compilation}
MDKP (MA-QAOA): circuit and compilation statistics. 
Logical metrics (qubits, depth, gates, two-qubit gates, parameters) and post-transpilation metrics per backend.}
\centering
\footnotesize
\setlength{\tabcolsep}{3pt}
\begin{tabular*}{\textwidth}{@{\extracolsep{\fill}}llccccccccc}
\toprule
 & & \multicolumn{5}{c}{\textbf{Logical}} & \multicolumn{3}{c}{\textbf{Transpiled}} \\
\cmidrule(lr){3-7} \cmidrule(lr){8-10}
\textbf{Inst.} & \textbf{Backend} & \textbf{Q} & \textbf{D} & \textbf{G} & \textbf{2Q} & \textbf{P} & \textbf{D} & \textbf{G} & \textbf{2Q} \\
\midrule
hp1 & ibm\_fez & 60 & 220 & 4245 & 3765 & 4125 & 13221 & 86775 & 22896 \\
hp1 & ibm\_torino & 60 & 220 & 4245 & 3765 & 4125 & 12323 & 85215 & 22019 \\
hp2 & ibm\_fez & 67 & 248 & 5552 & 5016 & 5418 & 16548 & 114826 & 30125 \\
hp2 & ibm\_torino & 67 & 248 & 5552 & 5016 & 5418 & 14728 & 114316 & 29687 \\
pb1 & ibm\_fez & 59 & 216 & 4060 & 3588 & 3942 & 12245 & 82684 & 21754 \\
pb1 & ibm\_torino & 59 & 216 & 4060 & 3588 & 3942 & 12517 & 82578 & 21758 \\
pb2 & ibm\_fez & 66 & 244 & 5346 & 4818 & 5214 & 13485 & 108642 & 28136 \\
pb2 & ibm\_torino & 66 & 244 & 5346 & 4818 & 5214 & 14160 & 110256 & 28448 \\
pb4 & ibm\_fez & 45 & 177 & 2316 & 1956 & 2226 & 7424 & 44664 & 11506 \\
pb4 & ibm\_torino & 45 & 177 & 2316 & 1956 & 2226 & 7190 & 44751 & 11769 \\
pb5 & ibm\_fez & 116 & 383 & 8443 & 7515 & 8211 & 23279 & 179959 & 46919 \\
pb5 & ibm\_torino & 116 & 383 & 8443 & 7515 & 8211 & 21192 & 178172 & 46025 \\
pet2 & ibm\_fez & 99 & 321 & 4581 & 3789 & 4383 & 16045 & 90960 & 23551 \\
pet2 & ibm\_torino & 99 & 321 & 4581 & 3789 & 4383 & 14688 & 90967 & 23581 \\
pet3 & ibm\_fez & 102 & 335 & 5958 & 5142 & 5754 & 17605 & 124917 & 32638 \\
pet3 & ibm\_torino & 102 & 335 & 5958 & 5142 & 5754 & 16966 & 122721 & 32058 \\
pet4 & ibm\_fez & 107 & 355 & 7405 & 6549 & 7191 & 20985 & 159546 & 41625 \\
pet4 & ibm\_torino & 107 & 355 & 7405 & 6549 & 7191 & 24080 & 160513 & 41971 \\
pet5 & ibm\_fez & 122 & 408 & 10777 & 9801 & 10533 & 29079 & 239213 & 62656 \\
pet5 & ibm\_marrakesh & 122 & 408 & 10777 & 9801 & 10533 & 27861 & 234668 & 60929 \\
pet5 & ibm\_torino & 122 & 408 & 10777 & 9801 & 10533 & 30942 & 244192 & 64103 \\
pet6 & ibm\_fez & 86 & 311 & 8329 & 7641 & 8157 & 23220 & 174830 & 45006 \\
pet6 & ibm\_marrakesh & 86 & 311 & 8329 & 7641 & 8157 & 20645 & 175473 & 45943 \\
pet6 & ibm\_torino & 86 & 311 & 8329 & 7641 & 8157 & 24555 & 181081 & 47282 \\
pet7 & ibm\_marrakesh & 100 & 364 & 11774 & 10974 & 11574 & 31645 & 256899 & 66556 \\
pet7 & ibm\_torino & 100 & 364 & 11774 & 10974 & 11574 & 27783 & 252845 & 65806 \\
\bottomrule
\end{tabular*}
\end{table*}
\begin{table*}[htbp]
\caption{\label{tab:mdkp_maqaoa_execution_robustness}
MDKP (MA-QAOA): execution robustness summary. 
Shots per evaluation, number of evaluations, job success counts, latency statistics, and wall-clock time.}
\centering
\footnotesize
\setlength{\tabcolsep}{4pt}
\begin{tabular*}{\textwidth}{@{\extracolsep{\fill}}lcccccc}
\toprule
\textbf{Inst.} & \textbf{Shots} & \textbf{Evals} & \textbf{Jobs (ok/fail)} & \textbf{Med.\ Lat.\ (s)} & \textbf{P95 Lat.\ (s)} & \textbf{Time (min)} \\
\midrule
hp1 & 1000 & 200 & 200/0 & 15.656 & 31.904 & 74.174 \\
hp2 & 1000 & 200 & 200/0 & 16.204 & 26.851 & 68.616 \\
pb1 & 1000 & 200 & 200/0 & 15.768 & 31.438 & 78.120 \\
pb2 & 1000 & 200 & 200/0 & 16.224 & 81.988 & 127.219 \\
pb4 & 1000 & 200 & 200/0 & 15.253 & 25.794 & 63.497 \\
pb5 & 1000 & 200 & 200/0 & 22.260 & 34.752 & 85.626 \\
pet2 & 1000 & 200 & 200/0 & 15.737 & 86.152 & 86.775 \\
pet3 & 1000 & 200 & 200/0 & 16.482 & 195.877 & 245.035 \\
pet4 & 1000 & 200 & 200/0 & 22.304 & 125.021 & 154.754 \\
pet5 & 1000 & 200 & 200/0 & 23.493 & 61.140 & 120.356 \\
pet6 & 1000 & 200 & 200/0 & 22.777 & 77.704 & 128.654 \\
pet7 & 1000 & 200 & 200/0 & 24.330 & 72.478 & 127.831 \\
\bottomrule
\end{tabular*}
\end{table*}
\begin{table*}[htbp]
\caption{\label{tab:mdkp_wsqaoa_circuit_compilation}
MDKP (WS-QAOA): circuit and compilation statistics. 
Logical metrics (qubits, depth, gates, two-qubit gates, parameters) and post-transpilation metrics per backend.}
\centering
\footnotesize
\setlength{\tabcolsep}{3pt}
\begin{tabular*}{\textwidth}{@{\extracolsep{\fill}}llccccccccc}
\toprule
 & & \multicolumn{5}{c}{\textbf{Logical}} & \multicolumn{3}{c}{\textbf{Transpiled}} \\
\cmidrule(lr){3-7} \cmidrule(lr){8-10}
\textbf{Inst.} & \textbf{Backend} & \textbf{Q} & \textbf{D} & \textbf{G} & \textbf{2Q} & \textbf{P} & \textbf{D} & \textbf{G} & \textbf{2Q} \\
\midrule
hp1 & ibm\_fez & 60 & 226 & 4605 & 3765 & 6 & 11913 & 82642 & 20658 \\
hp1 & ibm\_torino & 60 & 226 & 4605 & 3765 & 6 & 11945 & 88193 & 23114 \\
hp2 & ibm\_fez & 67 & 254 & 5954 & 5016 & 6 & 14710 & 114305 & 29780 \\
hp2 & ibm\_torino & 67 & 254 & 5954 & 5016 & 6 & 14804 & 112124 & 29111 \\
pb1 & ibm\_fez & 59 & 222 & 4414 & 3588 & 6 & 12646 & 83889 & 22192 \\
pb1 & ibm\_torino & 59 & 222 & 4414 & 3588 & 6 & 11390 & 82514 & 21601 \\
pb2 & ibm\_fez & 66 & 250 & 5742 & 4818 & 6 & 15456 & 110628 & 28862 \\
pb2 & ibm\_torino & 66 & 250 & 5742 & 4818 & 6 & 13743 & 109202 & 28131 \\
pb4 & ibm\_fez & 45 & 183 & 2586 & 1956 & 6 & 7172 & 44994 & 11808 \\
pb4 & ibm\_torino & 45 & 183 & 2586 & 1956 & 6 & 8043 & 45132 & 11698 \\
pb5 & ibm\_fez & 116 & 389 & 9139 & 7515 & 6 & 21330 & 181673 & 47353 \\
pb5 & ibm\_torino & 116 & 389 & 9139 & 7515 & 6 & 24725 & 181525 & 47419 \\
pet2 & ibm\_fez & 99 & 327 & 5175 & 3789 & 6 & 14412 & 91265 & 23747 \\
pet2 & ibm\_torino & 99 & 327 & 5175 & 3789 & 6 & 12887 & 91099 & 23458 \\
pet3 & ibm\_fez & 102 & 341 & 6570 & 5142 & 6 & 17378 & 124527 & 32589 \\
pet3 & ibm\_torino & 102 & 341 & 6570 & 5142 & 6 & 16250 & 122289 & 31727 \\
pet4 & ibm\_torino & 107 & 361 & 8047 & 6549 & 6 & 20281 & 159973 & 41282 \\
pet5 & ibm\_torino & 122 & 414 & 11509 & 9801 & 6 & 30377 & 245792 & 64628 \\
pet6 & ibm\_torino & 86 & 317 & 8845 & 7641 & 6 & 20991 & 178066 & 45559 \\
pet7 & ibm\_torino & 100 & 370 & 12374 & 10974 & 6 & 27930 & 254296 & 65953 \\
\bottomrule
\end{tabular*}
\end{table*}
\begin{table*}[htbp]
\caption{\label{tab:mdkp_wsqaoa_execution_robustness}
MDKP (WS-QAOA): execution robustness summary. 
Shots per evaluation, number of evaluations, job success counts, latency statistics, and wall-clock time.}
\centering
\footnotesize
\setlength{\tabcolsep}{4pt}
\begin{tabular*}{\textwidth}{@{\extracolsep{\fill}}lcccccc}
\toprule
\textbf{Inst.} & \textbf{Shots} & \textbf{Evals} & \textbf{Jobs (ok/fail)} & \textbf{Med.\ Lat.\ (s)} & \textbf{P95 Lat.\ (s)} & \textbf{Time (min)} \\
\midrule
hp1 & 1000 & 60 & 60/0 & 15.780 & 21.253 & 19.351 \\
hp2 & 1000 & 56 & 56/0 & 16.218 & 17.868 & 17.124 \\
pb1 & 1000 & 68 & 68/0 & 16.047 & 43.036 & 30.552 \\
pb2 & 1000 & 55 & 55/0 & 16.148 & 22.355 & 18.115 \\
pb4 & 1000 & 63 & 63/0 & 15.391 & 20.746 & 21.395 \\
pb5 & 1000 & 58 & 58/0 & 22.204 & 27.670 & 22.251 \\
pet2 & 1000 & 55 & 55/0 & 15.794 & 21.586 & 17.101 \\
pet3 & 1000 & 54 & 54/0 & 16.349 & 71.712 & 24.479 \\
pet4 & 1000 & 49 & 49/0 & 16.403 & 17.710 & 15.347 \\
pet5 & 1000 & 55 & 55/0 & 23.262 & 23.984 & 23.728 \\
pet6 & 1000 & 45 & 45/0 & 16.956 & 22.491 & 17.986 \\
pet7 & 1000 & 57 & 57/0 & 23.478 & 25.012 & 25.352 \\
\bottomrule
\end{tabular*}
\end{table*}

\newpage

\subsection{MDKP: encoding-based methods}
\label{sec:appendix_mdkp_encoding_family}

Tables~\ref{tab:mdkp_pce_circuit_compilation}--\ref{tab:mdkp_pce_execution_robustness}
report diagnostics for PCE, and
Tables~\ref{tab:mdkp_qrao_circuit_compilation}--\ref{tab:mdkp_qrao_execution_robustness}
for QRAO\@.
Despite their substantially lower circuit width, both encoding methods
produce non-trivial post-transpilation circuits.  In particular, PCE
circuits have higher circuit depth than VQE circuits on the same
instances, illustrating the trade-off between qubit compression and
circuit complexity that is central to the benchmark's findings.

\begin{table*}[htbp]
\caption{\label{tab:mdkp_pce_circuit_compilation}
MDKP (PCE): circuit and compilation statistics. 
Logical metrics (qubits, depth, gates, two-qubit gates, parameters) and post-transpilation metrics per backend.}
\centering
\footnotesize
\setlength{\tabcolsep}{3pt}
\begin{tabular*}{\textwidth}{@{\extracolsep{\fill}}llccccccccc}
\toprule
 & & \multicolumn{5}{c}{\textbf{Logical}} & \multicolumn{3}{c}{\textbf{Transpiled}} \\
\cmidrule(lr){3-7} \cmidrule(lr){8-10}
\textbf{Inst.} & \textbf{Backend} & \textbf{Q} & \textbf{D} & \textbf{G} & \textbf{2Q} & \textbf{P} & \textbf{D} & \textbf{G} & \textbf{2Q} \\
\midrule
hp1 & ibm\_torino & 7 & 57 & 287 & 84 & 182 & 393 & 1493 & 168 \\
hp2 & ibm\_torino & 8 & 65 & 376 & 112 & 240 & 449 & 1979 & 224 \\
pb1 & ibm\_torino & 7 & 57 & 287 & 84 & 182 & 393 & 1493 & 168 \\
pb2 & ibm\_torino & 8 & 65 & 376 & 112 & 240 & 449 & 1979 & 224 \\
pb4 & ibm\_torino & 6 & 49 & 210 & 60 & 132 & 337 & 1086 & 120 \\
pb5 & ibm\_fez & 10 & 81 & 590 & 180 & 380 & 561 & 3135 & 360 \\
pb5 & ibm\_torino & 10 & 81 & 590 & 180 & 380 & 561 & 3135 & 360 \\
pet2 & ibm\_fez & 9 & 73 & 477 & 144 & 306 & 505 & 2529 & 288 \\
pet2 & ibm\_torino & 9 & 73 & 477 & 144 & 306 & 505 & 2529 & 288 \\
pet3 & ibm\_fez & 9 & 73 & 477 & 144 & 306 & 505 & 2529 & 288 \\
pet3 & ibm\_torino & 9 & 73 & 477 & 144 & 306 & 505 & 2529 & 288 \\
pet4 & ibm\_fez & 9 & 73 & 477 & 144 & 306 & 505 & 2529 & 288 \\
pet4 & ibm\_torino & 9 & 73 & 477 & 144 & 306 & 505 & 2529 & 288 \\
pet5 & ibm\_fez & 10 & 81 & 590 & 180 & 380 & 561 & 3135 & 360 \\
pet5 & ibm\_torino & 10 & 81 & 590 & 180 & 380 & 561 & 3135 & 360 \\
pet6 & ibm\_fez & 9 & 73 & 477 & 144 & 306 & 505 & 2529 & 288 \\
pet6 & ibm\_torino & 9 & 73 & 477 & 144 & 306 & 505 & 2529 & 288 \\
pet7 & ibm\_fez & 9 & 73 & 477 & 144 & 306 & 505 & 2529 & 288 \\
pet7 & ibm\_torino & 9 & 73 & 477 & 144 & 306 & 505 & 2529 & 288 \\
\bottomrule
\end{tabular*}
\end{table*}
\begin{table*}[htbp]
\caption{\label{tab:mdkp_pce_execution_robustness}
MDKP (PCE): execution robustness summary. 
Shots per evaluation, number of evaluations, job success counts, latency statistics, and wall-clock time.}
\centering
\footnotesize
\setlength{\tabcolsep}{4pt}
\begin{tabular*}{\textwidth}{@{\extracolsep{\fill}}lcccccc}
\toprule
\textbf{Inst.} & \textbf{Shots} & \textbf{Evals} & \textbf{Jobs (ok/fail)} & \textbf{Med.\ Lat.\ (s)} & \textbf{P95 Lat.\ (s)} & \textbf{Time (min)} \\
\midrule
hp1 & 1000 & 200 & 200/0 & 8.66 & 14.32 & 40.42 \\
hp2 & 1000 & 200 & 200/0 & 8.61 & 13.98 & 37.01 \\
pb1 & 1000 & 200 & 200/0 & 8.62 & 19.32 & 43.53 \\
pb2 & 1000 & 200 & 200/0 & 8.62 & 20.14 & 42.26 \\
pb4 & 1000 & 200 & 200/0 & 8.62 & 14.74 & 41.06 \\
pb5 & 1000 & 200 & 200/0 & 8.72 & 46.52 & 72.70 \\
pet2 & 1000 & 200 & 200/0 & 8.75 & 30.59 & 47.60 \\
pet3 & 1000 & 200 & 200/0 & 8.81 & 127.30 & 147.46 \\
pet4 & 1000 & 200 & 200/0 & 8.95 & 73.63 & 71.73 \\
pet5 & 1000 & 200 & 200/0 & 8.87 & 14.78 & 40.72 \\
pet6 & 1000 & 200 & 200/0 & 8.96 & 30.31 & 45.87 \\
pet7 & 1000 & 200 & 200/0 & 9.07 & 58.03 & 77.65 \\
\bottomrule
\end{tabular*}
\end{table*}
\begin{table*}[htbp]
\caption{\label{tab:mdkp_qrao_circuit_compilation}
MDKP (QRAO): circuit and compilation statistics. 
Logical metrics (qubits, depth, gates, two-qubit gates, parameters) and post-transpilation metrics per backend.}
\centering
\footnotesize
\setlength{\tabcolsep}{3pt}
\begin{tabular*}{\textwidth}{@{\extracolsep{\fill}}llccccccccc}
\toprule
 & & \multicolumn{5}{c}{\textbf{Logical}} & \multicolumn{3}{c}{\textbf{Transpiled}} \\
\cmidrule(lr){3-7} \cmidrule(lr){8-10}
\textbf{Inst.} & \textbf{Backend} & \textbf{Q} & \textbf{D} & \textbf{G} & \textbf{2Q} & \textbf{P} & \textbf{D} & \textbf{G} & \textbf{2Q} \\
\midrule
hp1 & ibm\_fez & 38 & 49 & 415 & 111 & 304 & 61 & 871 & 111 \\
hp2 & ibm\_fez & 45 & 56 & 492 & 132 & 360 & 68 & 1032 & 132 \\
pb1 & ibm\_fez & 37 & 48 & 404 & 108 & 296 & 60 & 848 & 108 \\
pb2 & ibm\_marrakesh & 44 & 55 & 481 & 129 & 352 & 67 & 1009 & 129 \\
pb4 & ibm\_marrakesh & 32 & 43 & 349 & 93 & 256 & 55 & 733 & 93 \\
pb5 & ibm\_marrakesh & 57 & 68 & 624 & 168 & 456 & 80 & 1308 & 168 \\
pet2 & ibm\_torino & 44 & 55 & 481 & 129 & 352 & 162 & 1522 & 129 \\
pet3 & ibm\_marrakesh & 44 & 55 & 481 & 129 & 352 & 67 & 1009 & 129 \\
pet4 & ibm\_torino & 47 & 58 & 514 & 138 & 376 & 171 & 1627 & 138 \\
pet5 & ibm\_fez & 58 & 69 & 635 & 171 & 464 & 81 & 1331 & 171 \\
pet5 & ibm\_marrakesh & 58 & 69 & 635 & 171 & 464 & 81 & 1331 & 171 \\
pet6 & ibm\_fez & 52 & 63 & 569 & 153 & 416 & 75 & 1193 & 153 \\
pet6 & ibm\_marrakesh & 52 & 63 & 569 & 153 & 416 & 75 & 1193 & 153 \\
pet7 & ibm\_fez & 62 & 73 & 679 & 183 & 496 & 85 & 1423 & 183 \\
pet7 & ibm\_marrakesh & 62 & 73 & 679 & 183 & 496 & 85 & 1423 & 183 \\
\bottomrule
\end{tabular*}
\end{table*}
\begin{table*}[htbp]
\caption{\label{tab:mdkp_qrao_execution_robustness}
MDKP (QRAO): execution robustness summary. 
Shots per evaluation, number of evaluations, job success counts, latency statistics, and wall-clock time.}
\centering
\footnotesize
\setlength{\tabcolsep}{4pt}
\begin{tabular*}{\textwidth}{@{\extracolsep{\fill}}lcccccc}
\toprule
\textbf{Inst.} & \textbf{Shots} & \textbf{Evals} & \textbf{Jobs (ok/fail)} & \textbf{Med.\ Lat.\ (s)} & \textbf{P95 Lat.\ (s)} & \textbf{Time (min)} \\
\midrule
hp1 & 1000 & 200 & 200/0 & 30.95 & 66.41 & 237.44 \\
hp2 & 1000 & 200 & 200/0 & 32.80 & 91.78 & 227.31 \\
pb1 & 1000 & 200 & 200/0 & 29.60 & 57.16 & 159.12 \\
pb2 & 1000 & 200 & 200/0 & 31.00 & 69.14 & 138.50 \\
pb4 & 1000 & 200 & 200/0 & 24.50 & 37.49 & 95.23 \\
pb5 & 1000 & 200 & 200/0 & 31.85 & 65.62 & 188.67 \\
pet2 & 1000 & 200 & 200/0 & 24.80 & 30.16 & 96.92 \\
pet3 & 1000 & 200 & 200/0 & 31.95 & 80.71 & 157.92 \\
pet4 & 1000 & 200 & 200/0 & 28.85 & 50.72 & 115.51 \\
pet5 & 1000 & 200 & 200/0 & 36.55 & 165.74 & 258.83 \\
pet6 & 1000 & 200 & 200/0 & 35.30 & 113.99 & 229.84 \\
pet7 & 1000 & 200 & 200/0 & 42.80 & 128.37 & 221.51 \\
\bottomrule
\end{tabular*}
\end{table*}

\clearpage

\section{Hardware Diagnostics: MIS}
\label{app:hw_mis}

This appendix reports compilation and execution diagnostics for all
method families on the MIS benchmark.  These diagnostics are particularly
important for MIS because the main results
(Section~\ref{sec:results_mis}) reveal a sharp feasibility cliff as
instance size increases.  The tables below show how transpiled circuit
depth and two-qubit gate counts grow across instances and backends,
providing the hardware-level context for interpreting the observed
feasibility breakdown.

\subsection{MIS: VQE-style methods}
\label{sec:appendix_mis_vqe_family}

Tables~\ref{tab:mis_vqe_circuit_compilation}--\ref{tab:mis_vqe_execution_robustness}
report diagnostics for VQE, and
Tables~\ref{tab:mis_cvar_vqe_circuit_compilation}--\ref{tab:mis_cvar_vqe_execution_robustness}
for CVaR-VQE\@.

\begin{table*}[htbp]
\caption{\label{tab:mis_vqe_circuit_compilation}
MIS (VQE): circuit and compilation statistics. 
Logical metrics (qubits, depth, gates, two-qubit gates, parameters) and post-transpilation metrics per backend.}
\centering
\footnotesize
\setlength{\tabcolsep}{3pt}
\begin{tabular*}{\textwidth}{@{\extracolsep{\fill}}llccccccccc}
\toprule
 & & \multicolumn{5}{c}{\textbf{Logical}} & \multicolumn{3}{c}{\textbf{Transpiled}} \\
\cmidrule(lr){3-7} \cmidrule(lr){8-10}
\textbf{Inst.} & \textbf{Backend} & \textbf{Q} & \textbf{D} & \textbf{G} & \textbf{2Q} & \textbf{P} & \textbf{D} & \textbf{G} & \textbf{2Q} \\
\midrule
1tc.8 & ibm\_fez & 8 & 29 & 64 & 24 & 32 & 174 & 387 & 78 \\
1tc.8 & ibm\_torino & 8 & 29 & 64 & 24 & 32 & 167 & 328 & 60 \\
1tc.16 & ibm\_fez & 16 & 53 & 128 & 48 & 64 & 261 & 623 & 102 \\
1tc.16 & ibm\_torino & 16 & 53 & 128 & 48 & 64 & 257 & 611 & 102 \\
1tc.32 & ibm\_fez & 32 & 101 & 256 & 96 & 128 & 395 & 928 & 96 \\
1tc.32 & ibm\_torino & 32 & 101 & 256 & 96 & 128 & 395 & 928 & 96 \\
1tc.64 & ibm\_fez & 64 & 197 & 512 & 192 & 256 & 779 & 1856 & 192 \\
1tc.64 & ibm\_torino & 64 & 197 & 512 & 192 & 256 & 779 & 1856 & 192 \\
1et.64 & ibm\_fez & 64 & 197 & 512 & 192 & 256 & 779 & 1856 & 192 \\
1et.64 & ibm\_torino & 64 & 197 & 512 & 192 & 256 & 779 & 1856 & 192 \\
1dc.64 & ibm\_fez & 64 & 197 & 512 & 192 & 256 & 779 & 1856 & 192 \\
1dc.64 & ibm\_torino & 64 & 197 & 512 & 192 & 256 & 779 & 1856 & 192 \\
1dc.128 & ibm\_fez & 128 & 389 & 1024 & 384 & 512 & 1931 & 4489 & 657 \\
1dc.128 & ibm\_torino & 128 & 389 & 1024 & 384 & 512 & 2965 & 6581 & 1403 \\
\bottomrule
\end{tabular*}
\end{table*}
\begin{table*}[htbp]
\caption{\label{tab:mis_vqe_execution_robustness}
MIS (VQE): execution robustness summary. 
Shots per evaluation, number of evaluations, job success counts, latency statistics, and wall-clock time.}
\centering
\footnotesize
\setlength{\tabcolsep}{4pt}
\begin{tabular*}{\textwidth}{@{\extracolsep{\fill}}lcccccc}
\toprule
\textbf{Inst.} & \textbf{Shots} & \textbf{Evals} & \textbf{Jobs (ok/fail)} & \textbf{Med.\ Lat.\ (s)} & \textbf{P95 Lat.\ (s)} & \textbf{Time (min)} \\
\midrule
1tc.8 & 1000 & 200 & 200/0 & 8.489 & 14.052 & 33.394 \\
1tc.16 & 1000 & 200 & 200/0 & 8.467 & 24.764 & 40.188 \\
1tc.32 & 1000 & 200 & 200/0 & 8.522 & 14.119 & 36.399 \\
1tc.64 & 1000 & 200 & 200/0 & 8.654 & 14.244 & 38.898 \\
1et.64 & 1000 & 200 & 200/0 & 8.437 & 13.826 & 35.473 \\
1dc.64 & 1000 & 200 & 200/0 & 8.437 & 13.952 & 35.324 \\
1dc.128 & 1000 & 200 & 200/0 & 8.797 & 19.425 & 62.911 \\
\bottomrule
\end{tabular*}
\end{table*}
\begin{table*}[htbp]
\caption{\label{tab:mis_cvar_vqe_circuit_compilation}
MIS (CVaR-VQE): circuit and compilation statistics. 
Logical metrics (qubits, depth, gates, two-qubit gates, parameters) and post-transpilation metrics per backend.}
\centering
\footnotesize
\setlength{\tabcolsep}{3pt}
\begin{tabular*}{\textwidth}{@{\extracolsep{\fill}}llccccccccc}
\toprule
 & & \multicolumn{5}{c}{\textbf{Logical}} & \multicolumn{3}{c}{\textbf{Transpiled}} \\
\cmidrule(lr){3-7} \cmidrule(lr){8-10}
\textbf{Inst.} & \textbf{Backend} & \textbf{Q} & \textbf{D} & \textbf{G} & \textbf{2Q} & \textbf{P} & \textbf{D} & \textbf{G} & \textbf{2Q} \\
\midrule
1tc.8 & ibm\_marrakesh & 8 & 29 & 64 & 24 & 32 & 170 & 331 & 60 \\
1tc.8 & ibm\_torino & 8 & 29 & 64 & 24 & 32 & 176 & 337 & 60 \\
1tc.16 & ibm\_marrakesh & 16 & 53 & 128 & 48 & 64 & 255 & 622 & 105 \\
1tc.16 & ibm\_torino & 16 & 53 & 128 & 48 & 64 & 257 & 610 & 102 \\
1tc.32 & ibm\_marrakesh & 32 & 101 & 256 & 96 & 128 & 395 & 928 & 96 \\
1tc.32 & ibm\_torino & 32 & 101 & 256 & 96 & 128 & 395 & 928 & 96 \\
1tc.64 & ibm\_marrakesh & 64 & 197 & 512 & 192 & 256 & 779 & 1856 & 192 \\
1tc.64 & ibm\_torino & 64 & 197 & 512 & 192 & 256 & 779 & 1856 & 192 \\
1et.64 & ibm\_marrakesh & 64 & 197 & 512 & 192 & 256 & 779 & 1856 & 192 \\
1et.64 & ibm\_torino & 64 & 197 & 512 & 192 & 256 & 779 & 1856 & 192 \\
1dc.64 & ibm\_marrakesh & 64 & 197 & 512 & 192 & 256 & 779 & 1856 & 192 \\
1dc.64 & ibm\_torino & 64 & 197 & 512 & 192 & 256 & 779 & 1856 & 192 \\
1dc.128 & ibm\_torino & 128 & 389 & 1024 & 384 & 512 & 2855 & 6213 & 1284 \\
\bottomrule
\end{tabular*}
\end{table*}
\begin{table*}[htbp]
\caption{\label{tab:mis_cvar_vqe_execution_robustness}
MIS (CVaR-VQE): execution robustness summary. 
Shots per evaluation, number of evaluations, job success counts, latency statistics, and wall-clock time.}
\centering
\footnotesize
\setlength{\tabcolsep}{4pt}
\begin{tabular*}{\textwidth}{@{\extracolsep{\fill}}lcccccc}
\toprule
\textbf{Inst.} & \textbf{Shots} & \textbf{Evals} & \textbf{Jobs (ok/fail)} & \textbf{Med.\ Lat.\ (s)} & \textbf{P95 Lat.\ (s)} & \textbf{Time (min)} \\
\midrule
1tc.8 & 1000 & 200 & 200/0 & 8.723 & 46.214 & 60.812 \\
1tc.16 & 1000 & 200 & 200/0 & 8.558 & 24.632 & 41.127 \\
1tc.32 & 1000 & 200 & 200/0 & 13.898 & 41.342 & 98.269 \\
1tc.64 & 1000 & 200 & 200/0 & 8.748 & 30.126 & 47.822 \\
1et.64 & 1000 & 200 & 200/0 & 19.782 & 24.163 & 58.046 \\
1dc.64 & 1000 & 200 & 200/0 & 19.621 & 21.412 & 59.166 \\
1dc.128 & 1000 & 200 & 200/0 & 8.650 & 14.456 & 39.590 \\
\bottomrule
\end{tabular*}
\end{table*}

\subsection{MIS: QAOA-style methods}
\label{sec:appendix_mis_qaoa_family}

Tables~\ref{tab:appendix_mis_qaoa_circuit_compilation}--\ref{tab:mis_qaoa_execution_robustness}
report diagnostics for QAOA,
Tables~\ref{tab:appendix_mis_ma_qaoa_circuit_compilation}--\ref{tab:appendix_mis_ma_qaoa_execution_robustness}
for MA-QAOA, and
Tables~\ref{tab:appendix_mis_wsqaoa_circuit_compilation}--\ref{tab:appendix_mis_wsqaoa_execution_robustness}
for WS-QAOA\@.
For MIS, the QAOA-style circuits grow rapidly with graph size due to the
problem Hamiltonian's edge structure, and routing overhead on
sparse-connectivity backends further amplifies the effective depth.

\begin{table*}[htbp]
\caption{\label{tab:appendix_mis_qaoa_circuit_compilation}
MIS (QAOA): circuit and compilation statistics. 
Logical metrics (qubits, depth, gates, two-qubit gates, parameters) and post-transpilation metrics per backend.}
\centering
\footnotesize
\setlength{\tabcolsep}{3pt}
\begin{tabular*}{\textwidth}{@{\extracolsep{\fill}}llccccccccc}
\toprule
 & & \multicolumn{5}{c}{\textbf{Logical}} & \multicolumn{3}{c}{\textbf{Transpiled}} \\
\cmidrule(lr){3-7} \cmidrule(lr){8-10}
\textbf{Inst.} & \textbf{Backend} & \textbf{Q} & \textbf{D} & \textbf{G} & \textbf{2Q} & \textbf{P} & \textbf{D} & \textbf{G} & \textbf{2Q} \\
\midrule
1tc.8 & ibm\_torino & 8 & 17 & 82 & 18 & 6 & 124 & 390 & 60 \\
1tc.16 & ibm\_torino & 16 & 29 & 194 & 66 & 6 & 442 & 1233 & 241 \\
1tc.32 & ibm\_torino & 32 & 41 & 460 & 204 & 6 & 826 & 4141 & 935 \\
1tc.64 & ibm\_torino & 64 & 66 & 1088 & 576 & 6 & 1880 & 18387 & 4600 \\
1et.64 & ibm\_torino & 64 & 76 & 1304 & 792 & 6 & 1880 & 18387 & 4600 \\
1dc.64 & ibm\_fez & 64 & 155 & 2141 & 1629 & 6 & 7823 & 51710 & 14383 \\
1dc.64 & ibm\_torino & 64 & 155 & 2141 & 1629 & 6 & 7230 & 50389 & 13924 \\
1dc.128 & ibm\_fez & 128 & 242 & 5437 & 4413 & 6 & 16655 & 169152 & 48702 \\
1dc.128 & ibm\_torino & 128 & 242 & 5437 & 4413 & 6 & 17379 & 171179 & 49279 \\
\bottomrule
\end{tabular*}
\end{table*}
\begin{table*}[htbp]
\caption{\label{tab:mis_qaoa_execution_robustness}
MIS (QAOA): execution robustness summary. 
Shots per evaluation, number of evaluations, job success counts, latency statistics, and wall-clock time.}
\centering
\footnotesize
\setlength{\tabcolsep}{4pt}
\begin{tabular*}{\textwidth}{@{\extracolsep{\fill}}lcccccc}
\toprule
\textbf{Inst.} & \textbf{Shots} & \textbf{Evals} & \textbf{Jobs (ok/fail)} & \textbf{Med.\ Lat.\ (s)} & \textbf{P95 Lat.\ (s)} & \textbf{Time (min)} \\
\midrule
1tc.8 & 1000 & 50 & 50/0 & 13.72 & 14.67 & 13.10 \\
1tc.16 & 1000 & 62 & 62/0 & 8.75 & 14.09 & 11.91 \\
1tc.32 & 1000 & 56 & 56/0 & 8.74 & 19.46 & 12.35 \\
1tc.64 & 1000 & 59 & 59/0 & 8.85 & 12.41 & 10.38 \\
1et.64 & 1000 & 65 & 65/0 & 9.08 & 19.64 & 15.14 \\
1dc.64 & 1000 & 59 & 59/0 & 15.02 & 18.49 & 17.88 \\
1dc.128 & 1000 & 56 & 56/0 & 16.64 & 21.78 & 17.89 \\
\bottomrule
\end{tabular*}
\end{table*}
\begin{table*}[htbp]
\caption{\label{tab:appendix_mis_ma_qaoa_circuit_compilation}
MIS (MA-QAOA): circuit and compilation statistics. 
Logical metrics (qubits, depth, gates, two-qubit gates, parameters) and post-transpilation metrics per backend.}
\centering
\footnotesize
\setlength{\tabcolsep}{3pt}
\begin{tabular*}{\textwidth}{@{\extracolsep{\fill}}llccccccccc}
\toprule
 & & \multicolumn{5}{c}{\textbf{Logical}} & \multicolumn{3}{c}{\textbf{Transpiled}} \\
\cmidrule(lr){3-7} \cmidrule(lr){8-10}
\textbf{Inst.} & \textbf{Backend} & \textbf{Q} & \textbf{D} & \textbf{G} & \textbf{2Q} & \textbf{P} & \textbf{D} & \textbf{G} & \textbf{2Q} \\
\midrule
1tc.8 & ibm\_fez & 8 & 17 & 82 & 18 & 66 & 126 & 383 & 60 \\
1tc.8 & ibm\_torino & 8 & 17 & 82 & 18 & 66 & 126 & 383 & 60 \\
1tc.16 & ibm\_fez & 16 & 29 & 194 & 66 & 162 & 375 & 1226 & 237 \\
1tc.16 & ibm\_torino & 16 & 29 & 194 & 66 & 162 & 386 & 1232 & 232 \\
1tc.32 & ibm\_fez & 32 & 41 & 460 & 204 & 396 & 735 & 4159 & 931 \\
1tc.32 & ibm\_torino & 32 & 41 & 460 & 204 & 396 & 761 & 4112 & 915 \\
1tc.64 & ibm\_fez & 64 & 66 & 1088 & 576 & 960 & 1453 & 13369 & 3265 \\
1tc.64 & ibm\_torino & 64 & 66 & 1088 & 576 & 960 & 2305 & 13433 & 3274 \\
1et.64 & ibm\_fez & 64 & 76 & 1304 & 792 & 1176 & 1903 & 18247 & 4495 \\
1et.64 & ibm\_torino & 64 & 76 & 1304 & 792 & 1176 & 2160 & 18341 & 4548 \\
1dc.64 & ibm\_fez & 64 & 155 & 2141 & 1629 & 2013 & 7006 & 50752 & 14015 \\
1dc.64 & ibm\_torino & 64 & 155 & 2141 & 1629 & 2013 & 7354 & 50523 & 13926 \\
1dc.128 & ibm\_fez & 128 & 242 & 5437 & 4413 & 5181 & 15885 & 166696 & 47824 \\
1dc.128 & ibm\_torino & 128 & 242 & 5437 & 4413 & 5181 & 17098 & 167445 & 48272 \\
\bottomrule
\end{tabular*}
\end{table*}
\begin{table*}[htbp]
\caption{\label{tab:appendix_mis_ma_qaoa_execution_robustness}
MIS (MA-QAOA): execution robustness summary. 
Shots per evaluation, number of evaluations, job success counts, latency statistics, and wall-clock time.}
\centering
\footnotesize
\setlength{\tabcolsep}{4pt}
\begin{tabular*}{\textwidth}{@{\extracolsep{\fill}}lcccccc}
\toprule
\textbf{Inst.} & \textbf{Shots} & \textbf{Evals} & \textbf{Jobs (ok/fail)} & \textbf{Med.\ Lat.\ (s)} & \textbf{P95 Lat.\ (s)} & \textbf{Time (min)} \\
\midrule
1tc.8 & 1000 & 200 & 200/0 & 8.96 & 19.74 & 44.73 \\
1tc.16 & 1000 & 200 & 200/0 & 9.14 & 14.90 & 39.73 \\
1tc.32 & 1000 & 200 & 200/0 & 9.17 & 24.64 & 44.84 \\
1tc.64 & 1000 & 200 & 200/0 & 9.45 & 19.87 & 45.10 \\
1et.64 & 1000 & 200 & 200/0 & 14.51 & 67.91 & 86.82 \\
1dc.64 & 1000 & 200 & 200/0 & 15.25 & 20.68 & 58.58 \\
1dc.128 & 1000 & 200 & 200/0 & 17.31 & 44.04 & 89.61 \\
\bottomrule
\end{tabular*}
\end{table*}
\begin{table*}[htbp]
\caption{\label{tab:appendix_mis_wsqaoa_circuit_compilation}
MIS (WS-QAOA): circuit and compilation statistics. 
Logical metrics (qubits, depth, gates, two-qubit gates, parameters) and post-transpilation metrics per backend.}
\centering
\footnotesize
\setlength{\tabcolsep}{3pt}
\begin{tabular*}{\textwidth}{@{\extracolsep{\fill}}llccccccccc}
\toprule
 & & \multicolumn{5}{c}{\textbf{Logical}} & \multicolumn{3}{c}{\textbf{Transpiled}} \\
\cmidrule(lr){3-7} \cmidrule(lr){8-10}
\textbf{Inst.} & \textbf{Backend} & \textbf{Q} & \textbf{D} & \textbf{G} & \textbf{2Q} & \textbf{P} & \textbf{D} & \textbf{G} & \textbf{2Q} \\
\midrule
1tc.8 & ibm\_torino & 8 & 23 & 130 & 18 & 6 & 133 & 405 & 60 \\
1tc.16 & ibm\_torino & 16 & 35 & 290 & 66 & 6 & 456 & 1277 & 237 \\
1tc.32 & ibm\_torino & 32 & 47 & 652 & 204 & 6 & 849 & 4159 & 911 \\
1tc.64 & ibm\_torino & 64 & 72 & 1472 & 576 & 6 & 6986 & 50824 & 14029 \\
1et.64 & ibm\_torino & 64 & 82 & 1688 & 792 & 6 & 6986 & 50824 & 14029 \\
1dc.64 & ibm\_torino & 64 & 161 & 2525 & 1629 & 6 & 6986 & 50824 & 14029 \\
1dc.128 & ibm\_torino & 128 & 248 & 6205 & 4413 & 6 & 17168 & 169184 & 48544 \\
\bottomrule
\end{tabular*}
\end{table*}
\begin{table*}[htbp]
\caption{\label{tab:appendix_mis_wsqaoa_execution_robustness}
MIS (WS-QAOA): execution robustness summary. 
Shots per evaluation, number of evaluations, job success counts, latency statistics, and wall-clock time.}
\centering
\footnotesize
\setlength{\tabcolsep}{4pt}
\begin{tabular*}{\textwidth}{@{\extracolsep{\fill}}lcccccc}
\toprule
\textbf{Inst.} & \textbf{Shots} & \textbf{Evals} & \textbf{Jobs (ok/fail)} & \textbf{Med.\ Lat.\ (s)} & \textbf{P95 Lat.\ (s)} & \textbf{Time (min)} \\
\midrule
1tc.8 & 1000 & 64 & 64/0 & 8.56 & 13.77 & 10.59 \\
1tc.16 & 1000 & 60 & 60/0 & 8.89 & 41.30 & 15.88 \\
1tc.32 & 1000 & 72 & 72/0 & 8.78 & 22.29 & 15.08 \\
1tc.64 & 1000 & 59 & 59/0 & 14.79 & 69.14 & 22.13 \\
1et.64 & 1000 & 65 & 65/0 & 14.91 & 15.68 & 17.36 \\
1dc.64 & 1000 & 73 & 73/0 & 14.82 & 35.43 & 26.30 \\
1dc.128 & 1000 & 62 & 62/0 & 16.51 & 22.63 & 20.62 \\
\bottomrule
\end{tabular*}
\end{table*}

\subsection{MIS: encoding-based methods}
\label{sec:appendix_mis_encoding_family}

Tables~\ref{tab:mis_pce_circuit_compilation}--\ref{tab:mis_pce_execution_robustness}
report diagnostics for PCE, and
Tables~\ref{tab:mis_qrao_circuit_compilation}--\ref{tab:mis_qrao_execution_robustness}
for QRAO\@.
Although both methods compress the algorithmic-level circuit width substantially, PCE's
aggressive compression (e.g., 3 qubits for an 8-variable instance) comes
with high circuit depth, while QRAO uses more qubits but shallower
circuits.

\begin{table*}[htbp]
\caption{\label{tab:mis_pce_circuit_compilation}
MIS (PCE): circuit and compilation statistics. 
Logical metrics (qubits, depth, gates, two-qubit gates, parameters) and post-transpilation metrics per backend.}
\centering
\footnotesize
\setlength{\tabcolsep}{3pt}
\begin{tabular*}{\textwidth}{@{\extracolsep{\fill}}llccccccccc}
\toprule
 & & \multicolumn{5}{c}{\textbf{Logical}} & \multicolumn{3}{c}{\textbf{Transpiled}} \\
\cmidrule(lr){3-7} \cmidrule(lr){8-10}
\textbf{Inst.} & \textbf{Backend} & \textbf{Q} & \textbf{D} & \textbf{G} & \textbf{2Q} & \textbf{P} & \textbf{D} & \textbf{G} & \textbf{2Q} \\
\midrule
1tc.8 & ibm\_fez & 3 & 10 & 22 & 4 & 18 & 27 & 61 & 4 \\
1tc.8 & ibm\_marrakesh & 3 & 10 & 22 & 4 & 18 & 27 & 61 & 4 \\
1tc.16 & ibm\_torino & 4 & 11 & 30 & 6 & 24 & 30 & 84 & 6 \\
1tc.32 & ibm\_fez & 6 & 13 & 46 & 10 & 36 & 36 & 130 & 10 \\
1tc.32 & ibm\_marrakesh & 6 & 13 & 46 & 10 & 36 & 36 & 130 & 10 \\
1tc.64 & ibm\_torino & 8 & 15 & 62 & 14 & 48 & 42 & 176 & 14 \\
1et.64 & ibm\_fez & 8 & 15 & 62 & 14 & 48 & 42 & 176 & 14 \\
1et.64 & ibm\_marrakesh & 8 & 15 & 62 & 14 & 48 & 42 & 176 & 14 \\
1dc.64 & ibm\_fez & 8 & 15 & 62 & 14 & 48 & 42 & 176 & 14 \\
1dc.64 & ibm\_marrakesh & 8 & 15 & 62 & 14 & 48 & 42 & 176 & 14 \\
1dc.128 & ibm\_torino & 10 & 17 & 78 & 18 & 60 & 48 & 222 & 18 \\
\bottomrule
\end{tabular*}
\end{table*}
\begin{table*}[htbp]
\caption{\label{tab:mis_pce_execution_robustness}
MIS (PCE): execution robustness summary. 
Shots per evaluation, number of evaluations, job success counts, latency statistics, and wall-clock time.}
\centering
\footnotesize
\setlength{\tabcolsep}{4pt}
\begin{tabular*}{\textwidth}{@{\extracolsep{\fill}}lcccccc}
\toprule
\textbf{Inst.} & \textbf{Shots} & \textbf{Evals} & \textbf{Jobs (ok/fail)} & \textbf{Med.\ Lat.\ (s)} & \textbf{P95 Lat.\ (s)} & \textbf{Time (min)} \\
\midrule
1tc.8 & 1000 & 124 & 124/0 & 23.994 & 46.039 & 104.451 \\
1tc.16 & 1000 & 184 & 184/0 & 28.940 & 56.058 & 113.205 \\
1tc.32 & 1000 & 200 & 200/0 & 28.168 & 54.459 & 150.307 \\
1tc.64 & 1000 & 200 & 200/0 & 21.708 & 40.302 & 86.065 \\
1et.64 & 1000 & 200 & 200/0 & 27.535 & 53.285 & 136.793 \\
1dc.64 & 1000 & 200 & 200/0 & 22.786 & 111.521 & 167.134 \\
1dc.128 & 1000 & 200 & 200/0 & 21.618 & 30.380 & 75.589 \\
\bottomrule
\end{tabular*}
\end{table*}
\begin{table*}[htbp]
\caption{\label{tab:mis_qrao_circuit_compilation}
MIS (QRAO): circuit and compilation statistics. 
Logical metrics (qubits, depth, gates, two-qubit gates, parameters) and post-transpilation metrics per backend.}
\centering
\footnotesize
\setlength{\tabcolsep}{3pt}
\begin{tabular*}{\textwidth}{@{\extracolsep{\fill}}llccccccccc}
\toprule
 & & \multicolumn{5}{c}{\textbf{Logical}} & \multicolumn{3}{c}{\textbf{Transpiled}} \\
\cmidrule(lr){3-7} \cmidrule(lr){8-10}
\textbf{Inst.} & \textbf{Backend} & \textbf{Q} & \textbf{D} & \textbf{G} & \textbf{2Q} & \textbf{P} & \textbf{D} & \textbf{G} & \textbf{2Q} \\
\midrule
1tc.8 & statevector\_primitives & 4 & 15 & 41 & 9 & 32 & -- & -- & -- \\
1tc.16 & ibm\_fez & 6 & 17 & 63 & 15 & 48 & 48 & 192 & 15 \\
1tc.16 & ibm\_marrakesh & 6 & 17 & 63 & 15 & 48 & 48 & 192 & 15 \\
1tc.32 & ibm\_fez & 13 & 24 & 140 & 36 & 104 & 69 & 437 & 36 \\
1tc.32 & ibm\_marrakesh & 13 & 24 & 140 & 36 & 104 & 69 & 437 & 36 \\
1tc.64 & ibm\_torino & 23 & 34 & 250 & 66 & 184 & 99 & 787 & 66 \\
1et.64 & ibm\_fez & 23 & 34 & 250 & 66 & 184 & 99 & 787 & 66 \\
1et.64 & ibm\_marrakesh & 23 & 34 & 250 & 66 & 184 & 99 & 787 & 66 \\
1dc.64 & ibm\_torino & 25 & 36 & 272 & 72 & 200 & 105 & 857 & 72 \\
1dc.128 & ibm\_fez & 46 & 57 & 503 & 135 & 368 & 135 & 540 & 135 \\
1dc.128 & ibm\_marrakesh & 57 & 57 & 503 & 135 & 368 & 135 & 540 & 135 \\
\bottomrule
\end{tabular*}
\end{table*}
\begin{table*}[htbp]
\caption{\label{tab:mis_qrao_execution_robustness}
MIS (QRAO): execution robustness summary. 
Shots per evaluation, number of evaluations, job success counts, latency statistics, and wall-clock time.}
\centering
\footnotesize
\setlength{\tabcolsep}{4pt}
\begin{tabular*}{\textwidth}{@{\extracolsep{\fill}}lcccccc}
\toprule
\textbf{Inst.} & \textbf{Shots} & \textbf{Evals} & \textbf{Jobs (ok/fail)} & \textbf{Med.\ Lat.\ (s)} & \textbf{P95 Lat.\ (s)} & \textbf{Time (min)} \\
\midrule
1tc.8 & 1000 & 200 & 200/0 & -- & -- & 0.261 \\
1tc.16 & 1000 & 200 & 205/0 & 25.771 & 55.854 & 119.435 \\
1tc.32 & 1000 & 200 & 202/0 & 27.735 & 61.712 & 115.894 \\
1tc.64 & 1000 & 200 & 202/0 & 22.527 & 23.630 & 83.023 \\
1et.64 & 1000 & 200 & 202/0 & 28.560 & 51.449 & 115.114 \\
1dc.64 & 1000 & 200 & 202/0 & 26.062 & 58.875 & 114.982 \\
1dc.128 & 1000 & 200 & 201/0 & 31.121 & 52.935 & 160.077 \\
\bottomrule
\end{tabular*}
\end{table*}

\clearpage

\section{Hardware Diagnostics: QAP}
\label{app:hw_qap}

This appendix reports compilation and execution diagnostics for all
method families on the QAP benchmark.  QAP is the most demanding problem
in the study: its dense quadratic assignment structure produces the
largest encoded Hamiltonians and the highest post-transpilation gate
counts.  These diagnostics quantify the circuit burden that underlies
the complete infeasibility observed in Section~\ref{sec:results_qap}
and help explain why even qubit-efficient methods cannot recover
feasible solutions on current hardware for this problem class.

\subsection{QAP: VQE-style methods}
\label{sec:appendix_qap_vqe_family}

Tables~\ref{tab:qap_vqe_circuit_compilation}--\ref{tab:qap_vqe_execution_robustness}
report diagnostics for VQE, and
Tables~\ref{tab:qap_cvar_vqe_circuit_compilation}--\ref{tab:qap_cvar_vqe_execution_robustness}
for CVaR-VQE\@.
The 144-qubit circuit width required for the 12-variable QAP
instances already approaches the physical capacity of the tested
backends, and transpilation further amplifies circuit depth and
two-qubit gate counts.

\begin{table*}[htbp]
\caption{\label{tab:qap_vqe_circuit_compilation}
QAP (VQE): circuit and compilation statistics. 
Logical metrics (qubits, depth, gates, two-qubit gates, parameters) and post-transpilation metrics per backend.}
\centering
\footnotesize
\setlength{\tabcolsep}{3pt}
\begin{tabular*}{\textwidth}{@{\extracolsep{\fill}}llccccccccc}
\toprule
 & & \multicolumn{5}{c}{\textbf{Logical}} & \multicolumn{3}{c}{\textbf{Transpiled}} \\
\cmidrule(lr){3-7} \cmidrule(lr){8-10}
\textbf{Inst.} & \textbf{Backend} & \textbf{Q} & \textbf{D} & \textbf{G} & \textbf{2Q} & \textbf{P} & \textbf{D} & \textbf{G} & \textbf{2Q} \\
\midrule
chr12a & ibm\_fez & 144 & 437 & 1152 & 432 & 576 & 2995 & 6706 & 1323 \\
chr12b & ibm\_fez & 144 & 437 & 1152 & 432 & 576 & 2995 & 6706 & 1323 \\
chr12c & ibm\_fez & 144 & 441 & 1728 & 432 & 1152 & 2596 & 7380 & 1218 \\
nug12 & ibm\_fez & 144 & 441 & 1728 & 432 & 1152 & 2596 & 7380 & 1218 \\
had12 & ibm\_fez & 144 & 441 & 1728 & 432 & 1152 & 2596 & 7380 & 1218 \\
rou12 & ibm\_fez & 144 & 441 & 1728 & 432 & 1152 & 2596 & 7380 & 1218 \\
scr12 & ibm\_fez & 144 & 441 & 1728 & 432 & 1152 & 2596 & 7380 & 1218 \\
tai10a & ibm\_fez & 100 & 309 & 1200 & 300 & 800 & 919 & 3497 & 300 \\
tai10a & ibm\_torino & 100 & 309 & 1200 & 300 & 800 & 919 & 3497 & 300 \\
tai10b & ibm\_fez & 100 & 309 & 1200 & 300 & 800 & 919 & 3497 & 300 \\
tai10b & ibm\_torino & 100 & 309 & 1200 & 300 & 800 & 919 & 3497 & 300 \\
tai12a & ibm\_fez & 144 & 441 & 1728 & 432 & 1152 & 2596 & 7380 & 1218 \\
tai12b & ibm\_fez & 144 & 441 & 1728 & 432 & 1152 & 2458 & 7120 & 1134 \\
\bottomrule
\end{tabular*}
\end{table*}
\begin{table*}[htbp]
\caption{\label{tab:qap_vqe_execution_robustness}
QAP (VQE): execution robustness summary. 
Shots per evaluation, number of evaluations, job success counts, latency statistics, and wall-clock time.}
\centering
\footnotesize
\setlength{\tabcolsep}{4pt}
\begin{tabular*}{\textwidth}{@{\extracolsep{\fill}}lcccccc}
\toprule
\textbf{Inst.} & \textbf{Shots} & \textbf{Evals} & \textbf{Jobs (ok/fail)} & \textbf{Med.\ Lat.\ (s)} & \textbf{P95 Lat.\ (s)} & \textbf{Time (min)} \\
\midrule
chr12a & 1000 & 200 & 200/0 & 8.681 & 30.096 & 49.703 \\
chr12b & 1000 & 200 & 200/0 & 9.196 & 159.688 & 144.250 \\
chr12c & 1000 & 200 & 200/0 & 13.837 & 19.875 & 53.118 \\
nug12 & 1000 & 200 & 200/0 & 9.294 & 19.723 & 63.117 \\
had12 & 1000 & 200 & 200/0 & 14.300 & 41.324 & 89.017 \\
rou12 & 1000 & 200 & 200/0 & 9.535 & 79.171 & 109.948 \\
scr12 & 1000 & 200 & 200/0 & 9.242 & 74.311 & 133.229 \\
tai10a & 1000 & 200 & 200/0 & 9.362 & 31.067 & 62.898 \\
tai10b & 1000 & 200 & 200/0 & 9.286 & 20.099 & 60.371 \\
tai12a & 1000 & 200 & 200/0 & 14.304 & 41.058 & 74.305 \\
tai12b & 1000 & 200 & 200/0 & 14.039 & 30.613 & 100.887 \\
\bottomrule
\end{tabular*}
\end{table*}
\begin{table*}[htbp]
\caption{\label{tab:qap_cvar_vqe_circuit_compilation}
QAP (CVaR-VQE): circuit and compilation statistics. 
Logical metrics (qubits, depth, gates, two-qubit gates, parameters) and post-transpilation metrics per backend.}
\centering
\footnotesize
\setlength{\tabcolsep}{3pt}
\begin{tabular*}{\textwidth}{@{\extracolsep{\fill}}llccccccccc}
\toprule
 & & \multicolumn{5}{c}{\textbf{Logical}} & \multicolumn{3}{c}{\textbf{Transpiled}} \\
\cmidrule(lr){3-7} \cmidrule(lr){8-10}
\textbf{Inst.} & \textbf{Backend} & \textbf{Q} & \textbf{D} & \textbf{G} & \textbf{2Q} & \textbf{P} & \textbf{D} & \textbf{G} & \textbf{2Q} \\
\midrule
chr12a & ibm\_marrakesh & 144 & 437 & 1152 & 432 & 576 & 3026 & 6901 & 1389 \\
chr12b & ibm\_marrakesh & 144 & 437 & 1152 & 432 & 576 & 3015 & 6714 & 1329 \\
chr12c & ibm\_marrakesh & 144 & 437 & 1152 & 432 & 576 & 3026 & 6901 & 1389 \\
nug12 & ibm\_fez & 144 & 437 & 1152 & 432 & 576 & 3412 & 7473 & 1580 \\
had12 & ibm\_marrakesh & 144 & 437 & 1152 & 432 & 576 & 3026 & 6901 & 1389 \\
rou12 & ibm\_fez & 144 & 437 & 1152 & 432 & 576 & 3412 & 7473 & 1580 \\
scr12 & ibm\_fez & 144 & 437 & 1152 & 432 & 576 & 3412 & 7473 & 1580 \\
tai10a & ibm\_fez & 100 & 305 & 800 & 300 & 400 & 1211 & 2900 & 300 \\
tai10a & ibm\_torino & 100 & 305 & 800 & 300 & 400 & 1211 & 2900 & 300 \\
tai10b & ibm\_fez & 100 & 305 & 800 & 300 & 400 & 1211 & 2900 & 300 \\
tai10b & ibm\_torino & 100 & 305 & 800 & 300 & 400 & 1211 & 2900 & 300 \\
tai12a & ibm\_fez & 144 & 437 & 1152 & 432 & 576 & 3412 & 7473 & 1580 \\
tai12b & ibm\_fez & 144 & 437 & 1152 & 432 & 576 & 3412 & 7473 & 1580 \\
\bottomrule
\end{tabular*}
\end{table*}
\begin{table*}[htbp]
\caption{\label{tab:qap_cvar_vqe_execution_robustness}
QAP (CVaR-VQE): execution robustness summary. 
Shots per evaluation, number of evaluations, job success counts, latency statistics, and wall-clock time.}
\centering
\footnotesize
\setlength{\tabcolsep}{4pt}
\begin{tabular*}{\textwidth}{@{\extracolsep{\fill}}lcccccc}
\toprule
\textbf{Inst.} & \textbf{Shots} & \textbf{Evals} & \textbf{Jobs (ok/fail)} & \textbf{Med.\ Lat.\ (s)} & \textbf{P95 Lat.\ (s)} & \textbf{Time (min)} \\
\midrule
chr12a & 1000 & 200 & 200/0 & 8.507 & 14.091 & 40.771 \\
chr12b & 1000 & 200 & 200/0 & 8.848 & 30.082 & 54.166 \\
chr12c & 1000 & 200 & 200/0 & 8.924 & 73.387 & 119.847 \\
nug12 & 1000 & 200 & 200/0 & 8.680 & 24.723 & 63.019 \\
had12 & 1000 & 200 & 200/0 & 8.816 & 73.071 & 92.513 \\
rou12 & 1000 & 200 & 200/0 & 8.561 & 24.753 & 77.076 \\
scr12 & 1000 & 200 & 200/0 & 8.438 & 13.844 & 42.205 \\
tai10a & 1000 & 200 & 200/0 & 8.557 & 14.621 & 45.653 \\
tai10b & 1000 & 200 & 200/0 & 8.558 & 14.132 & 42.362 \\
tai12a & 1000 & 200 & 200/0 & 8.543 & 19.417 & 102.043 \\
tai12b & 1000 & 200 & 200/0 & 8.549 & 14.173 & 50.708 \\
\bottomrule
\end{tabular*}
\end{table*}

\subsection{QAP: QAOA-style methods}
\label{sec:appendix_qap_qaoa_family}

Tables~\ref{tab:qap_qaoa_circuit_compilation}--\ref{tab:qap_qaoa_execution_robustness}
report diagnostics for QAOA,
Tables~\ref{tab:qap_maqaoa_circuit_compilation}--\ref{tab:qap_maqaoa_execution_robustness}
for MA-QAOA, and
Tables~\ref{tab:qap_wsqaoa_circuit_compilation}--\ref{tab:qap_wsqaoa_execution_robustness}
for WS-QAOA\@.

\begin{table*}[htbp]
\caption{\label{tab:qap_qaoa_circuit_compilation}
QAP (QAOA): circuit and compilation statistics. 
Logical metrics (qubits, depth, gates, two-qubit gates, parameters) and post-transpilation metrics per backend.}
\centering
\footnotesize
\setlength{\tabcolsep}{3pt}
\begin{tabular*}{\textwidth}{@{\extracolsep{\fill}}llccccccccc}
\toprule
 & & \multicolumn{5}{c}{\textbf{Logical}} & \multicolumn{3}{c}{\textbf{Transpiled}} \\
\cmidrule(lr){3-7} \cmidrule(lr){8-10}
\textbf{Inst.} & \textbf{Backend} & \textbf{Q} & \textbf{D} & \textbf{G} & \textbf{2Q} & \textbf{P} & \textbf{D} & \textbf{G} & \textbf{2Q} \\
\midrule
chr12a & ibm\_fez & 144 & 350 & 10194 & 9042 & 6 & 33622 & 297432 & 84111 \\
chr12b & ibm\_fez & 144 & 383 & 10194 & 9042 & 6 & 33622 & 297432 & 84111 \\
chr12c & ibm\_fez & 144 & 350 & 10194 & 9042 & 6 & 33622 & 297432 & 84111 \\
nug12 & ibm\_marrakesh & 144 & 581 & 23724 & 22572 & 6 & 45119 & 586299 & 158163 \\
had12 & ibm\_fez & 144 & 581 & 32040 & 30888 & 6 & 33622 & 297432 & 84111 \\
rou12 & ibm\_marrakesh & 144 & 581 & 31644 & 30492 & 6 & 45119 & 586299 & 158163 \\
scr12 & ibm\_marrakesh & 144 & 504 & 16992 & 15840 & 6 & 45119 & 586299 & 158163 \\
tai10a & ibm\_fez & 100 & 405 & 15110 & 14310 & 6 & 21681 & 299467 & 74145 \\
tai10a & ibm\_marrakesh & 100 & 405 & 15110 & 14310 & 6 & 35011 & 312712 & 79041 \\
tai10a & ibm\_torino & 100 & 405 & 15110 & 14310 & 6 & 30214 & 308388 & 77265 \\
tai10b & ibm\_fez & 100 & 405 & 12950 & 12150 & 6 & 21681 & 299467 & 74145 \\
tai10b & ibm\_marrakesh & 100 & 405 & 12950 & 12150 & 6 & 35011 & 312712 & 79041 \\
tai12a & ibm\_fez & 144 & 581 & 31248 & 30096 & 6 & 53641 & 664051 & 168995 \\
tai12a & ibm\_marrakesh & 144 & 581 & 31248 & 30096 & 6 & 45119 & 586299 & 158163 \\
tai12b & ibm\_marrakesh & 144 & 570 & 27024 & 25872 & 6 & 45119 & 586299 & 158163 \\
\bottomrule
\end{tabular*}
\end{table*}
\begin{table*}[htbp]
\caption{\label{tab:qap_qaoa_execution_robustness}
QAP (QAOA): execution robustness summary. 
Shots per evaluation, number of evaluations, job success counts, latency statistics, and wall-clock time.}
\centering
\footnotesize
\setlength{\tabcolsep}{4pt}
\begin{tabular*}{\textwidth}{@{\extracolsep{\fill}}lcccccc}
\toprule
\textbf{Inst.} & \textbf{Shots} & \textbf{Evals} & \textbf{Jobs (ok/fail)} & \textbf{Med.\ Lat.\ (s)} & \textbf{P95 Lat.\ (s)} & \textbf{Time (min)} \\
\midrule
chr12a & 1000 & 60 & 60/0 & 23.779 & 52.064 & 34.534 \\
chr12b & 1000 & 63 & 63/0 & 23.834 & 133.728 & 50.833 \\
chr12c & 1000 & 64 & 64/0 & 23.821 & 39.282 & 32.922 \\
nug12 & 1000 & 53 & 53/0 & 40.163 & 200.741 & 59.933 \\
had12 & 1000 & 62 & 62/0 & 28.894 & 77.350 & 42.513 \\
rou12 & 1000 & 48 & 48/0 & 39.588 & 42.271 & 36.761 \\
scr12 & 1000 & 62 & 62/0 & 40.433 & 93.911 & 55.197 \\
tai10a & 1000 & 54 & 54/0 & 25.504 & 176.651 & 52.640 \\
tai10b & 1000 & 61 & 61/0 & 24.921 & 31.502 & 29.735 \\
tai12a & 1000 & 55 & 55/0 & 39.530 & 57.626 & 44.326 \\
tai12b & 1000 & 54 & 54/0 & 39.849 & 43.370 & 45.861 \\
\bottomrule
\end{tabular*}
\end{table*}
\begin{table*}[htbp]
\caption{\label{tab:qap_maqaoa_circuit_compilation}
QAP (MA-QAOA): circuit and compilation statistics. 
Logical metrics (qubits, depth, gates, two-qubit gates, parameters) and post-transpilation metrics per backend.}
\centering
\footnotesize
\setlength{\tabcolsep}{3pt}
\begin{tabular*}{\textwidth}{@{\extracolsep{\fill}}llccccccccc}
\toprule
 & & \multicolumn{5}{c}{\textbf{Logical}} & \multicolumn{3}{c}{\textbf{Transpiled}} \\
\cmidrule(lr){3-7} \cmidrule(lr){8-10}
\textbf{Inst.} & \textbf{Backend} & \textbf{Q} & \textbf{D} & \textbf{G} & \textbf{2Q} & \textbf{P} & \textbf{D} & \textbf{G} & \textbf{2Q} \\
\midrule
chr12a & ibm\_fez & 144 & 350 & 10194 & 9042 & 9906 & 32646 & 297939 & 83724 \\
chr12a & ibm\_marrakesh & 144 & 350 & 10194 & 9042 & 9906 & 31882 & 300715 & 84615 \\
chr12b & ibm\_fez & 144 & 383 & 10194 & 9042 & 9906 & 32378 & 298821 & 83936 \\
chr12b & ibm\_marrakesh & 144 & 383 & 10194 & 9042 & 9906 & 32937 & 298766 & 84276 \\
chr12c & ibm\_marrakesh & 144 & 350 & 10194 & 9042 & 9906 & 32937 & 298766 & 84276 \\
nug12 & ibm\_fez & 144 & 581 & 23724 & 22572 & 23436 & 50753 & 582411 & 158188 \\
nug12 & ibm\_marrakesh & 144 & 581 & 23724 & 22572 & 23436 & 49186 & 585337 & 159279 \\
had12 & ibm\_fez & 144 & 581 & 32040 & 30888 & 23436 & 58958 & 679545 & 173096 \\
had12 & ibm\_marrakesh & 144 & 581 & 32040 & 30888 & 23436 & 52379 & 670490 & 169151 \\
rou12 & ibm\_fez & 144 & 581 & 31644 & 30492 & 23436 & 58958 & 679545 & 173096 \\
rou12 & ibm\_marrakesh & 144 & 581 & 31644 & 30492 & 23436 & 52379 & 670490 & 169151 \\
scr12 & ibm\_fez & 144 & 504 & 16992 & 15840 & 16704 & 46354 & 464930 & 127688 \\
scr12 & ibm\_marrakesh & 144 & 504 & 16992 & 15840 & 16704 & 40375 & 461528 & 126628 \\
tai10a & ibm\_fez & 100 & 405 & 15110 & 14310 & 14910 & 38869 & 317872 & 81971 \\
tai10a & ibm\_marrakesh & 100 & 405 & 15110 & 14310 & 14910 & 26862 & 311241 & 79027 \\
tai10a & ibm\_torino & 100 & 405 & 15110 & 14310 & 14910 & 36950 & 312637 & 80095 \\
tai10b & ibm\_fez & 100 & 405 & 12950 & 12150 & 12750 & 28807 & 296936 & 78718 \\
tai10b & ibm\_marrakesh & 100 & 405 & 12950 & 12150 & 12750 & 27730 & 291243 & 77471 \\
tai12a & ibm\_fez & 144 & 581 & 31248 & 30096 & 23436 & 68253 & 664859 & 170894 \\
tai12a & ibm\_marrakesh & 144 & 581 & 31248 & 30096 & 23436 & 48566 & 654259 & 164264 \\
tai12b & ibm\_fez & 144 & 570 & 27024 & 25872 & 26736 & 59372 & 635760 & 167992 \\
tai12b & ibm\_marrakesh & 144 & 570 & 27024 & 25872 & 26736 & 55396 & 640370 & 168270 \\
\bottomrule
\end{tabular*}
\end{table*}
\begin{table*}[htbp]
\caption{\label{tab:qap_maqaoa_execution_robustness}
QAP (MA-QAOA): execution robustness summary. 
Shots per evaluation, number of evaluations, job success counts, latency statistics, and wall-clock time.}
\centering
\footnotesize
\setlength{\tabcolsep}{4pt}
\begin{tabular*}{\textwidth}{@{\extracolsep{\fill}}lcccccc}
\toprule
\textbf{Inst.} & \textbf{Shots} & \textbf{Evals} & \textbf{Jobs (ok/fail)} & \textbf{Med.\ Lat.\ (s)} & \textbf{P95 Lat.\ (s)} & \textbf{Time (min)} \\
\midrule
chr12a & 1000 & 200 & 200/0 & 24.858 & 35.665 & 102.356 \\
chr12b & 1000 & 200 & 200/0 & 25.002 & 65.119 & 203.732 \\
chr12c & 1000 & 200 & 200/0 & 24.384 & 41.098 & 102.559 \\
nug12 & 1000 & 200 & 200/0 & 43.779 & 103.283 & 232.456 \\
had12 & 1000 & 67 & 67/0 & 47.348 & 73.956 & 65.430 \\
rou12 & 1000 & 62 & 62/0 & 47.785 & 60.243 & 60.291 \\
scr12 & 1000 & 200 & 200/0 & 36.732 & 52.809 & 154.193 \\
tai10a & 1000 & 200 & 200/0 & 29.018 & 45.262 & 114.011 \\
tai10b & 1000 & 200 & 200/0 & 30.108 & 38.175 & 113.757 \\
tai12a & 1000 & 67 & 67/0 & 48.202 & 54.950 & 62.244 \\
tai12b & 1000 & 200 & 200/0 & 46.908 & 79.634 & 206.645 \\
\bottomrule
\end{tabular*}
\end{table*}
\begin{table*}[htbp]
\caption{\label{tab:qap_wsqaoa_circuit_compilation}
QAP (WS-QAOA): circuit and compilation statistics. 
Logical metrics (qubits, depth, gates, two-qubit gates, parameters) and post-transpilation metrics per backend.}
\centering
\footnotesize
\setlength{\tabcolsep}{3pt}
\begin{tabular*}{\textwidth}{@{\extracolsep{\fill}}llccccccccc}
\toprule
 & & \multicolumn{5}{c}{\textbf{Logical}} & \multicolumn{3}{c}{\textbf{Transpiled}} \\
\cmidrule(lr){3-7} \cmidrule(lr){8-10}
\textbf{Inst.} & \textbf{Backend} & \textbf{Q} & \textbf{D} & \textbf{G} & \textbf{2Q} & \textbf{P} & \textbf{D} & \textbf{G} & \textbf{2Q} \\
\midrule
chr12a & ibm\_fez & 144 & 356 & 11058 & 9042 & 6 & 30456 & 301381 & 84503 \\
chr12a & ibm\_marrakesh & 144 & 356 & 11058 & 9042 & 6 & 28717 & 296861 & 82789 \\
chr12b & ibm\_fez & 144 & 389 & 11058 & 9042 & 6 & 30456 & 301381 & 84503 \\
chr12b & ibm\_marrakesh & 144 & 389 & 11058 & 9042 & 6 & 28717 & 296861 & 82789 \\
chr12c & ibm\_fez & 144 & 356 & 11058 & 9042 & 6 & 30456 & 301381 & 84503 \\
chr12c & ibm\_marrakesh & 144 & 356 & 11058 & 9042 & 6 & 28717 & 296861 & 82789 \\
nug12 & ibm\_fez & 144 & 587 & 24588 & 22572 & 6 & 50864 & 589536 & 159188 \\
nug12 & ibm\_marrakesh & 144 & 587 & 24588 & 22572 & 6 & 49604 & 591422 & 159153 \\
had12 & ibm\_marrakesh & 144 & 587 & 32904 & 30888 & 6 & 28717 & 296861 & 82789 \\
rou12 & ibm\_fez & 144 & 587 & 32508 & 30492 & 6 & 50864 & 589536 & 159188 \\
rou12 & ibm\_marrakesh & 144 & 587 & 32508 & 30492 & 6 & 49604 & 591422 & 159153 \\
scr12 & ibm\_marrakesh & 144 & 510 & 17856 & 15840 & 6 & 49604 & 591422 & 159153 \\
tai10a & ibm\_marrakesh & 100 & 411 & 15710 & 14310 & 6 & 33740 & 309729 & 77609 \\
tai10a & ibm\_torino & 100 & 411 & 15710 & 14310 & 6 & 26709 & 307357 & 76532 \\
tai10b & ibm\_marrakesh & 100 & 411 & 13550 & 12150 & 6 & 33740 & 309729 & 77609 \\
tai10b & ibm\_torino & 100 & 411 & 13550 & 12150 & 6 & 26709 & 307357 & 76532 \\
\bottomrule
\end{tabular*}
\end{table*}
\begin{table*}[htbp]
\caption{\label{tab:qap_wsqaoa_execution_robustness}
QAP (WS-QAOA): execution robustness summary. 
Shots per evaluation, number of evaluations, job success counts, latency statistics, and wall-clock time.}
\centering
\footnotesize
\setlength{\tabcolsep}{4pt}
\begin{tabular*}{\textwidth}{@{\extracolsep{\fill}}lcccccc}
\toprule
\textbf{Inst.} & \textbf{Shots} & \textbf{Evals} & \textbf{Jobs (ok/fail)} & \textbf{Med.\ Lat.\ (s)} & \textbf{P95 Lat.\ (s)} & \textbf{Time (min)} \\
\midrule
chr12a & 1000 & 49 & 49/0 & 23.849 & 30.497 & 22.715 \\
chr12b & 1000 & 46 & 46/0 & 25.030 & 51.947 & 27.362 \\
chr12c & 1000 & 63 & 63/0 & 24.938 & 35.776 & 32.358 \\
nug12 & 1000 & 46 & 46/0 & 40.295 & 338.060 & 122.527 \\
had12 & 1000 & 50 & 50/0 & 23.870 & 29.658 & 25.958 \\
rou12 & 1000 & 59 & 59/0 & 40.639 & 55.351 & 61.584 \\
scr12 & 1000 & 66 & 66/0 & 40.469 & 64.525 & 53.502 \\
tai10a & 1000 & 50 & 50/0 & 23.919 & 46.100 & 25.578 \\
tai10b & 1000 & 58 & 58/0 & 24.067 & 34.706 & 29.370 \\
\bottomrule
\end{tabular*}
\end{table*}

\subsection{QAP: encoding-based methods}
\label{sec:appendix_qap_encoding_family}

Tables~\ref{tab:qap_pce_circuit_compilation}--\ref{tab:qap_pce_execution_robustness}
report diagnostics for PCE, and
Tables~\ref{tab:qap_qrao_circuit_compilation}--\ref{tab:qap_qrao_execution_robustness}
for QRAO\@.
PCE compresses the 144-variable QAP instances to just 11 qubits, but
the resulting circuits still fail to produce feasible solutions on
hardware, illustrating that extreme qubit compression does not
compensate for the dense coupling structure of QAP\@.

\begin{table*}[htbp]
\caption{\label{tab:qap_pce_circuit_compilation}
QAP (PCE): circuit and compilation statistics. 
Logical metrics (qubits, depth, gates, two-qubit gates, parameters) and post-transpilation metrics per backend.}
\centering
\footnotesize
\setlength{\tabcolsep}{3pt}
\begin{tabular*}{\textwidth}{@{\extracolsep{\fill}}llccccccccc}
\toprule
 & & \multicolumn{5}{c}{\textbf{Logical}} & \multicolumn{3}{c}{\textbf{Transpiled}} \\
\cmidrule(lr){3-7} \cmidrule(lr){8-10}
\textbf{Inst.} & \textbf{Backend} & \textbf{Q} & \textbf{D} & \textbf{G} & \textbf{2Q} & \textbf{P} & \textbf{D} & \textbf{G} & \textbf{2Q} \\
\midrule
chr12a & ibm\_fez & 11 & 89 & 715 & 220 & 462 & 617 & 3817 & 440 \\
chr12a & ibm\_torino & 11 & 89 & 715 & 220 & 462 & 617 & 3817 & 440 \\
chr12b & ibm\_fez & 11 & 89 & 715 & 220 & 462 & 617 & 3817 & 440 \\
chr12b & ibm\_torino & 11 & 89 & 715 & 220 & 462 & 617 & 3817 & 440 \\
chr12c & ibm\_fez & 11 & 89 & 715 & 220 & 462 & 617 & 3817 & 440 \\
chr12c & ibm\_torino & 11 & 89 & 715 & 220 & 462 & 617 & 3817 & 440 \\
nug12 & ibm\_fez & 11 & 18 & 86 & 20 & 66 & 51 & 245 & 20 \\
nug12 & ibm\_marrakesh & 11 & 18 & 86 & 20 & 66 & 51 & 245 & 20 \\
had12 & ibm\_fez & 11 & 18 & 86 & 20 & 66 & 51 & 245 & 20 \\
had12 & ibm\_marrakesh & 11 & 18 & 86 & 20 & 66 & 51 & 245 & 20 \\
rou12 & aer\_simulator\_mps & 11 & 18 & 86 & 20 & 66 & 51 & 245 & 20 \\
rou12 & ibm\_fez & 11 & 18 & 86 & 20 & 66 & 51 & 245 & 20 \\
rou12 & ibm\_marrakesh & 11 & 18 & 86 & 20 & 66 & 51 & 245 & 20 \\
\bottomrule
\end{tabular*}
\end{table*}
\begin{table*}[htbp]
\caption{\label{tab:qap_pce_execution_robustness}
QAP (PCE): execution robustness summary. 
Shots per evaluation, number of evaluations, job success counts, latency statistics, and wall-clock time.}
\centering
\footnotesize
\setlength{\tabcolsep}{4pt}
\begin{tabular*}{\textwidth}{@{\extracolsep{\fill}}lcccccc}
\toprule
\textbf{Inst.} & \textbf{Shots} & \textbf{Evals} & \textbf{Jobs (ok/fail)} & \textbf{Med.\ Lat.\ (s)} & \textbf{P95 Lat.\ (s)} & \textbf{Time (min)} \\
\midrule
chr12a & 1000 & 200 & 200/0 & 13.951 & 63.975 & 81.420 \\
chr12b & 1000 & 200 & 200/0 & 9.171 & 25.250 & 50.613 \\
chr12c & 1000 & 200 & 200/0 & 9.113 & 30.273 & 45.370 \\
nug12 & 1000 & 200 & 200/0 & 24.223 & 83.771 & 115.372 \\
had12 & 1000 & 200 & 200/0 & 23.536 & 42.756 & 99.672 \\
rou12 & 1000 & 200 & 200/0 & 23.337 & 113.295 & 125.164 \\
\bottomrule
\end{tabular*}
\end{table*}
\begin{table*}[htbp]
\caption{\label{tab:qap_qrao_circuit_compilation}
QAP (QRAO): circuit and compilation statistics. 
Logical metrics (qubits, depth, gates, two-qubit gates, parameters) and post-transpilation metrics per backend.}
\centering
\footnotesize
\setlength{\tabcolsep}{3pt}
\begin{tabular*}{\textwidth}{@{\extracolsep{\fill}}llccccccccc}
\toprule
 & & \multicolumn{5}{c}{\textbf{Logical}} & \multicolumn{3}{c}{\textbf{Transpiled}} \\
\cmidrule(lr){3-7} \cmidrule(lr){8-10}
\textbf{Inst.} & \textbf{Backend} & \textbf{Q} & \textbf{D} & \textbf{G} & \textbf{2Q} & \textbf{P} & \textbf{D} & \textbf{G} & \textbf{2Q} \\
\midrule
chr12a & ibm\_fez & 56 & 67 & 613 & 165 & 448 & 198 & 1942 & 165 \\
chr12a & ibm\_marrakesh & 56 & 67 & 613 & 165 & 448 & 198 & 1942 & 165 \\
chr12b & ibm\_fez & 62 & 73 & 679 & 183 & 496 & 216 & 2152 & 183 \\
chr12b & ibm\_marrakesh & 62 & 73 & 679 & 183 & 496 & 216 & 2152 & 183 \\
chr12c & ibm\_fez & 54 & 65 & 591 & 159 & 432 & 192 & 1872 & 159 \\
chr12c & ibm\_marrakesh & 54 & 65 & 591 & 159 & 432 & 192 & 1872 & 159 \\
nug12 & ibm\_fez & 63 & 74 & 690 & 186 & 504 & 219 & 2187 & 186 \\
had12 & ibm\_fez & 144 & 155 & 1581 & 429 & 1152 & 1195 & 7033 & 1110 \\
had12 & ibm\_marrakesh & 144 & 155 & 1581 & 429 & 1152 & 1195 & 7033 & 1110 \\
rou12 & ibm\_torino & 132 & 143 & 1449 & 393 & 1056 & 1737 & 7659 & 1413 \\
scr12 & ibm\_fez & 60 & 71 & 657 & 177 & 480 & 210 & 2082 & 177 \\
scr12 & ibm\_kingston & 60 & 71 & 657 & 177 & 480 & 210 & 2082 & 177 \\
scr12 & ibm\_marrakesh & 60 & 71 & 657 & 177 & 480 & 210 & 2082 & 177 \\
\bottomrule
\end{tabular*}
\end{table*}
\begin{table*}[htbp]
\caption{\label{tab:qap_qrao_execution_robustness}
QAP (QRAO): execution robustness summary. 
Shots per evaluation, number of evaluations, job success counts, latency statistics, and wall-clock time.}
\centering
\footnotesize
\setlength{\tabcolsep}{4pt}
\begin{tabular*}{\textwidth}{@{\extracolsep{\fill}}lcccccc}
\toprule
\textbf{Inst.} & \textbf{Shots} & \textbf{Evals} & \textbf{Jobs (ok/fail)} & \textbf{Med.\ Lat.\ (s)} & \textbf{P95 Lat.\ (s)} & \textbf{Time (min)} \\
\midrule
chr12a & 1000 & 200 & 202/0 & 38.992 & 260.216 & 255.781 \\
chr12b & 1000 & 200 & 202/0 & 36.777 & 50.377 & 139.077 \\
chr12c & 1000 & 200 & 202/0 & 39.006 & 157.909 & 192.154 \\
nug12 & 1000 & 200 & 202/0 & 67.249 & 390.702 & 489.529 \\
had12 & 1000 & 200 & 202/0 & 118.453 & 224.432 & 463.684 \\
rou12 & 1000 & 200 & 202/0 & 107.806 & 139.019 & 403.562 \\
scr12 & 1000 & 200 & 202/0 & 48.854 & 155.421 & 219.811 \\
\bottomrule
\end{tabular*}
\end{table*}




\section{Hardware Diagnostics: MSP}
\label{app:hw_msp}

This appendix reports compilation and execution diagnostics for all
method families on the Market Share Problem benchmark. Unlike QAP,
all methods produce feasible solutions for MSP, so the diagnostics
here help explain the variation in recovered solution quality rather
than feasibility. In particular, the tables reveal how the
balancing-constraint structure of MSP interacts differently with
full-width and compressed encodings after transpilation.

\subsection{MSP normalization reference bounds}
\label{app:msp_tdev_bounds}

The range-normalized approximation ratio
$\mathrm{AR}_{\mathrm{TDev}}$ defined in
Eq.~\eqref{eq:msp_ar_tdev} uses instance-specific reference bounds
computed over the original MSP feasible decision domain. For each
instance, we solve both the minimum and maximum total deviation,
$\mathrm{TDev}^{\star}$ and $\mathrm{TDev}^{\max}$, using a
classical mixed-integer optimization audit. Both bounds are certified
optimal. All revised MSP result
tables use the certified bounds reported in
Table~\ref{tab:msp_tdev_reference_bounds}.

\begin{table}[h]
  \centering
  \caption{Classically certified MSP reference bounds used to compute the
    range-normalized approximation ratio $\mathrm{AR}_{\mathrm{TDev}}$.
    The minimum and maximum are evaluated over the original feasible
  decision domain used for decoding and result evaluation.}
  \label{tab:msp_tdev_reference_bounds}
  \footnotesize
  \setlength{\tabcolsep}{4pt}
  \renewcommand{\arraystretch}{1.08}
  \begin{tabular}{lrrr}
    \toprule
    \textbf{Instance} &
    $\boldsymbol{\mathrm{TDev}^{\star}}$ &
    $\boldsymbol{\mathrm{TDev}^{\max}}$ &
    $\boldsymbol{\mathrm{TDev}^{\max}-\mathrm{TDev}^{\star}}$ \\
    \midrule
    ms20 & 3 & 491  & 488 \\
    ms30 & 2 & 1549 & 1547 \\
    ms40 & 1 & 2944 & 2943 \\
    ms50 & 0 & 4762 & 4762 \\
    ms21 & 3 & 515  & 512 \\
    ms31 & 3 & 1513 & 1510 \\
    ms41 & 1 & 3238 & 3237 \\
    ms51 & 1 & 5154 & 5153 \\
    \bottomrule
  \end{tabular}
\end{table}

\subsection{MSP: VQE-style methods}
\label{sec:appendix_ms_vqe_family}

Tables~\ref{tab:ms_vqe_circuit_compilation}--\ref{tab:ms_vqe_execution_robustness}
report diagnostics for VQE, and
Tables~\ref{tab:ms_cvar_vqe_circuit_compilation}--\ref{tab:ms_cvar_vqe_execution_robustness}
for CVaR-VQE\@.

\begin{table*}[htbp]
\caption{\label{tab:ms_vqe_circuit_compilation}
MSP (VQE): circuit and compilation statistics. 
Logical metrics (qubits, depth, gates, two-qubit gates, parameters) and post-transpilation metrics per backend.}
\centering
\footnotesize
\setlength{\tabcolsep}{3pt}
\begin{tabular*}{\textwidth}{@{\extracolsep{\fill}}llccccccccc}
\toprule
 & & \multicolumn{5}{c}{\textbf{Logical}} & \multicolumn{3}{c}{\textbf{Transpiled}} \\
\cmidrule(lr){3-7} \cmidrule(lr){8-10}
\textbf{Inst.} & \textbf{Backend} & \textbf{Q} & \textbf{D} & \textbf{G} & \textbf{2Q} & \textbf{P} & \textbf{D} & \textbf{G} & \textbf{2Q} \\
\midrule
ms\_seed0\_prod2 & ibm\_fez & 48 & 149 & 384 & 144 & 192 & 587 & 1392 & 144 \\
ms\_seed0\_prod2 & ibm\_torino & 48 & 149 & 384 & 144 & 192 & 587 & 1392 & 144 \\
ms\_seed0\_prod3 & ibm\_fez & 84 & 257 & 672 & 252 & 336 & 1019 & 2436 & 252 \\
ms\_seed0\_prod3 & ibm\_torino & 84 & 257 & 672 & 252 & 336 & 1019 & 2436 & 252 \\
ms\_seed0\_prod4 & ibm\_fez & 118 & 359 & 944 & 354 & 472 & 1532 & 3581 & 408 \\
ms\_seed0\_prod4 & ibm\_torino & 118 & 359 & 944 & 354 & 472 & 1979 & 4704 & 808 \\
ms\_seed0\_prod5 & ibm\_fez & 150 & 455 & 1200 & 450 & 600 & 3018 & 6901 & 1348 \\
ms\_seed1\_prod2 & ibm\_fez & 50 & 155 & 400 & 150 & 200 & 914 & 2069 & 369 \\
ms\_seed1\_prod2 & ibm\_torino & 50 & 155 & 400 & 150 & 200 & 796 & 1814 & 272 \\
ms\_seed1\_prod3 & ibm\_fez & 84 & 257 & 672 & 252 & 336 & 1019 & 2436 & 252 \\
ms\_seed1\_prod3 & ibm\_torino & 84 & 257 & 672 & 252 & 336 & 1019 & 2436 & 252 \\
ms\_seed1\_prod4 & ibm\_fez & 118 & 359 & 944 & 354 & 472 & 1532 & 3581 & 408 \\
ms\_seed1\_prod4 & ibm\_torino & 118 & 359 & 944 & 354 & 472 & 1979 & 4704 & 808 \\
ms\_seed1\_prod5 & ibm\_fez & 156 & 473 & 1248 & 468 & 624 & 3276 & 7592 & 1536 \\
\bottomrule
\end{tabular*}
\end{table*}
\begin{table*}[htbp]
\caption{\label{tab:ms_vqe_execution_robustness}
MSP (VQE): execution robustness summary. 
Shots per evaluation, number of evaluations, job success counts, latency statistics, and wall-clock time.}
\centering
\footnotesize
\setlength{\tabcolsep}{4pt}
\begin{tabular*}{\textwidth}{@{\extracolsep{\fill}}lcccccc}
\toprule
\textbf{Inst.} & \textbf{Shots} & \textbf{Evals} & \textbf{Jobs (ok/fail)} & \textbf{Med.\ Lat.\ (s)} & \textbf{P95 Lat.\ (s)} & \textbf{Time (min)} \\
\midrule
ms\_seed0\_prod2 & 1000 & 200 & 200/0 & 8.545 & 13.995 & 39.038 \\
ms\_seed0\_prod3 & 1000 & 200 & 200/0 & 8.529 & 13.990 & 41.030 \\
ms\_seed0\_prod4 & 1000 & 200 & 200/0 & 8.752 & 14.005 & 41.955 \\
ms\_seed0\_prod5 & 1000 & 200 & 200/0 & 8.606 & 26.120 & 51.288 \\
ms\_seed1\_prod2 & 1000 & 200 & 200/0 & 8.509 & 14.447 & 39.081 \\
ms\_seed1\_prod3 & 1000 & 200 & 200/0 & 8.570 & 14.011 & 40.675 \\
ms\_seed1\_prod4 & 1000 & 200 & 200/0 & 8.957 & 20.124 & 51.252 \\
ms\_seed1\_prod5 & 1000 & 200 & 200/0 & 8.505 & 14.008 & 45.000 \\
\bottomrule
\end{tabular*}
\end{table*}
\begin{table*}[htbp]
\caption{\label{tab:ms_cvar_vqe_circuit_compilation}
MSP (CVaR-VQE): circuit and compilation statistics. 
Logical metrics (qubits, depth, gates, two-qubit gates, parameters) and post-transpilation metrics per backend.}
\centering
\footnotesize
\setlength{\tabcolsep}{3pt}
\begin{tabular*}{\textwidth}{@{\extracolsep{\fill}}llccccccccc}
\toprule
 & & \multicolumn{5}{c}{\textbf{Logical}} & \multicolumn{3}{c}{\textbf{Transpiled}} \\
\cmidrule(lr){3-7} \cmidrule(lr){8-10}
\textbf{Inst.} & \textbf{Backend} & \textbf{Q} & \textbf{D} & \textbf{G} & \textbf{2Q} & \textbf{P} & \textbf{D} & \textbf{G} & \textbf{2Q} \\
\midrule
ms\_seed0\_prod2 & ibm\_marrakesh & 48 & 149 & 384 & 144 & 192 & 587 & 1392 & 144 \\
ms\_seed0\_prod2 & ibm\_torino & 48 & 149 & 384 & 144 & 192 & 587 & 1392 & 144 \\
ms\_seed0\_prod3 & ibm\_marrakesh & 84 & 257 & 672 & 252 & 336 & 1019 & 2436 & 252 \\
ms\_seed0\_prod3 & ibm\_torino & 84 & 257 & 672 & 252 & 336 & 1019 & 2436 & 252 \\
ms\_seed0\_prod4 & ibm\_marrakesh & 118 & 359 & 944 & 354 & 472 & 1532 & 3581 & 408 \\
ms\_seed0\_prod4 & ibm\_torino & 118 & 359 & 944 & 354 & 472 & 1923 & 4661 & 784 \\
ms\_seed0\_prod5 & ibm\_marrakesh & 150 & 455 & 1200 & 450 & 600 & 3239 & 7428 & 1534 \\
ms\_seed1\_prod2 & ibm\_marrakesh & 50 & 155 & 400 & 150 & 200 & 894 & 1943 & 318 \\
ms\_seed1\_prod2 & ibm\_torino & 50 & 155 & 400 & 150 & 200 & 818 & 1807 & 272 \\
ms\_seed1\_prod3 & ibm\_marrakesh & 84 & 257 & 672 & 252 & 336 & 1019 & 2436 & 252 \\
ms\_seed1\_prod3 & ibm\_torino & 84 & 257 & 672 & 252 & 336 & 1019 & 2436 & 252 \\
ms\_seed1\_prod4 & ibm\_marrakesh & 118 & 359 & 944 & 354 & 472 & 1532 & 3581 & 408 \\
ms\_seed1\_prod4 & ibm\_torino & 118 & 359 & 944 & 354 & 472 & 1923 & 4661 & 784 \\
ms\_seed1\_prod5 & ibm\_fez & 156 & 473 & 1248 & 468 & 624 & 3551 & 7727 & 1621 \\
\bottomrule
\end{tabular*}
\end{table*}
\begin{table*}[htbp]
\caption{\label{tab:ms_cvar_vqe_execution_robustness}
MSP (CVaR-VQE): execution robustness summary. 
Shots per evaluation, number of evaluations, job success counts, latency statistics, and wall-clock time.}
\centering
\footnotesize
\setlength{\tabcolsep}{4pt}
\begin{tabular*}{\textwidth}{@{\extracolsep{\fill}}lcccccc}
\toprule
\textbf{Inst.} & \textbf{Shots} & \textbf{Evals} & \textbf{Jobs (ok/fail)} & \textbf{Med.\ Lat.\ (s)} & \textbf{P95 Lat.\ (s)} & \textbf{Time (min)} \\
\midrule
ms\_seed0\_prod2 & 1000 & 200 & 200/0 & 13.987 & 19.851 & 51.716 \\
ms\_seed0\_prod3 & 1000 & 200 & 200/0 & 8.568 & 14.837 & 42.078 \\
ms\_seed0\_prod4 & 1000 & 200 & 200/0 & 8.948 & 19.892 & 46.589 \\
ms\_seed0\_prod5 & 1000 & 200 & 200/0 & 8.908 & 30.100 & 61.004 \\
ms\_seed1\_prod2 & 1000 & 200 & 200/0 & 13.978 & 46.324 & 99.710 \\
ms\_seed1\_prod3 & 1000 & 200 & 200/0 & 8.642 & 45.728 & 57.528 \\
ms\_seed1\_prod4 & 1000 & 200 & 200/0 & 9.033 & 91.113 & 79.357 \\
ms\_seed1\_prod5 & 1000 & 200 & 200/0 & 8.576 & 25.067 & 61.076 \\
\bottomrule
\end{tabular*}
\end{table*}

\subsection{MSP: QAOA-style methods}
\label{sec:appendix_ms_qaoa_family}

Tables~\ref{tab:appendix_ms_qaoa_circuit_compilation}--\ref{tab:ms_qaoa_execution_robustness}
report diagnostics for QAOA,
Tables~\ref{tab:ms_maqaoa_circuit_compilation}--\ref{tab:ms_maqaoa_execution_robustness}
for MA-QAOA, and
Tables~\ref{tab:appendix_ms_wsqaoa_circuit_compilation}--\ref{tab:appendix_ms_wsqaoa_execution_robustness}
for WS-QAOA\@.
The WS-QAOA circuits are by far the largest in the benchmark due to the
dense warm-start operator structure, with transpiled two-qubit gate
counts exceeding 100\,000 on several instances.

\begin{table*}[htbp]
\caption{\label{tab:appendix_ms_qaoa_circuit_compilation}
MSP (QAOA): circuit and compilation statistics. 
Logical metrics (qubits, depth, gates, two-qubit gates, parameters) and post-transpilation metrics per backend.}
\centering
\footnotesize
\setlength{\tabcolsep}{3pt}
\begin{tabular*}{\textwidth}{@{\extracolsep{\fill}}llccccccccc}
\toprule
 & & \multicolumn{5}{c}{\textbf{Logical}} & \multicolumn{3}{c}{\textbf{Transpiled}} \\
\cmidrule(lr){3-7} \cmidrule(lr){8-10}
\textbf{Inst.} & \textbf{Backend} & \textbf{Q} & \textbf{D} & \textbf{G} & \textbf{2Q} & \textbf{P} & \textbf{D} & \textbf{G} & \textbf{2Q} \\
\midrule
ms\_seed0\_prod2 & ibm\_fez & 48 & 177 & 2688 & 2304 & 6 & 7779 & 51252 & 13304 \\
ms\_seed0\_prod2 & ibm\_torino & 48 & 177 & 2688 & 2304 & 6 & 8056 & 51421 & 13467 \\
ms\_seed0\_prod3 & ibm\_fez & 84 & 299 & 6972 & 6300 & 6 & 17879 & 145758 & 38039 \\
ms\_seed0\_prod3 & ibm\_torino & 84 & 299 & 6972 & 6300 & 6 & 19016 & 148732 & 38936 \\
ms\_seed0\_prod4 & ibm\_fez & 118 & 411 & 12875 & 11931 & 6 & 30660 & 292910 & 76599 \\
ms\_seed0\_prod4 & ibm\_torino & 118 & 411 & 12875 & 11931 & 6 & 31394 & 292454 & 76264 \\
ms\_seed0\_prod5 & ibm\_fez & 150 & 517 & 20139 & 18939 & 6 & 48019 & 471852 & 123875 \\
ms\_seed1\_prod2 & ibm\_fez & 50 & 182 & 2845 & 2475 & 6 & 8795 & 56016 & 14614 \\
ms\_seed1\_prod2 & ibm\_torino & 50 & 182 & 2845 & 2475 & 6 & 7836 & 55147 & 14494 \\
ms\_seed1\_prod3 & ibm\_fez & 84 & 296 & 6918 & 6306 & 6 & 17879 & 145758 & 38039 \\
ms\_seed1\_prod3 & ibm\_torino & 84 & 296 & 6918 & 6306 & 6 & 19016 & 148732 & 38936 \\
ms\_seed1\_prod4 & ibm\_fez & 118 & 411 & 12875 & 11931 & 6 & 30660 & 292910 & 76599 \\
ms\_seed1\_prod4 & ibm\_torino & 118 & 411 & 12875 & 11931 & 6 & 31394 & 292454 & 76264 \\
ms\_seed1\_prod5 & ibm\_fez & 156 & 535 & 21174 & 19926 & 6 & 47838 & 504835 & 133151 \\
\bottomrule
\end{tabular*}
\end{table*}
\begin{table*}[htbp]
\caption{\label{tab:ms_qaoa_execution_robustness}
MSP (QAOA): execution robustness summary. 
Shots per evaluation, number of evaluations, job success counts, latency statistics, and wall-clock time.}
\centering
\footnotesize
\setlength{\tabcolsep}{4pt}
\begin{tabular*}{\textwidth}{@{\extracolsep{\fill}}lcccccc}
\toprule
\textbf{Inst.} & \textbf{Shots} & \textbf{Evals} & \textbf{Jobs (ok/fail)} & \textbf{Med.\ Lat.\ (s)} & \textbf{P95 Lat.\ (s)} & \textbf{Time (min)} \\
\midrule
ms\_seed0\_prod2 & 1000 & 51 & 51/0 & 15.372 & 172.062 & 32.583 \\
ms\_seed0\_prod3 & 1000 & 58 & 58/0 & 16.906 & 82.326 & 37.490 \\
ms\_seed0\_prod4 & 1000 & 55 & 55/0 & 24.950 & 54.760 & 37.729 \\
ms\_seed0\_prod5 & 1000 & 56 & 56/0 & 33.051 & 61.486 & 42.136 \\
ms\_seed1\_prod2 & 1000 & 59 & 59/0 & 15.314 & 69.292 & 25.101 \\
ms\_seed1\_prod3 & 1000 & 55 & 55/0 & 17.174 & 49.124 & 26.164 \\
ms\_seed1\_prod4 & 1000 & 48 & 48/0 & 25.237 & 83.587 & 60.450 \\
ms\_seed1\_prod5 & 1000 & 49 & 49/0 & 38.445 & 101.818 & 44.298 \\
\bottomrule
\end{tabular*}
\end{table*}
\begin{table*}[htbp]
\caption{\label{tab:ms_maqaoa_circuit_compilation}
MSP (MA-QAOA): circuit and compilation statistics. 
Logical metrics (qubits, depth, gates, two-qubit gates, parameters) and post-transpilation metrics per backend.}
\centering
\footnotesize
\setlength{\tabcolsep}{3pt}
\begin{tabular*}{\textwidth}{@{\extracolsep{\fill}}llccccccccc}
\toprule
 & & \multicolumn{5}{c}{\textbf{Logical}} & \multicolumn{3}{c}{\textbf{Transpiled}} \\
\cmidrule(lr){3-7} \cmidrule(lr){8-10}
\textbf{Inst.} & \textbf{Backend} & \textbf{Q} & \textbf{D} & \textbf{G} & \textbf{2Q} & \textbf{P} & \textbf{D} & \textbf{G} & \textbf{2Q} \\
\midrule
ms\_seed0\_prod2 & ibm\_torino & 48 & 177 & 2688 & 2304 & 2592 & 8265 & 52094 & 13644 \\
ms\_seed0\_prod3 & ibm\_fez & 84 & 299 & 6972 & 6300 & 6804 & 19539 & 151739 & 39882 \\
ms\_seed0\_prod3 & ibm\_torino & 84 & 299 & 6972 & 6300 & 6804 & 17352 & 150102 & 39302 \\
ms\_seed0\_prod4 & ibm\_fez & 118 & 411 & 12875 & 11931 & 12639 & 30936 & 291142 & 76248 \\
ms\_seed0\_prod4 & ibm\_torino & 118 & 411 & 12875 & 11931 & 12639 & 31327 & 290871 & 75551 \\
ms\_seed0\_prod5 & ibm\_fez & 150 & 517 & 20139 & 18939 & 19839 & 44337 & 460244 & 120201 \\
ms\_seed1\_prod2 & ibm\_fez & 50 & 182 & 2845 & 2475 & 2745 & 8267 & 55586 & 14473 \\
ms\_seed1\_prod2 & ibm\_torino & 50 & 182 & 2845 & 2475 & 2745 & 9113 & 55648 & 14473 \\
ms\_seed1\_prod3 & ibm\_fez & 84 & 296 & 6918 & 6306 & 6750 & 19222 & 149753 & 39159 \\
ms\_seed1\_prod3 & ibm\_torino & 84 & 296 & 6918 & 6306 & 6750 & 18957 & 152431 & 40562 \\
ms\_seed1\_prod4 & ibm\_fez & 118 & 411 & 12875 & 11931 & 12639 & 30936 & 291142 & 76248 \\
ms\_seed1\_prod4 & ibm\_torino & 118 & 411 & 12875 & 11931 & 12639 & 31327 & 290871 & 75551 \\
ms\_seed1\_prod5 & ibm\_fez & 156 & 535 & 21174 & 19926 & 20862 & 44655 & 495541 & 128487 \\
ms\_seed1\_prod5 & ibm\_marrakesh & 156 & 535 & 21174 & 19926 & 20862 & 50183 & 504651 & 132622 \\
\bottomrule
\end{tabular*}
\end{table*}
\begin{table*}[htbp]
\caption{\label{tab:ms_maqaoa_execution_robustness}
MSP (MA-QAOA): execution robustness summary. 
Shots per evaluation, number of evaluations, job success counts, latency statistics, and wall-clock time.}
\centering
\footnotesize
\setlength{\tabcolsep}{4pt}
\begin{tabular*}{\textwidth}{@{\extracolsep{\fill}}lcccccc}
\toprule
\textbf{Inst.} & \textbf{Shots} & \textbf{Evals} & \textbf{Jobs (ok/fail)} & \textbf{Med.\ Lat.\ (s)} & \textbf{P95 Lat.\ (s)} & \textbf{Time (min)} \\
\midrule
ms\_seed0\_prod2 & 1000 & 200 & 200/0 & 15.186 & 36.421 & 69.192 \\
ms\_seed0\_prod3 & 1000 & 200 & 200/0 & 16.914 & 27.327 & 71.063 \\
ms\_seed0\_prod4 & 1000 & 200 & 200/0 & 24.383 & 51.213 & 133.832 \\
ms\_seed0\_prod5 & 1000 & 200 & 200/0 & 36.655 & 128.320 & 187.643 \\
ms\_seed1\_prod2 & 1000 & 200 & 200/0 & 15.312 & 30.668 & 61.797 \\
ms\_seed1\_prod3 & 1000 & 200 & 200/0 & 16.550 & 70.541 & 86.375 \\
ms\_seed1\_prod4 & 1000 & 200 & 200/0 & 25.452 & 208.449 & 294.116 \\
ms\_seed1\_prod5 & 1000 & 200 & 200/0 & 37.311 & 79.805 & 161.180 \\
\bottomrule
\end{tabular*}
\end{table*}
\begin{table*}[htbp]
\caption{\label{tab:appendix_ms_wsqaoa_circuit_compilation}
MSP (WS-QAOA): circuit and compilation statistics. 
Logical metrics (qubits, depth, gates, two-qubit gates, parameters) and post-transpilation metrics per backend.}
\centering
\footnotesize
\setlength{\tabcolsep}{3pt}
\begin{tabular*}{\textwidth}{@{\extracolsep{\fill}}llccccccccc}
\toprule
 & & \multicolumn{5}{c}{\textbf{Logical}} & \multicolumn{3}{c}{\textbf{Transpiled}} \\
\cmidrule(lr){3-7} \cmidrule(lr){8-10}
\textbf{Inst.} & \textbf{Backend} & \textbf{Q} & \textbf{D} & \textbf{G} & \textbf{2Q} & \textbf{P} & \textbf{D} & \textbf{G} & \textbf{2Q} \\
\midrule
ms\_seed0\_prod2 & ibm\_fez & 48 & 183 & 2976 & 2304 & 6 & 8207 & 52335 & 13434 \\
ms\_seed0\_prod2 & ibm\_marrakesh & 48 & 183 & 2976 & 2304 & 6 & 8017 & 52247 & 13512 \\
ms\_seed0\_prod3 & ibm\_marrakesh & 84 & 305 & 7476 & 6300 & 6 & 19052 & 150195 & 39227 \\
ms\_seed0\_prod3 & ibm\_torino & 84 & 305 & 7476 & 6300 & 6 & 18206 & 151219 & 39165 \\
ms\_seed0\_prod4 & ibm\_marrakesh & 118 & 417 & 13583 & 11931 & 6 & 30844 & 290228 & 75459 \\
ms\_seed0\_prod4 & ibm\_torino & 118 & 417 & 13583 & 11931 & 6 & 31861 & 290794 & 76190 \\
ms\_seed0\_prod5 & ibm\_fez & 150 & 523 & 21039 & 18939 & 6 & 46370 & 472119 & 123414 \\
ms\_seed0\_prod5 & ibm\_marrakesh & 150 & 523 & 21039 & 18939 & 6 & 47791 & 468318 & 122758 \\
ms\_seed1\_prod2 & ibm\_fez & 50 & 188 & 3145 & 2475 & 6 & 8079 & 55016 & 14116 \\
ms\_seed1\_prod2 & ibm\_marrakesh & 50 & 188 & 3145 & 2475 & 6 & 8356 & 56799 & 14801 \\
ms\_seed1\_prod3 & ibm\_fez & 84 & 302 & 7422 & 6306 & 6 & 20804 & 152636 & 40149 \\
ms\_seed1\_prod3 & ibm\_marrakesh & 84 & 302 & 7422 & 6306 & 6 & 19052 & 150195 & 39227 \\
ms\_seed1\_prod3 & ibm\_torino & 84 & 302 & 7422 & 6306 & 6 & 18206 & 151219 & 39165 \\
ms\_seed1\_prod4 & ibm\_marrakesh & 118 & 417 & 13583 & 11931 & 6 & 30844 & 290228 & 75459 \\
ms\_seed1\_prod4 & ibm\_torino & 118 & 417 & 13583 & 11931 & 6 & 31861 & 290794 & 76190 \\
ms\_seed1\_prod5 & ibm\_marrakesh & 156 & 541 & 22110 & 19926 & 6 & 44420 & 501025 & 130345 \\
\bottomrule
\end{tabular*}
\end{table*}
\begin{table*}[htbp]
\caption{\label{tab:appendix_ms_wsqaoa_execution_robustness}
MSP (WS-QAOA): execution robustness summary. 
Shots per evaluation, number of evaluations, job success counts, latency statistics, and wall-clock time.}
\centering
\footnotesize
\setlength{\tabcolsep}{4pt}
\begin{tabular*}{\textwidth}{@{\extracolsep{\fill}}lcccccc}
\toprule
\textbf{Inst.} & \textbf{Shots} & \textbf{Evals} & \textbf{Jobs (ok/fail)} & \textbf{Med.\ Lat.\ (s)} & \textbf{P95 Lat.\ (s)} & \textbf{Time (min)} \\
\midrule
ms\_seed0\_prod2 & 1000 & 47 & 47/0 & 15.588 & 17.731 & 20.871 \\
ms\_seed0\_prod3 & 1000 & 55 & 55/0 & 17.915 & 36.054 & 21.653 \\
ms\_seed0\_prod4 & 1000 & 55 & 55/0 & 29.904 & 107.441 & 40.353 \\
ms\_seed0\_prod5 & 1000 & 61 & 61/0 & 33.791 & 314.956 & 69.344 \\
ms\_seed1\_prod2 & 1000 & 52 & 52/0 & 15.670 & 16.473 & 15.024 \\
ms\_seed1\_prod3 & 1000 & 53 & 53/0 & 16.765 & 22.371 & 18.412 \\
ms\_seed1\_prod4 & 1000 & 63 & 63/0 & 24.157 & 30.182 & 29.032 \\
ms\_seed1\_prod5 & 1000 & 50 & 50/0 & 34.046 & 41.184 & 34.045 \\
\bottomrule
\end{tabular*}
\end{table*}

\subsection{MSP: encoding-based methods}
\label{sec:appendix_ms_encoding_family}

Tables~\ref{tab:ms_pce_circuit_compilation}--\ref{tab:ms_pce_execution_robustness}
report diagnostics for PCE, and
Tables~\ref{tab:ms_qrao_circuit_compilation}--\ref{tab:ms_qrao_execution_robustness}
for QRAO\@.
MSP provides the most extreme width compression in the benchmark:
PCE and QRAO reduce the 48--156 qubit full-width formulations to
7--11 qubits.  However, the main results show that this compression
does not translate into better recovered solution quality, and the
compilation diagnostics here help explain why.

\begin{table*}[htbp]
\caption{\label{tab:ms_pce_circuit_compilation}
MSP (PCE): circuit and compilation statistics. 
Logical metrics (qubits, depth, gates, two-qubit gates, parameters) and post-transpilation metrics per backend.}
\centering
\footnotesize
\setlength{\tabcolsep}{3pt}
\begin{tabular*}{\textwidth}{@{\extracolsep{\fill}}llccccccccc}
\toprule
 & & \multicolumn{5}{c}{\textbf{Logical}} & \multicolumn{3}{c}{\textbf{Transpiled}} \\
\cmidrule(lr){3-7} \cmidrule(lr){8-10}
\textbf{Inst.} & \textbf{Backend} & \textbf{Q} & \textbf{D} & \textbf{G} & \textbf{2Q} & \textbf{P} & \textbf{D} & \textbf{G} & \textbf{2Q} \\
\midrule
ms\_seed0\_prod2 & ibm\_torino & 7 & 57 & 287 & 84 & 182 & 393 & 1493 & 168 \\
ms\_seed0\_prod3 & ibm\_torino & 8 & 65 & 376 & 112 & 240 & 449 & 1979 & 224 \\
ms\_seed0\_prod4 & ibm\_torino & 10 & 81 & 590 & 180 & 380 & 561 & 3135 & 360 \\
ms\_seed0\_prod5 & ibm\_torino & 11 & 89 & 715 & 220 & 462 & 617 & 3817 & 440 \\
ms\_seed1\_prod2 & ibm\_torino & 7 & 57 & 287 & 84 & 182 & 393 & 1493 & 168 \\
ms\_seed1\_prod3 & ibm\_fez & 8 & 65 & 376 & 112 & 240 & 449 & 1979 & 224 \\
ms\_seed1\_prod3 & ibm\_torino & 8 & 65 & 376 & 112 & 240 & 449 & 1979 & 224 \\
ms\_seed1\_prod4 & ibm\_fez & 10 & 81 & 590 & 180 & 380 & 561 & 3135 & 360 \\
ms\_seed1\_prod4 & ibm\_torino & 10 & 81 & 590 & 180 & 380 & 561 & 3135 & 360 \\
ms\_seed1\_prod5 & ibm\_fez & 11 & 89 & 715 & 220 & 462 & 617 & 3817 & 440 \\
ms\_seed1\_prod5 & ibm\_torino & 11 & 89 & 715 & 220 & 462 & 617 & 3817 & 440 \\
\bottomrule
\end{tabular*}
\end{table*}
\begin{table*}[htbp]
\caption{\label{tab:ms_pce_execution_robustness}
MSP (PCE): execution robustness summary. 
Shots per evaluation, number of evaluations, job success counts, latency statistics, and wall-clock time.}
\centering
\footnotesize
\setlength{\tabcolsep}{4pt}
\begin{tabular*}{\textwidth}{@{\extracolsep{\fill}}lcccccc}
\toprule
\textbf{Inst.} & \textbf{Shots} & \textbf{Evals} & \textbf{Jobs (ok/fail)} & \textbf{Med.\ Lat.\ (s)} & \textbf{P95 Lat.\ (s)} & \textbf{Time (min)} \\
\midrule
ms\_seed0\_prod2 & 1000 & 200 & 200/0 & 8.792 & 14.989 & 41.288 \\
ms\_seed0\_prod3 & 1000 & 200 & 200/0 & 8.527 & 14.136 & 37.539 \\
ms\_seed0\_prod4 & 1000 & 200 & 200/0 & 8.560 & 19.221 & 42.853 \\
ms\_seed0\_prod5 & 1000 & 200 & 200/0 & 8.771 & 19.698 & 41.875 \\
ms\_seed1\_prod2 & 1000 & 200 & 200/0 & 8.690 & 14.561 & 40.588 \\
ms\_seed1\_prod3 & 1000 & 200 & 200/0 & 8.768 & 52.741 & 69.656 \\
ms\_seed1\_prod4 & 1000 & 200 & 200/0 & 8.778 & 20.076 & 46.427 \\
ms\_seed1\_prod5 & 1000 & 200 & 200/0 & 8.941 & 31.443 & 56.768 \\
\bottomrule
\end{tabular*}
\end{table*}
\begin{table*}[htbp]
\caption{\label{tab:ms_qrao_circuit_compilation}
MSP (QRAO): circuit and compilation statistics. 
Logical metrics (qubits, depth, gates, two-qubit gates, parameters) and post-transpilation metrics per backend.}
\centering
\footnotesize
\setlength{\tabcolsep}{3pt}
\begin{tabular*}{\textwidth}{@{\extracolsep{\fill}}llccccccccc}
\toprule
 & & \multicolumn{5}{c}{\textbf{Logical}} & \multicolumn{3}{c}{\textbf{Transpiled}} \\
\cmidrule(lr){3-7} \cmidrule(lr){8-10}
\textbf{Inst.} & \textbf{Backend} & \textbf{Q} & \textbf{D} & \textbf{G} & \textbf{2Q} & \textbf{P} & \textbf{D} & \textbf{G} & \textbf{2Q} \\
\midrule
ms\_seed0\_prod2 & ibm\_fez & 30 & 41 & 327 & 87 & 240 & 120 & 1032 & 87 \\
ms\_seed0\_prod2 & ibm\_marrakesh & 30 & 41 & 327 & 87 & 240 & 120 & 1032 & 87 \\
ms\_seed0\_prod3 & ibm\_torino & 42 & 53 & 459 & 123 & 336 & 156 & 1452 & 123 \\
ms\_seed0\_prod4 & ibm\_torino & 73 & 84 & 800 & 216 & 584 & 249 & 2537 & 216 \\
ms\_seed0\_prod5 & ibm\_torino & 84 & 95 & 921 & 249 & 672 & 282 & 2922 & 249 \\
ms\_seed1\_prod2 & ibm\_fez & 30 & 41 & 327 & 87 & 240 & 120 & 1032 & 87 \\
ms\_seed1\_prod3 & ibm\_fez & 42 & 53 & 459 & 123 & 336 & 156 & 1452 & 123 \\
ms\_seed1\_prod3 & ibm\_marrakesh & 42 & 53 & 459 & 123 & 336 & 156 & 1452 & 123 \\
ms\_seed1\_prod4 & ibm\_fez & 73 & 84 & 800 & 216 & 584 & 249 & 2537 & 216 \\
ms\_seed1\_prod4 & ibm\_marrakesh & 73 & 84 & 800 & 216 & 584 & 249 & 2537 & 216 \\
ms\_seed1\_prod5 & ibm\_fez & 85 & 96 & 932 & 252 & 680 & 285 & 2957 & 252 \\
ms\_seed1\_prod5 & ibm\_marrakesh & 85 & 96 & 932 & 252 & 680 & 285 & 2957 & 252 \\
\bottomrule
\end{tabular*}
\end{table*}
\begin{table*}[htbp]
\caption{\label{tab:ms_qrao_execution_robustness}
MSP (QRAO): execution robustness summary. 
Shots per evaluation, number of evaluations, job success counts, latency statistics, and wall-clock time.}
\centering
\footnotesize
\setlength{\tabcolsep}{4pt}
\begin{tabular*}{\textwidth}{@{\extracolsep{\fill}}lcccccc}
\toprule
\textbf{Inst.} & \textbf{Shots} & \textbf{Evals} & \textbf{Jobs (ok/fail)} & \textbf{Med.\ Lat.\ (s)} & \textbf{P95 Lat.\ (s)} & \textbf{Time (min)} \\
\midrule
ms\_seed0\_prod2 & 1000 & 200 & 202/0 & 23.565 & 33.261 & 91.169 \\
ms\_seed0\_prod3 & 1000 & 200 & 202/0 & 24.884 & 50.117 & 100.389 \\
ms\_seed0\_prod4 & 1000 & 200 & 202/0 & 43.719 & 97.333 & 186.746 \\
ms\_seed0\_prod5 & 1000 & 200 & 202/0 & 52.110 & 99.067 & 210.891 \\
ms\_seed1\_prod2 & 1000 & 200 & 202/0 & 23.350 & 157.532 & 171.703 \\
ms\_seed1\_prod3 & 1000 & 200 & 202/0 & 29.245 & 346.852 & 471.645 \\
ms\_seed1\_prod4 & 1000 & 200 & 202/0 & 41.341 & 131.894 & 238.789 \\
ms\_seed1\_prod5 & 1000 & 200 & 202/0 & 59.588 & 304.199 & 401.842 \\
\bottomrule
\end{tabular*}
\end{table*}

\section{Matched Uniform-Random Baseline for Low-Fidelity QAOA-Family Runs}
\label{app:random_baseline}

This appendix evaluates whether the decoded quality of low-fidelity
QAOA-family hardware runs exceeds a matched uniform-random best-shot control.
The purpose is to test the practical interpretation of the strongly
noise-dominated regime without assuming that a small gate-count fidelity proxy
implies a fully uniform measured output distribution.

\subsection{Scope and matched-control construction}

The analysis includes all 106 saved QAOA, MA-QAOA, and warm-start-QAOA hardware
artifacts whose reconstructed gate-count fidelity proxy satisfies
\begin{equation}
  F_{\mathrm{est}} < 10^{-3}.
\end{equation}
PCE and QRAO are excluded because their compressed encodings use distinct
native reconstruction or rounding procedures; the present control addresses
the low-fidelity QAOA-family interpretation specifically.

For each hardware run, the random control matches the decision-variable
dimension, optimizer-trajectory batch count, shots per batch, batch-selection
rule, within-batch tie-breaking rule, and shared one-round local-swap
refinement used by the corresponding hardware artifact. The saved hardware
path stores counts from the optimizer-evaluation batch with the lowest
expectation objective and does not retain an independent final-sampling batch.
Each random replicate therefore generates the same number of uniformly random
bitstrings as the corresponding hardware run, grouped into the same
optimizer-evaluation batches. The replicate selects the batch with the lowest
mean QUBO energy, selects the lowest-QUBO-energy bitstring within that batch,
and then applies the same one-round local-swap refinement used by the hardware
pipeline.

Each hardware run is compared with 300 independent matched uniform-random
replicates. For MDKP, MIS, and QAP, the outcome metric is optimality gap. For
MSP, the outcome metric is total deviation, TDev. QAP is reported as a
feasibility-only comparison because all selected hardware QAP outputs are
infeasible and valid random permutations are astronomically unlikely at the
available candidate budgets.

\subsection{Run-level classification}

For a feasible hardware outcome with a finite matched random-quality
distribution, let $Q_{\mathrm{HW}}$ denote the final hardware metric and let
$Q_{\mathrm{rand}}^{(b)}$ denote the final metric from matched random replicate
$b$. Since lower values are better, we define the empirical random tail
fraction
\begin{equation}
  \tau =
  \frac{
    1+\sum_{b=1}^{300}
    \mathbb{I}\left[
      Q_{\mathrm{rand}}^{(b)}
      \leq
      Q_{\mathrm{HW}}
    \right]
  }{301}.
  \label{eq:random_baseline_tail}
\end{equation}
A hardware outcome is described as within the empirical random range when
$0.05 < \tau < 0.95$ and as worse than the empirical random range when
$\tau \geq 0.95$. The figure below displays the central $2.5$–$97.5\%$
random interval for visualization, whereas the run-level classification uses
the stated empirical tail-fraction rule.

A separate category is required when the hardware pipeline returns a feasible
candidate but none of the 300 matched random replicates does so. Such a result
does not have a finite random-quality distribution against which a standard
optimality-gap comparison can be made. We therefore report it as a
finite-sample feasibility exception rather than as a finite-gap outperformance
or a quantum-advantage claim.

\subsection{Results}

Table~\ref{tab:random_baseline_summary} summarizes the matched control. Of the
106 low-fidelity QAOA-family hardware runs, 43 are hardware-infeasible and are
therefore treated as feasibility-only comparisons. Among the 62 feasible
hardware runs for which the matched random control produces a finite
quality distribution, 51 lie within the empirical random range and 11 are
worse than the random baseline. One additional warm-start-QAOA MIS run,
\texttt{1tc.64}, is a finite-sample feasibility exception: the hardware
pipeline returned a feasible candidate, whereas none of the 300 matched random
replicates did.

\begin{figure*}[t]
  \centering
  \includegraphics[width=\textwidth]{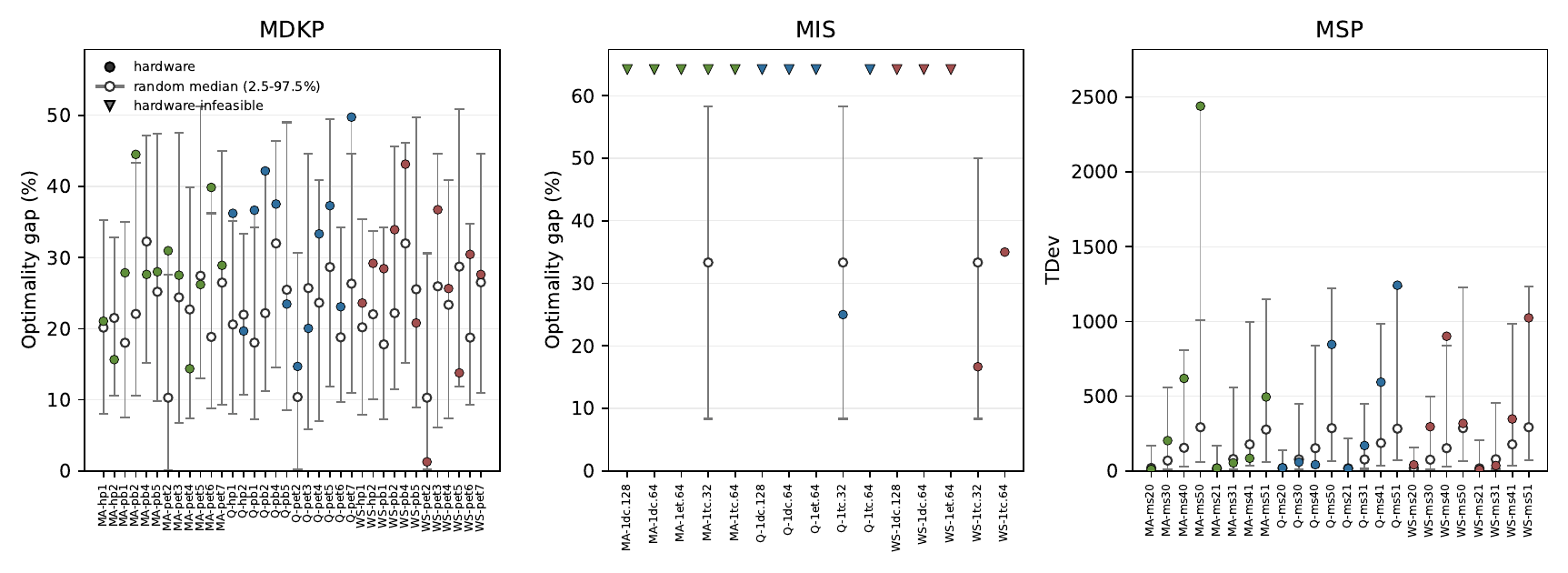}
  \caption{Matched uniform-random baseline for low-fidelity QAOA-family hardware
    runs. Filled circles show final hardware outcomes. Open circles show random
    medians, and vertical intervals show the central $2.5$--$97.5\%$ range across
    300 matched random replicates. Each random pool matches the original run’s
    decision-variable dimension, optimizer-trajectory batch structure,
    candidate-selection rule, and one-round local refinement. MDKP and MIS report
    optimality gap; MSP reports TDev. Downward triangles indicate hardware-infeasible
    runs. QAP is omitted because all hardware outputs are infeasible and valid
    random permutations are astronomically unlikely at the sampled candidate
  budgets.}
  \label{fig:low_fidelity_random_baseline}
\end{figure*}

The dominant empirical pattern is therefore not broad evidence of optimization
beyond the matched random baseline. Instead, most feasible low-fidelity
QAOA-family outcomes are compatible with best-of-budget random candidate
selection under the same trajectory-level selection and common local-refinement
protocol, or are worse than that baseline. At the same time, the
\texttt{1tc.64} feasibility exception shows why the gate-count fidelity proxy
must not be interpreted as proof that every low-fidelity output distribution
is exactly uniform random. Figure~\ref{fig:low_fidelity_random_baseline}
shows the run-level hardware outcomes relative to the matched-random medians
and central empirical intervals.

\begin{table*}[t]
  \centering
  \caption{Matched uniform-random control for low-fidelity QAOA-family hardware
    runs. Within range'' and Worse’’ apply only to feasible hardware runs for
    which the matched random control yields a finite quality distribution.
    ``Feasibility exception’’ denotes a feasible hardware result with zero feasible
    matched random draws. QAP is feasibility-only because all selected hardware
  outputs are infeasible.}
  \label{tab:random_baseline_summary}
  \footnotesize
  \setlength{\tabcolsep}{3.3pt}
  \renewcommand{\arraystretch}{1.08}
  \begin{tabular}{llrrrrrr}
    \toprule
    Problem & Method & Runs & HW feasible & Within range & Worse &
    Feasibility exception & HW infeasible \\
    \midrule
    MDKP & MA-QAOA & 12 & 12 & 9 & 3 & 0 & 0 \\
    MDKP & QAOA    & 12 & 12 & 8 & 4 & 0 & 0 \\
    MDKP & WS-QAOA & 12 & 12 & 12 & 0 & 0 & 0 \\
    MIS  & MA-QAOA & 5  & 0  & 0 & 0 & 0 & 5 \\
    MIS  & QAOA    & 5  & 1  & 1 & 0 & 0 & 4 \\
    MIS  & WS-QAOA & 5  & 2  & 1 & 0 & 1 & 3 \\
    MSP  & MA-QAOA & 8  & 8  & 7 & 1 & 0 & 0 \\
    MSP  & QAOA    & 8  & 8  & 7 & 1 & 0 & 0 \\
    MSP  & WS-QAOA & 8  & 8  & 6 & 2 & 0 & 0 \\
    QAP  & MA-QAOA & 11 & 0  & – & – & – & 11 \\
    QAP  & QAOA    & 11 & 0  & – & – & – & 11 \\
    QAP  & WS-QAOA & 9  & 0  & – & – & – & 9 \\
    \midrule
    Total & – & 106 & 63 & 51 & 11 & 1 & 43 \\
    \bottomrule
  \end{tabular}
\end{table*}

The common local-refinement stage is materially active in the matched
pipeline. Its scale is reported separately by problem family because MDKP and
MIS use percentage optimality gap whereas MSP uses TDev. Across matched random
replicates, the median pre-to-post refinement improvement is $20.56$ percentage
points for MDKP, $8.33$ percentage points for MIS, and $6675$ TDev units for
MSP. These values describe the shared hybrid post-processing stage and should
not be compared across problem families as a single pooled quantity.

The complete selection ledger, all 31,800 random-baseline replicates,
corrected run-level classifications, and problem-level aggregate summaries are
provided in the reproducibility artifact.

\clearpage

\section{Reproducibility details}
\label{app:reproducibility}

\subsection{Software stack and experimental window}

\begin{table}[htbp]
  \centering
  \caption{Software stack used for all experiments in this study.}
  \label{tab:software_stack}
  \small
  \begin{tabular}{ll}
    \toprule
    Component & Version \\
    \midrule
    Python & 3.10.16 \\
    Qiskit & 2.3.0 \\
    \texttt{qiskit-ibm-runtime} & 0.45.1 \\
    \texttt{qiskit-aer} & 0.17.2 \\
    SciPy & 1.15.3 \\
    NumPy & 2.2.6 \\
    CPLEX Python API & 22.1.2.0 \\
    Gurobi (classical baseline) & 13.0.1 \\
    \bottomrule
  \end{tabular}
\end{table}

\subsection{Execution and Runtime mitigation scope}
\label{app:mitigation_provenance}

The execution contains 255 method–instance records:
84 MDKP, 49 MIS, 56 MSP, and 66 QAP records. All Estimator-based
objective evaluations use \texttt{resilience\_level=2} with 1,000 requested
shots and no manually configured mitigation sub-options. In particular, the metadata does not retain custom TREX randomization settings, ZNE
noise factors, extrapolator choices, twirling parameters, or
dynamical-decoupling settings. These settings are therefore not reported as
historical protocol facts.

Final candidate bitstrings are generated by separate SamplerV2 jobs with
raw/default options and no custom sampler noise-management configuration.
The ledger records raw/default final sampling for all 255 archived
records. Most use 1,000 final sampling shots; a small subset follows
the recorded one-shot final-sampling pathway. In all cases, final bitstrings
are decoded and checked classically rather than inferred from Estimator
expectation values.

\begin{table}[ht]
  \centering
  \caption{Execution summary. The table distinguishes
  Estimator mitigation from final SamplerV2 decoding samples.}
  \label{tab:mitigation_provenance_summary}
  \footnotesize
  \setlength{\tabcolsep}{3.0pt}
  \renewcommand{\arraystretch}{1.08}
  \begin{tabularx}{\textwidth}{@{}>{\raggedright\arraybackslash}p{0.43\textwidth}>{\raggedright\arraybackslash}X@{}}
    \toprule
    Provenance field & Archived result \\
    \midrule
    Archived method–instance records & 255 \\
    Estimator resilience level & 2 for all records \\
    Estimator requested shots & 1,000 for all records \\
    Custom Estimator mitigation options & None recorded \\
    Final sampling primitive & Separate SamplerV2 \\
    Custom Sampler noise-management options & None recorded \\
    Final bitstrings from raw/default SamplerV2 & Yes, all records \\
    Final sampling shots & 1,000 for 235 records; 1 for 20 records \\
    Paired level-0 mitigation-overhead study & Not performed \\
    \bottomrule
  \end{tabularx}
\end{table}

The data does not contain paired executions of the same circuit
at \texttt{resilience\_level=0} and \texttt{resilience\_level=2}. We therefore
do not infer a circuit-specific mitigation-overhead multiplier
retrospectively. Wall-clock durations are retained where available, but they
combine queueing, backend availability, calibration-time variation,
compilation, managed Runtime work, and ordinary execution time; they are not
interpreted as a controlled mitigation-cost measurement.

Transpiler \texttt{optimization\_level=3} was used for hardware-aware circuit
compilation. Compilation and Runtime mitigation are distinct pipeline stages:
the former determines the executable circuit, whereas the latter affects only
Estimator-based expectation-value evaluation in the protocol.

\subsection{Master fidelity table}

Table~\ref{tab:master_fidelity} (on the following pages) reports the
estimated process fidelity, hardware outcome, and simulator outcome for every
method--instance--backend combination evaluated in this study.

\begin{landscape}
\footnotesize
\begin{xltabular}{\linewidth}{p{1.2cm}p{1.8cm}p{1.4cm}Xrrrrc}
\caption{Complete process-fidelity estimates for every method--instance--backend combination evaluated on hardware in this benchmark. $N_{2Q}$: transpiled two-qubit gate count. $F_{\mathrm{est}}$: estimated process fidelity under the independent-depolarizing model of Section~4.11 using the median backend $\varepsilon_{2Q}$ during the experimental window. Sim and HW gap are reported as percentage optimality gap (or TDev for MSP) where available; `--' indicates not evaluated. Feas: Y if the hardware run returned a feasible solution, N otherwise.}\label{tab:master_fidelity}\\
\toprule
Problem & Method & Instance & Backend & $N_{2Q}$ & $F_{\mathrm{est}}$ & Sim gap \% & HW gap \% & Feas \\
\midrule
\endfirsthead
\multicolumn{9}{c}{\tablename\ \thetable{} -- continued from previous page} \\
\toprule
Problem & Method & Instance & Backend & $N_{2Q}$ & $F_{\mathrm{est}}$ & Sim gap \% & HW gap \% & Feas \\
\midrule
\endhead
\midrule
\multicolumn{9}{r}{\textit{Continued on next page}} \\
\endfoot
\bottomrule
\endlastfoot
MDKP & CVaR-VQE & hp1 & \texttt{ibm\_torino} & 180 & 0.486 & 20.59 & 33.85 & Y \\
MDKP & CVaR-VQE & hp2 & \texttt{ibm\_torino} & 279 & 0.327 & 21.78 & 15.88 & Y \\
MDKP & CVaR-VQE & pb1 & \texttt{ibm\_torino} & 314 & 0.284 & 23.43 & 14.89 & Y \\
MDKP & CVaR-VQE & pb2 & \texttt{ibm\_torino} & 295 & 0.307 & 22.78 & 11.93 & Y \\
MDKP & CVaR-VQE & pb4 & \texttt{ibm\_torino} & 309 & 0.29 & 39.38 & 34.97 & Y \\
MDKP & CVaR-VQE & pb5 & \texttt{ibm\_torino} & 876 & 0.03 & 33.34 & 18.14 & Y \\
MDKP & CVaR-VQE & pet2 & \texttt{ibm\_torino} & 347 & 0.249 & 30.37 & 1.50 & Y \\
MDKP & CVaR-VQE & pet3 & \texttt{ibm\_torino} & 324 & 0.273 & 34.74 & 36.61 & Y \\
MDKP & CVaR-VQE & pet4 & \texttt{ibm\_torino} & 459 & 0.159 & 48.44 & 41.91 & Y \\
MDKP & CVaR-VQE & pet5 & \texttt{ibm\_torino} & 1227 & 0.00731 & 53.15 & 13.83 & Y \\
MDKP & CVaR-VQE & pet6 & \texttt{ibm\_torino} & 332 & 0.264 & 40.41 & 13.88 & Y \\
MDKP & CVaR-VQE & pet7 & \texttt{ibm\_torino} & 300 & 0.3 & 46.47 & 24.52 & Y \\
MDKP & MA-QAOA & hp1 & \texttt{ibm\_fez} & 22896 & 1.3299999999999999e-30 & -- & 21.06 & Y \\
MDKP & MA-QAOA & hp2 & \texttt{ibm\_fez} & 30125 & 4.9199999999999996e-40 & -- & 15.66 & Y \\
MDKP & MA-QAOA & pb1 & \texttt{ibm\_fez} & 21754 & 4.1200000000000003e-29 & -- & 27.86 & Y \\
MDKP & MA-QAOA & pb2 & \texttt{ibm\_fez} & 28136 & 1.94e-37 & -- & 44.51 & Y \\
MDKP & MA-QAOA & pb4 & \texttt{ibm\_fez} & 11506 & 9.69e-16 & -- & 27.63 & Y \\
MDKP & MA-QAOA & pb5 & \texttt{ibm\_fez} & 46919 & 6e-62 & -- & 28.00 & Y \\
MDKP & MA-QAOA & pet2 & \texttt{ibm\_fez} & 23551 & 1.8600000000000002e-31 & -- & 30.97 & Y \\
MDKP & MA-QAOA & pet3 & \texttt{ibm\_fez} & 32638 & 2.5900000000000003e-43 & -- & 27.52 & Y \\
MDKP & MA-QAOA & pet4 & \texttt{ibm\_fez} & 41625 & 4.85e-55 & -- & 14.38 & Y \\
MDKP & MA-QAOA & pet5 & \texttt{ibm\_fez} & 62656 & 1.75e-82 & -- & 26.21 & Y \\
MDKP & MA-QAOA & pet6 & \texttt{ibm\_fez} & 45006 & 1.8800000000000002e-59 & -- & 39.87 & Y \\
MDKP & MA-QAOA & pet7 & \texttt{ibm\_marrakesh} & 66556 & 1.43e-87 & -- & 28.91 & Y \\
MDKP & PCE & hp1 & \texttt{ibm\_torino} & 168 & 0.51 & 20.88 & 13.25 & Y \\
MDKP & PCE & hp2 & \texttt{ibm\_torino} & 224 & 0.407 & 38.38 & 14.56 & Y \\
MDKP & PCE & pb1 & \texttt{ibm\_torino} & 168 & 0.51 & 14.04 & 22.78 & Y \\
MDKP & PCE & pb2 & \texttt{ibm\_torino} & 224 & 0.407 & 14.37 & 21.37 & Y \\
MDKP & PCE & pb4 & \texttt{ibm\_torino} & 120 & 0.618 & 32.91 & 21.30 & Y \\
MDKP & PCE & pb5 & \texttt{ibm\_fez} & 360 & 0.339 & 11.87 & 17.72 & Y \\
MDKP & PCE & pet2 & \texttt{ibm\_fez} & 288 & 0.421 & 28.19 & 1.69 & Y \\
MDKP & PCE & pet3 & \texttt{ibm\_fez} & 288 & 0.421 & 15.56 & 7.97 & Y \\
MDKP & PCE & pet4 & \texttt{ibm\_fez} & 288 & 0.421 & 50.98 & 20.42 & Y \\
MDKP & PCE & pet5 & \texttt{ibm\_fez} & 360 & 0.339 & 22.74 & 37.50 & Y \\
MDKP & PCE & pet6 & \texttt{ibm\_fez} & 288 & 0.421 & 33.84 & 19.92 & Y \\
MDKP & PCE & pet7 & \texttt{ibm\_fez} & 288 & 0.421 & 15.87 & 25.27 & Y \\
MDKP & QAOA & hp1 & \texttt{ibm\_fez} & 22501 & 4.36e-30 & -- & 36.22 & Y \\
MDKP & QAOA & hp2 & \texttt{ibm\_fez} & 28967 & 1.59e-38 & -- & 19.68 & Y \\
MDKP & QAOA & pb1 & \texttt{ibm\_fez} & 21821 & 3.37e-29 & -- & 36.67 & Y \\
MDKP & QAOA & pb2 & \texttt{ibm\_fez} & 29320 & 5.52e-39 & -- & 42.18 & Y \\
MDKP & QAOA & pb4 & \texttt{ibm\_fez} & 11522 & 9.24e-16 & -- & 37.53 & Y \\
MDKP & QAOA & pb5 & \texttt{ibm\_fez} & 46275 & 4.1499999999999996e-61 & -- & 23.47 & Y \\
MDKP & QAOA & pet2 & \texttt{ibm\_fez} & 24261 & 2.2e-32 & -- & 14.70 & Y \\
MDKP & QAOA & pet3 & \texttt{ibm\_fez} & 32232 & 8.760000000000001e-43 & -- & 20.05 & Y \\
MDKP & QAOA & pet4 & \texttt{ibm\_fez} & 41325 & 1.1999999999999999e-54 & -- & 33.33 & Y \\
MDKP & QAOA & pet5 & \texttt{ibm\_fez} & 63249 & 2.95e-83 & -- & 37.30 & Y \\
MDKP & QAOA & pet6 & \texttt{ibm\_fez} & 47056 & 3.9799999999999995e-62 & -- & 23.09 & Y \\
MDKP & QAOA & pet7 & \texttt{ibm\_fez} & 65365 & 5.1200000000000004e-86 & -- & 49.76 & Y \\
MDKP & QRAO & hp1 & \texttt{ibm\_fez} & 111 & 0.716 & -- & 5.85 & Y \\
MDKP & QRAO & hp2 & \texttt{ibm\_fez} & 132 & 0.673 & -- & 16.67 & Y \\
MDKP & QRAO & pb1 & \texttt{ibm\_fez} & 108 & 0.723 & -- & 15.40 & Y \\
MDKP & QRAO & pb2 & \texttt{ibm\_marrakesh} & 129 & 0.679 & -- & 13.65 & Y \\
MDKP & QRAO & pb4 & \texttt{ibm\_marrakesh} & 93 & 0.756 & -- & 21.71 & Y \\
MDKP & QRAO & pb5 & \texttt{ibm\_marrakesh} & 168 & 0.604 & -- & 18.00 & Y \\
MDKP & QRAO & pet2 & \texttt{ibm\_torino} & 129 & 0.596 & -- & 0.64 & Y \\
MDKP & QRAO & pet3 & \texttt{ibm\_marrakesh} & 129 & 0.679 & -- & 34.12 & Y \\
MDKP & QRAO & pet4 & \texttt{ibm\_torino} & 138 & 0.575 & -- & 18.63 & Y \\
MDKP & QRAO & pet5 & \texttt{ibm\_fez} & 171 & 0.598 & -- & 25.52 & Y \\
MDKP & QRAO & pet6 & \texttt{ibm\_fez} & 153 & 0.631 & -- & 15.23 & Y \\
MDKP & QRAO & pet7 & \texttt{ibm\_fez} & 183 & 0.577 & -- & 16.79 & Y \\
MDKP & VQE & hp1 & \texttt{ibm\_fez} & 180 & 0.582 & 39.76 & 26.95 & Y \\
MDKP & VQE & hp2 & \texttt{ibm\_fez} & 338 & 0.362 & 12.34 & 17.92 & Y \\
MDKP & VQE & pb1 & \texttt{ibm\_fez} & 363 & 0.336 & 19.94 & 22.62 & Y \\
MDKP & VQE & pb2 & \texttt{ibm\_fez} & 344 & 0.356 & 19.49 & 26.77 & Y \\
MDKP & VQE & pb4 & \texttt{ibm\_fez} & 270 & 0.444 & -- & 33.63 & Y \\
MDKP & VQE & pb5 & \texttt{ibm\_fez} & 348 & 0.351 & 4.25 & 29.73 & Y \\
MDKP & VQE & pet2 & \texttt{ibm\_fez} & 326 & 0.376 & 41.07 & 16.98 & Y \\
MDKP & VQE & pet3 & \texttt{ibm\_fez} & 347 & 0.353 & 4.98 & 32.38 & Y \\
MDKP & VQE & pet4 & \texttt{ibm\_fez} & 410 & 0.292 & 66.58 & 44.69 & Y \\
MDKP & VQE & pet5 & \texttt{ibm\_fez} & 384 & 0.315 & 33.23 & 25.65 & Y \\
MDKP & VQE & pet6 & \texttt{ibm\_fez} & 390 & 0.31 & 12.50 & 20.24 & Y \\
MDKP & VQE & pet7 & \texttt{ibm\_fez} & 300 & 0.406 & 43.46 & 23.28 & Y \\
MDKP & WS-QAOA & hp1 & \texttt{ibm\_fez} & 20658 & 1.11e-27 & -- & 23.61 & Y \\
MDKP & WS-QAOA & hp2 & \texttt{ibm\_fez} & 29780 & 1.39e-39 & -- & 29.19 & Y \\
MDKP & WS-QAOA & pb1 & \texttt{ibm\_fez} & 22192 & 1.1000000000000001e-29 & -- & 28.45 & Y \\
MDKP & WS-QAOA & pb2 & \texttt{ibm\_fez} & 28862 & 2.19e-38 & -- & 33.93 & Y \\
MDKP & WS-QAOA & pb4 & \texttt{ibm\_fez} & 11808 & 3.91e-16 & -- & 43.13 & Y \\
MDKP & WS-QAOA & pb5 & \texttt{ibm\_fez} & 47353 & 1.63e-62 & -- & 20.80 & Y \\
MDKP & WS-QAOA & pet2 & \texttt{ibm\_fez} & 23747 & 1.03e-31 & -- & 1.28 & Y \\
MDKP & WS-QAOA & pet3 & \texttt{ibm\_fez} & 32589 & 3e-43 & -- & 36.74 & Y \\
MDKP & WS-QAOA & pet4 & \texttt{ibm\_torino} & 41282 & 1.3900000000000001e-72 & -- & 25.65 & Y \\
MDKP & WS-QAOA & pet5 & \texttt{ibm\_torino} & 64628 & 3.2e-113 & -- & 13.79 & Y \\
MDKP & WS-QAOA & pet6 & \texttt{ibm\_torino} & 45559 & 4.98e-80 & -- & 30.45 & Y \\
MDKP & WS-QAOA & pet7 & \texttt{ibm\_torino} & 65953 & 1.58e-115 & -- & 27.62 & Y \\
\midrule
MIS & CVaR-VQE & 1dc.128 & \texttt{ibm\_torino} & 1284 & 0.00582 & -- & -- & N \\
MIS & CVaR-VQE & 1dc.64 & \texttt{ibm\_torino} & 192 & 0.463 & -- & -- & N \\
MIS & CVaR-VQE & 1et.64 & \texttt{ibm\_torino} & 192 & 0.463 & -- & -- & N \\
MIS & CVaR-VQE & 1tc.16 & \texttt{ibm\_torino} & 102 & 0.664 & 0.00 & 12.50 & Y \\
MIS & CVaR-VQE & 1tc.32 & \texttt{ibm\_marrakesh} & 96 & 0.749 & 8.33 & 16.67 & Y \\
MIS & CVaR-VQE & 1tc.64 & \texttt{ibm\_torino} & 192 & 0.463 & -- & -- & N \\
MIS & CVaR-VQE & 1tc.8 & \texttt{ibm\_torino} & 60 & 0.786 & 0.00 & 0.00 & Y \\
MIS & MA-QAOA & 1dc.128 & \texttt{ibm\_fez} & 47824 & 3.96e-63 & -- & -- & N \\
MIS & MA-QAOA & 1dc.64 & \texttt{ibm\_fez} & 14015 & 5.16e-19 & -- & -- & N \\
MIS & MA-QAOA & 1et.64 & \texttt{ibm\_fez} & 4495 & 1.36e-06 & -- & -- & N \\
MIS & MA-QAOA & 1tc.16 & \texttt{ibm\_fez} & 237 & 0.491 & -- & 12.50 & Y \\
MIS & MA-QAOA & 1tc.32 & \texttt{ibm\_fez} & 931 & 0.061 & -- & -- & N \\
MIS & MA-QAOA & 1tc.64 & \texttt{ibm\_fez} & 3265 & 5.49e-05 & -- & -- & N \\
MIS & MA-QAOA & 1tc.8 & \texttt{ibm\_fez} & 60 & 0.835 & -- & 0.00 & Y \\
MIS & PCE & 1dc.128 & \texttt{ibm\_torino} & 18 & 0.93 & -- & -- & N \\
MIS & PCE & 1dc.64 & \texttt{ibm\_fez} & 14 & 0.959 & -- & -- & N \\
MIS & PCE & 1et.64 & \texttt{ibm\_fez} & 14 & 0.959 & -- & -- & N \\
MIS & PCE & 1tc.16 & \texttt{ibm\_torino} & 6 & 0.976 & 50.00 & 62.50 & Y \\
MIS & PCE & 1tc.32 & \texttt{ibm\_fez} & 10 & 0.97 & 41.67 & -- & N \\
MIS & PCE & 1tc.64 & \texttt{ibm\_torino} & 14 & 0.945 & -- & -- & N \\
MIS & PCE & 1tc.8 & \texttt{ibm\_fez} & 4 & 0.988 & 25.00 & 25.00 & Y \\
MIS & QAOA & 1dc.128 & \texttt{ibm\_fez} & 48702 & 2.83e-64 & -- & -- & N \\
MIS & QAOA & 1dc.64 & \texttt{ibm\_fez} & 14383 & 1.71e-19 & -- & -- & N \\
MIS & QAOA & 1et.64 & \texttt{ibm\_torino} & 4600 & 9.84e-09 & -- & -- & N \\
MIS & QAOA & 1tc.16 & \texttt{ibm\_torino} & 241 & 0.381 & -- & 25.00 & Y \\
MIS & QAOA & 1tc.32 & \texttt{ibm\_torino} & 935 & 0.024 & -- & 25.00 & Y \\
MIS & QAOA & 1tc.64 & \texttt{ibm\_torino} & 4600 & 9.84e-09 & -- & -- & N \\
MIS & QAOA & 1tc.8 & \texttt{ibm\_torino} & 60 & 0.786 & -- & 0.00 & Y \\
MIS & QRAO & 1dc.128 & \texttt{ibm\_fez} & 135 & 0.667 & -- & -- & N \\
MIS & QRAO & 1dc.64 & \texttt{ibm\_torino} & 72 & 0.749 & -- & -- & N \\
MIS & QRAO & 1et.64 & \texttt{ibm\_fez} & 66 & 0.82 & -- & -- & N \\
MIS & QRAO & 1tc.16 & \texttt{ibm\_fez} & 15 & 0.956 & -- & 12.50 & Y \\
MIS & QRAO & 1tc.32 & \texttt{ibm\_fez} & 36 & 0.897 & -- & 25.00 & Y \\
MIS & QRAO & 1tc.64 & \texttt{ibm\_torino} & 66 & 0.768 & -- & -- & N \\
MIS & QRAO & 1tc.8 & \texttt{sim} & 9 & 0.969 & -- & 0.00 & Y \\
MIS & VQE & 1dc.128 & \texttt{ibm\_fez} & 657 & 0.139 & -- & -- & N \\
MIS & VQE & 1dc.64 & \texttt{ibm\_fez} & 192 & 0.562 & -- & -- & N \\
MIS & VQE & 1et.64 & \texttt{ibm\_fez} & 192 & 0.562 & -- & -- & N \\
MIS & VQE & 1tc.16 & \texttt{ibm\_fez} & 102 & 0.736 & 0.00 & 0.00 & Y \\
MIS & VQE & 1tc.32 & \texttt{ibm\_fez} & 96 & 0.749 & 8.33 & -- & N \\
MIS & VQE & 1tc.64 & \texttt{ibm\_fez} & 192 & 0.562 & -- & -- & N \\
MIS & VQE & 1tc.8 & \texttt{ibm\_fez} & 78 & 0.791 & 0.00 & 0.00 & Y \\
MIS & WS-QAOA & 1dc.128 & \texttt{ibm\_torino} & 48544 & 3.17e-85 & -- & -- & N \\
MIS & WS-QAOA & 1dc.64 & \texttt{ibm\_torino} & 14029 & 3.8e-25 & -- & -- & N \\
MIS & WS-QAOA & 1et.64 & \texttt{ibm\_torino} & 14029 & 3.8e-25 & -- & -- & N \\
MIS & WS-QAOA & 1tc.16 & \texttt{ibm\_torino} & 237 & 0.387 & -- & 0.00 & Y \\
MIS & WS-QAOA & 1tc.32 & \texttt{ibm\_torino} & 911 & 0.026 & -- & 16.67 & Y \\
MIS & WS-QAOA & 1tc.64 & \texttt{ibm\_torino} & 14029 & 3.8e-25 & -- & 35.00 & Y \\
MIS & WS-QAOA & 1tc.8 & \texttt{ibm\_torino} & 60 & 0.786 & -- & 0.00 & Y \\
\midrule
MSP & CVaR-VQE & ms20 & \texttt{ibm\_marrakesh} & 144 & 0.649 & -- & 27.00 & Y \\
MSP & CVaR-VQE & ms21 & \texttt{ibm\_marrakesh} & 318 & 0.385 & -- & 9.00 & Y \\
MSP & CVaR-VQE & ms30 & \texttt{ibm\_marrakesh} & 252 & 0.469 & -- & 347.00 & Y \\
MSP & CVaR-VQE & ms31 & \texttt{ibm\_marrakesh} & 252 & 0.469 & -- & 51.00 & Y \\
MSP & CVaR-VQE & ms40 & \texttt{ibm\_marrakesh} & 408 & 0.294 & -- & 150.00 & Y \\
MSP & CVaR-VQE & ms41 & \texttt{ibm\_marrakesh} & 408 & 0.294 & -- & 48.00 & Y \\
MSP & CVaR-VQE & ms50 & \texttt{ibm\_marrakesh} & 1534 & 0.00996 & -- & 476.00 & Y \\
MSP & CVaR-VQE & ms51 & \texttt{ibm\_fez} & 1621 & 0.00767 & -- & 565.00 & Y \\
MSP & PCE & ms20 & \texttt{ibm\_torino} & 168 & 0.51 & -- & 45.00 & Y \\
MSP & PCE & ms21 & \texttt{ibm\_torino} & 168 & 0.51 & -- & 52.00 & Y \\
MSP & PCE & ms30 & \texttt{ibm\_torino} & 224 & 0.407 & -- & 167.00 & Y \\
MSP & PCE & ms31 & \texttt{ibm\_fez} & 224 & 0.51 & -- & 185.00 & Y \\
MSP & PCE & ms40 & \texttt{ibm\_torino} & 360 & 0.236 & -- & 296.00 & Y \\
MSP & PCE & ms41 & \texttt{ibm\_fez} & 360 & 0.339 & -- & 317.00 & Y \\
MSP & PCE & ms50 & \texttt{ibm\_torino} & 440 & 0.171 & -- & 822.00 & Y \\
MSP & PCE & ms51 & \texttt{ibm\_fez} & 440 & 0.267 & -- & 845.00 & Y \\
MSP & QAOA & ms20 & \texttt{ibm\_fez} & 13304 & 4.37e-18 & -- & 21.00 & Y \\
MSP & QAOA & ms21 & \texttt{ibm\_fez} & 14614 & 8.53e-20 & -- & 15.00 & Y \\
MSP & QAOA & ms30 & \texttt{ibm\_fez} & 38039 & 2.32e-50 & -- & 59.00 & Y \\
MSP & QAOA & ms31 & \texttt{ibm\_fez} & 38039 & 2.32e-50 & -- & 170.00 & Y \\
MSP & QAOA & ms40 & \texttt{ibm\_fez} & 76599 & 1.1200000000000001e-100 & -- & 42.00 & Y \\
MSP & QAOA & ms41 & \texttt{ibm\_fez} & 76599 & 1.1200000000000001e-100 & -- & 594.00 & Y \\
MSP & QAOA & ms50 & \texttt{ibm\_fez} & 123875 & 2.31e-162 & -- & 846.00 & Y \\
MSP & QAOA & ms51 & \texttt{ibm\_fez} & 133151 & 1.8199999999999999e-174 & -- & 1242.00 & Y \\
MSP & QRAO & ms20 & \texttt{ibm\_fez} & 87 & 0.77 & -- & 81.00 & Y \\
MSP & QRAO & ms21 & \texttt{ibm\_fez} & 87 & 0.77 & -- & 99.00 & Y \\
MSP & QRAO & ms30 & \texttt{ibm\_torino} & 123 & 0.611 & -- & 291.00 & Y \\
MSP & QRAO & ms31 & \texttt{ibm\_fez} & 123 & 0.691 & -- & 310.00 & Y \\
MSP & QRAO & ms40 & \texttt{ibm\_torino} & 216 & 0.421 & -- & 485.00 & Y \\
MSP & QRAO & ms41 & \texttt{ibm\_fez} & 216 & 0.523 & -- & 509.00 & Y \\
MSP & QRAO & ms50 & \texttt{ibm\_torino} & 249 & 0.369 & -- & 1120.00 & Y \\
MSP & QRAO & ms51 & \texttt{ibm\_fez} & 252 & 0.469 & -- & 1150.00 & Y \\
MSP & VQE & ms20 & \texttt{ibm\_fez} & 144 & 0.649 & -- & 4.00 & Y \\
MSP & VQE & ms21 & \texttt{ibm\_fez} & 369 & 0.33 & -- & 215.00 & Y \\
MSP & VQE & ms30 & \texttt{ibm\_fez} & 252 & 0.469 & -- & 127.00 & Y \\
MSP & VQE & ms31 & \texttt{ibm\_fez} & 252 & 0.469 & -- & 12.00 & Y \\
MSP & VQE & ms40 & \texttt{ibm\_fez} & 408 & 0.294 & -- & 188.00 & Y \\
MSP & VQE & ms41 & \texttt{ibm\_fez} & 408 & 0.294 & -- & 273.00 & Y \\
MSP & VQE & ms50 & \texttt{ibm\_fez} & 1348 & 0.017 & -- & 644.00 & Y \\
MSP & VQE & ms51 & \texttt{ibm\_fez} & 1536 & 0.0099 & -- & 439.00 & Y \\
MSP & WS-QAOA & ms20 & \texttt{ibm\_fez} & 13434 & 2.96e-18 & -- & 41.00 & Y \\
MSP & WS-QAOA & ms21 & \texttt{ibm\_fez} & 14116 & 3.81e-19 & -- & 9.00 & Y \\
MSP & WS-QAOA & ms30 & \texttt{ibm\_marrakesh} & 39227 & 6.529999999999999e-52 & -- & 296.00 & Y \\
MSP & WS-QAOA & ms31 & \texttt{ibm\_fez} & 40149 & 4.09e-53 & -- & 36.00 & Y \\
MSP & WS-QAOA & ms40 & \texttt{ibm\_marrakesh} & 75459 & 3.45e-99 & -- & 901.00 & Y \\
MSP & WS-QAOA & ms41 & \texttt{ibm\_marrakesh} & 75459 & 3.45e-99 & -- & 348.00 & Y \\
MSP & WS-QAOA & ms50 & \texttt{ibm\_fez} & 123414 & 9.21e-162 & -- & 318.00 & Y \\
MSP & WS-QAOA & ms51 & \texttt{ibm\_marrakesh} & 130345 & 8.33e-171 & -- & 1024.00 & Y \\
\midrule
QAP & CVaR-VQE & chr12a & \texttt{ibm\_marrakesh} & 1389 & 0.015 & -- & -- & N \\
QAP & CVaR-VQE & chr12b & \texttt{ibm\_marrakesh} & 1329 & 0.018 & -- & -- & N \\
QAP & CVaR-VQE & chr12c & \texttt{ibm\_marrakesh} & 1389 & 0.015 & -- & -- & N \\
QAP & CVaR-VQE & had12 & \texttt{ibm\_marrakesh} & 1389 & 0.015 & -- & -- & N \\
QAP & CVaR-VQE & nug12 & \texttt{ibm\_fez} & 1580 & 0.00868 & -- & -- & N \\
QAP & CVaR-VQE & rou12 & \texttt{ibm\_fez} & 1580 & 0.00868 & -- & -- & N \\
QAP & CVaR-VQE & scr12 & \texttt{ibm\_fez} & 1580 & 0.00868 & -- & -- & N \\
QAP & CVaR-VQE & tai10a & \texttt{ibm\_fez} & 300 & 0.406 & -- & -- & N \\
QAP & CVaR-VQE & tai10b & \texttt{ibm\_fez} & 300 & 0.406 & -- & -- & N \\
QAP & CVaR-VQE & tai12a & \texttt{ibm\_fez} & 1580 & 0.00868 & -- & -- & N \\
QAP & CVaR-VQE & tai12b & \texttt{ibm\_fez} & 1580 & 0.00868 & -- & -- & N \\
QAP & MA-QAOA & chr12a & \texttt{ibm\_fez} & 83724 & 5.670000000000001e-110 & -- & -- & N \\
QAP & MA-QAOA & chr12b & \texttt{ibm\_fez} & 83936 & 3.0000000000000004e-110 & -- & -- & N \\
QAP & MA-QAOA & chr12c & \texttt{ibm\_marrakesh} & 84276 & 1.0800000000000001e-110 & -- & -- & N \\
QAP & MA-QAOA & had12 & \texttt{ibm\_fez} & 173096 & 1.3700000000000001e-226 & -- & -- & N \\
QAP & MA-QAOA & nug12 & \texttt{ibm\_fez} & 158188 & 3.8899999999999996e-207 & -- & -- & N \\
QAP & MA-QAOA & rou12 & \texttt{ibm\_fez} & 173096 & 1.3700000000000001e-226 & -- & -- & N \\
QAP & MA-QAOA & scr12 & \texttt{ibm\_fez} & 127688 & 2.44e-167 & -- & -- & N \\
QAP & MA-QAOA & tai10a & \texttt{ibm\_fez} & 81971 & 1.1e-107 & -- & -- & N \\
QAP & MA-QAOA & tai10b & \texttt{ibm\_fez} & 78718 & 1.93e-103 & -- & -- & N \\
QAP & MA-QAOA & tai12a & \texttt{ibm\_fez} & 170894 & 1.02e-223 & -- & -- & N \\
QAP & MA-QAOA & tai12b & \texttt{ibm\_fez} & 167992 & 6.2699999999999996e-220 & -- & -- & N \\
QAP & PCE & chr12a & \texttt{ibm\_fez} & 440 & 0.267 & 234.75 & -- & N \\
QAP & PCE & chr12b & \texttt{ibm\_fez} & 440 & 0.267 & 15.47 & -- & N \\
QAP & PCE & chr12c & \texttt{ibm\_fez} & 440 & 0.267 & 66.82 & -- & N \\
QAP & PCE & had12 & \texttt{ibm\_fez} & 20 & 0.942 & -- & -- & N \\
QAP & PCE & nug12 & \texttt{ibm\_fez} & 20 & 0.942 & 35.98 & -- & N \\
QAP & PCE & rou12 & \texttt{ibm\_fez} & 20 & 0.942 & 26.30 & -- & N \\
QAP & QAOA & chr12a & \texttt{ibm\_fez} & 84111 & 1.77e-110 & -- & -- & N \\
QAP & QAOA & chr12b & \texttt{ibm\_fez} & 84111 & 1.77e-110 & -- & -- & N \\
QAP & QAOA & chr12c & \texttt{ibm\_fez} & 84111 & 1.77e-110 & -- & -- & N \\
QAP & QAOA & had12 & \texttt{ibm\_fez} & 84111 & 1.77e-110 & -- & -- & N \\
QAP & QAOA & nug12 & \texttt{ibm\_marrakesh} & 158163 & 4.1899999999999994e-207 & -- & -- & N \\
QAP & QAOA & rou12 & \texttt{ibm\_marrakesh} & 158163 & 4.1899999999999994e-207 & -- & -- & N \\
QAP & QAOA & scr12 & \texttt{ibm\_marrakesh} & 158163 & 4.1899999999999994e-207 & -- & -- & N \\
QAP & QAOA & tai10a & \texttt{ibm\_fez} & 74145 & 1.79e-97 & -- & -- & N \\
QAP & QAOA & tai10b & \texttt{ibm\_fez} & 74145 & 1.79e-97 & -- & -- & N \\
QAP & QAOA & tai12a & \texttt{ibm\_fez} & 168995 & 3.08e-221 & -- & -- & N \\
QAP & QAOA & tai12b & \texttt{ibm\_marrakesh} & 158163 & 4.1899999999999994e-207 & -- & -- & N \\
QAP & QRAO & chr12a & \texttt{ibm\_fez} & 165 & 0.609 & -- & -- & N \\
QAP & QRAO & chr12b & \texttt{ibm\_fez} & 183 & 0.577 & -- & -- & N \\
QAP & QRAO & chr12c & \texttt{ibm\_fez} & 159 & 0.62 & -- & -- & N \\
QAP & QRAO & had12 & \texttt{ibm\_fez} & 1110 & 0.036 & -- & -- & N \\
QAP & QRAO & nug12 & \texttt{ibm\_fez} & 186 & 0.572 & -- & -- & N \\
QAP & QRAO & rou12 & \texttt{ibm\_torino} & 1413 & 0.00347 & -- & -- & N \\
QAP & QRAO & scr12 & \texttt{ibm\_fez} & 177 & 0.588 & -- & -- & N \\
QAP & VQE & chr12a & \texttt{ibm\_fez} & 1323 & 0.019 & -- & -- & N \\
QAP & VQE & chr12b & \texttt{ibm\_fez} & 1323 & 0.019 & -- & -- & N \\
QAP & VQE & chr12c & \texttt{ibm\_fez} & 1218 & 0.026 & -- & -- & N \\
QAP & VQE & had12 & \texttt{ibm\_fez} & 1218 & 0.026 & -- & -- & N \\
QAP & VQE & nug12 & \texttt{ibm\_fez} & 1218 & 0.026 & -- & -- & N \\
QAP & VQE & rou12 & \texttt{ibm\_fez} & 1218 & 0.026 & -- & -- & N \\
QAP & VQE & scr12 & \texttt{ibm\_fez} & 1218 & 0.026 & -- & -- & N \\
QAP & VQE & tai10a & \texttt{ibm\_fez} & 300 & 0.406 & -- & -- & N \\
QAP & VQE & tai10b & \texttt{ibm\_fez} & 300 & 0.406 & -- & -- & N \\
QAP & VQE & tai12a & \texttt{ibm\_fez} & 1218 & 0.026 & -- & -- & N \\
QAP & VQE & tai12b & \texttt{ibm\_fez} & 1134 & 0.033 & -- & -- & N \\
QAP & WS-QAOA & chr12a & \texttt{ibm\_fez} & 84503 & 5.46e-111 & -- & -- & N \\
QAP & WS-QAOA & chr12b & \texttt{ibm\_fez} & 84503 & 5.46e-111 & -- & -- & N \\
QAP & WS-QAOA & chr12c & \texttt{ibm\_fez} & 84503 & 5.46e-111 & -- & -- & N \\
QAP & WS-QAOA & had12 & \texttt{ibm\_marrakesh} & 82789 & 9.410000000000001e-109 & -- & -- & N \\
QAP & WS-QAOA & nug12 & \texttt{ibm\_fez} & 159188 & 1.9300000000000002e-208 & -- & -- & N \\
QAP & WS-QAOA & rou12 & \texttt{ibm\_fez} & 159188 & 1.9300000000000002e-208 & -- & -- & N \\
QAP & WS-QAOA & scr12 & \texttt{ibm\_marrakesh} & 159153 & 2.14e-208 & -- & -- & N \\
QAP & WS-QAOA & tai10a & \texttt{ibm\_marrakesh} & 77609 & 5.4e-102 & -- & -- & N \\
QAP & WS-QAOA & tai10b & \texttt{ibm\_marrakesh} & 77609 & 5.4e-102 & -- & -- & N \\
\end{xltabular}
\end{landscape}

\section{Classical Baseline Performance}
\label{app:classical_baselines}

This section reports the classical baseline performance for all benchmark
families considered in the paper: the Multi-Dimensional Knapsack Problem
(MDKP), Maximum Independent Set (MIS), Quadratic Assignment Problem (QAP),
and Market Share Problem (MSP). These results provide a classical reference
for interpreting the hardware and simulator outcomes reported in the main
text.

For MDKP, MIS, and QAP, we report the performance of exact Gurobi-based
integer programming formulations, including the best incumbent found at
termination, the final best bound, the relative optimality gap, solver
status, and runtime. For MSP, we report the performance of CPLEX using the
same summary metrics adopted in the main experimental workflow, namely the
search effort in terms of branch-and-bound nodes and simplex/barrier
iterations, together with the final gap, solution value, and runtime.

Overall, the classical baselines confirm that the selected benchmark
instances span a meaningful range of difficulty. Many MDKP and MIS instances
are solved to proven optimality almost immediately, providing a clean
reference point for evaluating quantum solution quality. In contrast, several
QAP and larger MSP instances remain substantially more challenging: some QAP
instances terminate at the time limit with nonzero optimality gaps, while for
MSP the larger cases reach the one-hour limit with gaps close to~1. These
results support the role of the chosen benchmark families as classically
meaningful testbeds rather than trivial small-scale examples.

\begin{table*}[htbp]
  \caption{\label{tab:mdkp_classical}
    Classical Gurobi performance on MDKP instances.
    \textbf{Size}: items $\times$ dimensions.
    \textbf{BKS}: known optimal value from the benchmark file.
    \textbf{Incumbent}: best feasible solution at termination.
    \textbf{Bound}: final best bound.
    \textbf{Gap}: final relative optimality gap.
    \textbf{Status}: solver termination status.
  All listed instances were solved to proven optimality.}
  \centering
  \scriptsize
  \setlength{\tabcolsep}{2pt}
  \begin{tabular*}{\textwidth}{@{\extracolsep{\fill}}lccccccc@{}}
    \toprule
    \textbf{Inst.} & \textbf{Size} & \textbf{BKS} & \textbf{Inc.} &
    \textbf{Bound} & \textbf{Gap (\%)} & \textbf{Status} & \textbf{Time (s)} \\
    \midrule
    hp1  & $28{\times}4$  & 3\,418  & 3\,418  & 3\,418  & 0.0 & OPT & 0.01 \\
    hp2  & $35{\times}4$  & 3\,186  & 3\,186  & 3\,186  & 0.0 & OPT & 0.00 \\
    pb1  & $27{\times}4$  & 3\,090  & 3\,090  & 3\,090  & 0.0 & OPT & 0.00 \\
    pb2  & $34{\times}4$  & 3\,186  & 3\,186  & 3\,186  & 0.0 & OPT & 0.00 \\
    pb4  & $29{\times}2$  & 95\,168 & 95\,168 & 95\,168 & 0.0 & OPT & 0.00 \\
    pb5  & $20{\times}10$ & 2\,139  & 2\,139  & 2\,139  & 0.0 & OPT & 0.02 \\
    pet2 & $10{\times}10$ & 87\,061 & 87\,061 & 87\,061 & 0.0 & OPT & 0.00 \\
    pet3 & $15{\times}10$ & 4\,015  & 4\,015  & 4\,015  & 0.0 & OPT & 0.00 \\
    pet4 & $20{\times}10$ & 6\,120  & 6\,120  & 6\,120  & 0.0 & OPT & 0.00 \\
    pet5 & $28{\times}10$ & 12\,400 & 12\,400 & 12\,400 & 0.0 & OPT & 0.01 \\
    pet6 & $39{\times}5$  & 10\,618 & 10\,618 & 10\,618 & 0.0 & OPT & 0.00 \\
    pet7 & $50{\times}5$  & 16\,537 & 16\,537 & 16\,537 & 0.0 & OPT & 0.09 \\
    \bottomrule
  \end{tabular*}
\end{table*}

\begin{table*}[htbp]
  \caption{\label{tab:mis_classical}
    Classical Gurobi performance on MIS instances.
    \textbf{$|V|$}: number of vertices.
    \textbf{$|E|$}: number of edges.
    \textbf{BKS}: known optimal independent-set size.
    \textbf{Incumbent}: best feasible solution at termination.
    \textbf{Bound}: final best bound.
    \textbf{Gap}: final relative optimality gap.
  All listed instances were solved to proven optimality.}
  \centering
  \footnotesize
  \setlength{\tabcolsep}{4pt}
  \begin{tabular*}{\textwidth}{@{\extracolsep{\fill}}lcccccccc}
    \toprule
    \textbf{Instance} & \textbf{$|V|$} & \textbf{$|E|$} & \textbf{BKS} &
    \textbf{Incumbent} & \textbf{Bound} & \textbf{Gap (\%)} &
    \textbf{Status} & \textbf{Time (s)} \\
    \midrule
    1dc.64  & 64  & 543  & 10 & 10 & 10 & 0.0 & OPT & 0.00 \\
    1dc.128 & 128 & 1\,471 & 16 & 16 & 16 & 0.0 & OPT & 0.02 \\
    1et.64  & 64  & 264  & 18 & 18 & 18 & 0.0 & OPT & 0.00 \\
    1tc.8   & 8   & 6    & 4  & 4  & 4  & 0.0 & OPT & 0.00 \\
    1tc.16  & 16  & 22   & 8  & 8  & 8  & 0.0 & OPT & 0.00 \\
    1tc.32  & 32  & 68   & 12 & 12 & 12 & 0.0 & OPT & 0.00 \\
    1tc.64  & 64  & 192  & 20 & 20 & 20 & 0.0 & OPT & 0.00 \\
    \bottomrule
  \end{tabular*}
\end{table*}

\begin{table*}[htbp]
  \caption{\label{tab:qap_classical}
    Classical Gurobi performance on QAP instances.
    \textbf{$n$}: number of facilities/locations.
    \textbf{BKS}: known optimal objective value.
    \textbf{Incumbent}: best feasible solution at termination.
    \textbf{Bound}: final best bound.
    \textbf{Gap}: final relative optimality gap.
    \textbf{Status}: solver termination status.
  Entries with status TL reached the imposed time limit (100 seconds) before proving optimality.}
  \centering
  \footnotesize
  \setlength{\tabcolsep}{4pt}
  \begin{tabular*}{\textwidth}{@{\extracolsep{\fill}}lccccccc}
    \toprule
    \textbf{Instance} & \textbf{$n$} & \textbf{BKS} & \textbf{Incumbent} &
    \textbf{Bound} & \textbf{Gap (\%)} & \textbf{Status} & \textbf{Time (s)} \\
    \midrule
    chr12a & 12 & 9\,552     & 9\,552        & 9\,552        & 0.0  & OPT & 0.36 \\
    chr12b & 12 & 9\,742     & 9\,742        & 9\,742        & 0.0  & OPT & 0.21 \\
    chr12c & 12 & 11\,156    & 11\,156       & 11\,156       & 0.0  & OPT & 2.00 \\
    had12  & 12 & 1\,652     & 1\,652        & 1\,260        & 23.7 & TL  & 100.00 \\
    nug12  & 12 & 578        & 578           & 578           & 0.0  & OPT & 98.41 \\
    rou12  & 12 & 235\,528   & 235\,528      & 131\,514      & 44.2 & TL  & 100.00 \\
    scr12  & 12 & 31\,410    & 31\,410       & 31\,410       & 0.0  & OPT & 0.89 \\
    tai10a & 10 & 135\,028   & 135\,028      & 135\,028      & 0.0  & OPT & 5.82 \\
    tai10b & 10 & 1\,183\,760 & 1\,183\,760  & 1\,183\,760   & 0.0  & OPT & 0.62 \\
    tai12a & 12 & 224\,416   & 224\,416      & 142\,796      & 36.4 & TL  & 100.00 \\
    tai12b & 12 & 39\,464\,925 & 39\,464\,894.17 & 39\,464\,894.17 & 0.0 & OPT & 11.17 \\
    \bottomrule
  \end{tabular*}
\end{table*}

\begin{table*}[htbp]
  \caption{\label{tab:cplex_performance_metrics}
    CPLEX solver performance on Market Share Problem instances.
    \textbf{Gap}: relative difference between incumbent and lower bound
    (values near~$1$ indicate the solver could not close the gap within
    the time limit).
    \textbf{Nodes}: branch-and-bound nodes explored.
    \textbf{Iters}: simplex/barrier iterations.
  For sizes $6{\times}50$ and above, a one-hour time limit was imposed.}
  \centering
  \footnotesize
  \setlength{\tabcolsep}{4pt}
  \begin{tabular*}{\textwidth}{@{\extracolsep{\fill}}lccccc}
    \toprule
    \textbf{Size} & \textbf{Nodes} & \textbf{Iters} &
    \textbf{Gap} & \textbf{Sol.} & \textbf{Time (s)} \\
    \midrule
    $3{\times}20$ & 7\,552    & 9\,776         & 0.0         & 3 & 2.97 \\
    $3{\times}20$ & 7\,270    & 9\,035         & 0.0         & 2 & 1.70 \\
    $3{\times}20$ & 7\,559    & 10\,287        & 0.0         & 3 & 2.05 \\
    $3{\times}20$ & 9\,205    & 11\,446        & 0.0         & 3 & 1.13 \\
    $3{\times}20$ & 8\,301    & 11\,358        & 0.0         & 2 & 1.47 \\
    \midrule
    $4{\times}30$ & 765\,655  & 1\,622\,772    & 0.0         & 1 & 5.64 \\
    $4{\times}30$ & 72\,436   & 138\,158       & 0.0         & 0 & 5.92 \\
    $4{\times}30$ & 439\,294  & 960\,557       & 0.0         & 2 & 7.33 \\
    $4{\times}30$ & 538\,857  & 1\,197\,837    & 0.0         & 1 & 9.28 \\
    $4{\times}30$ & 772\,561  & 1\,687\,113    & 0.0         & 1 & 7.33 \\
    \midrule
    $5{\times}40$ & 50\,288\,061  & 117\,363\,318   & 0.0         & 1 & 777 \\
    $5{\times}40$ & 72\,691\,565  & 173\,591\,038   & 0.0         & 1 & 1\,076 \\
    $5{\times}40$ & 65\,010\,492  & 153\,229\,808   & 0.0         & 1 & 777 \\
    $5{\times}40$ & 41\,687\,999  & 98\,461\,473    & 0.0         & 1 & 292 \\
    $5{\times}40$ & 14\,683\,912  & 36\,843\,546    & 0.0         & 0 & 95 \\
    \midrule
    $6{\times}50$ & 263\,122\,381 & 682\,375\,105   & ${\approx}1$ & 1 & 3\,607 \\
    $6{\times}50$ & 433\,287\,296 & 1\,046\,486\,269 & ${\approx}1$ & 2 & 3\,600 \\
    $6{\times}50$ & 426\,949\,570 & 1\,037\,766\,414 & ${\approx}1$ & 2 & 3\,600 \\
    $6{\times}50$ & 544\,066\,780 & 1\,327\,364\,155 & ${\approx}1$ & 2 & 3\,600 \\
    $6{\times}50$ & 477\,967\,274 & 1\,147\,025\,106 & ${\approx}1$ & 2 & 3\,600 \\
    \midrule
    $7{\times}60$ & 448\,161\,278 & 1\,166\,369\,595 & ${\approx}1$ & 4 & 3\,617 \\
    $7{\times}60$ & 344\,532\,350 & 910\,970\,829    & ${\approx}1$ & 6 & 3\,617 \\
    $7{\times}60$ & 409\,426\,464 & 1\,086\,559\,026 & ${\approx}1$ & 5 & 3\,617 \\
    $7{\times}60$ & 491\,680\,008 & 1\,327\,295\,105 & ${\approx}1$ & 5 & 3\,619 \\
    $7{\times}60$ & 426\,054\,792 & 1\,120\,381\,222 & ${\approx}1$ & 6 & 3\,617 \\
    \bottomrule
  \end{tabular*}
\end{table*}

\clearpage

\section{Simulator Resource Usage by Problem and Method}
\label{app:resource_usage}

This appendix reports the computational resource usage for each
method--instance combination under simulator-based evaluation.  These
tables complement the hardware diagnostics by providing the algorithmic-level
circuit baseline before transpilation and backend-specific effects are
introduced.

\subsection{MDKP resource usage}
\label{app:resource_mdkp}

\begin{table*}[htbp]
  \caption{\label{tab:resource_usage_mdkp}
  MDKP simulator resource usage for VQE, CVaR-VQE, and PCE}
  \centering
  \footnotesize
  \setlength{\tabcolsep}{3pt}
  \begin{tabular*}{\textwidth}{@{\extracolsep{\fill}}llcccccr}
    \toprule
    \textbf{Inst.} & \textbf{Method} & \textbf{Q} & \textbf{D} & \textbf{G} & \textbf{2Q} & \textbf{P} & \textbf{Time (min)} \\
    \midrule
    hp1  & VQE      & 60  & 75  & 836  & 236 & 600  & 351.9 \\
    & CVaR-VQE & 60  & 71  & 557  & 177 & 480  & 246.8 \\
    & PCE      & 7   & 336 & 1568 & 336 & 364  & 72.5  \\
    \midrule
    hp2  & VQE      & 67  & 82  & 934  & 264 & 670  & 395.0 \\
    & CVaR-VQE & 67  & 78  & 734  & 198 & 538  & 384.3 \\
    & PCE      & 8   & 384 & 2080 & 448 & 480  & 125.8  \\
    \midrule
    pb1  & VQE      & 59  & 74  & 822  & 232 & 590  & 344.1 \\
    & CVaR-VQE & 59  & 70  & 646  & 174 & 472  & 241.6 \\
    & PCE      & 7   & 336 & 1568 & 336 & 364  & 71.0  \\
    \midrule
    pb2  & VQE      & 66  & 81  & 920  & 260 & 660  & 385.2 \\
    & CVaR-VQE & 66  & 77  & 723  & 195 & 528  & 312.2 \\
    & PCE      & 8   & 384 & 2080 & 448 & 480  & 119.0  \\
    \midrule
    pb4  & VQE      & 45  & 60  & 626  & 176 & 450  & 261.9 \\
    & CVaR-VQE & 45  & 56  & 492  & 132 & 360  & 174.4 \\
    & PCE      & 6   & 288 & 1128 & 240 & 264  & 28.1  \\
    \midrule
    pb5  & VQE      & 116 & 131 & 1620 & 460 & 1160 & 678.6 \\
    & CVaR-VQE & 116 & 127 & 1273 & 345 & 928  & 593.7 \\
    & PCE      & 10  & 480 & 3320 & 720 & 760  & 336.5  \\
    \midrule
    pet2 & VQE      & 99  & 114 & 1382 & 392 & 990  & 578.8 \\
    & CVaR-VQE & 99  & 110 & 1086 & 294 & 792  & 467.2 \\
    & PCE      & 9   & 432 & 2664 & 576 & 612  & 151.3  \\
    \midrule
    pet3 & VQE      & 102 & 117 & 1424 & 404 & 1020 & 596.4 \\
    & CVaR-VQE & 102 & 113 & 1119 & 303 & 816  & 565.9 \\
    & PCE      & 9   & 432 & 2664 & 576 & 612  & 187.1 \\
    \midrule
    pet4 & VQE      & 107 & 122 & 1494 & 424 & 1070 & 625.8 \\
    & CVaR-VQE & 107 & 118 & 1174 & 318 & 856  & 517.3 \\
    & PCE      & 9   & 432 & 2664 & 576 & 612  & 175.3  \\
    \midrule
    pet5 & VQE      & 122 & 137 & 1704 & 484 & 1220 & 713.8 \\
    & CVaR-VQE & 122 & 133 & 1339 & 363 & 976  & 576.8 \\
    & PCE      & 10  & 480 & 3320 & 720 & 760  & 345.5 \\
    \midrule
    pet6 & VQE      & 86  & 101 & 1210 & 340 & 860  & 502.5 \\
    & CVaR-VQE & 86  & 97  & 943  & 255 & 688  & 477.5 \\
    & PCE      & 9   & 432 & 2664 & 576 & 612  & 198.3  \\
    \midrule
    pet7 & VQE      & 100 & 115 & 1396 & 396 & 1000 & 584.7 \\
    & CVaR-VQE & 100 & 111 & 1097 & 297 & 800  & 550.1 \\
    & PCE      & 9   & 432 & 2664 & 576 & 612  & 198.6  \\
    \bottomrule
  \end{tabular*}
\end{table*}

\subsection{MIS resource usage}
\label{app:resource_mis}

\begin{table*}[!htbp]
  \caption{\label{tab:resource_usage_mis}
    MIS simulator resource usage for VQE, CVaR-VQE, PCE, QRAO, QAOA, and
    MA-QAOA\@.  Execution time is reported in seconds.  M.E.\ denotes
  memory exhaustion.}
  \centering
  \footnotesize
  \setlength{\tabcolsep}{3pt}
  \begin{tabular*}{\textwidth}{@{\extracolsep{\fill}}llcccccr}
    \toprule
    \textbf{Inst.} & \textbf{Method} & \textbf{Q} & \textbf{D} & \textbf{G} & \textbf{2Q} & \textbf{P} & \textbf{Time (s)} \\
    \midrule
    1tc.8   & VQE      & 8  & 19  & 85   & 21  & 64  & 3.1 \\
    & CVaR-VQE & 8  & 15  & 62   & 14  & 48  & 1.4 \\
    & PCE      & 3  & 144 & 240  & 48  & 60  & 0.6 \\
    & QRAO     & 4  & 11  & 30   & 6   & 24  & 1.4 \\
    & QAOA     & 8  & 12  & 42   & 12  & 30  & 15.2 \\
    & MA-QAOA  & 8  & 6   & 30   & 6   & 22  & 1.5 \\
    \midrule
    1tc.16  & VQE      & 16 & 27  & 173  & 45  & 128 & 17.4 \\
    & CVaR-VQE & 16 & 23  & 126  & 30  & 96  & 14.3 \\
    & PCE      & 4  & 192 & 464  & 96  & 112 & 1.9 \\
    & QRAO     & 6  & 13  & 54   & 10  & 34  & 4.3 \\
    & QAOA     & 16 & 30  & 114  & 44  & 70  & 52.4 \\
    & MA-QAOA  & 16 & 12  & 70   & 22  & 54  & 4.3 \\
    \midrule
    1tc.32  & VQE      & 32 & 43  & 349  & 93  & 256 & 68.2 \\
    & CVaR-VQE & 32 & 39  & 254  & 62  & 192 & 54.4 \\
    & PCE      & 6  & 288 & 1128 & 240 & 264 & 9.9 \\
    & QRAO     & 13 & 20  & 102  & 24  & 78  & 111.2 \\
    & QAOA     & 32 & 54  & 300  & 136 & 164 & 158.7 \\
    & MA-QAOA  & 32 & 20  & 164  & 68  & 132 & 11.8 \\
    \midrule
    1tc.64  & VQE      & 64 & 75  & 701  & 189 & 512 & 248.2 \\
    & CVaR-VQE & 64 & 71  & 510  & 126 & 384 & 186.6 \\
    & PCE      & 8  & 384 & 2080 & 448 & 480 & 34.7 \\
    & QRAO     & 23 & 30  & 182  & 44  & 138 & 310.1 \\
    & QAOA     & 64 & 105 & 768  & 384 & 384 & 702.4 \\
    & MA-QAOA  & 64 & 32  & 350  & 145 & 286 & 47.9 \\
    \midrule
    1et.64  & VQE      & 62 & 73  & 679  & 183 & 496 & 218.5 \\
    & CVaR-VQE & 62 & 69  & 494  & 122 & 372 & 156.2 \\
    & PCE      & 7  & 336 & 1568 & 336 & 364 & 26.8 \\
    & QRAO     & 24 & 31  & 190  & 46  & 144 & 516.6 \\
    & QAOA     & 62 & 105 & 978  & 528 & 450 & 812.8 \\
    & MA-QAOA  & 62 & 37  & 414  & 210 & 352 & 2\,668.7 \\
    \midrule
    1dc.64  & VQE      & 50 & 61  & 547  & 147 & 400 & 231.4 \\
    & CVaR-VQE & 50 & 57  & 398  & 98  & 300 & 184.1 \\
    & PCE      & 7  & 336 & 1568 & 336 & 364 & 31.9 \\
    & QRAO     & 18 & 25  & 142  & 34  & 108 & 133.0 \\
    & QAOA     & 50 & 252 & 1377 & 818 & 559 & M.E. \\
    & MA-QAOA  & 50 & 86  & 559  & 409 & 509 & 3\,135.2 \\
    \bottomrule
  \end{tabular*}
\end{table*}

\subsection{QAP resource usage}
\label{app:resource_qap}

\begin{table*}[htbp]
  \caption{\label{tab:resource_usage_qap}
  QAP simulator resource usage for VQE, CVaR-VQE, and PCE}
  \centering
  \footnotesize
  \setlength{\tabcolsep}{3pt}
  \begin{tabular*}{\textwidth}{@{\extracolsep{\fill}}llcccccr}
    \toprule
    \textbf{Inst.} & \textbf{Method} & \textbf{Q} & \textbf{D} & \textbf{G} & \textbf{2Q} & \textbf{P} & \textbf{Time (min)} \\
    \midrule
    chr12a & VQE      & 144 & 163 & 2443 & 715 & 1728 & 1\,917.5 \\
    & CVaR-VQE & 144 & 163 & 2443 & 715 & 1728 & 777.4 \\
    & PCE      & 11  & 528 & 4048 & 880 & 924  & 537.6 \\
    \midrule
    chr12b & VQE      & 144 & 163 & 2443 & 715 & 1728 & 1\,927.9 \\
    & CVaR-VQE & 144 & 163 & 2443 & 715 & 1728 & 772.5 \\
    & PCE      & 11  & 528 & 4048 & 880 & 924  & 592.7  \\
    \midrule
    chr12c & VQE      & 144 & 163 & 2443 & 715 & 1728 & 1\,910.4 \\
    & CVaR-VQE & 144 & 163 & 2443 & 715 & 1728 & 760.2 \\
    & PCE      & 11  & 528 & 4048 & 880 & 924  & 930.5  \\
    \midrule
    nug12  & VQE      & 144 & 163 & 2443 & 715 & 1728 & 1\,947.8 \\
    & CVaR-VQE & 144 & 163 & 2443 & 715 & 1728 & 765.6 \\
    & PCE      & 11  & 528 & 4048 & 880 & 924  & 1\,169.1  \\
    \midrule
    rou12  & VQE      & 144 & 163 & 2443 & 715 & 1728 & 1\,961.6 \\
    & CVaR-VQE & 144 & 163 & 2443 & 715 & 1728 & 796.5 \\
    & PCE      & 11  & 528 & 4048 & 880 & 924  & 1\,500.4  \\
    \midrule
    scr12  & VQE      & 144 & 163 & 2443 & 715 & 1728 & 1\,906.6 \\
    & CVaR-VQE & 144 & 163 & 2443 & 715 & 1728 & 783.4 \\
    & PCE      & 11  & 528 & 4048 & 880 & 924  & 831.4  \\
    \midrule
    tai12a & VQE      & 144 & 163 & 2443 & 715 & 1728 & 1\,883.8 \\
    & CVaR-VQE & 144 & 163 & 2443 & 715 & 1728 & 755.5 \\
    & PCE      & 11  & 528 & 4048 & 880 & 924  & 1\,545.3  \\
    \midrule
    tai12b & VQE      & 144 & 163 & 2443 & 715 & 1728 & 1\,911.3 \\
    & CVaR-VQE & 144 & 163 & 2443 & 715 & 1728 & 789.0 \\
    & PCE      & 11  & 528 & 4048 & 880 & 924  & 1\,206.2  \\
    \bottomrule
  \end{tabular*}
\end{table*}

\subsection{MSP resource usage}
\label{app:resource_msp}

\begin{table*}[htbp]
  \caption{\label{tab:resource_usage_ms}
  MSP simulator resource usage for VQE, CVaR-VQE, and PCE}
  \centering
  \footnotesize
  \setlength{\tabcolsep}{3pt}
  \begin{tabular*}{\textwidth}{@{\extracolsep{\fill}}llcccccr}
    \toprule
    \textbf{Inst.} & \textbf{Method} & \textbf{Q} & \textbf{D} & \textbf{G} & \textbf{2Q} & \textbf{P} & \textbf{Time (min)} \\
    \midrule
    $2\times 10$ S0 & VQE      & 50  & 69  & 845  & 245 & 600  & 2\,135.7 \\
    & CVaR-VQE & 50  & 69  & 845  & 245 & 600  & 1\,727.8 \\
    & PCE      & 7   & 336 & 1568 & 336 & 364  & 55.4  \\
    \midrule
    $2\times 10$ S1 & VQE      & 46  & 65  & 777  & 225 & 552  & 1\,680.0 \\
    & CVaR-VQE & 46  & 65  & 777  & 225 & 552  & 1\,343.8 \\
    & PCE      & 7   & 336 & 1568 & 336 & 364  & 47.4  \\
    \midrule
    $3\times 20$ S0 & VQE      & 84  & 103 & 1423 & 415 & 1008 & 2\,050.6 \\
    & CVaR-VQE & 84  & 103 & 1423 & 415 & 1008 & 1\,358.7 \\
    & PCE      & 8   & 384 & 2080 & 448 & 480  & 164.9  \\
    \midrule
    $3\times 20$ S1 & VQE      & 80  & 99  & 1355 & 395 & 960  & 2\,494.1 \\
    & CVaR-VQE & 80  & 99  & 1355 & 395 & 960  & 1\,857.6 \\
    & PCE      & 8   & 384 & 2080 & 448 & 480  & 155.6 \\
    \midrule
    $4\times 30$ S0 & VQE      & 118 & 137 & 2001 & 585 & 1416 & 4\,283.2 \\
    & CVaR-VQE & 118 & 137 & 2001 & 585 & 1416 & 3\,629.8 \\
    & PCE      & 10  & 480 & 3320 & 720 & 760  & 504.6  \\
    \midrule
    $4\times 30$ S1 & VQE      & 118 & 137 & 2001 & 585 & 1416 & 4\,663.9 \\
    & CVaR-VQE & 118 & 137 & 2001 & 585 & 1416 & 3\,712.6 \\
    & PCE      & 10  & 480 & 3320 & 720 & 760  & 512.2  \\
    \midrule
    $5\times 40$ S0 & VQE      & 154 & 165 & 1691 & 459 & 1232 & 6\,956.9 \\
    & CVaR-VQE & 154 & 165 & 1691 & 459 & 1232 & 5\,958.5 \\
    & PCE      & 11  & 528 & 4048 & 880 & 924  & 1\,029.2 \\
    \midrule
    $5\times 40$ S1 & VQE      & 154 & 165 & 1691 & 459 & 1232 & 6\,693.6 \\
    & CVaR-VQE & 154 & 165 & 1691 & 459 & 1232 & 5\,719.3 \\
    & PCE      & 11  & 528 & 4048 & 880 & 924  & 1\,020.9 \\
    \bottomrule
  \end{tabular*}
\end{table*}





\bibliographystyle{iopart-num}   
\bibliography{reference}          

\end{document}